\documentstyle[epsfig]{myelstart}
\renewcommand{\thefigure}{\thesection.\arabic{figure}}
\renewcommand{\theequation}{\thesection.\arabic{equation}}
\makeatletter\@addtoreset{equation}{section}\makeatother

\newcommand{\pmb}[1]{%
        \setbox0=\hbox{#1}%
        \kern-.02em\copy0\kern-\wd0
        \kern+.04em\copy0\kern-\wd0
        \kern-.02em\raise.0217em\box0}
\renewcommand{\vec}[1]{
        \mathchoice{\mbox{\boldmath$#1$}}%
        {\mbox{\boldmath$#1$}}%
        {\pmb{$\scriptstyle#1$}}%
        {\pmb{$\scriptscriptstyle#1$}}}

\newcommand{\llongrightarrow}{\relbar\joinrel\relbar\joinrel\longrightarrow}
\newcommand{\xpom}{x_{_{\rm I\!P}}}
\newcommand{\pom}{\rm I\!P}
\newcommand{\reg}{\rm I\!R}
\newcommand{\areg}{\alpha_{_{\rm I\!R}}}
\newcommand{\apom}{\alpha_{_{\rm I\!P}}}

\newcommand{\Slash}[1]{{#1}\!\!\!/}

\newcommand{\lsim}{
 \mathrel{\setbox0=\hbox{$<$}\raise0.6ex\copy0\kern-\wd0
 \lower0.65ex\hbox{$\sim$}}}

\newcommand{\gsim}{
 \mathrel{\setbox0=\hbox{$>$}\raise0.6ex\copy0\kern-\wd0
 \lower0.65ex\hbox{$\sim$}}}
 
\newcommand{\T}[1]{{\mathrm{#1}}}

\begin{document}
\begin{frontmatter}
\begin{flushright}
\bf 
TUM/T39-99-11
\end{flushright}

\vspace*{2cm}

\title{Nuclear Deep-Inelastic Lepton Scattering \\ 
and Coherence Phenomena$^{*)}$} 

\author {G. Piller},
\author {W. Weise} 
\address {Physik Department, Technische Universit\"{a}t M\"{u}nchen,
      D-85747 Garching, Germany}

\bigskip
\bigskip
\bigskip
\bigskip

\begin{abstract} 
\noindent
This review outlines our present experimental knowledge and 
theoretical understanding of deep-inelastic scattering on nuclear 
targets. 
The emphasis is primarily on nuclear coherence phenomena, 
such as shadowing, where the key physics issue 
is the exploration of 
hadronic and quark-gluon fluctuations of a high-energy 
virtual photon and their passage through the nuclear 
medium. 
New developments in polarized deep-inelastic scattering 
on nuclei are also discussed, and more  
conventional binding and Fermi motion effects are 
summarized. 
The report closes with a brief outlook  on vector meson 
electroproduction, nuclear shadowing at very large $Q^2$ 
and the physics of high parton densities in QCD. 
\end{abstract}

\end{frontmatter}
%
\vspace*{\fill}
\noindent $^{*}$) Work supported  in part by BMBF.

     \newpage
\tableofcontents \newpage

\section{Introduction}
This review is written with the intention to summarize 
and discuss nuclear phenomena observed in the deep-inelastic 
scattering (DIS) of leptons (mostly muons and electrons) on 
nuclear targets. Experimental developments in the last 
decade have brought several such effects into 
focus (the EMC effect; shadowing, etc.). 
This first came as a surprise. 
At the high energy and momentum transfers 
characteristic of DIS one did not expect to see 
sizable nuclear effects which usually occur on 
length scales of order $1$ fm or larger, 
governed by the inverse Fermi momentum of nucleons in nuclei. 

Today such nuclear effects are well established by a large 
amount of high-quality experimental data. 
Also, their theoretical understanding has progressed in 
recent years, so that an updated review of these developments 
appears justified. 
Our presentation, however, does not aim for completeness in all 
details. 
We wish to emphasize those effects in which two 
or more nucleons act coherently to produce significant deviations 
from the incoherent sum of individual nucleon structure functions. 
The most prominent effect of this kind is shadowing.
Its close relationship with diffraction in 
high-energy hadronic processes is now quite well understood, 
which  points to the significance  of optical analogues when 
dealing with the interaction of high-energy, virtual photons 
in a nuclear medium. 

Other, less pronounced nuclear effects such as binding and 
Fermi motion will  also be discussed.
Some overlap with previous  
reviews 
\cite{Frankfurt:1988nt,Jaffe:1985je,Bickerstaff:1989ch,Arneodo:1994wf,%
Geesaman:1995yd} 
is not unwanted for reasons of continuity. 
At the same time, this report incorporates plenty of more recent 
material, including polarization observables in DIS on nuclei, 
a field in which experimental activities progress rapidly and 
forcefully. 

Before turning to our main subject it is necessary and useful 
to summarize, in the following Section 2, our  knowledge 
on free nucleon structure functions. Deep-inelastic 
scattering probes the substructure of the nucleon with very 
high resolution down to length and time scales of order 
$10^{-2}$ fm. The QCD analysis of the structure functions 
gives detailed insights into the composition of nucleons in 
terms of quarks and gluons, and their momentum 
and spin distributions. 

A fundamental question from a nuclear physics perspective 
is then the following: how do the quark 
and gluon distributions of the nucleon change in a nuclear 
many-body environment? 
What are the mechanisms responsible for such changes? 
These issues will be addressed starting from Section 3 in 
which the basic observations and phenomenology of DIS from nuclear 
targets will be described. 
A particularly instructive way of illustrating the physics content 
of nuclear structure functions is provided by a 
space-time (rather than momentum space) analysis to which we refer 
in a separate Section 4. 
Shadowing is discussed in detail in Section 5. 
Binding effects,  Fermi motion and pionic 
contributions are dealt with in Section 6. 
Section 7 turns to a discussion of more recent work on polarized 
DIS from nuclei. 
A status summary and further perspectives 
follow in Section 8. 

We close this introduction with a remark on references to the literature.
As usual, aiming for completeness is an impossible task. 
What we hope to be a reasonable compromise is a combination of 
references to previous reviews in which earlier references can 
be found, together with selected original references 
to data and theory whenever they are of direct relevance in the 
text.

\section{Structure functions of free nucleons}

\subsection{Deep-inelastic scattering: kinematics and structure functions}

\begin{figure}[h]
\bigskip
\begin{center} 
\epsfig{file=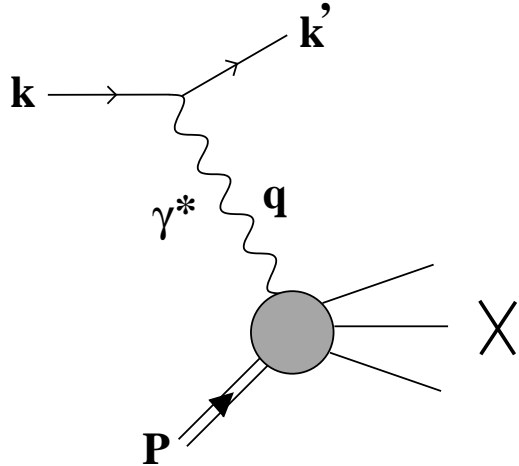,height=60mm}
\end{center}
\caption[...]{Inclusive deep-inelastic lepton-nucleon scattering.}
\label{fig:Feynman_DIS}
\bigskip
\end{figure}

Consider the scattering of an electron or muon with four-momentum 
$k^{\mu}=(E, \vec k)$ and invariant mass $m$ from 
a nucleon carrying the four-momentum 
$P^{\mu} = (E_{\T p},\vec P)$  and 
mass $M$. 
Inclusive measurements observe only the scattered lepton with momentum 
${k'}^{\mu} = (E',\vec k')$ as indicated in Fig.\ref{fig:Feynman_DIS}.  

Neglecting weak interactions which are relevant at very high 
energies only, the differential cross section is 
given by:\footnote{For introductions to deep-inelastic lepton scattering 
see e.g. Refs.\cite{Roberts:1990ww,Muta:1987mz,Cheng:1984}.}
\begin{equation} \label{eq:crossDIS}
\frac{d^2\sigma}{d\Omega d E'} 
= 
\frac{\alpha^2}{Q^4} \frac{E'}{E} 
L_{\mu\nu} W^{\mu\nu},
\end{equation}
to  leading order in the electromagnetic coupling constant 
$\alpha = e^2/4\pi \simeq 1/137$.   
Here 
\begin{equation}
q^{\mu}=k^{\mu}-{k'}^{\mu} = (\nu,\vec q)
\end{equation}
is the four-momentum of the exchanged virtual photon, 
and $Q^2 = - q^2$. 
The lepton-photon interaction is described by the lepton tensor 
$L_{\mu\nu}$. 
Let the spin projections of the initial and final lepton be 
$s$ and $s'$. 
After summing over  $s'$ the lepton tensor 
can be split  into pieces which are symmetric and antisymmetric   
with respect to the Lorentz indices $\mu$ and $\nu$:
\begin{equation} 
L_{\mu\nu}(k,s;k') = L_{\mu\nu}^S(k;k') + i \,L_{\mu\nu}^{A}(k,s;k'),   
\end{equation}
with:
%
\begin{eqnarray} 
\label{eq:lepton_tensor_s_as}
&&L_{\mu\nu}^S (k;k') \, = \, 
2 \left(k_{\mu} k'_{\nu} + k_{\nu} k'_{\mu} \right)
+ g_{\mu\nu} \,q^2, 
\\
&&L_{\mu\nu}^A(k,s;k')\, = \,2 \,m \,\epsilon_{\mu\nu\alpha\beta} \,
s^{\alpha} \,q^{\beta}, 
\end{eqnarray}
%
where the lepton spin vector is defined by 
$2 m \,s^{\alpha} = \bar u \gamma^{\alpha} \gamma_5 u$. 
For unpolarized lepton scattering the average over the 
initial lepton polarization is carried out. 
In this case only the symmetric term, $L_{\mu\nu}^S$,  
remains.

The complete information about the target response 
is in the hadronic tensor $W_{\mu\nu}$. 
We denote the nucleon spin by $S$.  
Gauge invariance and symmetry  properties allow a  
parametrization  of the hadronic tensor, 
\begin{equation} \label{eq:hadronic_tensor}
W_{\mu\nu}(q;P,S) = W_{\mu\nu}^S(q;P) + i \,W_{\mu\nu}^A(q;P,S), 
\end{equation}
in terms of four structure functions.
The symmetric part is 
\begin{eqnarray} \label{eq:hadronic_tensor_sym}
W_{\mu\nu}^S(q;P) &=& \left( \frac{q_{\mu}q_{\nu}}{q^2} - g_{\mu\nu} \right) 
W_1(P\cdot q,q^2)  
\nonumber \\
&+&
\left(P_{\mu} - \frac{P\cdot q}{q^2}q_{\mu}\right) 
\left(P_{\nu} - \frac{P\cdot q}{q^2}q_{\nu}\right) 
\frac{W_2(P\cdot q,q^2)}{M^2}, 
\end{eqnarray}
and the antisymmetric part can be written
\begin{eqnarray} \label{eq:hadronic_tensor_asym}
W_{\mu\nu}^A(q;P,S) &=& \epsilon_{\mu\nu\alpha\beta} \,q^{\alpha}
\left[
S^{\beta} M G_1(P\cdot q,q^2) + 
\left( P\cdot q \,S^{\beta} - S\cdot q \, P^{\beta} \right) 
\frac{G_2(P\cdot q,q^2)}{M}\right]. \nonumber \\
\end{eqnarray}
Here the nucleon spin vector $S^{\mu}$, with 
$2 M S^{\mu} = \overline U(P,S)\,  \gamma^{\mu} \gamma_5 \,U(P,S)$, 
has been introduced.
In the conventions used in this paper nucleon Dirac spinors $U$ 
are normalized according to  $\bar U(P) \gamma^{\mu} U(P) = 2 P^{\mu}$.  
It is evident that  $W_{1,2}$ can be  
measured in unpolarized scattering processes, whereas the complete 
investigation of $G_{1,2}$ requires both 
beam and target to be polarized.

It is convenient to introduce dimensionless structure functions 
%
\begin{eqnarray} \label{eq:F1_F2}
F_1(x,Q^2) &=& M \,W_1(P\cdot q,q^2), 
\\
F_2(x,Q^2) &=& \frac{P\cdot q}{M}\, W_2(P\cdot q,q^2), 
\end{eqnarray}
%
which depend on the Bjorken scaling variable, 
\begin{equation}
x = \frac{Q^2}{2 P\cdot q}. 
\end{equation} 
In terms of $F_{1,2}$ the charged lepton scattering 
cross section (\ref{eq:crossDIS}) for an unpolarized lepton and 
nucleon is:  
\begin{equation}
\frac{d^2\sigma}{dx \,dQ^2} = 
\frac{4 \pi \alpha^2}{Q^4} 
\left[
\left(1-y-\frac{M xy}{2 E}\right) \frac{F_2}{x} 
+ y^2 F_1\right],
\end{equation}
with
\begin{equation}
y = \frac{P\cdot q}{P\cdot k}.
\end{equation}

Let us recall the behavior of the structure functions in 
the Bjorken limit, i.e. at large momentum and 
energy transfers, 
\begin{equation}
Q^2 = - q^2\rightarrow \infty, \quad P\cdot q \rightarrow \infty,
\end{equation} 
but fixed ratio $Q^2/P\cdot q$.
Here the unpolarized structure functions 
\begin{eqnarray} 
&F_1(x,Q^2) &\stackrel{Q^2\rightarrow \infty}{\llongrightarrow} F_1(x), 
\\
&F_2(x,Q^2) &\stackrel{Q^2\rightarrow \infty}{\llongrightarrow} F_2(x) 
\end{eqnarray}
are observed to depend in good approximation only on the dimensionless 
Bjorken scaling variable $x$.
Variations of the structure functions with $Q^2$ at 
fixed $x$ turn out to be small.

A similar scaling behavior is expected for the spin-dependent 
structure functions: 
\begin{eqnarray}  \label{eq:g1_g2}
g_1(x,Q^2) &=& M P\cdot q \,G_1(P\cdot q,q^2),\\
g_2(x,Q^2) &=& \frac{(P\cdot q)^2}{M} \,G_2(P\cdot q,q^2), 
\end{eqnarray}
which  likewise reduce to functions of $x$ only when 
the limit $Q^2 \rightarrow \infty$ is taken. 

\subsection{Parton model}

The approximate $Q^2$-independence of nucleon structure functions 
at large $Q^2$ has led to the conclusion that 
the virtual photon sees point-like constituents in the nucleon.
This is the basis of the naive parton model  
which gives a simple interpretation of nucleon structure functions. 
In this picture the nucleon is composed of free pointlike 
constituents, the partons, identified with quarks and gluons.
Introducing distributions $q_f(x)$ and $\bar q_f(x)$ of 
quarks and antiquarks with flavor $f$ and fractional electric charge 
$e_f$, one finds:
\begin{eqnarray} \label{eq:F12_parton}
F_1(x) &=& \frac{1}{2}\sum_f e_f^2 \left( q_f(x) + \bar q_f(x) \right), 
\\
\label{eq:F2_parton}
F_2(x) &=& 2 x \, F_1(x). 
\end{eqnarray} 
The Bjorken variable $x$ coincides with the 
fraction of the target light-cone momentum carried by 
the interacting quark with momentum $l$: 
\begin{equation} 
x=\frac{Q^2}{2 P\cdot q} = \frac{l\cdot q}{P\cdot q}.
\end{equation}
The Callan-Gross relation (\ref{eq:F2_parton}) 
connecting $F_1$ and $F_2$ reflects 
the spin-$1/2$ nature of the quarks. 

For the spin structure functions the naive parton model gives:   
\begin{eqnarray} \label{eq:g12_parton}
g_1(x) &=& \frac{1}{2}\sum_f e_f^2 \left[ \Delta q_f(x) + 
\Delta \bar q_f(x) \right],\\ 
g_2(x) &=& 0.
\end{eqnarray} 
The helicity distributions 
$\Delta q_f(x) = q_f^{\uparrow}(x) - q_f^{\downarrow}(x)$ 
and 
$\Delta \bar q_f(x) = \bar q_f^{\uparrow}(x) - \bar q_f^{\downarrow}(x)$
involve  the differences of quark or antiquark distributions  
with helicities parallel and 
antiparallel with respect to the helicity of the target nucleon.

\subsection{Virtual Compton scattering}

The hadronic tensor (\ref{eq:hadronic_tensor}) can be expressed 
as the Fourier transform of a correlation function of  
electromagnetic currents, 
with its expectation value taken for the nucleon ground state 
$|P,S\rangle$ normalized as 
$\langle P',S'|P,S\rangle = 2 E_{\T p} \,(2\pi)^3 \,\delta^3(\vec P - \vec P') 
\,\delta_{SS'}$ \cite{Roberts:1990ww,Muta:1987mz,Cheng:1984}:
\begin{equation} \label{eq:WJJ}
W_{\mu\nu} (q;P,S) = \frac{1}{4\pi\,M} 
\int {d}^4 z \,e^{i q\cdot z} \, 
\langle P,S| J_{\mu}(z) J_{\nu}(0)|P,S \rangle .
\end{equation}
It is related to the forward virtual Compton scattering amplitude:
\begin{equation} \label{eq:TJJ}
T_{\mu\nu} (q;P,S) = i 
\int {d}^4 z \,e^{i q\cdot z} \, 
\langle P,S |{\cal T}\left(J_{\mu}(z) J_{\nu}(0)\right)|P,S \rangle, 
\end{equation}
where ${\cal T}$ denotes the time-ordered product.
By comparison of Eqs.(\ref{eq:WJJ}) and (\ref{eq:TJJ}) one finds 
the optical theorem:
\begin{equation} \label{eq:Opt}
2 \pi M\,W_{\mu\nu} = {\T {Im}}\,T_{\mu\nu}.
\end{equation}
Consequently, nucleon structure functions can be represented  
in terms of virtual photon-nucleon helicity amplitudes,
\begin{equation}
{\cal A}_{h H,h'H'} =  
e^2 \epsilon^{\mu *}_{h'} T_{\mu\nu}(H,H') \,\epsilon_{h}^{\nu}.
\end{equation} 
Here $\epsilon_{h}$ and $\epsilon_{h'}$ are the polarization vectors 
of the incoming and scattered photon with helicities 
$h$ and $h'$, respectively. 
They have values $+1,-1, 0$ (abbreviated as $ +,-,0$).
Helicities of the initial and final nucleon are denoted 
by $H$ and $H'$. 
Their values are $\pm 1$, symbolically denoted by $\uparrow, \downarrow$.
Choosing the $z$-axis in space to coincide with $\vec q/|\vec q|$, 
the direction of the propagating virtual photon, 
and quantizing the angular momentum of the target and photon along 
this axis yields the following relations:
\begin{center}
\begin{eqnarray} 
F_1&=&\frac{1}{4\pi e^2}
\left({\T {Im}} \,{\cal A}_{+\downarrow,+\downarrow}+
{\T {Im}} \,{\cal A}_{+\uparrow,+\uparrow}
\right),\label{f1helamp}\\
F_2&=&
\frac{x}{2 \pi e^2 \kappa}
\left({\T {Im}} \,{\cal A}_{+\downarrow,+\downarrow}+
{\T {Im}} \,{\cal A}_{+\uparrow,+\uparrow}
+ 2 \,{\T {Im}} \,{\cal A}_{0\uparrow,0\uparrow}\right),
\end{eqnarray}\end{center}
where $\kappa = 1 + (2 M x/Q)^2$. 
For the spin-dependent structure functions one finds:
\begin{eqnarray} \label{eq:g1_helicity}
g_1 &=&\frac{1}{4\pi e^2 \kappa}
\left({\T {Im}} \,{\cal A}_{+ \downarrow,+ \downarrow}-
{\T {Im}} \,{\cal A}_{+\uparrow,+ \uparrow}+\sqrt{2(\kappa-1)} \,
{\T {Im}} \,{\cal A}_{+\downarrow,0\uparrow}\right),
\\
\label{eq:g2_helicity}
g_2 &=&\frac{1}{4\pi e^2\kappa}
\left({\T {Im}} \,{\cal A}_{+ \uparrow,+ \uparrow}-
{\T {Im}} \,{\cal A}_{+\downarrow,+\downarrow}
+ \frac{2}{\sqrt{2(\kappa-1)}} \,
{\T {Im}} \,{\cal A}_{+\downarrow,0\uparrow}
\right).
\end{eqnarray}
In the scaling limit the structure functions $F_1$, $F_2$ and 
$g_1$ are determined by helicity conserving amplitudes. 
It is therefore possible to express them through 
virtual photon-nucleon cross sections defined as:
\begin{equation} \label{eq:sigma_hH}
\sigma_{h H} = \frac{1}{2 M K} 
{\T {Im}} \,{\cal A}_{hH,hH},
\end{equation}
with the virtual photon flux $K= (2 P\cdot q - Q^2)/2M$. 
For example, the structure function $F_2$  reads:
\begin{equation}  \label{eq:F2_sig}
F_2 = \frac{1-x}{1 + (2 M x/Q)^2} \frac{Q^2}{4 \pi^2 \alpha}
\left( \sigma_L +  \sigma_T \right),
\end{equation}
where the longitudinal and transverse photon-nucleon cross sections 
$\sigma_{L,T}(\nu,Q^2)$ are given by: 
\begin{eqnarray}
\sigma_L &=& 
\frac{1}{2}\left(
\sigma_{0 \uparrow} + \sigma_{0\downarrow}
\right),
\\  
\sigma_T &=& 
\frac{1}{4}
\left(
\sigma_{+\uparrow} + \sigma_{+\downarrow} +  
\sigma_{-\uparrow} + \sigma_{-\downarrow} \right).
\end{eqnarray}
An interesting quantity is their ratio: 
\begin{equation} \label{eq:R_L_T}
R = \frac{\sigma_L}{\sigma_T} = 
\frac{F_2 ( 1  + (2 M x/Q)^2)} {2x F_1} - 1. 
\end{equation}
In the simple parton model the Callan-Gross relation 
(\ref{eq:F2_parton}) implies $R=0$ as $Q^2\rightarrow \infty$. 
Due to their interaction with gluons, quarks receive 
momentum components transverse to the photon direction. 
Then they can  absorb also longitudinally polarized photons. 
This leads  to  $R\ne 0$.

\subsection{QCD-improved parton model}
\label{ssec:AP_eq}

Nucleon structure functions systematically exhibit a weak 
$Q^2$-dependence, even at large $Q^2$.  
These scaling violations can be described within the framework 
of the QCD-improved parton model 
which incorporates the interaction between quarks and gluons in 
the nucleon in a perturbative way 
(see e.g. \cite{Roberts:1990ww,Muta:1987mz,Cheng:1984}). 
The scale at which this interaction is resolved is determined 
by the momentum transfer. 
The $Q^2$-dependence of parton distributions, e.g. 
\begin{eqnarray} \label{eq:parton_QCD}
F_2(x,Q^2) &=&  \sum_f e_f^2 \,x\,\left[ q_f(x,Q^2) + \bar q_f(x,Q^2) \right], 
\\
g_1(x,Q^2) &=& \frac{1}{2}\sum_f e_f^2 \left[ \Delta q_f(x,Q^2) + 
\Delta \bar q_f(x,Q^2) \right], 
\end{eqnarray}
is described 
by the Dokshitzer-Gribov-Lipatov-Altarelli-Parisi (DGLAP) evolution equations. 
They are different for flavor non-singlet and  
singlet distribution functions. 
Typical examples of non-singlet combinations are the difference of quark 
and antiquark distribution functions, or the difference of up and down 
quark distributions. 
The difference of the proton and neutron structure function, 
$F_2^{\T p} - F_2^{\T n}$, also behaves as a  flavor non-singlet, 
whereas  the deuteron structure function $F_2^{\T d}$ 
is an almost  pure flavor singlet combination.
For the flavor non-singlet quark distribution, 
$q^{\T{NS}}$, and the flavor-singlet quark and gluon 
distributions, $q^{\T{S}}$ and $g$, the DGLAP evolution equations read 
as follows:
\begin{eqnarray} \label{eq:DGLAP}
\frac{{d}q^{\T{NS}}(x,Q^2)}{{d} \ln Q^2}
&=&
\frac{\alpha_s(Q^2)}{2\pi}\int_x^1\frac{{d} y}{y} 
q^{\T{NS}}(y,Q^2) P_{qq}\left(\frac{x}{y}\right), 
\\
\label{eq:DGLAP_s}
\frac{d}{{d}\ln Q^2} 
\left( \begin{array}{c} q^{\T S}(x,Q^2) \\ g(x,Q^2) \end{array} \right)
&=& \frac{\alpha_s(Q^2)}{2\pi}\int_x^1\frac{{d}y}{y} 
\left( \begin{array}{rr}
 P_{qq}(\frac{x}{y}) & P_{qg}(\frac{x}{y}) \\
 P_{gq}(\frac{x}{y}) & P_{gg}(\frac{x}{y}) \end{array} \right)
 \left( \begin{array}{c} q^{\T S}(y,Q^2) \\ g(y,Q^2) \end{array} \right).
\end{eqnarray}
Here $\alpha_s(Q^2)$ is the running QCD coupling strength.
The splitting function $P_{qq}(x/y)$ determines the probability 
for a quark to radiate a gluon such that the quark momentum 
is reduced by a fraction $x/y$. 
Similar interpretations hold for the remaining splitting functions. 
For further details we refer the reader to one of the many textbooks 
on applications of QCD, 
e.g. \cite{Roberts:1990ww,Muta:1987mz,Cheng:1984}.

\subsection{Light-cone dominance of deep-inelastic scattering}
\label{ssec:OPE}

The QCD analysis of deep-inelastic scattering has generated its own 
terminology and specialized jargon. 
In this section we summarize some of the basic notions. 
The general framework is Wilson's operator product expansion 
applied to the current-current correlation function. 
A detailed investigation reveals that the 
hadronic tensor 
\begin{equation}
2 \pi\,M\,W_{\mu\nu} = {\T {Im}}\left[\, i 
\int {d}^4z \, e^{i q\cdot z}
\left \langle P \right| {\cal T}\left( J_{\mu}(z) J_{\nu}(0)\right) 
\left|P\right\rangle\right],
\end{equation}
at $Q^2\rightarrow \infty$ but fixed Bjorken $x$, is dominated 
by contributions from near the light-cone, $z^2 = t^2 - \vec z^2 
\simeq 0$ \cite{Roberts:1990ww,Muta:1987mz,Cheng:1984}. 
The operator product expansion makes use of this fact 
by expanding the time-ordered product of currents 
around the singularity at $z^2 = 0$: 
\begin{equation} \label{eq:OPE1}
{\cal T}\left(J (z) J (0)\right) \sim \sum_{n=0}^{\infty} 
c^{\cal O}_n(z^2;\mu^2) \,z^{\mu_1}\cdots z^{\mu_n}\, 
{\cal O}_{\mu_1\cdots \mu_n}(\mu^2),
\end{equation}
where the ${\cal O}_{\mu_1\cdots \mu_n}$ are local operators 
involving quark and gluon fields. 
The coefficient functions $c_n$ are singular at $z^2 = 0$. 
They are grouped according to the order of their singularity. 
Both the operators ${\cal O}$ and the c-number coefficient 
functions $c_n$ depend on the renormalization point $\mu^2$.

The operators ${\cal O}$ can be organized according to the 
irreducible representation of the Lorentz group to 
which they belong. 
Each operator has a characteristic dimensionality, $d$, 
in powers of mass or momentum. 
For example, the symmetric traceless Lorentz tensors of 
rank $n$ with minimum dimensionality are the operators 
\begin{eqnarray} \label{eq:op_quark}
{\cal O}_{\mu_1\cdots \mu_n}^q &=& 
\left\{
\overline{\psi} \gamma_{\mu_1} i D_{\mu_2} \cdots i D_{\mu_n} \psi
\right\}_{\cal S},
\\
\label{eq:op_glue}
{\cal O}_{\mu_1\cdots \mu_n}^g  &=& 
\left\{
G_{\mu_1\nu} i D_{\mu_2} \cdots i D_{\mu_{n-1}} G_{\mu_n}^{\nu}
\right \}_{\cal S},
\end{eqnarray}
local bilinears in the quark field $\psi$ and the gluon 
field tensor $G_{\mu\nu}$, 
with any number of gauge-covariant derivatives 
$D_\mu$ inserted between them.
The brackets $\{\}_{\cal S}$ indicate symmetrization 
with respect to Lorentz indices and subtraction of 
trace terms. 
The operators ${\cal O}^q$ and ${\cal O}^g$ 
have dimensionality $d = 3 + (n-1) = 4 + (n-2) = 2 + n$. 
The difference $\tau = d-n$ is called ``twist'' 
($\tau = 2$ in our example), a useful bookkeeping device to 
classify the light-cone ($z^2 \rightarrow 0$) singularity of 
the coefficient function $c_n$. 
Comparing dimensions in Eq.(\ref{eq:OPE1}) one finds that, 
for each given operator on the right-hand side, 
the coefficient behaves as $c_n \sim (1/z^2)^{(2 d_J - \tau )/2}$ 
when $z^2 \rightarrow 0$, where $d_J = 3$ is the dimensionality of each 
of the currents on the left-hand side of Eq.(\ref{eq:OPE1}).

Matrix elements of the operators $\cal O$ between nucleon states 
are of genuinely non-perturbative origin. For 
spin-averaged quantities they must be of the form 
\begin{eqnarray} \label{eq:reduced1}
\langle P|{\cal O}_{\mu_1\cdots \mu_n}^q (\mu^2)|P\rangle 
&=& a^q_n(\mu^2) 
P_{\mu_1} \cdots P_{\mu_n},
\\
\label{eq:reduced2}
\langle P|{\cal O}_{\mu_1 \cdots \mu_n}^g (\mu^2)|P\rangle 
&=& a^g_n(\mu^2) 
P_{\mu_1} \cdots P_{\mu_n}, 
\end{eqnarray}
since Lorentz-covariant tensorial functions of the nucleon 
four-momentum $P^{\mu}$, with $P^2 = M^2$ fixed, 
are proportional to the symmetric tensors 
$P_{\mu_1} \cdots P_{\mu_n}$. 
Trace terms have been subtracted in Eqs.(\ref{eq:reduced1},\ref{eq:reduced2}).
The constants $a_n^q$ and $a_n^g$ are fixed at a given 
renormalization scale $\mu^2$ and represent the non-perturbative 
quark and gluon dynamics of the nucleon. 

We can now make contact with observables. 
Since the Fourier transform of 
\linebreak
$\langle P | T(J_{\mu}(z) J_{\nu}(0)|P\rangle$ 
is proportional to the forward virtual Compton 
scattering amplitude and its imaginary part determines 
the structure functions $F_1$ and $F_2$, it is clear that 
the $a_n$ represent moments of those structure functions,  
with $Q^2$-dependent coefficients. 
Consider as an example the structure function $F_1(x,Q^2)$ 
in the flavor singlet channel. 
One finds, 
\begin{equation}  \label{eq:OPE_F1}
\int_0^1 {d}x\,x^{n-1} F_1(x,Q^2)  = 
C_{n}^{q}(Q^2;\mu^2)  \, a^{q}_{n}(\mu^2) +
C_{n}^{g}(Q^2;\mu^2)  \, a^{g}_{n}(\mu^2), 
\end{equation}
where crossing symmetry implies a restriction to 
even orders $n=2,4,\dots$. 
The momentum space coefficient functions $C_n(Q^2;\mu^2)$ 
are related to the c-number functions 
$c_n(z^2;\mu^2)$ of Eq.(\ref{eq:OPE1}) by Fourier transformation. 
The important point is  that the $C_n(Q^2;\mu^2)$ 
can be calculated perturbatively at large $Q^2$. 
Their $Q^2$-dependence is determined by renormalization group 
equations equivalent to the DGLAP equations in (\ref{eq:DGLAP_s}).

It is often useful to express the structure functions in a factorized 
form, by a convolution of ``hard'' effective cross sections 
$\hat \sigma_q$ and $\hat \sigma_g$ for the scattering of the 
virtual photon from quarks and gluons in the nucleon, 
with ``soft'' quark and gluon distributions of the target. 
For example, 
\begin{eqnarray} \label{eq:F1_conv}
F_1(x,Q^2) &=& \int_x^1 \frac{{d}y}{y}\, 
\left\{
\hat {\sigma}_{q}(y,Q^2;\mu^2) \,
\left[ q(x/y,\mu^2) + \bar q(x/y,\mu^2)\right] 
\right.
\nonumber \\
&& \hspace{2cm}+\left. \hat{\sigma}_{g}(y,Q^2;\mu^2) 
\,g(x/y,\mu^2)\right\}.
\end{eqnarray}
(Here we have generically used only 
one quark flavor with unit electric charge.)
The perturbatively calculable functions $C_n$ 
then find a simple physical interpretation in terms of 
moments of the ``hard'' cross sections:
\begin{equation}
\label{eq:sig_q}
C^{q,g}_{n}(Q^2;\mu^2) = \int_0^1 {d} x\,x^{n-1} 
\hat \sigma_{q,g}(x,Q^2;\mu^2), 
\end{equation}
while the quark and gluon distributions are related to the 
``soft'' reduced matrix elements (\ref{eq:reduced1},\ref{eq:reduced2}) 
by 
\begin{eqnarray} \label{eq:OPE_quark}
a^q_{n}(\mu^2) &=& \int_0^1 {d}x\,x^{n-1} 
\left[
q(x,\mu^2) + (-1)^n {\bar q}(x,\mu^2)
\right],
\\ \label{eq:OPE_gluon}
a^g_{n}(\mu^2) &=& \int_0^1 dx\,x^{n-1} 
g(x,\mu^2), 
\end{eqnarray} 
where Eq.(\ref{eq:OPE_gluon}) holds  only for even $n$. 
To lowest (zeroth) order in the running coupling strength 
$\alpha_s(\mu^2)$, 
the ``hard'' cross sections are simply 
$\hat \sigma_{q} \sim \delta(1-x)$ and $\hat \sigma_{g} = 0$, 
so that only quarks contribute to $F_1$. 
Gluons first enter at order $\alpha_s$. 
We mention that, in general, the representation 
of a given structure function in terms of separate 
quark and gluon contributions is a matter of definition. 
It is unique only in leading order and depends on the 
renormalization scheme at higher orders in $\alpha_s$ 
\cite{Roberts:1990ww,Muta:1987mz,Cheng:1984}. 
The measured structure functions are, of course, free of such 
ambiguities.

\subsection{Facts about free nucleon structure functions}

In this section we briefly review the present  experimental status  
on  free nucleon structure functions as measured in deep-inelastic 
lepton scattering. We focus on those aspects which are of direct relevance 
for our further discussion of nuclear deep-inelastic scattering.

\subsubsection{Spin independent structure functions}
\label{ssec:spin_ind_strfns}

Unpolarized deep-inelastic scattering has been explored in recent 
years over a wide kinematic range in fixed target experiments 
at CERN, FNAL and SLAC, and at the HERA collider  
at DESY. 
Reviews can be found e.g. in 
Refs.\cite{Cooper-Sarkar:1997jk,Badelek:1996ss,Badelek:1996rmp}.

{\bigskip\noindent \it The proton structure function $F_{2}^{\T p}$}
\bigskip

Accurate $F_2^{\T p}$ data are available from fixed target measurements 
at SLAC, at CERN (BCDMS, NMC) and at Fermilab (E665). 
They cover the  kinematic range $10^{-3} < x < 0.8$ and 
$0.2 \,{\T {GeV}}^2 < Q^2 < 260$ GeV$^2$ \cite{Cooper-Sarkar:1997jk}.
Due to experimental constraints  fixed target studies at small 
$x$ are possible only at low $Q^2$. 
For example, in the E665 measurements at Fermilab the smallest 
values of the Bjorken variable, $x \simeq 0.8\cdot 10^{-3}$, 
are measured 
typically at $Q^2 \simeq 0.2$ GeV$^2$ \cite{Adams:1996gu}. 
This is different at the HERA collider where 
the kinematic range  $3 \cdot 10^{-6} < x < 0.5$ and 
$0.16 \,{\T {GeV}}^2 < Q^2 < 5000$ GeV$^2$ is explored. 
In these experiments the region of small $x$ is accessible 
also at large $Q^2$.

\begin{figure}[b]
\bigskip
\begin{center} 
\epsfig{file=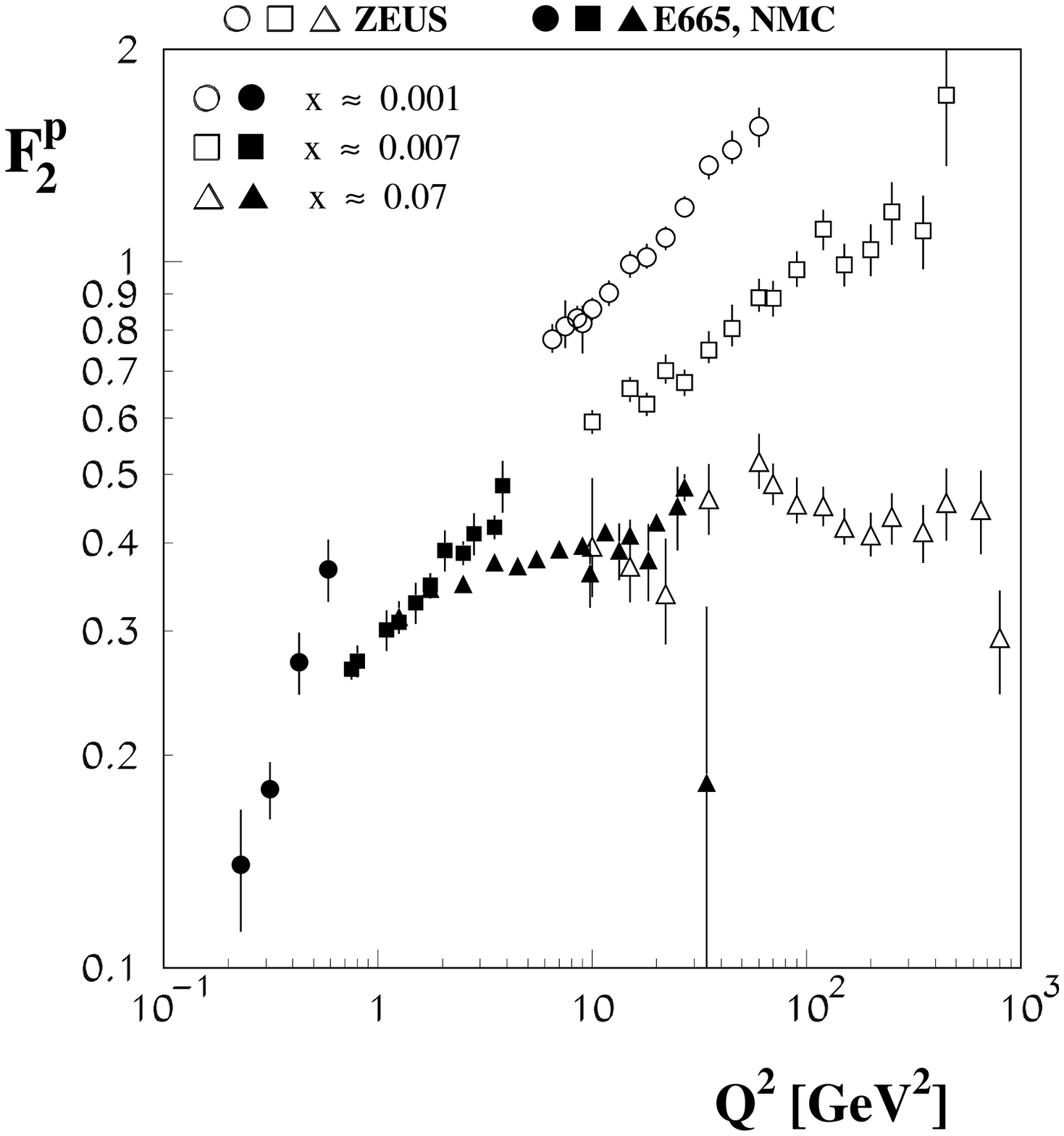,height=90mm}
\end{center}
\caption[...]{The $Q^2$-dependence of the proton  structure 
function $F_2^{\T p}$ for $x < 0.1$. 
The data are taken from ZEUS \cite{Derrick:1996hn,Breitweg:1998dz}, 
E665 \cite{Adams:1996gu}, and  NMC \cite{Arneodo:1997qe}.
}
\label{fig:Fq2}
\end{figure}

The data 
\nocite{Aid:1996au,Adloff:1997mf,Derrick:1996ef,Derrick:1996hn,%
Breitweg:1998dz,
Adams:1996gu,Arneodo:1997qe,Whitlow:1990dr,Benvenuti:1989rh}
summarized in Figures \ref{fig:Fpx} and \ref{fig:Fq2} 
display several important features 
(for references see e.g. 
\cite{Cooper-Sarkar:1997jk,Badelek:1996ss,Badelek:1996rmp}):

\begin{figure}[t]
\begin{center} 
\epsfig{file=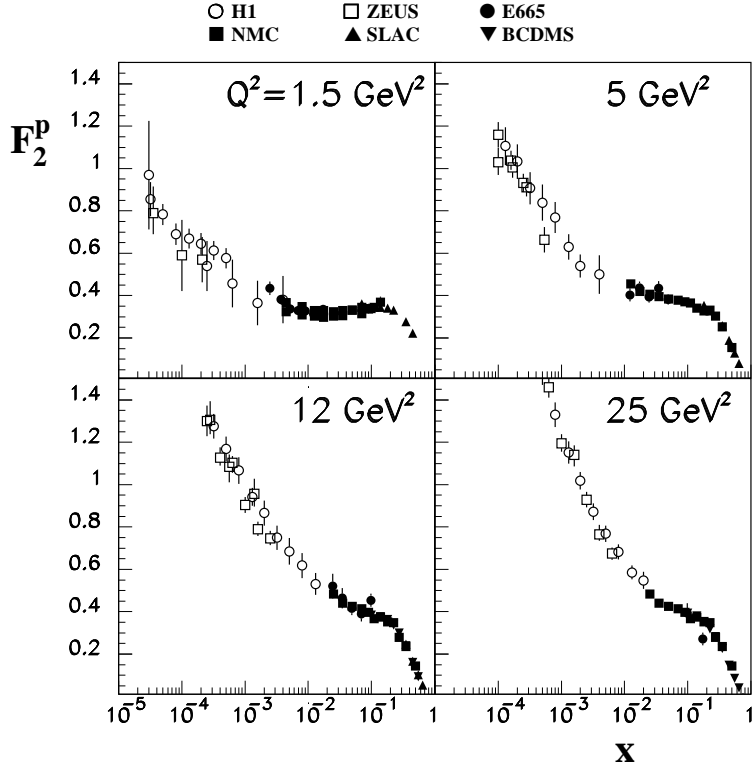,height=120mm}
\end{center}
\caption[...]{The proton structure function $F_2^{\T p}$ as a function 
of $x$ for various $Q^2$. 
The data are taken from H1 \cite{Aid:1996au,Adloff:1997mf}, 
ZEUS \cite{Derrick:1996ef,Derrick:1996hn,Breitweg:1998dz}, 
E665 \cite{Adams:1996gu}, 
NMC \cite{Arneodo:1997qe}, 
SLAC \cite{Whitlow:1990dr}, 
and BCDMS \cite{Benvenuti:1989rh}.
}
\label{fig:Fpx}
\bigskip
\end{figure}

\begin{itemize}

\item [$\bullet$]
At small Bjorken-$x$ ($x\ll 0.1$) but large $Q^2$ a strong increase of 
$F_2^{\T p}$ with decreasing $x$ has been found at HERA.
This behavior is commonly interpreted in terms of the dominant 
role of gluons at small $x$, the density 
of which rises strongly with decreasing $x$. 
This increase becomes weaker at low $Q^2$. 
Here only a minor $x$-dependence has been observed in  
fixed target experiments, which  is nevertheless
enhanced at very small $x\ll 0.1$ as recently 
explored at HERA \cite{Adloff:1997mf,Breitweg:1998dz}. 
Note that a rise of $F_2^{\T p}$ with decreasing  $x$ reflects a 
growing virtual photon-proton cross section as 
the photon-nucleon center-of-mass energy $W=\sqrt{s}$ increases. 
For example, at $Q^2 \simeq 100$ GeV$^2$ one observes a characteristic 
behavior \cite{Aid:1996au}:  
\begin{equation} \label{eq:sig_Q2}
\sigma_{\gamma^* {\T p}} \sim \left( W^2\right)^{\Delta},  
\quad \mbox{with} \, \Delta \approx 0.3. 
\end{equation}
For the real photon-nucleon cross section at high 
energies,  on the other hand,  
one has $\Delta \approx 0.08$ \cite{Donnachie:1992ny}. 
The dynamical origin of the observed  variation of the energy 
dependence of $\sigma_{\gamma^* \T N}$ 
with $Q^2$ is an important issue of 
ongoing investigations 
(see for example Refs.\cite{Cooper-Sarkar:1997jk,Badelek:1996ss}). 

\medskip
\noindent
Hadron-hadron interaction cross sections have  an energy dependence 
similar to that observed in photon-nucleon scattering. 
It is often parametrized using  
Regge phenomenology \cite{Donnachie:1992ny,Collins:1977jy}.
In Regge theory the dependence of cross sections 
on the center-of-mass energy is determined by 
the $t$-channel exchange of families of particles permitted 
by the conservation of all relevant quantum numbers. 
Each group of particles is characterized by a 
Regge trajectory, $\alpha(t) \approx \alpha(0) +  t\, \alpha'$,  
which relates their spin with  their invariant 
mass. 
The resulting dependence of  hadron-hadron total cross sections on the 
squared center-of-mass energy $s$ is:
\begin{equation}\label{eq:tot_regge}
\sigma_{tot} \sim s^{\alpha(0) - 1}. 
\end{equation}
The rising hadron-hadron cross sections at  high 
energies are well described by the so-called pomeron exchange. 
It corresponds to multi-gluon exchange with vacuum quantum numbers 
and it is characterized by the trajectory 
\cite{Donnachie:1992ny,Abe:1994xx}
\begin{equation} \label{eq:apom}
\apom (t) \approx  \apom(0) + t\,\apom' \approx 
1.08 +  t \, 0.26\,\mbox{GeV}^{-2}\,.
\end{equation}
Note that the fast growth of the interaction 
cross section (\ref{eq:tot_regge}) with energy 
as implied by Eq.(\ref{eq:apom}) 
cannot persist up to  arbitrarily high energies 
because of limitations imposed  by unitarity. 
At asymptotic energies the Froissart bound 
does not permit total hadronic cross sections to 
rise faster than $(\ln s/s_0)^2$ with some constant scale 
$s_0$ \cite{Collins:1977jy}. 

\medskip
\noindent
The slow decrease of hadron cross sections at 
moderate energies is described by an exchange 
made up from a set of reggeons which lie on the approximately 
degenerate trajectory \cite{Donnachie:1992ny,Apel:1979sp}
\begin{equation}
\areg(t) \approx \areg (0) + t \, \areg' \approx 
0.5 + t \, 0.9 \,\mbox{GeV}^{-2}\,, 
\end{equation}
and which carry the quantum numbers of the 
$\rho, \omega, a_2$ and $f_2$ mesons, respectively. 
At large energies these so-called subleading contributions 
are exceeded by pomeron exchange (\ref{eq:apom}).

\medskip
\item[$\bullet$]
At  small values of $Q^2$ (i.e. $Q^2 < 1$ GeV$^2$) the structure function 
$F_2^{\T p}$ drops.
This is quite natural in view of the fact that 
$F_2^{\T p}$ has to vanish linearly with $Q^2$  
in the limit $Q^2 \rightarrow 0$ as a consequence 
of  current conservation (see e.g. \cite{Badelek:1996rmp}). 
Bjorken scaling must break down in this 
kinematic regime. 
In particular, at small $x<0.1$ or large photon energy, 
$\nu > 5$ GeV, vector meson dominance is expected 
to play an important role. 
It describes  (virtual) photon-nucleon scattering 
via the interaction of vector meson fluctuations  
of the photon. The  contribution to $F_2^{\T p}$ 
from the three lightest vector mesons reads 
(see e.g. \cite{Bauer:1978iq}):
\begin{equation} \label{eq:VMD}
F_2^{{\T p} (VMD)} (x,Q^2) = 
\frac{Q^2}{4 \pi} \sum_{\T V={\rho,\omega,\phi}} 
\left(\frac{m_\T V^2}{g_{\T V}}\right)^2 
\left(\frac{1}{m_{\T V}^2 + Q^2}\right)^2 \sigma_{\T{Vp}}.  
\end{equation}
The sum is taken over 
$\rho$, $\omega$, and $\phi$ mesons with their invariant masses $m_\T V$. 
The vector meson-proton cross sections are denoted by $\sigma_{\T{Vp}}$.
The vector meson-photon coupling constants $g_{\T V}$ can be 
deduced from electron-positron annihilation 
into those vector mesons.   
One observes that $F_2^{{\T p} (VMD)} \sim Q^2$ at small $Q^2$. 
At large $Q^2$, however, the vector meson contribution (\ref{eq:VMD}) 
vanishes as $1/Q^2$. 
Then the scattering from parton  constituents in the target 
takes over and leads to Bjorken scaling.

\medskip
\item[$\bullet$] 

Finally, at large values of $x$ one observes a rapid decrease of 
the structure function. 
This can be understood within the framework of perturbative QCD.
In the limit $x\rightarrow 1$, a single valence quark struck by the virtual 
photon carries all of the nucleon momentum. 
The only way for such 
a configuration to evolve from a bound state wave 
function which is centered around low parton momenta, 
is through the exchange of hard gluons. A perturbative 
description of this process leads to 
$F_2^{{\T p}}(x \rightarrow 1) \sim (1-x)^3$ 
\cite{Brodsky:1995kg}. 
\end{itemize}

\subsubsection{The ratio of longitudinal and transverse cross sections}

\begin{figure}[t]
\begin{center} 
\epsfig{file=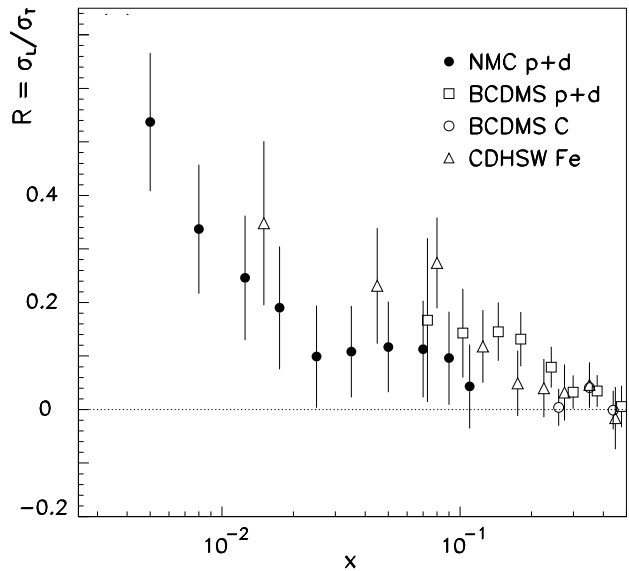,height=80mm}
\end{center}
\caption[...]{
The ratio $R=\sigma_L/\sigma_T$ as a function of $x$. 
The data are taken from  NMC \cite{Arneodo:1997qe}, 
BCDMS \cite{Benvenuti:1989rh,Benvenuti:1990fm,Benvenuti:1987zj}, 
and CDHSW \cite{Berge:1991hr}.
}
\label{fig:R_N}
\bigskip
\end{figure}

Extracting the structure function $F_2$ from lepton 
scattering data requires information on the 
ratio of the total cross section for  
longitudinally and transversely polarized photons,   
$R = \frac{\sigma_L}{\sigma_T}$ from Eq.(\ref{eq:R_L_T}). 
Previous data from SLAC and CERN cover the 
region $0.1 < x < 0.9$ and 
$0.6 \,{\T {GeV}}^2< Q^2 < 80 \,{\T {GeV}^2}$ \cite{Whitlow:1990gk}.
In this region  $R$ is small. 
New data from the NMC collaboration are  available for 
$0.002 < x < 0.12$ \cite{Arneodo:1997qe}. 
A rise of $R$ with decreasing $x$ has been observed 
\nocite{Benvenuti:1989rh,Benvenuti:1990fm,Benvenuti:1987zj,%
Berge:1991hr,Arneodo:1997qe}
as shown in Fig.\ref{fig:R_N}.  
This behavior can be understood within the framework of 
perturbative QCD \cite{Altarelli:1978tq}. 
Helicity conservation implies that 
a high-$Q^2$ longitudinally polarized photon 
cannot be absorbed by a quark moving in longitudinal direction: 
a non-zero transverse momentum is necessary for this process to 
occur. 
In the QCD-improved parton model such transverse quark 
momenta result from  gluon bremsstrahlung which is important  
for low parton momenta, i.e. at small $x$. 
Further studies  of $R$ in the domain of  small $x$ are currently 
performed at HERA. 
A first analysis gives  $R\simeq 0.5$ at $x=2.4\cdot 10^{-4}$ and 
$Q^2 = 15$ GeV$^2$ \cite{Adloff:1997yz}. 

\subsubsection{Spin dependent structure functions}

In recent years polarized deep-inelastic scattering experiments 
have become a major activity at all high-energy lepton 
beam facilities. 
They aim primarily at the exploration  of the spin structure 
of nucleons. 
Detailed investigations have been carried out at 
CERN (SMC), 
SLAC (E142/143/154/155) and DESY (HERMES). 
For references see [33 -- 41].
\nocite{Adeva:1998vv,Adeva:1998vw,Abe:1997qk,%
Airapetian:1998wi,Ackerstaff:1997ws,%
Anthony:1999rm,Anthony:1999py,Abe:1997dp,Abe:1997cx}

While the proton spin structure functions 
$g_{1}^{\T p}$ and $g_2^{\T p}$ have been measured 
directly using hydrogen targets, 
neutron structure functions have  been extracted 
from measurements using  deuterons 
and $^3\T{He}$ targets 
with corrections for nuclear effects.
In  the data analysis such corrections have commonly 
been done in terms of  
effective proton and neutron polarizations  obtained 
from realistic deuteron and $^3$He wave functions. 
They account for the fact that bound nucleons carry orbital 
angular momenta. As a consequence  their polarization vectors 
need not  be aligned with the total polarization of the target.
At the present level of accuracy the use of effective 
polarizations turns out to 
be a reasonable approximation as discussed at length in 
Section \ref{sec:Pol_DIS}. 

\begin{figure}[t]
\begin{center} 
\epsfig{file=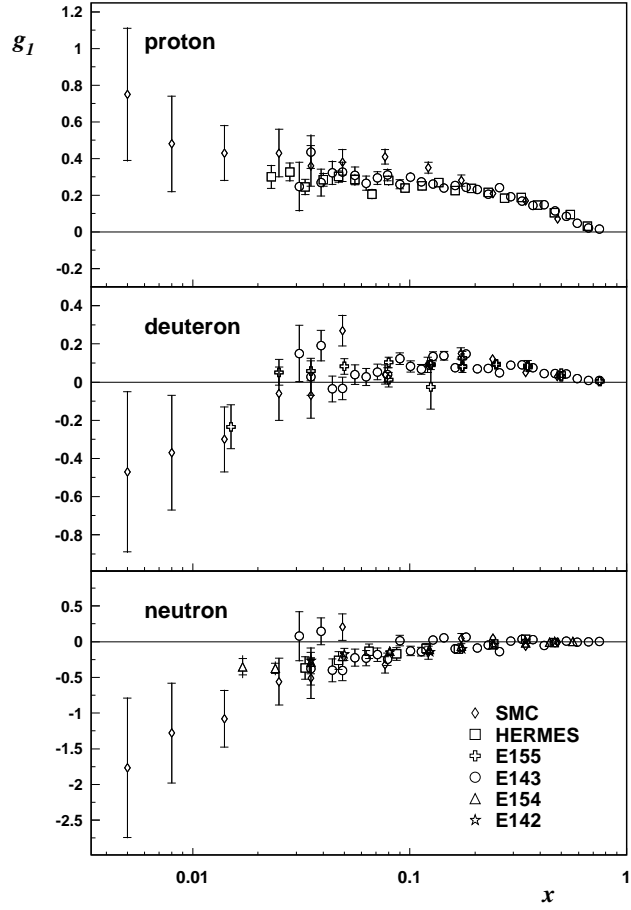,height=120mm}
\end{center}
\caption[...]{A compilation of data of the proton, deuteron, and neutron 
spin structure functions $g_1$ from 
Refs.\cite{Adeva:1998vv,Airapetian:1998wi,Ackerstaff:1997ws,Anthony:1999rm,%
Abe:1997dp,Anthony:1996mw,Abe:1998wq}. 
(We thank U. Stoesslein for the preparation of this figure.) 
}
\label{fig:g1}
\bigskip
\end{figure}

In Fig.\ref{fig:g1} we show a collection of data for $g_1$. 
The behavior of the proton, deuteron and neutron 
structure functions turns out to be quite different, especially 
in the region of  small $x$. 
This is in contrast to the unpolarized case where proton and neutron 
structure functions show a qualitatively similar behavior.

The moments 
\begin{equation} 
\label{eq:moments_g1}
\Gamma_1^{\T{p,n}}(Q^2) \equiv \int_0^1 {d} x \,g_1^{\T{p,n}} (x,Q^2)  
\end{equation}
of the proton and neutron spin structure functions are of fundamental 
importance. They can be decomposed in terms of  
proton matrix elements of SU$(3)$ axial currents, 
as follows (for a review see e.g. \cite{Anselmino:1995gn,Lampe:1998eu}):
\begin{equation} \label{eq:Gamma_1}
\Gamma_1^{\T{p,n}}  = \frac{1}{12} 
\left( \frac{4}{3}  \Delta q_0 + 
                    \frac{1}{\sqrt{3}} \Delta q_8 \pm \Delta q_3  
\right),
\end{equation} 
with the axial vector matrix elements:
\begin{equation}
M S_{\mu} \Delta q_a 
= \langle P,S | \bar \psi \gamma_{\mu} \gamma_5 
\frac{\lambda_a}{2}\,\psi | P,S \rangle,
\end{equation} 
where $ \psi = \left(\psi_u,\psi_d,\psi_s\right)$ is the quark field. 
Here $\lambda_a$ $(a = 1,\dots ,8)$ denote SU$(3)$ flavor matrices 
and the singlet $\lambda_0$ is the $3\times 3$ unit matrix. 
In Eq.(\ref{eq:Gamma_1}) and below we suppress  QCD corrections 
which are currently known  up to order $\alpha_s^3$. 
Current algebra and isospin symmetry 
equate the non-singlet matrix element 
$\Delta q_3 = \Delta u - \Delta d$ 
with the axial vector coupling constant $g_A = 1.26$ 
measured in neutron $\beta$-decay. 
One thus arrives at the fundamental Bjorken sum rule:
\begin{equation} \label{eq:Bj-SR}
\Gamma_1^{\T{p}} - \Gamma_1^{\T n} = 
\frac{1}{6} \Delta q_3 = \frac{1}{6} g_A . 
\end{equation}
Furthermore, assuming SU$(3)$ flavor symmetry, 
$\Delta q_8 = (\Delta u + \Delta d - 2 \Delta s)/\sqrt{3}$  
is determined by hyperon $\beta$-decays. 
The non-singlet matrix elements $\Delta q_{3,8}$ 
involve conserved currents, hence they are scale independent. 
This is different for the singlet term 
$\Delta q_0  = \Delta u + \Delta d + \Delta s$
which receives a $Q^2$-dependence through the QCD axial anomaly. 
Note that in next-to-leading order both quarks and gluons contribute  
to $\Delta q_0$. 
However, the detailed separation into quark and gluon parts 
depends on the factorization scheme used to separate 
perturbative and non-perturbative parts of the spin-dependent 
cross section.

An evaluation of the structure function moments 
$\Gamma_1^{\T{p,n}}(Q^2)$ from Eq.(\ref{eq:moments_g1})
requires knowledge of $g_1$ in the entire interval $0 \leq x \leq 1$. 
Since measurements cover only a limited kinematic range, 
data for  $g_1$ have to be extrapolated to $x\rightarrow 0$ and 
$x\rightarrow 1$.
The large-$x$ extrapolation is not critical since  
$g_1$ becomes small and ultimately vanishes as $x\rightarrow 1$.
The situation at small $x$ is, however, not yet well understood  
(for a review and references see Ref.\cite{Badelek:1996ss}). 
The common approach is to assume Regge behavior 
which implies  
that $g_1 \sim x^\alpha$ with $0\leq \alpha \leq 0.5$ 
for $x\rightarrow 0$.

A status review of the analysis 
of spin structure functions  and their moments 
can be found in Refs.\cite{Adeva:1998vw,Abe:1997dp}. 
All current studies arrive at the conclusion that 
the flavor singlet contribution to the nucleon spin 
is small. 
At $Q^2 = 1$ GeV$^2$ one finds (in the AB scheme) \cite{Adeva:1998vw}: 
\begin{equation}
\Delta q_0 = 0.23 \pm 0.07(\T{stat}) \pm 0.19(\T{sys}).
\end{equation}
This would imply that only 
about one third of the 
nucleon spin is carried by the quark spins alone. 
The missing two thirds probably involve gluon spin contributions and 
orbital angular momentum of quark, antiquark and gluon constituents.
Finally we note that  the Bjorken sum rule  (\ref{eq:Bj-SR}), 
with inclusion of QCD corrections,  is fulfilled at the $5\%$ 
level \cite{Adeva:1998vw}.

Measurements of $g_2$ have been performed at 
CERN \cite{Adams:1994id} and 
SLAC \cite{Abe:1997qk,Abe:1996dc}. 
For the neutron case  
$^3\T{He}$ \cite{Abe:1997qk} and deuteron \cite{Abe:1996dc} 
targets have been used again.
Within large experimental errors the data for $g_2$ 
are consistent  with the twist-$2$ prediction 
$g_2(x,Q^2) = \int_x^1 \frac{dy}{y}
    \left( 1 - \delta(1 - x/y) \right) g_1(y,Q^2)$ 
of Ref.\cite{Wandzura:1977qf}.

\subsubsection{Diffraction}  
\label{ssec:Diffraction}

\begin{figure}[t]
\begin{center} 
\epsfig{file=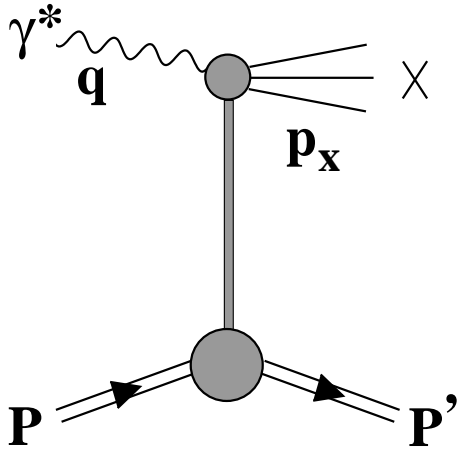,height=60mm}
\end{center}
\caption[...]{Diffractive scattering from a nucleon.
}
\label{fig:diff}
\bigskip
\end{figure}

A subclass of photon-nucleon processes, namely diffractive 
lepto- and photoproduction, plays a prominent role also 
in the interaction of real and virtual photons with complex 
nuclei at high energies. 
We focus here on so-called single diffractive processes. 
They are characterized 
by the proton emerging intact and well separated 
in rapidity from the hadronic state $X$ produced in 
the dissociation of the (virtual) photon (see Fig.\ref{fig:diff}):
\begin{equation}
\gamma^{(*)} + {\T p} \rightarrow  {\T X} + {\T p}'. 
\end{equation}
As in diffractive hadron-hadron collisions such processes 
are important at small momentum transfer. 
Their cross sections drop exponentially 
with the squared four-momentum transfered by the colliding 
particles. 
Furthermore, they generally exhibit  a weak energy dependence. 

\pagebreak 

\underline{Diffractive leptoproduction}
\medskip

\begin{figure}[b]
\bigskip
\begin{center} 
\hspace*{-1cm}
\epsfig{file=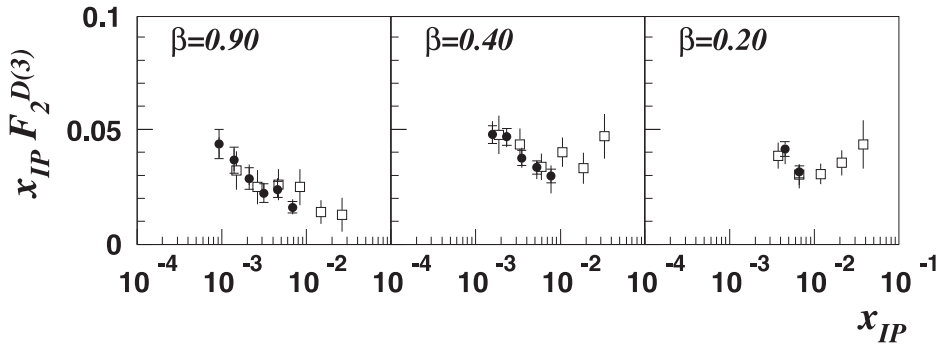,height=45mm}
\end{center}
\caption[...]{
The diffractive structure function 
$\xpom F_2^{D(3)}$ for different values 
of $\beta = \xpom/x$ and ${Q^2} \approx 28$ GeV$^2$.
Data from H1 \cite{Adloff:1997sc} (open squares)  
and ZEUS \cite{Breitweg:1998gc} (solid points). 
}
\label{fig:xF3D}
\end{figure}

In deep-inelastic scattering experiments at HERA 
approximately $10\%$ of the (virtual) photon-proton cross section 
result from  diffractive events (for a review see e.g. 
\cite{Abramowicz:1998ii}).
Their cross section is parametrized in terms of  
two structure functions, analogous  to the inclusive case.
One has:
\begin{equation} \label{eq:diff_lepto_cross}
\frac{d \sigma}{dx dQ^2  d\xpom dt}= 
\frac{4\pi\alpha^2}{Q^4} 
\left\{\frac{1-y}{x} + 
\frac{y^2}{2 x \left[1 + R^{\T {D}(4)}(x,Q^2;\xpom,t) \right]}
\right\}\,F_2^{\T {D}(4)}(x,Q^2;\xpom,t).
\end{equation}
The diffractive structure functions depend on $x$ and  $Q^2$, on   
the squared momentum transfer $t$ to the proton, 
$t = (P-P')^2 = (q - p_{\T X})^2$, and on the variable
\begin{equation}
\xpom =  \frac{(P-P')\cdot q}{P\cdot q} = 
\frac{Q^2 + M_{\T X}^2 - t}{Q^2 + W^2  - M^2}
\approx  
\frac{Q^2 + M_{\T X}^2}{Q^2 + W^2}.  
\end{equation}
Here $M_{\T X}$ is the invariant mass of the diffractively produced 
system $\T X$  in the final state. 
The diffractive structure function, 
conventionally denoted by $F_2^{D(4)}$ indicating its dependence 
on four kinematic variables,  
is directly related to the diffractive (virtual) photoproduction cross 
section. 
At small $x$ one finds in analogy with Eq.(\ref{eq:F2_sig}):
\begin{equation}
F_2^{\T {D}(4)}(x,Q^2;\xpom,t) \approx 
\frac{Q^2}{4\pi^2\alpha}\,
\frac{d\sigma_{\gamma^* \T p}^{\T diff}}{d\xpom  dt}.
\end{equation}
Most of the data have so far been obtained for the $t$-integrated 
structure function 
\begin{equation}
F_2^{{ \T D} (3)} (x,Q^2;\xpom) = \int_{-\infty}^{0} d t \, 
                            F_2^{{\T{D}} (4)} (x,Q^2;\xpom,t). 
\end{equation}   
Measurements by the H1 \cite{Adloff:1997sc,Ahmed:1995ns} and 
ZEUS 
\cite{Breitweg:1998gc,Breitweg:1998aa,Derrick:1996ma,Derrick:1995wv} groups 
cover the range  
$4.5 < Q^2 < 140$ GeV$^{2}$, $2\cdot 10^{-4} < \xpom <0.04$ and 
$0.02 < x/\xpom <0.9$. 
No substantial $Q^2$-dependence of $F_2^{{\T D} (3)}$ has 
been found.
Over most of the explored kinematic region, $\xpom F_2^{{ \T D} (3)}$ 
is either decreasing or approximately constant  
as a function of increasing $\xpom$. 
However at small  $x/\xpom$ there is a 
tendency for $\xpom F_2^{{ \T D} (3)}$ to increase at the highest values of 
$\xpom$. 
A typical collection of data is shown in Fig.\ref{fig:xF3D}.

A reasonably successful 
description of this behavior has been achieved 
within Regge phenomenology which assumes 
that the interaction proceeds in two steps: 
the emission of a pomeron or subleading reggeon from the 
proton, and the subsequent hard scattering of the virtual photon 
from the partons in the pomeron or reggeon, respectively. 
This picture leads to a factorization of the diffractive 
structure function \cite{Ingelman:1993qf}: 
\begin{equation} \label{eq:F_2D4_regge}
F_2^{{\T{D}}(4)} (x,Q^2;\xpom,t)  
= f_{_{\pom}} (\xpom,t) \,F_2^{^{\pom}}(x/\xpom, Q^2) 
+ f_{_{\reg}} (\xpom,t) \,F_2^{^{\reg}}(x/\xpom, Q^2), 
\end{equation}
where $F_2^{^{\pom(\reg)}}$ is 
commonly interpreted as the ``structure function'' of 
the pomeron (reggeon) and 
\begin{equation}
f_{i}(\xpom,t) = 
\frac{e^{B_{i} \, t}}
{\xpom^{2 \alpha_{i} (t)- 1}},
\end{equation}
with $i=\pom,\reg$, 
denotes the  pomeron (reggeon) distribution in the proton.

The H1 analysis \cite{Adloff:1997sc} 
gives $\apom (0) \approx 1.2$ 
and $\areg (0) \approx 0.5$. 
The slope parameters $B_{_{\pom(\reg)}}$,  
$\areg'$ and $\apom'$ were taken to reproduce hadron-hadron data. 
While $\apom (0)$ is found to be slightly  
larger than the value obtained from parametrizations of hadronic 
cross sections,  
$\areg(0)$ agrees well with the Regge phenomenology of 
hadron--hadron collisions \cite{Donnachie:1992ny} .

\begin{figure}[t]
\begin{center} 
\epsfig{file=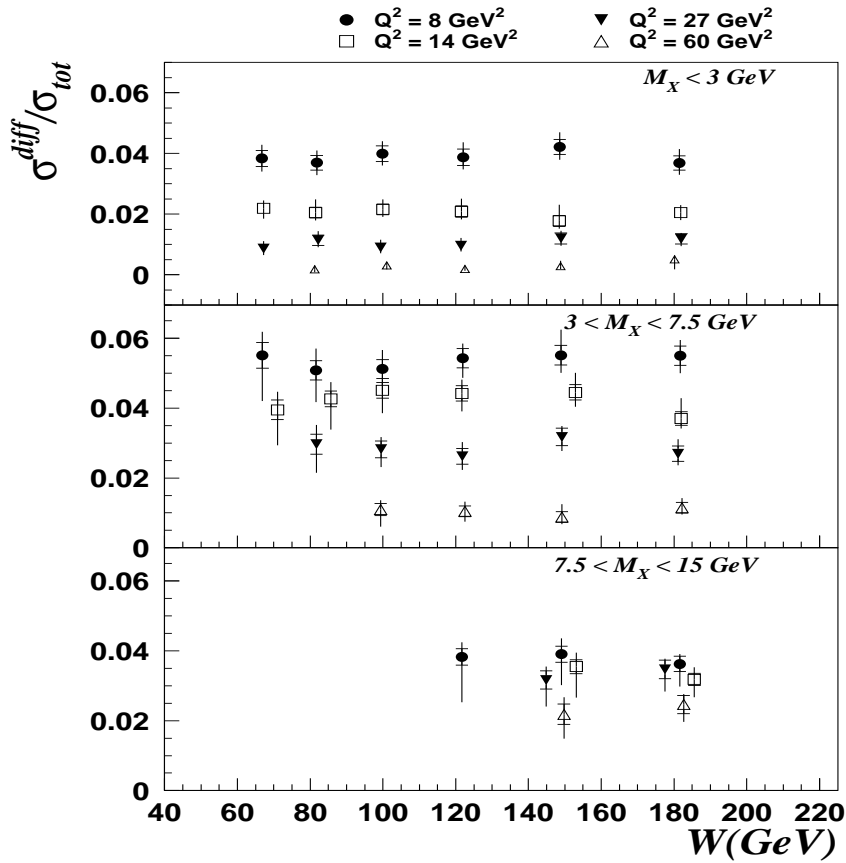,height=130mm}
\end{center}
\caption[...]{
ZEUS data \cite{Breitweg:1998gc} for the ratio of diffractive 
and  total photon-nucleon cross sections. 
The diffractive production cross section has been integrated over 
different intervals of $M_{\T X}$.
}
\label{fig:diff_tot}
\bigskip
\end{figure}

In Fig.(\ref{fig:diff_tot}) we show recent ZEUS data 
\cite{Breitweg:1998gc} on the ratio of diffractive 
and total photon-nucleon cross sections for different 
regions of $M_{\T X}$. 
The data show a similar energy dependence of both  
total and diffractive cross sections. 
Furthermore, the observed $Q^2$-dependence of the cross 
section ratio for different regions of $M_{\T X}$ 
suggests that, as $Q^2$ increases, 
diffractive states with large mass  
become important.

ZEUS measurements \cite{Breitweg:1998aa} have investigated  
the $t$-dependence of the diffractive leptoproduction cross section. 
In the range $5 < Q^2 < 20$ GeV$^2$ and 
$50 < W < 270$ GeV the $t$-dependence of the diffractive virtual 
photoproduction cross section is described 
for $0.07 < |t| < 0.4$ GeV$^2$ by the exponential form, 
$d\sigma_{\gamma^* p}^{diff} /dt \sim e^{B\, t}$, 
with  $B \approx 7$ GeV$^2$. 
This value is compatible with results from 
high-energy hadron-hadron scattering 
(see e.g. \cite{Goulianos:1995wy}).

\bigskip
\underline{Diffractive photoproduction}
\label{ssec:Diffractive_photoproduction}
\medskip

Diffractive dissociation of real photons, 
$\gamma + {\T N} \rightarrow {\T X} + {\T N}$, has been 
explored  with  fixed target and collider experiments. 
At FNAL \cite{Chapin:1985} photon-proton center of mass 
energies up to  $W \simeq 15$ GeV  were used to produce 
diffractive states with an invariant mass up to 
$M_{\T X} \simeq 5$ GeV. 
Recent experiments at HERA 
\cite{Breitweg:1997eh,Breitweg:1997za,Adloff:1997mi,Aid:1995bz,Derrick:1994dt} 
were carried out at $W \simeq 200$ GeV and $M_{\T X} < 30$ GeV. 
The diffractive cross section  amounts to 
approximately $20\%$ of the total 
photon-proton cross section. 
Around half of these events come from the production of 
the light vector mesons $\rho,\omega$ and $\phi$. 
This is contrary to diffractive leptoproduction 
at large $Q^2$ where 
vector meson contributions  are suppressed roughly as  
$1/Q^4$ \cite{Crittenden:1997yz}. 

\begin{figure}[t]
\begin{center} 
\epsfig{file=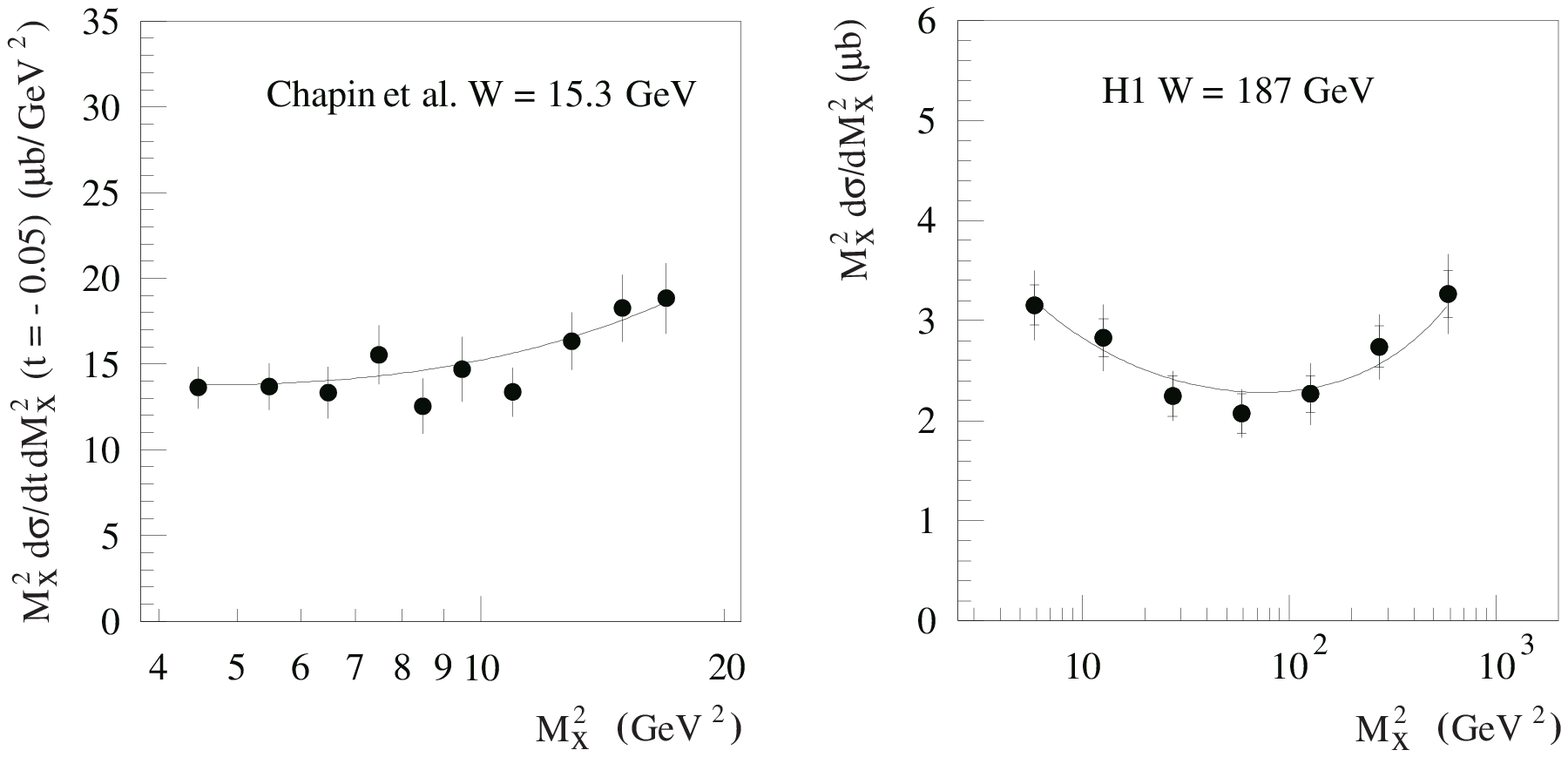,height=60mm}
\end{center}
\caption[...]{
Differential cross section for 
diffractive photoproduction off nucleons 
from  FNAL \cite{Chapin:1985}  and H1 \cite{Adloff:1997mi}  
for different center of mass energies $W$. 
The  curves corresponds to a Regge fit \cite{Adloff:1997mi}. 
}
\label{fig:diff_ph}
\bigskip
\end{figure}

At sufficiently large mass $M_{\T X}$ 
of the diffractively produced system $\T X$, 
the measured cross sections  
drop approximately as $1/M_{\T X}^2$, as shown in Fig.\ref{fig:diff_ph}. 
This is in accordance with Regge phenomenology. 
In the limit 
$W^2/M_{\T X}^2 \rightarrow \infty$ with  
$M_{\T X}^2 \rightarrow \infty$ only pomeron 
exchange is important and leads to \cite{Goulianos:1983vk}:
\begin{equation} \label{eq:DD_MX}
\frac{d \sigma_{\gamma \T N}^{diff}}{d M_{\T X}^2 dt}  \sim 
\frac{W^{4(\apom(0)-1)}}{M_{\T X}^{2 \apom (0)}} \,
\exp\left[
t \cdot \left(
{B + 2 \apom' \ln\left(\frac{W^2}{M_{\T X}^2}\right)}\right)\right],
\end{equation}
with a slope parameter $B$. 
Equation (\ref{eq:DD_MX}) implies 
that at energies $W = (15 - 30)$ GeV typical 
for fixed target experiments at CERN and FNAL,   
the relative amount of diffraction in deep-inelastic scattering 
is reduced to $(10 - 15)\%$ \cite{Piller:1998cy}.
Observed deviations from the  simple behavior (\ref{eq:DD_MX})
have been associated with contributions involving subleading Regge 
trajectories \cite{Breitweg:1997za,Adloff:1997mi}.

\section{Deep-inelastic scattering from nuclear systems} 

\setcounter{section}{3}
\setcounter{figure}{0}

\subsection{Introduction and motivation}

We now enter into the central topic of this review: 
an exploration of new phenomena specific to deep-inelastic 
lepton scattering from {\it nuclear} (rather than free nucleon) 
targets. 

Nuclei represent systems with a natural, built-in length scale. 
The baryon density in the center of a typical heavy nucleus is 
$\rho_0 \simeq 0.15$ fm$^{-3}$. 
The average distance between two nucleons at this density is
\begin{equation}  \label{eq:average_NN_distance}
 d \simeq 1.9 \,{\T{fm}}.
\end{equation}
The nucleons have a momentum distribution 
characterized by their Fermi momentum, 
\begin{equation}
p_F = \left(\frac{3 \pi^2}{2}\,\rho_0\right)^{1/3} 
\simeq 1.3\,\T{fm}^{-1} \simeq 0.26\,\T{GeV}.
\end{equation}
A high energy virtual photon which scatters from this system  
can expect to see two sorts of genuine nuclear effects:

\begin{enumerate}

\item[i)] 
Incoherent scattering from $A$ nucleons, but with their structure functions 
modified in the presence of the nuclear medium. Such modifications 
are expected to arise, for example, from the mean field that a 
nucleon experiences in the presence of other nucleons, and from its 
Fermi motion inside the nucleus;

\medskip
\item[ii)] 
Coherent scattering processes involving more than 
one nucleon at a time. Such effects can occur when hadronic 
excitations (or fluctuations) produced by the 
high energy photon propagate over distances (in the laboratory frame) 
which are comparable to or larger than the characteristic length 
scale $d\sim 2$ fm of Eq.(\ref{eq:average_NN_distance}). 
A typical example of a coherence effect is shadowing.

\end{enumerate}

It turns out, as we will demonstrate, that incoherent scattering 
takes place primarily in the range $0.1 < x < 1$ of the Bjorken variable. 
Strong coherence effects are observed at $x<0.1$. 
Cooperative phenomena in which several nucleons participate can also 
occur at $x>1$.
(In fact, the Bjorken variable can extend, in principle, up to 
$x\leq A$ in a nucleus with $A$ nucleons.) 

The aim of this section is to prepare the facts and 
phenomenology of nuclear DIS. An important 
subtopic in this context deals with the deuteron. 
While this is not a typical nucleus, it serves two purposes: 
first, as a convenient neutron target, and secondly, as the simplest 
prototype system in which coherence effects, involving proton and neutron 
simultaneously, can be investigated quite accurately. For this 
purpose we need to introduce the hadronic tensor and structure 
functions for spin-$1$ targets as well. 
Once the nuclear structure functions are at hand we 
will present a survey of nuclear DIS data and give first, qualitative 
interpretations. 
The more detailed understanding is then developed in subsequent 
sections.

\subsection{Nuclear structure functions}
\label{ssec:nucl_str_fns}

The deep-inelastic scattering cross sections  for 
free nucleons and nuclei have basically the same form 
as given by Eq.(\ref{eq:crossDIS}). 
All information about the target and its response to the 
interaction is included in the corresponding hadronic tensor. 
For nuclei with spin $1/2$ the hadronic tensor formally 
coincides with the one for free nucleons 
given in 
Eqs.(\ref{eq:hadronic_tensor},%
\ref{eq:hadronic_tensor_sym},\ref{eq:hadronic_tensor_asym}). 
In this case nuclei are characterized by four structure functions, 
$F_{1,2}^{\T A}$ and $g_{1,2}^{\T A}$. 
For spin-$0$ targets, only the  symmetric 
tensor (\ref{eq:hadronic_tensor_sym}) with the structure 
functions $F_{1,2}^{\T A}$ is present.
In the case of spin-$1$ targets 
the situation is more complex. 
Here the hadronic tensor is composed of  eight 
independent structure functions 
\cite{Hoodbhoy:1989am,Sather:1990bq}:\footnote{  
We omit  terms proportional to $q_{\mu}$ or $q_{\nu}$ which 
do not contribute to the cross section (\ref{eq:crossDIS}) due to 
electromagnetic gauge invariance.}
\begin{eqnarray} \label{eq:hadten_A}
M_{\T A}\,W^{\T A}_{\mu\nu} &=&
- g_{\mu\nu} \,F_1^{\T A} + \frac{P_{\mu}P_{\nu}}{P\cdot q}\,F_2^{\T A}
+i\frac{M_{\T A}}{P\cdot q}\,\varepsilon_{\mu\nu\alpha\beta}
\,q^{\alpha}
\left( S^{\beta}\,(g_1^{\T A} + g_2^{\T A}) 
- \frac{S\cdot q}{P\cdot q} \,P^{\beta}\,g_2^{\T A}
\right)   
\nonumber \\
&+& 
r_{\mu\nu} \,b_1^{\T A} + s_{\mu\nu} \,b_2^{\T A} 
+ t_{\mu\nu}\, \Delta^{\T A} + u_{\mu\nu}\, b_3^{\T A},
\end{eqnarray} 
with the Lorentz tensors:  
\begin{eqnarray} 
\label{eq:W1_tensors}
r_{\mu\nu} &=&  - g_{\mu\nu}
\left(\frac{M_{\T A}^2}{\kappa_{\T A}  (P\cdot q)^2} \,q\cdot {\cal E}  
\,q\cdot {\cal E}^* - \frac{1}{3}\right),
\\
s_{\mu\nu} &=&   \frac{P_{\mu}P_{\nu}}{P\cdot q} 
\left(\frac{M_{\T A}^2}{\kappa_{\T A}  (P\cdot q)^2} \,q\cdot {\cal E}  
\,q\cdot {\cal E}^* - \frac{1}{3}\right),
\nonumber \\
t_{\mu\nu} &=& 
\frac{1}{2} \left\{
\left(-g_{\mu\nu} + 
\frac{2x_{\T A}}{\kappa_{\T A}\,P\cdot q} P_{\mu}P_{\nu} \right)  
\left(\frac{M_{\T A}^2}{\kappa_{\T A}\,(P\cdot q)^2} \,q\cdot {\cal E}  
\,q\cdot {\cal E}^* - 1\right)  
\right.\nonumber\\
&+&
\left.
\left[
\left({\cal E}_{\mu} - \frac{q\cdot {\cal E}}{\kappa_{\T A} \,P\cdot q} 
P_{\mu}\right)
\left({\cal E}^*_{\nu} - 
\frac{q\cdot {\cal E}^*}{\kappa_{\T A} \,P\cdot q}P_{\nu}\right)
+ (\mu\leftrightarrow \nu)
\right]
\right\},
\nonumber\\
u_{\mu\nu} &=& 
\frac{\kappa_{\T A}-1}{\sqrt{\kappa_{\T A}}\,P\cdot q} 
\left[P_{\mu} \,q\cdot {\cal E}^* \left({\cal E}_{\nu} - 
\frac{q\cdot {\cal E}}{\kappa_{\T A} \,P\cdot q} P_{\nu} \right) 
+ P_{\mu} \,q\cdot {\cal E} 
\left({\cal E}^*_{\nu} - 
\frac{q\cdot {\cal E}^*}{\kappa_{\T A}\, P\cdot q} P_{\nu}\right)
+ (\mu\leftrightarrow \nu)
\right].
\nonumber 
\end{eqnarray}
The tensors (\ref{eq:W1_tensors}) are functions of the photon and target 
four-momenta 
$q^{\mu}$ and $P^{\mu}$, the target polarization vector ${\cal E}$, 
and the spin vector $S_{\alpha}
= -i \varepsilon_{\alpha\beta\gamma\delta}
   \,{\cal E}^{\beta *}\,{\cal E}^{\gamma} P^{\delta}/M_{\T A}$.
Furthermore we have used the notation  
$\kappa_{\T A}=1+{M_{\T A}^2Q^2}/{(P\cdot q)^2}$ where $M_{\T A}$ 
denotes the nuclear mass. 

The nuclear structure functions in Eq.(\ref{eq:hadten_A})
depend on the Bjorken scaling variable of the 
target, $x_{\T A} = Q^2/2 P\cdot q$ with $0\leq x_{\T A}\leq 1$, 
and on the momentum transfer $Q^2$. 
Note however that these functions 
are frequently expressed 
in terms of the Bjorken variable of the free nucleon 
which is $x=Q^2/2 M \nu = x_{\T A} M_{\T A}/M $ in the lab frame, 
and which can extend over the interval $0 \leq x \leq M_{\T A}/M \simeq A$. 
The first four structure functions in Eq.(\ref{eq:hadten_A})
are  proportional to  Lorentz structures 
already present  in the case of free nucleons 
(\ref{eq:hadronic_tensor}) or spin-$1/2$ nuclei. 
The new structure functions can be measured in the scattering of 
unpolarized leptons from polarized targets. 
By analogy with the Callan-Gross relation (\ref{eq:F2_parton}) 
one finds $b_2^{\T A} = 2 x_{\T A} b_1^{\T A}$ in the scaling limit.
The deuteron structure function $b_1^{\T d}$  
is subject of investigations at HERMES \cite{Jackson94}.

For spin-$1/2$ nuclei the relations between nuclear structure 
functions and photon-nucleus helicity amplitudes 
${\cal A}^{\gamma^* {\T A}}_{hH,h'H'}$ 
are analogous to the ones for free nucleons in 
Eqs.(\ref{f1helamp}--\ref{eq:g2_helicity}). 
For spin-$1$ targets with helicity $H,H'=+,-,0$ one obtains 
\cite{Hoodbhoy:1989am,Sather:1990bq,Edelmann:1997ik}: 
%
\begin{eqnarray}
F_1^{\T A} &=& \frac{1}{6\pi e^2} 
\left(
{\T {Im}}\,{\cal A}^{\gamma^* {\T A}}_{++,++} +  
{\T {Im}}\,{\cal A}^{\gamma^* {\T A}}_{+-,+-}  
+ 
{\T {Im}}\,{\cal A}^{\gamma^* {\T A}}_{+0,+0} 
\right),  
\\
F_2^{\T A} &=& \frac{x_{\T A}}{3\pi e^2\kappa_{\T A}} 
\left(
\!{\T {Im}}\,{\cal A}^{\gamma^*{\T A} }_{++,++} \!+\!  
{\T {Im}}\,{\cal A}^{\gamma^*{\T A}}_{+-,+-} + 
{\T {Im}}\,{\cal A}^{\gamma^*{\T A}}_{+0,+0} 
+ 2 \,{\T {Im}}\,{\cal A}^{\gamma^*{\T A}}_{0+,0+} + 
{\T {Im}}\,{\cal A}^{\gamma^*{\T A}}_{00,00} 
\right)\!,  
\\
\label{eq:g1A_hel}
g_1^{\T A} &=& \frac{1}{4\pi e^2\kappa_{\T A}} 
\left(
{\T {Im}}\,{\cal A}^{\gamma^*{\T A}}_{+-,+-} -  
{\T {Im}}\,{\cal A}^{\gamma^*{\T A}}_{++,++}  
+ \sqrt{\kappa_{\T A}-1}({\T {Im}}\,{\cal A}^{\gamma^*{\T A}}_{+0,0+} + 
{\T {Im}}\,{\cal A}^{\gamma^*{\T A}}_{+-,00}) \right),  
\\
g_2^{\T A} &=& \frac{1}{4\pi e^2\kappa_{\T A}} 
\left(
{\T {Im}}\,{\cal A}^{\gamma^*{\T A}}_{++,++} -  
{\T {Im}}\,{\cal A}^{\gamma^*{\T A}}_{+-,+-}  
+ \frac{1}{\sqrt{\kappa_{\T A}-1}} 
({\T {Im}}\,{\cal A}^{\gamma^*{\T A}}_{+0,0+} + 
{\T {Im}}\,{\cal A}^{\gamma^*{\T A}}_{+-,00})
\right),
\\  
\label{eq:b1_hel}
b_1^{\T A} &=& -\frac{1}{4\pi e^2} 
\left(
{\T {Im}}\,{\cal A}^{\gamma^*{\T A}}_{++,++} +  
{\T {Im}}\,{\cal A}^{\gamma^*{\T A}}_{+-,+-}  - 
2 {\T {Im}}\,{\cal A}^{\gamma^*{\T A}}_{+0,+0} 
\right).  
\end{eqnarray}
%
Corresponding relations for the remaining structure functions 
can be found for example in Ref.\cite{Sather:1990bq,Edelmann:1997ik}.

\subsection{Data on nuclear structure functions}
\label{Sec:nucl_dat}

In this section we summarize the existing  experimental 
information on nuclear effects in structure functions. 
Their systematic investigation for light and heavy nuclei 
has been carried out so far only in unpolarized scattering experiments.
Most of the data come  from deep-inelastic lepton scattering. 
Modifications of nuclear parton distributions have also been 
studied in other high-energy processes. 
We mention, in particular,  heavy quark production 
and Drell-Yan experiments.

\subsubsection{Nuclear effects in $F_2^{\T{A}}$}
\label{subs:Nucl_F2}

Experiments on 
deep-inelastic scattering from nuclei are reviewed in 
\cite{Arneodo:1994wf,Geesaman:1995yd}.
For a discussion of the data it is convenient to use 
structure functions which depend on 
the Bjorken scaling variable for a free nucleon, 
$x = Q^2/(2 M \nu)$. 
In charged lepton scattering from unpolarized nuclear targets 
these structure functions 
are defined by the  differential cross section per nucleon: 
\begin{equation}
\frac{d^2\sigma^{\T A}}{dx \,dQ^2} = 
\frac{4 \pi \alpha^2}{Q^4} 
\left[
\left(1-y-\frac{M xy}{2 E}\right) \frac{F_2^{\T A}(x,Q^2)}{x} 
+ y^2 F_1^{\T A}(x,Q^2)\right]. 
\end{equation}
%

\begin{figure}[t]
\bigskip
\begin{center} 
\epsfig{file=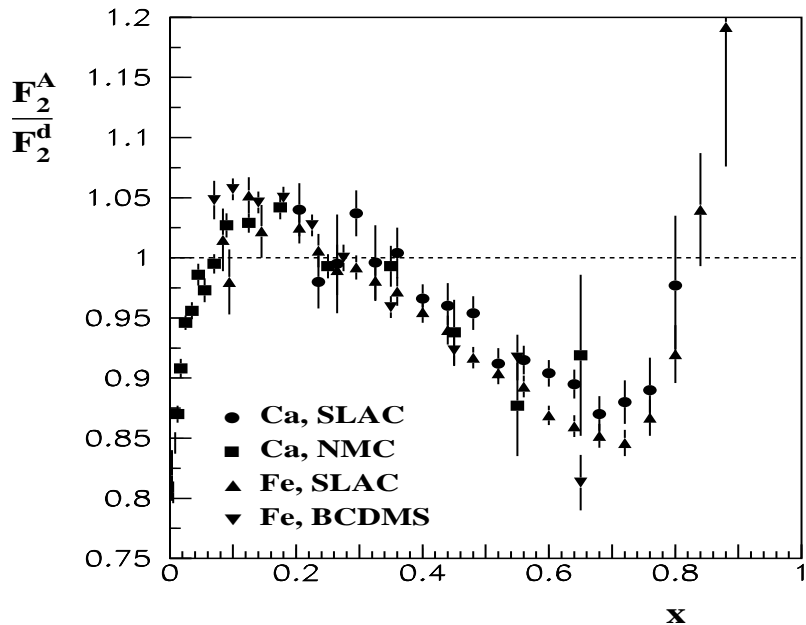,height=90mm,width=120mm}
\end{center}
\caption[...]{
The structure function ratio $F_2^{\T A} / F_2^{\T d}$
for $^{40}$Ca and $^{56}$Fe. 
The data are taken from NMC \cite{Amaudruz:1995tq}, 
SLAC \cite{Gomez:1994ri}, and 
BCDMS \cite{Benvenuti:1987az}.
}
\label{fig:RAd}
\bigskip
\end{figure}
Some time ago the EMC collaboration  discovered that the structure 
function $F_2$ for iron differs
substantially from the corresponding deuteron structure function 
\cite{Aubert:1983}, 
far beyond trivial Fermi motion corrections.  
Since then many experiments dedicated to a study of nuclear 
effects in unpolarized deep-inelastic scattering have been 
carried out at CERN, SLAC  and FNAL. 
The primary aim was to explore 
the difference of nuclear and deuterium structure functions.

Figure \ref{fig:RAd}
\nocite{Gomez:1994ri,Amaudruz:1995tq,Benvenuti:1987az}
presents a compilation of data for the 
structure function ratio $F_2^{\T A} / F_2^{\T d}$ over the  
range $0 \leq x \leq 1$. 
Here $F_2^{\T A}$ is the structure function 
per nucleon of a nucleus with mass number $A$, 
and $F_2^{\T d}$ refers to deuterium. 
In the absence of nuclear effects the ratios $F_2^{\T A} / F_2^{\T d}$ 
are thus  normalized to one. 
Neglecting small nuclear effects in the deuteron, $F_2^{\T d}$ 
can approximately stand for the isospin averaged nucleon 
structure function,  $F_2^{\T N}$. 
However, the more detailed analysis must include two-nucleon 
effects in the deuteron. 
Several distinct regions with characteristic nuclear effects can 
be identified: 
at $x < 0.1$ one observes a systematic reduction 
of $F_2^{\T A} / F_2^{\T d}$, the so-called nuclear shadowing.
A small enhancement is seen at $0.1 < x < 0.2$. 
The dip at $0.3 < x < 0.8$ is often referred to as 
the traditional ``EMC effect''. 
For $x>0.8$ the observed  enhancement of the nuclear structure function 
is associated with nuclear Fermi motion. 
Finally, note again that nuclear structure functions can extend 
beyond $x=1$, the kinematic limit for scattering from free nucleons. 
\begin{figure}[b]
\bigskip
\begin{center} 
\epsfig{file=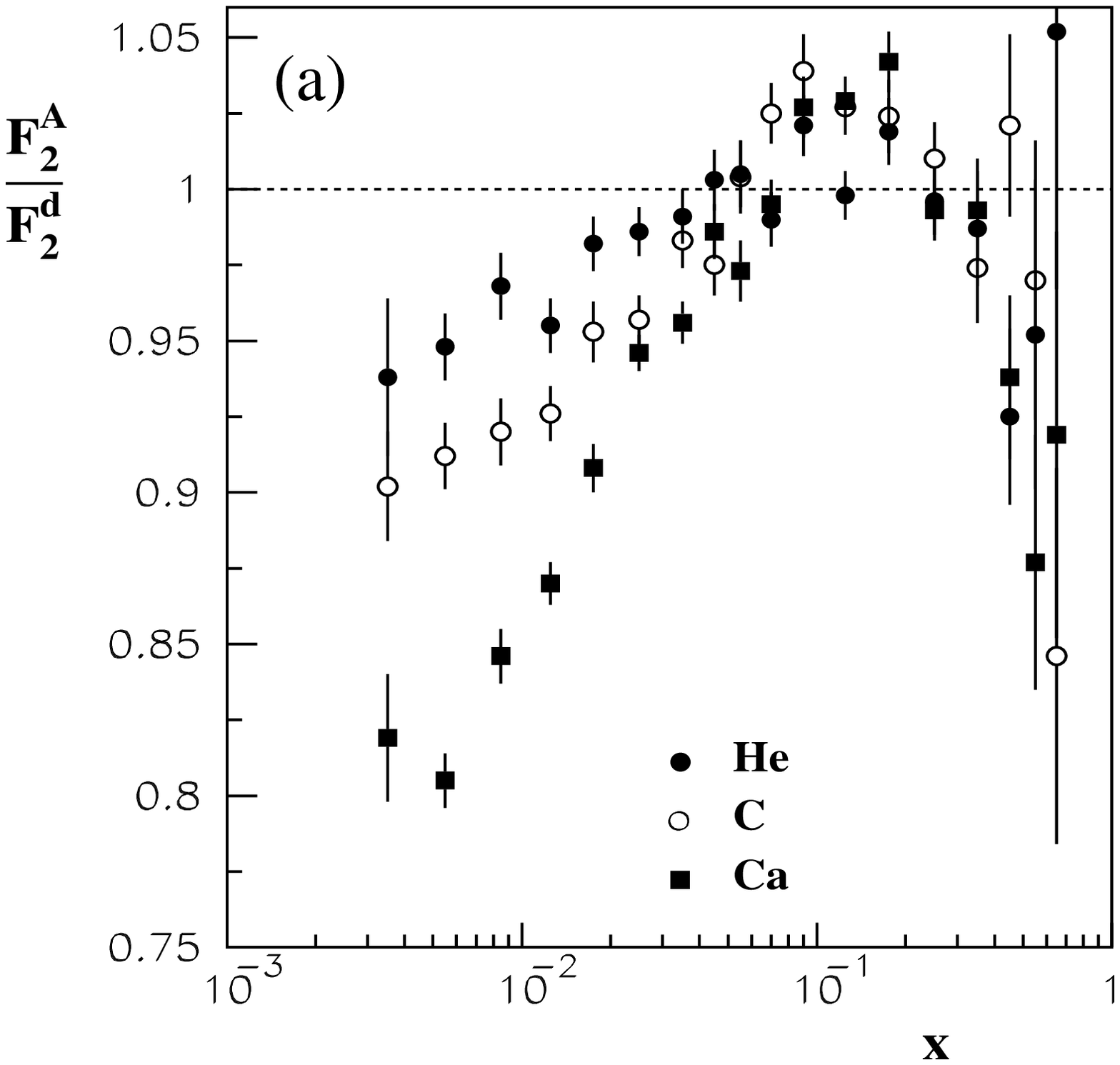,height=80mm,width=110mm}
\end{center}
\begin{center} 
\epsfig{file=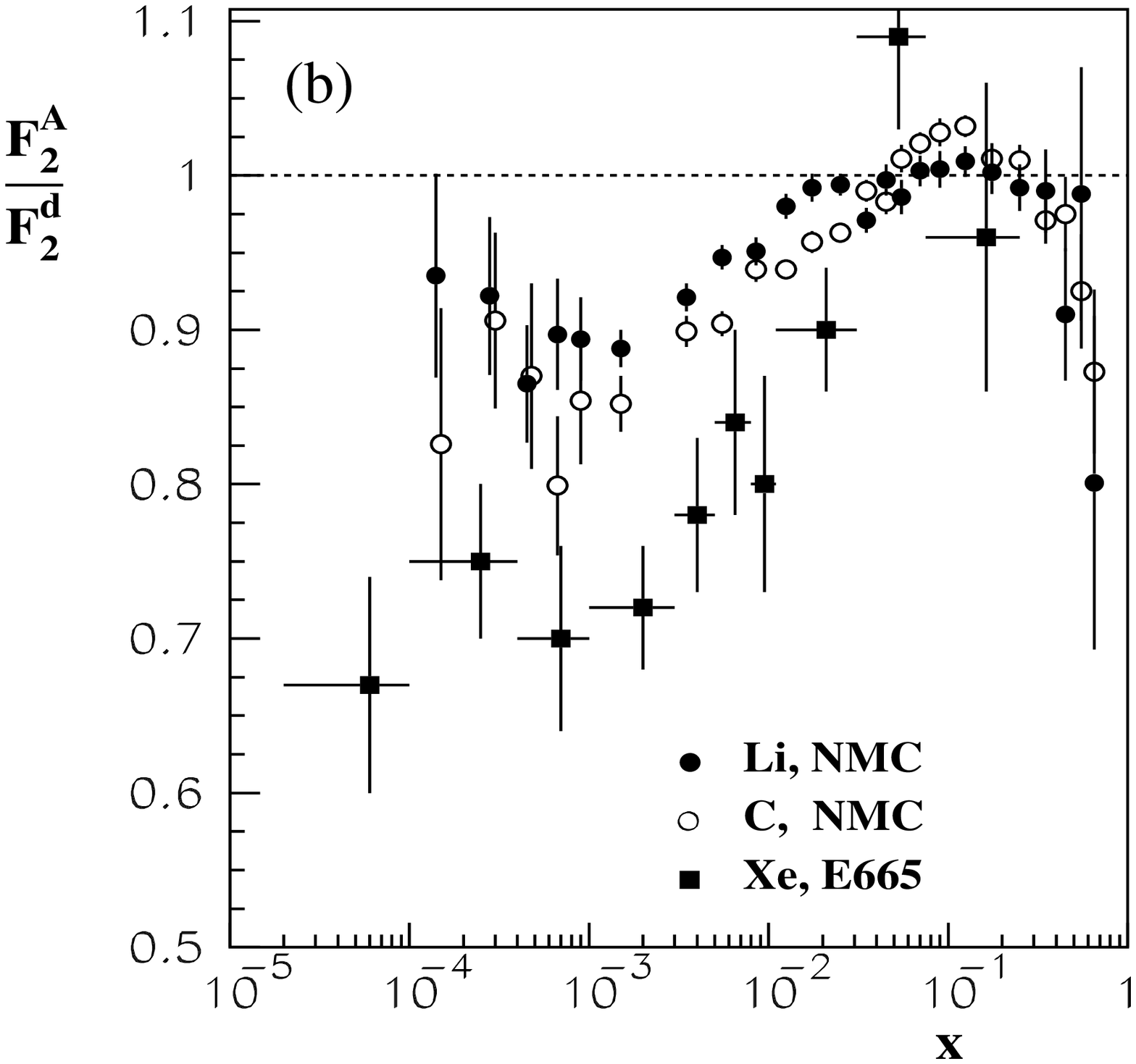,height=80mm,width=110mm}
\end{center}
\caption[...]{
(a) NMC data \cite{Amaudruz:1995tq} for the  
structure function ratio $F_2^{\T A} / F_2^{\T d}$ for  
$^{4}$He, $^{12}$C, and $^{40}$Ca.  
(b) The ratio $F_2^{\T A} / F_2^{\T d}$ for  
$^{6}$Li, $^{12}$C \cite{Arneodo:1995cs}, 
and $^{131}$Xe \cite{Adams:1992nf}.
}
\label{fig:RAd_shad2}
\bigskip
\end{figure}
\begin{figure}[t]
\begin{center} 
\epsfig{file=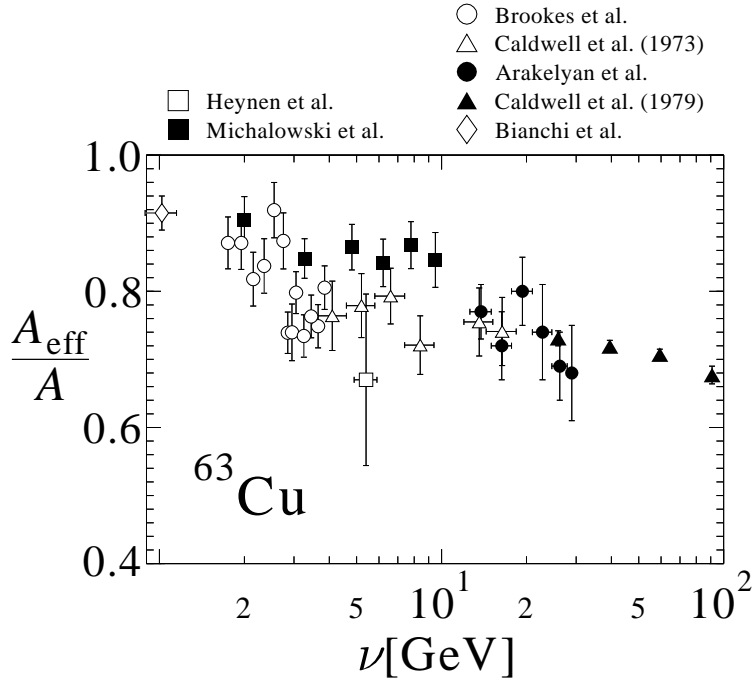,height=90mm}
\end{center}
\caption[...]{The shadowing ratio 
$A_{eff}/A = \sigma_{\gamma \T A}/{A \sigma_{\gamma\T N}}$ 
for $^{63}$Cu as a function of the photon energy $\nu$. 
The date are taken from 
Refs.\cite{Heynen71,Brookes73,Caldwell:1973bu,Michalowski:1977eg,%
Arakelian:1978rc,Caldwell:1979ik,Bianchi:1994ax}.
}
\label{fig:shad_photon}
\bigskip
\end{figure}
\begin{figure}[b]
\begin{center}
\begin{tabular}[htb]{lcr}
\epsfig{file=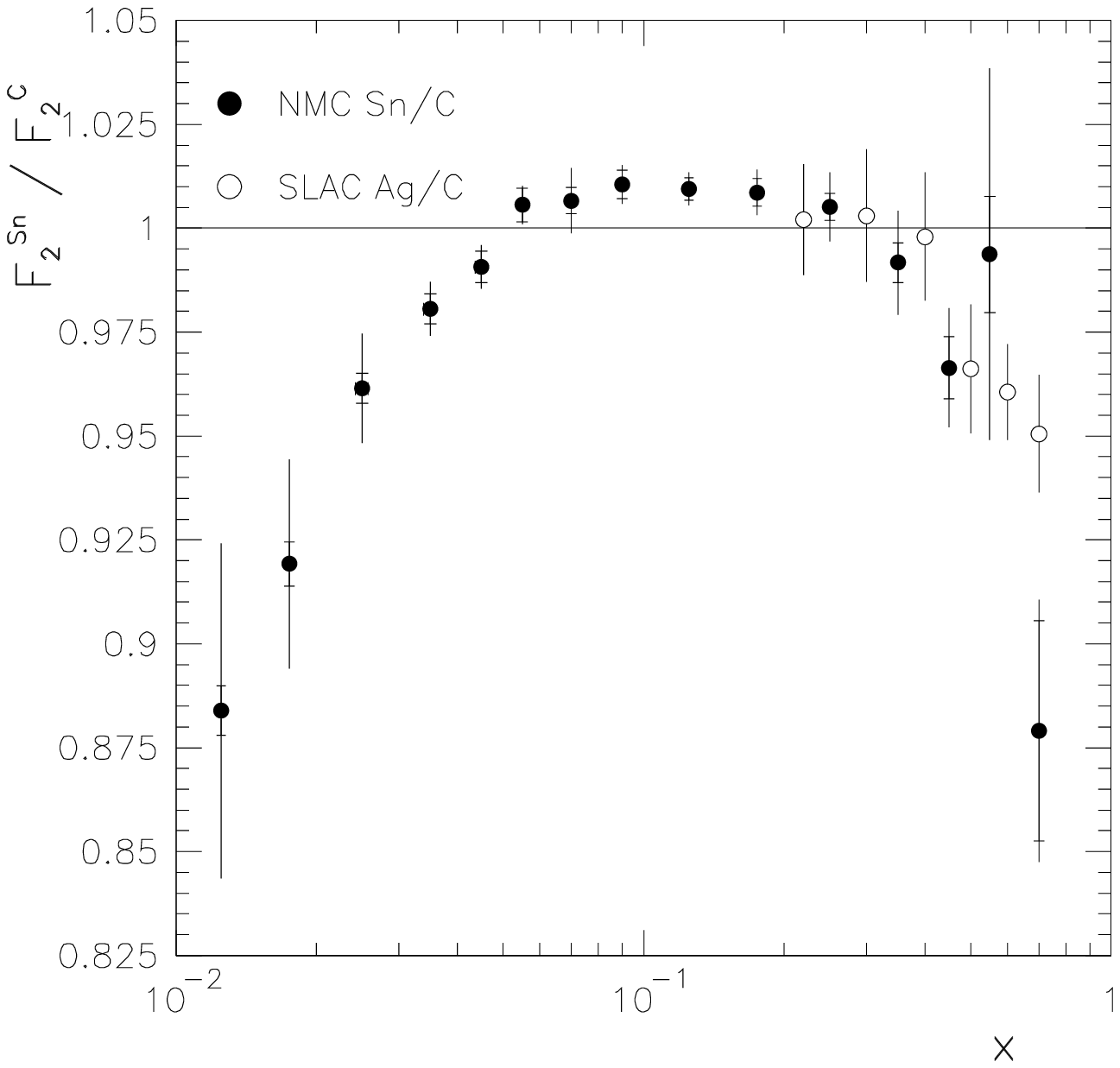,width=.40\textwidth}
& &
\epsfig{file=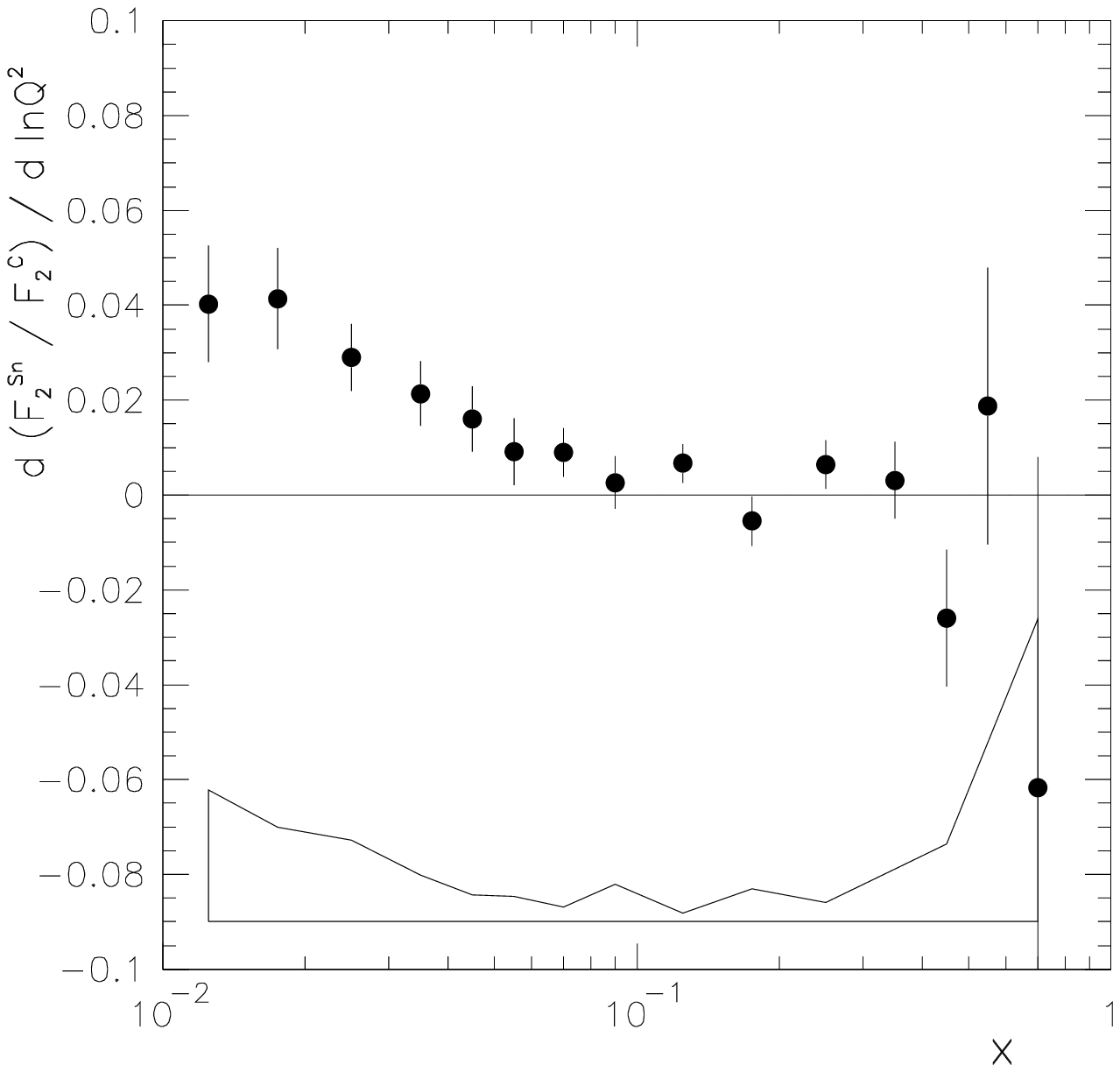,width=.40\textwidth} 
\end{tabular}
\end{center}
\caption{Left:
NMC data \cite{Arneodo:1996ru} for the ratio 
$F_2^{\T {Sn}} / F_2^{\T C}$ as a function of $x$ averaged over 
$Q^2$. 
At large $x$ SLAC  data \cite{Gomez:1994ri} for the ratio  
$F_2^{\T {Ag}} / F_2^{\T C}$ are added. 
Right: Results for  
the logarithmic slope  ${d}(F_2^{\T {Sn}}/F_2^{\T C})/{d}\ln Q^2$ 
from NMC \cite{Arneodo:1996ru}.
The error bars represent statistical uncertainties. 
The band indicates  the size of the systematic errors.
}
\label{fig:sn/c_a}
\label{emk_fig_sn}
\bigskip
\end{figure}

\begin{itemize} 

\item [$\bullet$]
{\bf Shadowing region}

\noindent
Measurements of 
E665 \cite{Adams:1992nf,Adams:1995is,Adams:1992vm} 
at Fermilab and NMC 
\cite{Amaudruz:1995tq,Arneodo:1995cs,Amaudruz:1991cc,Amaudruz:1992dj,%
Arneodo:1996rv,Arneodo:1996ru}
at CERN provide detailed and systematic 
information about  the $x$- and $A$-dependence of 
the structure function ratios $F_2^{\T A}/ F_2^{\T d}$. 
Nuclear targets ranging from He to Pb have been used.  
A sample of data for several  nuclei is shown in 
Fig.\ref{fig:RAd_shad2}.
While most experiments cover the region $x > 10^{-4}$,  
the E665 collaboration provides data for 
$F_2^{\T Xe}/F_2^{\T d}$ \cite{Adams:1992nf}
down to $x \simeq 2 \cdot 10^{-5}$.   
Given the kinematic constraints in fixed target experiments,  
the small $x$-region has been explored at low $Q^2$ only. 
For example, at $x \simeq 5 \cdot 10^{-3} $ the typical 
momentum transfers are $Q^2 \simeq 1$ GeV$^2$ \cite{Arneodo:1995cs}.
At extremely small values, $x \simeq 6 \cdot 10^{-5}$, one has 
$Q^2 \simeq 0.03$ GeV$^2$ \cite{Adams:1992nf}.

\medskip
\noindent
In the region $5\cdot 10^{-3} < x < 0.1$ the structure function 
ratios systematically decrease with decreasing $x$.
At still smaller $x$ one enters the range of small 
momentum transfers,  $Q^2 \simeq 0.5$ GeV$^2$, approaching the limit of  
high-energy photon-nucleus interactions  with real photons. 
As an example we show in Fig.\ref{fig:shad_photon} data on 
shadowing for real photon scattering from $^{63}$Cu. 
\nocite{Heynen71,Brookes73,Caldwell:1973bu,Michalowski:1977eg,%
Arakelian:1978rc,Caldwell:1979ik,Bianchi:1994ax}

\medskip
\noindent
Shadowing systematically increases with the nuclear mass
number $A$. For example, at $x \approx 0.01$ one finds   
$F_2^{\T A}/ F_2^{\T C} \sim A^{\alpha -1}$ with 
$\alpha \approx 0.95$ \cite{Arneodo:1996rv}. 
A similar behavior has been observed  in  
high-energy photonuclear cross sections \cite{Weise:1993}:
their $A$-dependence  is roughly 
$\sigma_{\gamma \T A} \approx A^{0.92} \sigma_{\gamma \T N}$ 
where $\sigma_{\gamma \T N}$ is the free photon-nucleon 
cross section averaged over proton and neutron. 
  
\medskip
\noindent
The shadowing effect depends only weakly on 
the momentum transfer $Q^2$. 
The most precise investigation of this issue has been 
performed for the ratio of Sn and carbon structure functions 
presented in Fig.\ref{fig:sn/c_a} 
\cite{Arneodo:1996ru}.  
It reveals that shadowing decreases at most linearly with 
$\ln Q^2$ for $x< 0.1$. 
The rate of this decrease becomes smaller with rising $x$. 
At $x > 0.1$ no significant $Q^2$-dependence of 
$F_2^{\T{Sn}}/F_2^{\T C}$ is found.

\medskip
\noindent
Shadowing has also been observed in deep-inelastic 
scattering from deuterium, the lightest and most weakly bound 
nucleus. 
In Fig.\ref{fig:deut_shad} we show data from E665 \cite{Adams:1995sh} and 
NMC \cite{Arneodo:1996kd}  
for the ratio $F_2^{\T d}/F_2^{\T p}$ of the deuteron and 
proton structure functions. 
At $x<0.1$ this ratio is systematically smaller than one. 
\begin{figure}[t]
\bigskip
\vspace*{-1cm}
\begin{center} 
\epsfig{file=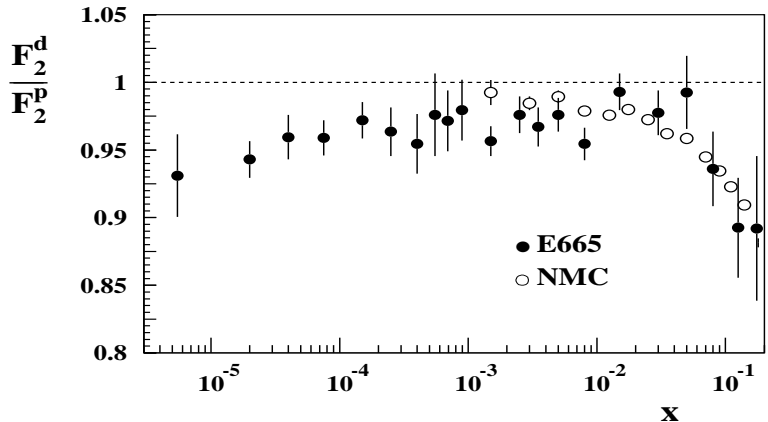,height=90mm,width=110mm}
\end{center}
\vspace*{-1cm}
\caption[...]{
The structure function ratio $F_2^{\T d} / F_2^{\T p}$. 
Data  from E665 \cite{Adams:1995sh} and NMC \cite{Arneodo:1996kd}.
}
\label{fig:deut_shad}
\bigskip
\end{figure}

\medskip
\item [$\bullet$]
{\bf Enhancement region}

\noindent
The NMC data have established a small but statistically significant 
enhancement of the structure function ratio at 
$0.1 <  x < 0.2$.
The observed enhancement is of the order of a few percent. 
For carbon and calcium it amounts to typically $2\%$ \cite{Arneodo:1996ru}. 
The most precise measurement of this enhancement has 
been obtained for $F_2^{\T{Sn}}/F_2^{\T{C}}$ shown in Fig.\ref{fig:sn/c_a}.
Within the accuracy of the data no significant $Q^2$-dependence 
of this effect has been found in this region.

\medskip
\item [$\bullet$]
{\bf Region of ``EMC effect''}

\noindent
The region of intermediate $0.2 < x < 0.8$ has been 
explored extensively at CERN and SLAC.
In the range  
$2 \,\T{GeV}^2  < Q^2 < 15$ GeV$^2$, 
data were taken 
by the E139 collaboration \cite{Gomez:1994ri}
for a large sample of nuclear targets 
between  deuterium and  gold.
The measured structure function ratios 
decrease with rising $x$ and have a minimum at $x \approx 0.6$. 
The magnitude of this depletion grows approximately 
logarithmically with the nuclear mass number. 
The observed effect agrees well with data 
for the ratios of iron and nitrogen to deuterium 
structure functions from BCDMS 
taken at large $Q^2$ values, $14 \,\T{GeV}^2 < Q^2 < 200$ 
GeV$^2$ \cite{Benvenuti:1987az,Bari:1985ga}. 
These data imply that a strong $Q^2$-dependence of 
the structure function ratios is excluded.

\medskip
\item [$\bullet$]
{\bf Fermi motion region}
 
\noindent
At $x>0.8$  the structure function ratios rise  
above unity \cite{Gomez:1994ri}, but experimental information is 
rather scarce. 
The free nucleon structure function 
$F_2^{\T N}$ is known to 
drop as  $(1-x)^3$ when approaching its kinematic 
limit at $x=1$.
Clearly, even minor nuclear effects 
appear artificially enhanced in this kinematic range 
when presented in the form of the ratio 
$F_2^{\T A}/F_2^{\T N}$.

\medskip
\item [$\bullet$]
{\bf The region $x > 1$}  

\noindent
Data at large Bjorken $x$ and large momentum transfer, 
$0.7 < x < 1.3$ and $50 \,\T{GeV}^2  <  Q^2 <  200$ GeV$^2$, 
have been taken for carbon and iron by the BCDMS \cite{Benvenuti:1994bb}
and CCFR \cite{Vakili:1999qt} collaborations, respectively.  
The results disagree with  model calculations  at $x \sim 1$ 
which account for Fermi motion effects only. 
For $Q^2 < 10$ GeV$^2$ data have been taken at SLAC 
for various nuclei 
\cite{Arrington:1996hs,Rock:1992jy,Bosted:1992fy,Filippone:1992iz,Day:1987az}. 
Both  quasielastic scattering from nucleons as well as inelastic 
scattering turns out to be  important here.

\end{itemize}

\subsection{Moments of nuclear structure functions}
\label{ssec:moments_str_fns}

Given data for the ratio  $F_2^{\T A}/F_2^{\T d}$ 
together with the measured deuteron structure function $F_2^{\T d}$, 
the difference $F_2^{\T A} - F_2^{\T d}$ can be evaluated. 
Its integral 
\begin{equation} \label{eq:F_2_momfrac}
M_2^{\T A} - M_2^{\T d}  = 
\int_0^1 d x_{\T A} \, F_2^{\T A} (x_{\T A}) 
- 
\int_0^1 d x_{\T d}  \,F_2^{\T d} (x_{\T d})
\approx  \int_0^2  d x \left(\frac{F_2^{\T A}(x)}{F_2^{\T d}(x)} - 1 
+ f_M \right) F_2^{\T d}(x)
\end{equation} 
represents the difference of the integrated momentum fraction carried 
by quarks in a nucleus relative to that for deuterium. 
The constant $f_M = (A M/M_{\T A} - 2 M/M_{\T d})$ 
corrects for the different mass defects of bound systems. 
Note that in Eq.(\ref{eq:F_2_momfrac}) we have omitted 
QCD target mass corrections \cite{Nachtmann:1974aj}. 
An analysis based on the NMC \cite{Amaudruz:1991cc} and 
SLAC \cite{Gomez:1994ri} data 
has been performed for $\T{He}$, $\T{C}$ and $\T{Ca}$ \cite{Arneodo:1994wf}. 
In the kinematic range covered by these experiments,  
$3.5\cdot 10^{-3} < x < 0.8$, 
the difference of the structure function moments 
$M_2^{\T A} - M_2^{\T d}$ turns out to be compatible with zero.  
Together with the well established result of the momentum sum 
rule for the proton \cite{Roberts:1990ww}, one can therefore conclude that,  
within the accuracy of present  
data,  quarks carry about half of the total momentum, 
in nuclei as well as in free nucleons.

\subsection{Ratios of longitudinal and transverse cross sections}

\begin{figure}[t]
\bigskip
\begin{center} 
\epsfig{file=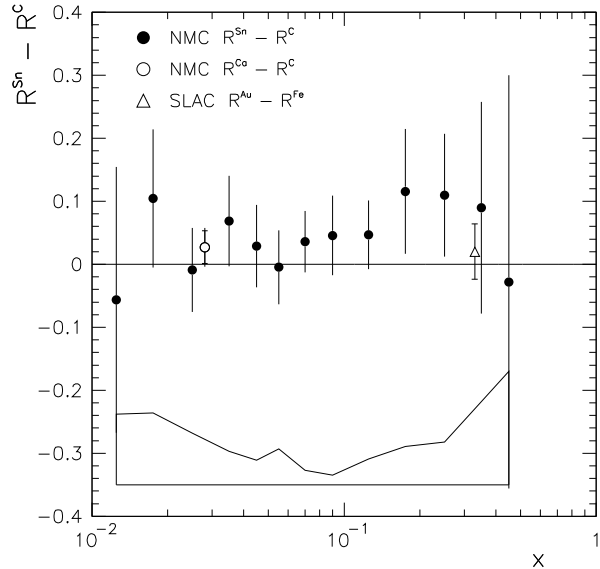,height=80mm}
\end{center}
\caption[...]{
NMC data \cite{Arneodo:1996ru} for 
$R^{\T{Sn}} - R^{\T{C}}$ as a function of $x$ for  
$\overline Q^2 \approx 10$ GeV$^2$. 
The average values for 
$R^{\T{Ca}} - R^{\T{C}}$ \cite{Amaudruz:1992wn}, and 
$R^{\T{Au}} - R^{\T{Fe}}$ \cite{Dasu:1988ru} 
are also shown.
}
\bigskip
\label{fig:R_nuclear}
\end{figure}

Investigations of the 
differences between the longitudinal-to-transverse cross section ratios 
$R = \sigma_L/\sigma_T$ (\ref{eq:R_L_T}) for different nuclei have 
been performed at SLAC for moderate and large values of $x$,
while  the region of small $x$ has been investigated by NMC.
The difference $R^{\T d} - R^{\T p}$ is found to 
be compatible with zero \cite{Whitlow:1990gk,Arneodo:1996kd,Tao:1996uh}. 
Similar observations have been made for heavier targets 
\cite{Arneodo:1996ru,Tao:1996uh,Dasu:1988ru,Amaudruz:1992wn,Dasu:1994vk}. 
In Fig.\ref{fig:R_nuclear} 
we show  NMC data \cite{Arneodo:1996ru} for 
$R^{\T{Sn}} - R^{\T{C}}$ as a function of $x$ for an average  $Q^2$ 
of about $10$ GeV$^2$. 
In addition we present the average values from the NMC measurement for 
$R^{\T{Ca}} - R^{\T{C}}$ \cite{Amaudruz:1992wn}, and for 
$R^{\T{Au}} - R^{\T{Fe}}$ from SLAC E140 \cite{Dasu:1988ru}. 
All  measurements are consistent with only marginal 
nuclear dependence of $R$. 
This implies that nuclear effects influence both structure functions 
$F_1$ and $F_2$ in a similar way, and that 
the ratio of  nuclear cross sections 
directly measures  the ratio of the corresponding structure 
functions $F_2$. 

\subsection{Other measurements of nuclear parton distributions}

Nuclear deep-inelastic scattering is sensitive only to the sum of 
valence and sea quark distributions (see e.g. Eq.(\ref{eq:parton_QCD})), 
weighted by their respective electric charges.
In order to separate nuclear effects in the valence and sea quark 
sectors, and directly measure nuclear gluon distributions,  
other types of processes are required which we briefly summarize 
in the following. 

\subsubsection{Drell-Yan lepton pair production}
\label{sssec:DY}

In the Drell-Yan production of lepton pairs  
(mostly $\mu^+\mu^-$) in hadron-nucleus collisions,  
the underlying partonic sub-process is  the annihilation of a quark 
and antiquark  from beam and target into a time-like 
high energy photon, which 
subsequently converts into the observed dilepton. The Drell-Yan 
cross section reads (see e.g. \cite{Field:1989uq}): 
\begin{equation} \label{eq:DY} 
\frac{d^2\sigma}{dx_T dx_B} 
= \frac{4 \pi \alpha^2}{9 \,m_{l}^2} 
\, K\, \sum_f e_f^2 \left[ q_f^B(x_B,Q^2)\,\bar q_f^T(x_T,Q^2) + 
\bar q_f^B(x_B,Q^2)\, q_f^T(x_T,Q^2) \right],
\end{equation}
where $m_l$ is the invariant mass of the produced lepton pair.
The flavor dependent quark distributions of the projectile and target
are denoted by $q_f^B$ and $q_f^T$, respectively. 
Seen from the center-of-mass frame the active quarks carry   
fractions  $x_B$ and $x_T$ of the beam and target momenta.
They are determined by the momentum component $q_L$ 
of the produced dilepton parallel 
to the beam, its  invariant mass $m_{l}$ 
and  the squared center-of-mass 
\linebreak
energy $s$: 
\begin{equation}
x_T\,x_B = \frac{m_{l}^2}{s}, \quad 
x_F = \frac{2 q_L}{\sqrt{s}} = x_B - x_T.  
\end{equation}
Higher order QCD corrections to the production cross section 
(\ref{eq:DY}) turn out to be significant. 
They are absorbed in the so-called ``$K$-factor'' and 
effectively double the leading order cross section.

\begin{figure}[t]
\bigskip
\vspace*{-1cm}
\begin{center} 
\epsfig{file=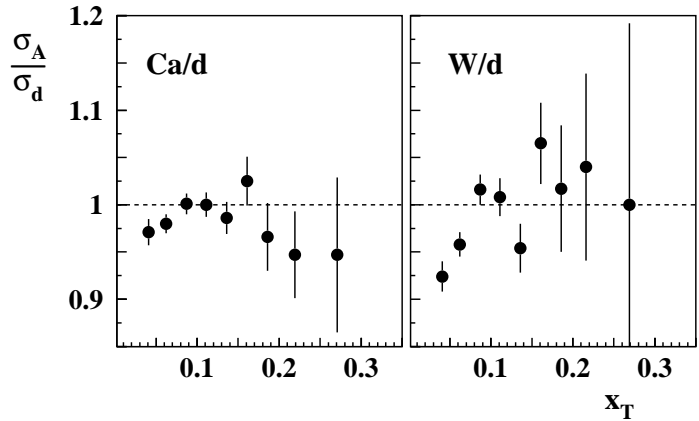,height=110mm}
\end{center}
\vspace*{-2cm}
\caption[...]{
Drell-Yan dimuon yields per nucleon for  $^{40}$Ca and  $^{184}$W 
as a function of $x_T$ for $x_F > 0$
\cite{Alde:1990im}.
}
\label{fig:DY}
\bigskip
\end{figure}

The E772 experiment at FNAL \cite{Alde:1990im} has investigated Drell-Yan 
dilepton production in proton-nucleus collisions at 
$s = 1600$ GeV$^2$.
At $x_F >  0.2$ the production process 
is dominated by the annihilation of projectile quarks with 
target antiquarks.  
Outside  the domain  of quarkonium resonances, 
i.e. for $4 \,{\T{GeV}} <  m_l <  9$ GeV and $m_l >  11$ GeV, 
this experiment explores possible modifications  of nuclear  
sea quark distributions. 
In Fig.\ref{fig:DY} we show ratios of  dimuon yields 
for nuclear targets and deuterium taken at $x_F > 0$. 
At $x_T>0.1$ 
no significant nuclear effects have been  observed within 
admittedly large experimental errors. 
This indicates the absence of strong modifications 
of nuclear sea quark distributions, as compared to those of free 
nucleons. 
At $x_T<0.1$, on the other hand, the observed attenuation for heavy 
nuclei implies a substantial  reduction 
of nuclear sea quarks, in qualitative agreement with 
the shadowing effects observed in nuclear deep-inelastic 
scattering at $x <0.1$.  
The detailed comparison of shadowing in Drell-Yan versus 
DIS requires, of course, a careful separation of valence 
and sea quark effects as well as their $Q^2$ evolution 
\cite{Frankfurt:1990xz}.

\subsubsection{Lepton-induced production  of heavy quarks}

The intrinsic heavy-quark ($c$- or $b$-quark) distributions in nucleons 
or nuclei 
are expected to be very small. 
Inelastic heavy-quark production is therefore assumed to 
receive its  major contributions from photon-gluon fusion, 
i.e. the coupling of the exchanged virtual photon 
to a heavy quark pair which is attached to a gluon out of the target. 
This mechanism is a basic ingredient of the so-called color-singlet model 
\cite{Berger:1981ni}.
In this model the cross section for heavy quark pair production 
is proportional to the gluon distribution of the target. 
A comparison of these  cross sections 
for nucleons and nuclei can then be directly translated 
into a difference of the corresponding gluon distributions. 

In this context  NMC has analyzed  $J/\psi$ production 
data from Sn and carbon nuclei  \cite{Amaudruz:1992sr}. 
The average ratio of the corresponding inelastic $J/\psi$ production cross 
sections was found slightly larger than one: 
\begin{equation}
\frac{\sigma(\gamma^* + {\T{Sn}} \rightarrow J/\psi + {\T X})}
 {\sigma(\gamma^* + {\T{C}} \rightarrow J/\psi + {\T X})}
= 1.13 \pm 0.08.
\end{equation}
Within the color singlet model this implies an  
enhancement by about $10\%$ of the gluon distribution in Sn 
as compared to carbon in the region $x\sim 0.1$, 
though with large errors.

\subsubsection{Neutrino scattering from nuclei}

Deep-inelastic neutrino 
scattering permits one to separate valence and sea quark distributions.
It is therefore a promising tool to investigate 
modifications of the different components of 
quark distributions in nuclei. 
The observed nuclear effects in neutrino 
experiments are qualitatively similar to 
the results from charged lepton scattering discussed previously
\cite{Guy:1987us,Allport:1989vf,Kitagaki:1988wc,Guy:1989iz}, 
although their statistical significance is poor, given the   
large experimental uncertainties.

\section{Space-time description of deep-inelastic scattering} 
\label{Sec:space_time}
\setcounter{section}{4}
\setcounter{figure}{0}

So far our picture of deep-inelastic scattering has been developed 
in momentum space. The partonic interpretation of structure 
functions is particularly transparent in the infinite momentum 
frame in which the nucleon (or nucleus) moves with (longitudinal) 
momentum $P \rightarrow \infty$. In this frame the Bjorken variable $x$ 
has a simple meaning as the fraction of the nucleon momentum carried 
by a parton when it is struck by the virtual photon.\footnote{
A simple interpretation is also possible in the laboratory 
frame using light-front dynamics. In this description, the scattering 
cross section is determined by the square of the target ground 
state wave function 
(for a review and references see e.g. \cite{Brodsky:1997de}). }

For an investigation of nuclear effects in DIS the infinite momentum 
frame is not always optimal. Instead, it is often preferable 
to describe the scattering process in the laboratory frame where 
the target is at rest. 
Only in that frame the detailed knowledge about nuclear structure in 
terms of many-body wave functions, meson exchange currents etc. can 
be used efficiently. Also, the physical effects implied by characteristic 
nuclear scales (the nuclear radius $R_{\T A} \sim A^{1/3}$ and the 
average nucleon-nucleon distance $d \simeq 2$ fm) 
are best discussed in the lab frame.

In this section we elaborate on several aspects relevant to 
deep-inelastic scattering as viewed in coordinate space. 
We first discuss the coordinate space resolution of the DIS probe. 
Then we introduce coordinate space distribution functions 
(so-called Ioffe-time distributions) of quarks and gluons and 
summarize results for free protons. 
A detailed discussion of nuclear effects in coordinate space 
distributions follows next. 
In the final part we comment on the relationship between 
lab frame and infinite momentum frame pictures.

\subsection{Deep-inelastic scattering in coordinate space}

We follow here essentially the discussion in  Ref.\cite{Vanttinen:1998iz}
(see also \cite{Frankfurt:1988nt,Ioffe:1969kf,LlewellynSmith:1985pv,%
Hoyer:1996nr} and references therein). 
Consider the scattering from a
free nucleon with momentum $P^{\mu} = (M,\vec 0)$ 
and invariant mass $M$ in the
laboratory frame. The four-momentum transfer $q^{\mu} = (\nu,\vec q)$, 
carried by the exchanged virtual photon, is taken 
to be in the (longitudinal) 
$z$-direction, $\vec q = (\vec 0_{\perp}, q_3)$ with 
$q_3 = \sqrt{\nu^2 + Q^2}$ and $Q^2 = -q^2$. 
In the Bjorken limit, $\nu^2 \gg Q^2 \gg M^2$ with $x=Q^2/(2 M \nu)$
fixed, the light-cone components of the photon momentum 
($q^{\pm}=\nu \pm q_3$) are $q^+ \simeq 2 \nu$ and $q^- \simeq - Mx$.
All information about the response of the target to the high-energy
virtual photon is in the hadronic tensor
\begin{equation} \label{eq:WJJ_ST}
  W_{\mu\nu} (q,P) \sim 
  \int d^4 y \,e^{i q\cdot y} \, 
  \langle P| J_{\mu}(y) J_{\nu}(0)|P\rangle,
\end{equation}
(see Eq.(\ref{eq:WJJ})). Using 
\begin{equation}
  q\cdot y = \frac{1}{2}\left(q^+ y^- + q^- y^+\right) - 
             \vec q_{\perp} \cdot \vec y_{\perp}
            \simeq \nu \,y^- - \frac{Mx}{2} \,y^+ 
            - \vec q_{\perp} \cdot \vec y_{\perp},
\end{equation}
one obtains the following coordinate-space resolutions along the 
light-cone distances $y^{\pm} = t \pm y_3$:
\begin{equation} \label{eq:typical_y}
  \delta y^- \sim \frac{1}{\nu}
  \quad\mbox{and} \quad 
  \delta y^+ \sim  \frac{1}{Mx}.
\end{equation}
At $y^-=0$ the current correlation function in Eq.(\ref{eq:WJJ_ST}) 
is not analytic since it vanishes for
$y^+ y^- - {(\vec y_{\perp})}^2 < 0$ because of causality
(see e.g.\ \cite{Muta:1987mz}). Indeed in perturbation theory it
turns out to be singular at $y^-=0$. Assuming that the integrand
in (\ref{eq:WJJ_ST}) is an analytic function of $y^-$ elsewhere, 
this implies that $W_{\mu\nu}$ is dominated for
$q^+ \rightarrow \infty$  by contributions from $y^-= 0$. 
Causality implies that, in the transverse plane, only 
contributions from ${(\vec y_{\perp})}^2 \simeq 1/Q^2$ are relevant: 
deep-inelastic scattering  is dominated 
by contributions from the light cone, i.e.\ $y^2 = 0$.

Furthermore, Eq.(\ref{eq:typical_y}) suggests that 
one probes increasing  distances along the light cone 
as $x$ is decreased.
Such a behavior is consistent with approximate 
Bjorken scaling \cite{Ioffe:1969kf}. 
The coordinate space analysis of nucleon structure
functions in Section \ref{section:CS-free-nucleon}
confirms this  conjecture. In the
Bjorken limit the dominant contributions to the hadronic tensor at
small $x$ come from light-like separations of order $y^+ \sim 1/(Mx)$
between the electromagnetic currents in (\ref{eq:WJJ_ST}).

In the laboratory frame these considerations imply that deep-inelastic
scattering involves a longitudinal correlation length  
\begin{equation} \label{eq:l_z}
  y_3 \simeq \frac{y^+}{2} \equiv l 
\end{equation}
of the virtual photon. Consequently, large longitudinal distances 
are important in the scattering process at small $x$. This can also
be deduced in the framework of time-ordered perturbation theory
(see Section \ref{ssec:DIS_SPTH}), where $l$ determines  
the typical propagation length of hadronic configurations present 
in the interacting photon.

\begin{figure}[t]
\bigskip
\centerline{
\epsfig{figure=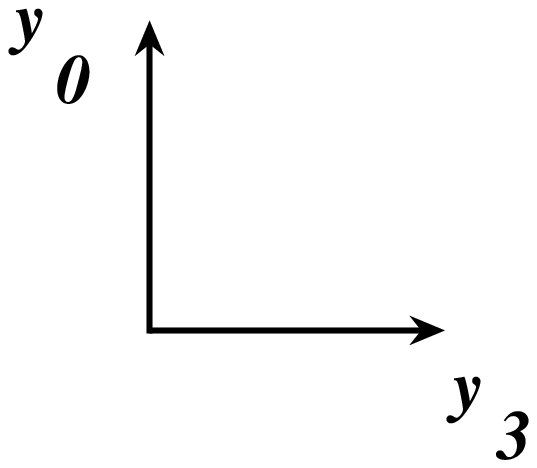,width=3cm}
\epsfig{figure=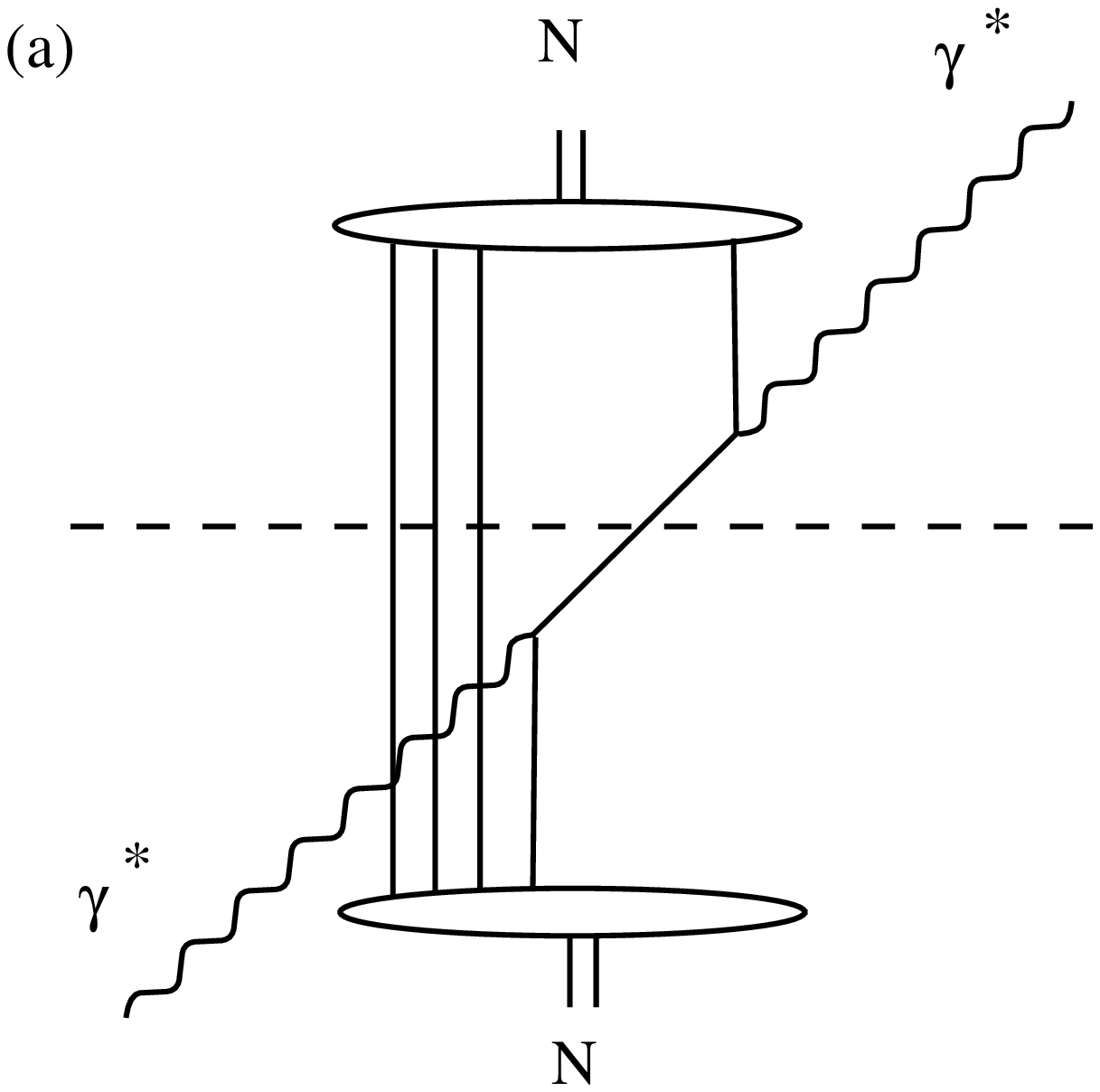,width=6cm}
\epsfig{figure=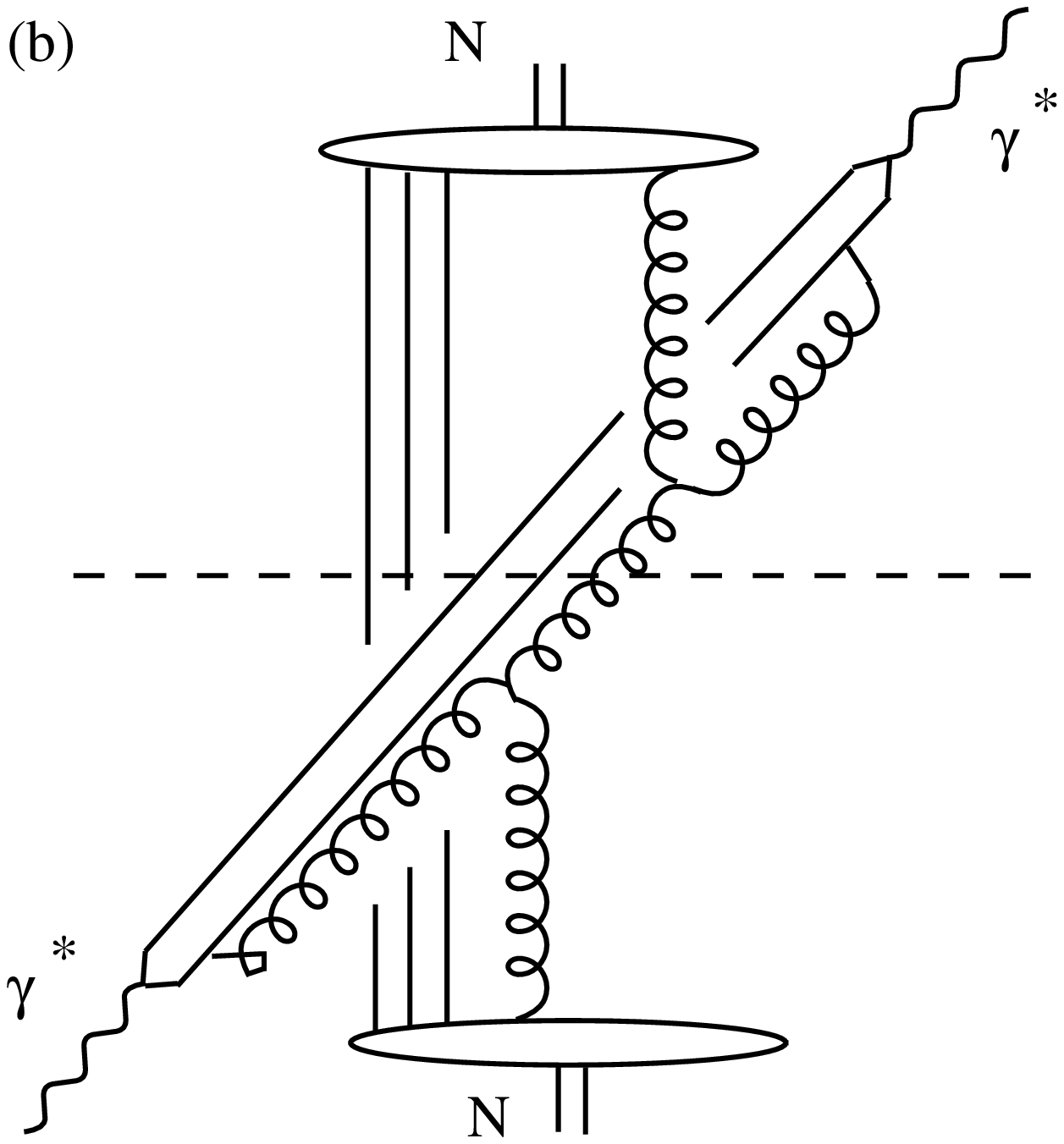,width=6cm}
}
\caption{
Two examples of diagrams 
illustrating the space-time pattern of deep inelastic scattering.} 
\label{fig:diagrams}
\bigskip
\end{figure}

The space-time pattern of deep-inelastic scattering is illustrated
in Fig.\ref{fig:diagrams} in terms of the imaginary part of the forward
Compton amplitude: the virtual photon interacts with
partons which propagate a distance $y^+$ along the light cone. 
The characteristic laboratory frame correlation length $l$ is one 
half of that distance.

\subsection{Coordinate-space distribution functions}
\label{ssec:CSDF}

Especially when it comes to the discussion 
of the relevant space-time scales which govern nuclear effects 
in deep-inelastic scattering, it is instructive to look at quark and gluon 
distribution functions in coordinate rather than in momentum space.
In this section we prepare the facts and return to the underlying 
dynamics at a later stage.

It is useful to express coordinate-space distributions in terms of
a suitable dimensionless variable. 
For this purpose let us introduce the
light-like vector $n^{\mu}$ with $n^2 = 0$ and $P\cdot n = P_0 - P_3$.
The hadronic tensor receives its dominant contributions  
from the vicinity of the light cone, where
$y$ is approximately parallel to $n$. The dimensionless variable
$z = y \cdot P$ then plays the role of a coordinate conjugate to
Bjorken $x$. It is helpful to bear in mind that the value $z = 5$
corresponds to a light-cone distance 
$y^+ =  2 z/M \approx 2$ fm in the laboratory  frame or, 
equivalently, to a longitudinal
distance $l \equiv y^+/2 \approx 1$ fm.  

In accordance with the charge conjugation ($C$) properties of
momentum-space quark and gluon distributions, one defines
coordinate-space distributions by \cite{Braun:1995jq}:
\begin{eqnarray}
  \label{eq:Coordinate_1}
  {\cal Q}(z,Q^2) &\equiv& \int_0^1 {d}x
  \, \left [q(x,Q^2) + \bar q(x,Q^2) \right]\,\sin (z \,x),   
  \\
  \label{eq:Coordinate_2}
  {\cal Q}_{v}(z,Q^2) &\equiv& \int_0^1 {d}x\, 
  \left[q(x,Q^2) - \bar q(x,Q^2) \right]\, \cos (z \,x),
  \\ 
  \label{eq:Coordinate_3}
  {\cal G}(z,Q^2) &\equiv& \int_0^1 {d}x\, 
  x\,g(x,Q^2)\, \cos (z \,x), 
\end{eqnarray}
where $q$, $\bar q$ and $g$ are the momentum-space quark, antiquark 
and gluon distributions, respectively. Flavor indices 
are suppressed here for simplicity.

At leading twist accuracy, the coordinate-space distributions 
(\ref{eq:Coordinate_1}--\ref{eq:Coordinate_3}) 
are related to forward matrix elements of non-local QCD operators
on the light cone \cite{Collins:1982uw,Balitskii:1988/89}:
\begin{eqnarray} 
  \label{eq:Coord_Op_Q}
  {\cal Q}(z,Q^2) 
  &=& \frac{1}{4 i P\cdot n} \,
  \langle P| \overline \psi(y) \,
  \Slash{n} 
  \Gamma(y) \,\psi(0)|P\rangle_{Q^2}
  - (y \leftrightarrow -y),
  \\
  \label{eq:Coord_Op_Qv}
  {\cal Q}_{v}(z,Q^2) 
  &=& \frac{1}{4 P\cdot n} \,
  \langle P| \overline {\psi}(y) \,
  \Slash{n} 
  \Gamma(y) \,\psi(0)
  |P\rangle_{Q^2}
  + (y \leftrightarrow -y), 
  \\
  \label{eq:Coord_Op_G}
  {\cal G}(z,Q^2) 
  &=& 
  n^{\mu} n^{\nu} 
  \frac{1}{2 (P\cdot n)^2}
  \langle P| G_{\mu\lambda}(y) \,\Gamma(y)\,
  G^{\lambda}_{\,\,\nu}(0)|P\rangle_{Q^2}.
\end{eqnarray}
Here $\psi$ denotes the quark field and $G_{\mu\nu}$ the gluon
field strength tensor. The path-ordered exponential
\begin{equation}
  \Gamma(y) = {\T P}\exp
  \left[ ig\,y^{\mu} \int_0^1 {d} \lambda  \,A_{\mu}(\lambda y)  
  \right] \, ,
\end{equation}
where $g$ denotes the strong coupling constant and $A^{\mu}$ the
gluon field, ensures gauge invariance of the parton distributions.
Note that an expansion of the right-hand side of 
Eqs.(\ref{eq:Coordinate_1}--\ref{eq:Coordinate_3}) and
(\ref{eq:Coord_Op_Q}--\ref{eq:Coord_Op_G}) around $y = 0$ 
(and hence $z=y\cdot P=0$) leads
to the conventional operator product expansion for parton
distributions \cite{Roberts:1990ww,Muta:1987mz,Cheng:1984}.

The functions ${\cal Q}(z)$, ${\cal Q}_v(z)$ and ${\cal G}(z)$ 
describe the mobility of partons in coordinate space. Consider, for
example, the valence quark distribution ${\cal Q}_{v}(z)$.
The matrix element in (\ref{eq:Coord_Op_Qv}) has
an obvious physical interpretation: as illustrated in
Fig.\ref{fig:diagrams}a, it measures the overlap between
the nucleon ground state and a state in which one quark has been
displaced along the light cone from $0$ to $y$.
A different sequence is shown in Fig.\ref{fig:diagrams}b. There the
photon converts into a beam of partons which propagates along
the light cone and interacts with partons of the target nucleon,
probing primarily its sea quark and gluon content.

\subsection{Coordinate-space distributions of free nucleons 
\label{section:CS-free-nucleon}}

\begin{figure}[t]
\bigskip
\hspace*{1cm}
\input{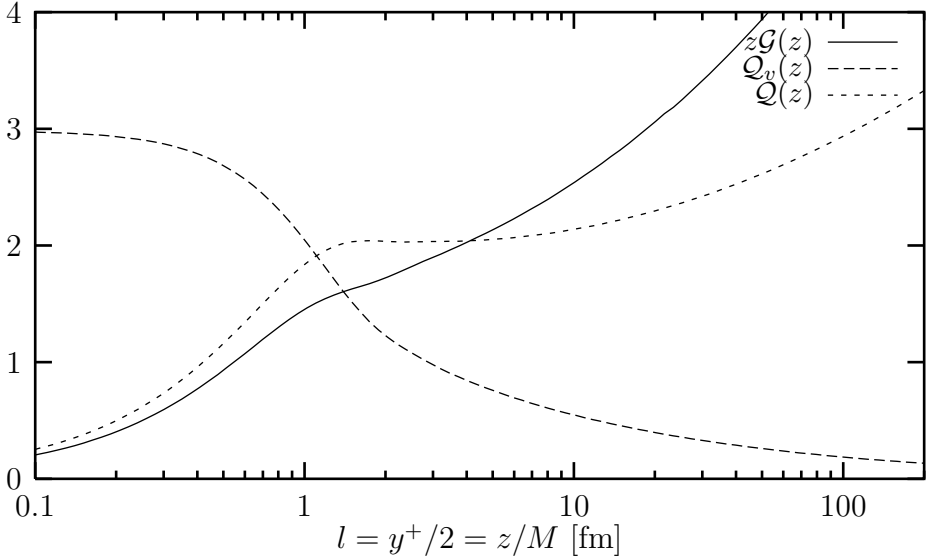}
\caption{Coordinate-space quark and gluon distributions resulting
from the CTEQ4L parametrization of momentum-space distributions,
taken at a momentum transfer $Q^2 = 4$ GeV$^2$. A sum over the $u$
and $d$ quarks is implied in the 
functions ${\cal Q}_v$ and ${\cal Q}$ \cite{Vanttinen:1998iz}.}
\label{fig:freenucleon}
\bigskip
\end{figure}

In this section we discuss the properties of coordinate-space
distribution functions of free nucleons. Examples of the distributions 
(\ref{eq:Coord_Op_Q}--\ref{eq:Coord_Op_G}) using
the CTEQ4L parametrization \cite{Lai:1997mg} of momentum-space quark
and gluon distributions taken at a momentum scale $Q^2 = 4$ GeV$^2$,
are shown in Fig.~\ref{fig:freenucleon}.

Some general features can be observed: the $C$-even quark distribution
${\cal Q}(z)$ rises at small
values of $z$, develops a plateau at $z \gsim 5$, and then exhibits a
slow rise at very large $z$.
At $z \lsim 5$, the gluon distribution 
function $z \,{\cal G}(z)$ behaves similarly as
${\cal Q}(z)$. For $z \gsim 5$, $z \,{\cal G}(z)$ rises somewhat
faster than ${\cal Q}(z)$. The $C$-odd (or valence) quark distribution
${\cal Q}_v(z)$ starts with  a finite value at small $z$, then begins to
fall at $z \simeq 3$  and vanishes at large $z$. Recall that in the
laboratory frame, the scale $z \simeq 5$ at which a significant change
in the behavior of coordinate-space distributions occurs, represents 
a longitudinal distance comparable to the typical size of a nucleon. 
\begin{figure}[b]
\bigskip
\hspace*{1cm}
\input{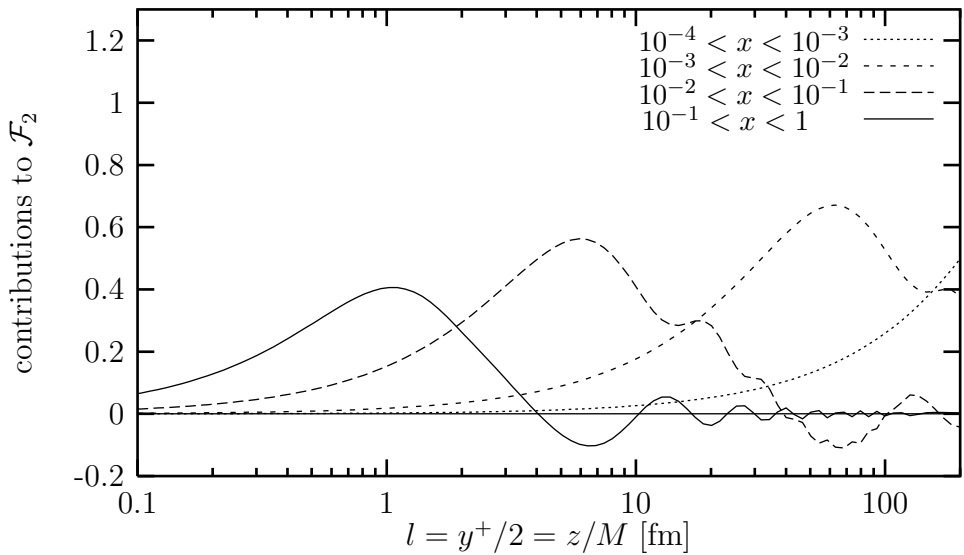}
\caption{Contributions from different regions in $x$ to the ${\cal F}_2$
combination of coordinate-space quark and antiquark distributions 
at $Q^2 = 4$ GeV$^2$ \cite{Vanttinen:1998iz}. 
}
\bigskip
\label{fig:xregions}
\end{figure}

At $z < 5$ the coordinate-space distributions are determined by average
properties of the corresponding momentum-space distribution functions
as expressed by their first few moments 
\cite{Mankiewicz:1996ep,Weigl:1996ii}. For example,
the derivative of the $C$-even quark distribution ${\cal Q}(z)$ taken
at $z = 0$ equals the fraction of the nucleon light-cone momentum
carried by quarks. The same is true for the gluon distribution
$z \,{\cal G}(z)$ (the momentum fractions carried by quarks and by
gluons are in fact approximately equal, a well-known experimental fact).  
At $z > 10$ the coordinate-space distributions are determined by the 
small-$x$ behavior of the corresponding momentum space distributions. 
Assuming, for example, $q(x) \sim x^{\beta}$  for $x < 0.05$ implies
${\cal Q}(z) \sim z^{-\beta - 1}$ at $z > 10$. Similarly, the 
small-$x$ behavior $g(x) \sim x^{\beta}$ leads to  
$z \,{\cal G}(z) \sim z^{ - \beta - 1}$ at large $z$. For typical
values  of $\beta$ as suggested by Regge phenomenology 
\cite{Collins:1977jy}
one obtains ${\cal Q}_{v} \sim z^{-0.5}$ while ${\cal Q}(z)$ and
$z \,{\cal G}(z)$ become constant at very large $z$. 

The fact that ${\cal Q}(z)$ and $z \,{\cal G}(z)$ extend over
large distances has a natural interpretation in the laboratory
frame. At correlation lengths $l$ much larger than the nucleon
size, both ${\cal Q}(z)$ and $z \,{\cal G}(z)$ reflect primarily
the partonic structure of the photon which 
behaves like a high-energy beam of gluons and quark-antiquark 
pairs incident on the nucleon.
For similar reasons,
the valence quark distribution ${\cal Q}_v(z)$ defined in
Eq.(\ref{eq:Coord_Op_Qv}) has a pronounced tail which extends
to distances beyond the nucleon radius. 
An antiquark in the ``beam'' can annihilate with a valence 
quark of the target nucleon, giving rise to  long distance 
contributions in  ${\cal Q}_v$.
A detailed and
instructive discussion of this frequently ignored feature
can be found in Ref.\cite{Brodsky:1991gn}.

Finally we illustrate the relevance of large distances in deep-inelastic
scattering at small $x$. In
Fig.~\ref{fig:xregions} we show contributions to the 
nucleon structure function $F_2$ in coordinate space,  
\begin{equation}
{\cal F}_2 (z,Q^2) 
= \int_0^1 \frac{d x}{x}  \,F_{2}^{\T N}(x,Q^2) \sin(z \,x),
\end{equation}
which result from different windows of Bjorken $x$. 
This confirms once more that contributions from large 
distances $\sim 1/(Mx)$ dominate at small $x$.

\subsection{Coordinate-space distributions of nuclei}
\label{ssec:Co_spa_nuclei}

The implications for scattering from nuclear targets, 
especially for coherence phenomena, are now obvious. 
If one compares, in the laboratory frame, 
the longitudinal correlation length $l$ from
Eq.(\ref{eq:l_z}) with the average nucleon-nucleon distance
in the nucleus, $d \simeq 2$ fm, one can clearly distinguish two
separate regions:
\begin{enumerate}
\item[(i)] 
At small distances, $l < d$, the virtual photon scatters incoherently
from the individual hadronic constituents of the target nucleus. 
Possible  modifications of the coordinate distribution functions 
(\ref{eq:Coordinate_1} -- \ref{eq:Coordinate_3})
in this region are 
caused by bulk nuclear effects such as binding and Fermi motion.

\medskip

\item[(ii)] 
At larger distances, $l > d$, it is likely that several nucleons 
participate collectively in the interaction. Modifications of
the coordinate distribution functions 
are now expected to 
come from the coherent scattering on 
at least two  nucleons in the target. 
Using $l \sim 1/(2 M x)$, this region corresponds to 
$x \lsim 0.05$. 
\end{enumerate}
This suggests that the nuclear modifications seen in coordinate-space
distributions will be  quite different in the regions $l > 2$ fm and
$l < 2$ fm. 
This is best demonstrated by studying  
the ratios of  nuclear and 
nucleon coordinate space distribution functions: 
\begin{eqnarray}
  {\cal R}_{F_2}(z,Q^2) &=& 
  \frac{
  \int_0^A \frac{d x}{x} \,F_{2}^{\T A}(x,Q^2) \sin(z \,x)
  }
  {\int_0^1 \frac{d x}{x}  \,F_{2}^{\T N}(x,Q^2) \sin(z \,x)
  }
  = 
  \frac{\sum _f e_f^2 \,{\cal Q}^{\T A}_f(z,Q^2)}
  {\sum _f e_f^2 \,{\cal Q}^{\T N}_f(z,Q^2)}, 
  \\
  {\cal R}_{v}(z,Q^2) &=& 
  \frac{{\cal Q}^{\T A}_{v}(z,Q^2)}
  {{\cal Q}^{\T N}_{v}(z,Q^2)},
  \\
  {\cal R}_{\cal G}(z,Q^2) &=& 
  \frac{z {\cal G}^{\T A}(z,Q^2)}{z {\cal G}^{\T N}(z,Q^2)}.  
\end{eqnarray}
The ratios ${\cal R}_{F_2}$ have been obtained  
for different nuclei from an analysis of the measured momentum space 
structure functions \cite{Hoyer:1996nr}. 
Furthermore, the ratios of valence quark and gluon distributions 
have been calculated in \cite{Vanttinen:1998iz}
as sine and cosine 
Fourier transforms
(\ref{eq:Coordinate_1} -- \ref{eq:Coordinate_3}) 
of momentum space distribution functions which result from an 
analysis of nuclear DIS and Drell-Yan data   \cite{Eskola:1998iy}
(see also Section \ref{ssec:nuclear_parton_distr}). 
\begin{figure}[t]
\bigskip
\hspace*{1cm}
\input{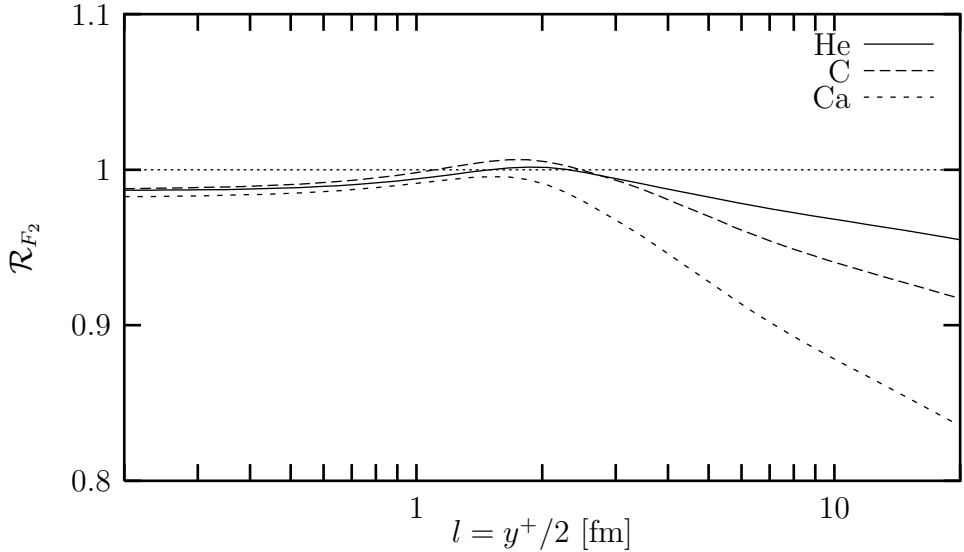}
\caption{
The coordinate space ratio ${\cal R}_{F_2}$ at $Q^2 = 5$ GeV$^2$  
for $^4$He, $^{12}$C, and $^{40}$Ca from Ref.\cite{Hoyer:1996nr}.
} 
\label{fig:CS_F2}
\bigskip
\end{figure}

In Fig.\ref{fig:CS_F2} we show the ratio 
${\cal R}_{F_2}$ for $Q^2 = 5$ GeV$^2$ 
taken from \cite{Hoyer:1996nr}. The most prominent feature is the pronounced 
depletion of ${\cal R}_{F_2}$  at $l > 2$ fm 
caused by nuclear shadowing. 
At $l \lsim 1$ fm, nuclear modifications of ${\cal R}_{F_2}$ 
are small, and  deep-inelastic scattering proceeds incoherently from the
hadronic constituents of the target nucleus. The intrinsic
structure of individual nucleons is evidently not much affected
by nuclear mean fields.
In momentum space, on the other hand, the pronounced nuclear dependence 
of the structure function $F_{2}^{\T A}$ at  $x>0.1$  evidently  
results from a superposition of long and short distance 
contributions as seen in Fig.\ref{fig:xregions}.  
(For a detailed discussion see Ref.\cite{Hoyer:1996nr}.)

\begin{figure}[t]
\bigskip
\hspace*{1cm}
\input{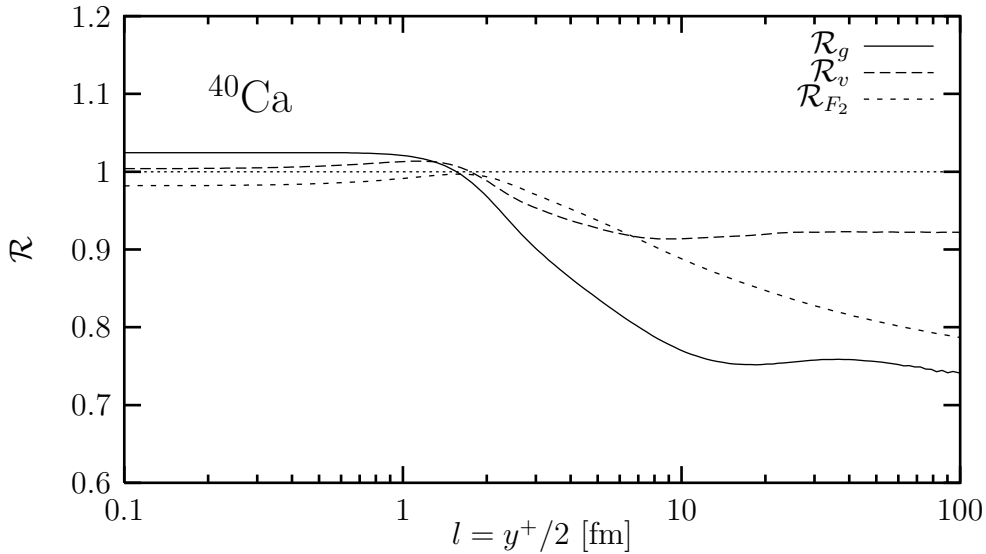}
\caption{
Coordinate-space ratios at $Q^2 = 4$ GeV$^2$ for gluon
distributions, valence-quark distributions, and the $F_2$ structure
function in $^{40}$Ca \cite{Vanttinen:1998iz}.
}
\label{fig:CS_CA}
\bigskip
\end{figure}

In Fig.\ref{fig:CS_CA}
we show the valence quark and gluon ratios 
${\cal R}_v$ and ${\cal R}_{\cal G}$ for $^{40}$Ca 
from Ref.\cite{Vanttinen:1998iz}.  
They behave similarly as the structure function ratio 
${\cal R}_{F_2}$, where the depletion of gluons at large 
distances is most pronounced. 
It is interesting to observe that in coordinate space,
shadowing sets in at approximately the same value of $l$ for all
sorts of partons. In momentum space, shadowing is found to start at 
different values of $x$ for different distributions \cite{Eskola:1998iy}.
Finally note that the shadowing effect continues to increase 
for distances larger than the nuclear diameter.

The results shown in Fig.\ref{fig:CS_CA} clearly emphasize 
the important role of gluons in the shadowing process. 
Of course the incident virtual photon does not directly 
``see'' the gluons. In the primary step the photon 
converts into a quark-antiquark pair. 
At small Bjorken-$x$, the subsequent QCD evolution of this pair 
rapidly induces a cascade of gluons. 
This cascade propagates along the light cone over distances 
which can exceed typical nuclear diameters by far: 
the high energy, high $Q^2$ photon behaves in part like a gluon 
beam which scatters coherently from the nucleus. 
This offers interesting new physics. The detailed QCD 
analysis of nuclear shadowing can in fact give information 
on the ``cross section'' $\sigma_{g \T N}$ for gluons 
incident on nucleons, and a simple eikonal estimate 
using ${\cal R}_g$ at asymptotic distances $l$ suggests 
that this $\sigma_{g \T N}$ is indeed large, comparable 
to typical hadronic cross sections 
(see also Refs.\cite{Frankfurt:1998ym,Alvero:1998bz}).

In summary,  a coordinate space representation  
which selects contributions from different longitudinal distances, 
lucidly demonstrates that 
nuclear effects of the structure function $F_2$ 
and parton distributions are by far 
dominated by shadowing and have a surprisingly 
simple geometric interpretation.

\subsection{Deep-inelastic scattering in standard perturbation 
            theory}

\label{ssec:DIS_SPTH}

\begin{figure}[t]
\bigskip
\begin{center} 
\epsfig{file=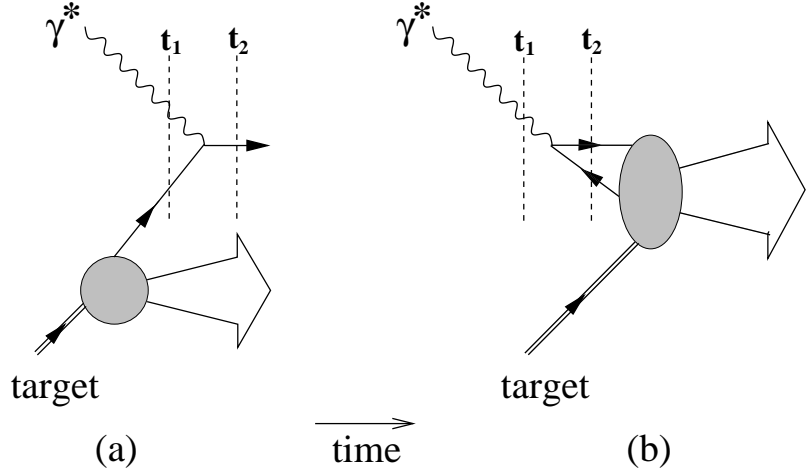,height=60mm}
\end{center}
\caption[...]{
The two possible time orderings for the interaction of a 
(virtual) photon with a nucleon or nuclear target: 
(a) the photon hits a quark in the target, (b) 
the photon creates a $q \bar q$ pair that subsequently 
interacts with the target.
}
\label{fig:tpth} 
\bigskip
\end{figure}

It is instructive to illustrate the previous results by 
looking at the lab frame space-time pattern of the  (virtual) 
photon-nucleon  interaction from the point of view of standard 
time-ordered perturbation theory.
The two basic  time orderings are shown in Figs.\ref{fig:tpth}a 
and \ref{fig:tpth}b: 

\begin{enumerate}

\item[(a)] 
the photon hits a quark or antiquark in the
target which picks up the large energy and
momentum transfer;

\medskip

\item[(b)]
the photon converts into a quark-antiquark 
pair which propagates and  subsequently interacts with the target.

\end{enumerate}

For small Bjorken-$x$ the pair production process
(b) dominates the scattering amplitude, as already mentioned.  
This can also be easily  seen in time-ordered perturbation
theory as follows (see e.g. \cite{Bauer:1978iq} and references therein): 
the amplitudes
${\cal{A}}_a$ and ${\cal A}_b$ of processes
(a) and (b) are roughly proportional to
the inverse of their corresponding energy
denominators $\Delta E_a$ and $\Delta
E_b$.  For large energy transfers $\nu \gg
M$ one finds:
\begin{eqnarray}
  \Delta E_a &=& E_a(t_2)-E_a(t_1) \approx
 -\left<p_q^2\right>^{1/2}
 +\frac{\left<p_q^2\right>+Q^2}{2\nu},
\label{eq:deltaEa}\\ 
\Delta E_b &=& E_b(t_2)-E_b(t_1)
\approx\frac{\mu^2+Q^2}{2\nu}\;,
\label{eq:deltaEb}
\end{eqnarray}
where $\left<p_q^2\right>^{1/2}$ is the
average quark momentum in a nucleon and
$\mu$ is the invariant mass of the
quark-antiquark pair.  We then obtain for
the ratio of these amplitudes:
\begin{equation} \label{amplitudes}
\left|\frac{{\cal A}_a}{{\cal A}_b}\right|
\sim
\left|\frac{\Delta E_b}{\Delta E_a}\right|
\approx
 \frac{M x}{\left<p_q^2\right>^{1/2}}
\left(1 + \frac{\mu^2}{Q^2}\right).
\end{equation}
When analyzing the spectral representation of the scattering 
amplitude one observes that the bulk 
contribution to process (b) results  from those 
hadronic components  in the photon wave function which have  
a squared mass $\mu^2 \sim Q^2$ (see Section \ref{subs:AJM}).  
The ratio in Eq.(\ref{amplitudes}) is evidently small
for $x\ll 0.1$.  Hence pair production,
Fig.\ref{fig:tpth}b, 
is the leading lab frame process in the small-$x$ region.
On the other hand, at $x>0.1$, both mechanisms 
(a) and (b) contribute. 
\begin{figure}[t]
\bigskip
\begin{center} 
\epsfig{file=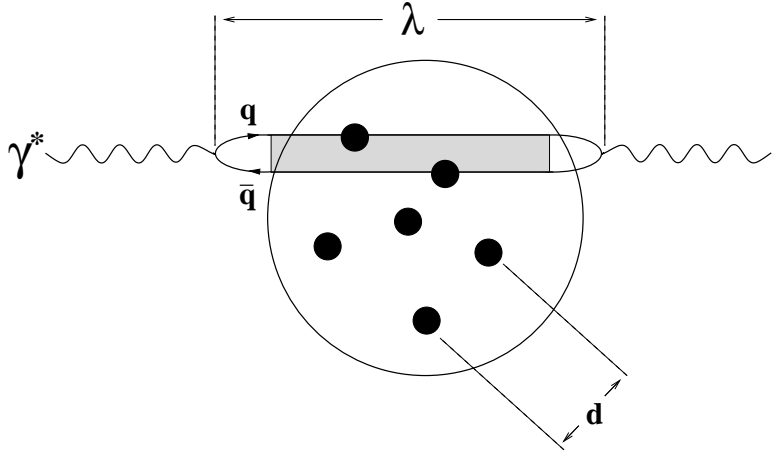,height=60mm}
\end{center}
\caption[...]{
Deep-inelastic scattering at small $x\ll 1$ in the laboratory 
frame proceeds via hadronic fluctuation present in the photon 
wave function.
}
\label{fig:gN_Lab}
\bigskip
\end{figure}

In process (b) the photon couples to a quark pair 
which can form  a complex 
(hadronic or quark-gluon) intermediate state and then scatters 
from the target. 
At small $x$ deep-inelastic scattering 
can therefore be described in the laboratory frame in terms of 
the interaction of quark-gluon components present in the wave 
function of the virtual photon (Fig.\ref{fig:gN_Lab}). 
The longitudinal propagation length $\lambda$ of a specific 
photon-induced quark-gluon fluctuation with mass $\mu$ is given by 
the inverse of the energy denominator (\ref{eq:deltaEb}):
\begin{equation} 
\lambda \sim \frac{1}{\Delta E_b} 
= 
\frac{2\nu}{\mu^2+Q^2} 
\stackrel{\mu^2\sim Q^2}{\llongrightarrow}
\frac {1}{2 x M}, 
\label{eq:coherence}
\end{equation}
which coincides  with the longitudinal correlation length $l$ 
of Eq.(\ref{eq:l_z}).
For $x<0.05$ the propagation length $\lambda$  exceeds
the average distance between nucleons in
nuclei, $\lambda > d \simeq 2\,\rm{fm}$.  
For a nuclear target,  
coherent multiple scattering of  quark-gluon fluctuations of 
the photon from several nucleons in the nucleus  can then occur, 
and this is clearly seen in the coordinate space analysis 
discussed in the previous section.  

For larger values of the Bjorken variable,  
$x > 0.2$, the propagation  length of
intermediate hadronic  states is small,
$\lambda < d$.  At the same time the process in 
Fig.\ref{fig:tpth}a
becomes prominent, i.e.\ the virtual
photon is absorbed directly by a quark or
antiquark in the target.  
Now the incoherent scattering from the hadronic 
constituents of the nucleus dominates.

\subsection{Nuclear deep-inelastic scattering in the infinite 
momentum frame}
\label{ssec:IMF_spati}

Let us finally view the deep-inelastic scattering process in the 
so-called infinite momentum frame where the target momentum is large. 
In this frame the standard parton model  applies in which a snapshot of  
the target at the short time scale of the 
interaction reveals an ensemble of almost non-interacting partons, 
i.e. quarks and gluons.

Consider the scattering from a nucleus which moves  with large 
longitudinal momentum 
$P_{\T A} \approx A P_{\T N}\rightarrow \infty$, where   
$P_{\T N}$ is the average longitudinal momentum of the bound nucleons 
\cite{Nikolaev:1975vy,Close:1988xw,Kumano:1992ef}. 
The average nucleon-nucleon distance in nuclei is now Lorentz contracted  
as compared to the lab frame:
$d^{inf} \approx 
d\,M_{\T A}/P_{\T A}\approx 2\,{\rm{fm}} \,M/P_{\T N}$. 
On the other hand the delocalization of a parton  with 
longitudinal momentum fraction $x$ in the nucleon is given  according 
to the 
Heisenberg uncertainty principle by $\delta l \approx 1/xP_{\T N}$. 
At small Bjorken-$x$, $x < 1/d^{inf} P_{\T N} \approx 0.1$,   
the wave functions  of partons from different nucleons 
have a chance to overlap, i.e. 
$d^{inf} < \delta l$. 
Therefore, at $x\ll 0.1$  we expect an  enhanced interaction between  
partons coming from different nucleons.
One can anticipate that, at $x\ll 0.1$, the parton delocalization 
extends over the whole nucleus. 
This is where the quark and gluon fluctuations of the 
photon interact simultaneously  with 
the parton content of several nucleons.

\section{Shadowing in unpolarized deep-inelastic scattering}
\label{sec:Shad}

\setcounter{section}{5}
\setcounter{figure}{0}

As outlined in Section \ref{Sec:nucl_dat}, the most pronounced 
nuclear effect in lepton-nucleus DIS is shadowing. For small 
values of the Bjorken variable ($x<0.1$), the nuclear 
structure functions  $F_{2}^{\T A}$ are significantly reduced 
as compared to the free nucleon structure function $F_{2}^{\T N}$.
Equivalently, the virtual photon-nucleus 
cross section is less than $A$ times 
the one for free nucleons, 
$\sigma_{\gamma^* \T A} < A \,\sigma_{\gamma^* \T N}$. 
The analogous behavior is observed for real 
photons at large energies ($\nu > 3\,\rm{GeV}$).

This reduction of nuclear cross sections is reminiscent of the 
features seen in high-energy hadron-nucleus collisions. 
For example, total cross sections for nucleon-nucleus 
scattering behave as $\sigma_{\T {N A}} \sim A^{0.8} \,\sigma_{\T{NN}}$ 
at center-of-mass energies $\sqrt{s} \sim (10$ -- $25)\,{\rm GeV}^2$ 
\cite{RamanaMurthy:1975qb}. 
A simple geometric picture interprets this effect as 
the hadron projectile interacting mainly with nucleons at the 
nuclear surface,  leading to $\sigma_{\T{N A}} \sim A^{2/3}$. 

The quantum mechanical description of shadowing in DIS 
explains this phenomenon by the 
destructive interference of  single and multiple scattering amplitudes.
Multiple scattering becomes important as soon as the lab frame coherence 
length for the hadronic fluctuations of the photon propagator exceeds the 
average distance between two nucleons in the nuclear target. We have 
seen in our space-time discussion of 
Section \ref{Sec:space_time} that this is precisely 
what happens in the region $x<0.1$ of the Bjorken variable. 

The physics 
issue of nuclear DIS at small $x$ is therefore, roughly speaking, 
the optics of hadronic or quark-gluon fluctuations of the virtual 
photon in the 
nuclear medium. Diffractive phenomena play an important role in this context, 
as we shall demonstrate. 

At extremely small $x$ (i.e. for $x < 10^{-3}$) in combination 
with large $Q^2$, the measured free nucleon structure functions 
indicate a rapidly growing number of partons (mostly gluons). 
This is the domain of ``high density QCD'' 
where individual partons interact perturbatively, at large $Q^2$, 
but their number increases so strongly that effective cross 
sections can become large 
(for references see e.g. 
\cite{McLerran:1994ni,Jalilian-Marian:1997xn,Ayala:1997em,Kovchegov:1997dm,%
Mueller:1999wm,Kovchegov:1999yj}). 
It is of great interest to investigate the transition of the observed 
shadowing phenomena into this new domain, 
accessible by collider experiments, but so far unexplored for nuclear 
systems.
 
In this section we first concentrate on  the relationship 
between diffractive photo- and leptoproduction from nucleons and 
shadowing in high-energy photon- and lepton-nucleus interactions.
Then we 
investigate perturbative and non-perturbative QCD aspects  
of  shadowing.
After that we summarize existing models 
which successfully describe data.
Finally we outline implications of shadowing for nuclear parton distributions.

\subsection{Diffractive production and nuclear shadowing}
\label{ssec:diff_shad}

In the shadowing region, diffractive photo- and leptoproduction 
of high energy  hadrons gives a substantial contribution to the (virtual) 
photon-nucleon cross section as discussed  in Section \ref{ssec:Diffraction}. 
This suggests that the diffractive excitation of hadronic states, 
$\gamma^* {\T N} \rightarrow {\T{X N}}$,  
and their coherent interaction with several nucleons inside the target 
plays an important role for shadowing in high energy  photon-nucleus 
scattering, in a similar way as for hadron-nucleus collisions.
For this effect to be significant,  the following 
two conditions have to be met  in the laboratory frame: 
\begin{enumerate} 

\item[(i)] 
The longitudinal propagation length, or coherence length, 
\begin{equation}  
 \lambda =\frac{2\nu}{M_{\T X}^2 + Q^2} 
\end{equation}
of the diffractively produced hadronic state of invariant 
mass $M_{\T X}$, Eq.(\ref{eq:coherence}), 
must exceed the average nucleon-nucleon distance in 
nuclei:
\begin{equation} \label{eq:cond_i}
\lambda > d \simeq 2\,\mbox{fm}.
\end{equation}

\medskip 

\item [(ii)] 
In addition, the mean free path $l_{\T X} = \left(\rho
\,\sigma_{\T{XN}}\right)^{-1}$ of the diffractively produced system 
in the nuclear medium must be sufficiently short, at least smaller than 
the nuclear radius

\end{enumerate}
Note that the mean free path of photons in a nucleus with 
density $\rho$ amounts to 
$l_{\gamma} \approx (\rho \,\sigma_{\gamma \T N})^{-1} 
\approx 550\,{\rm fm}$, 
which is much larger than any nuclear scale. 
Consequently ``bare'' photons do not scatter coherently from several nucleons 
and therefore do not contribute to shadowing. 

Shadowing results from 
the coherent scattering of a hadronic fluctuation  from 
at least two nucleons in the target, i.e. for  $\lambda >  d$. 
Since the longitudinal propagation length 
$\lambda$ of a diffractively produced hadronic state  ${\T X}$ 
decreases with its mass $M_{\T X}$, 
low mass excitations with  $M_{ \T X}\lsim 1\,{\rm GeV}$ are relevant  
for the onset of shadowing. 
Equation (\ref{eq:cond_i}) tells again 
that shadowing in deep-inelastic 
scattering at $Q^2\gg 1\,\rm{GeV^2}$ should start at  
$x \approx 0.1$, in accordance with the observed effect 
and in close correspondence with the space-time picture 
described in Section \ref{Sec:space_time}.

For real photons   diffractive processes at low mass are 
dominated by the excitation of the $\rho$- and $\omega$-meson. 
Significant contributions to double scattering and hence to 
shadowing  are therefore expected if the photon energy $\nu$ 
exceeds about $3$ GeV, 
in line with the experimental data. 

Consider now the scattering process in the laboratory frame. 
Realistic nuclear wave functions are well established only in this frame 
(with the exception of recent efforts  to construct 
relativistic nuclear model wave functions on the light front, 
see e.g. [140 -- 146]).
\nocite{Miller:1997xh,Miller:1997cr,Burkardt:1998bt,Miller:1998tp,%
Blunden:1999hy,Miller:1999ap,Blunden:1999gq} 
Later, in Section \ref{ssec:shad_IMF}, 
we comment on nuclear shadowing as seen in the Breit frame. 
We first neglect effects due to nuclear binding, Fermi-motion and 
non-nucleonic degrees of freedom in nuclei. 
They are relevant at moderate and large 
values of the Bjorken variable, $x > 0.1$, as discussed in 
Section \ref{sec:EMC}.

The (virtual) photon-nucleus cross section can be separated into 
a piece which accounts for the incoherent scattering 
from individual nucleons, and a correction from the 
coherent interaction with  several nucleons:
\begin{equation}
\sigma_{\gamma^*{\T A}} = 
Z \,\sigma_{\gamma^* {\T p}} + (A-Z) \,\sigma_{\gamma^* {\T n}} 
+ \delta \sigma_{\gamma^* {\T A}}. 
\end{equation}
The single scattering part  is the incoherent sum of 
photon-nucleon  cross sections, where $Z$ is the nuclear charge.
The multiple scattering correction can be expanded in 
contributions  which account for the scattering from $n\geq 2$ nucleons. 
Expressed in terms of the corresponding multiple scattering 
amplitudes ${\cal A}^{(n)}_{\gamma^* \T {A}}$ we have:  
\begin{equation} \label{eq:in+ds}
\delta \sigma_{\gamma^* {\T A}} = 
\frac{1}{2 M_{\T A} \nu} 
\sum_{n=2}^A {\T {Im}} \,{\cal A}_{\gamma^* {\T A}}^{(n)},
\end{equation}
where the photon 
flux (\ref{eq:sigma_hH}) is taken in the limit $x \ll 1$. 
The leading contribution to nuclear shadowing 
comes from double scattering.
Its mechanism is  best illustrated for a deuterium target 
on which we focus next.

\subsubsection{Shadowing in deuterium}
\label{ssec:shad_deu}

\begin{figure}[t]
\bigskip
\begin{center} 
\epsfig{file=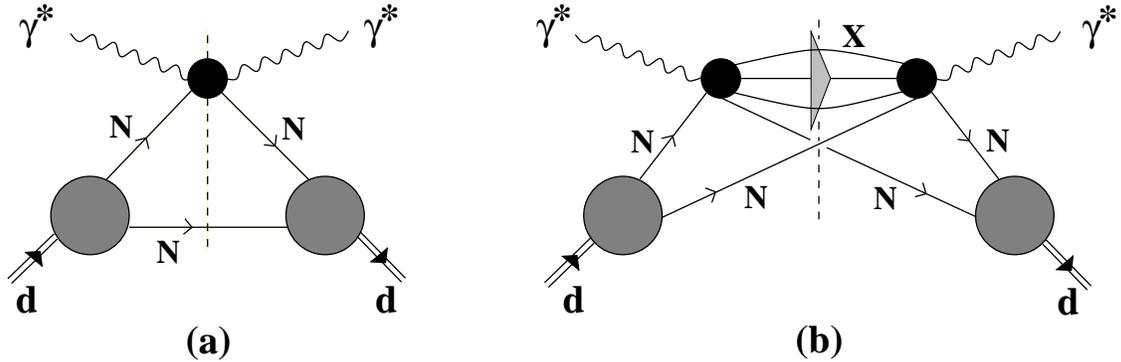,height=50mm}
\end{center}
\caption[...]{
Single (a) and double (b) scattering contribution to   
virtual photon-deuteron  scattering. 
The corresponding cross sections are obtained from the 
imaginary part of the forward scattering amplitude 
indicated by the dashed line.
}
\label{fig:mult_deut}
\bigskip
\end{figure}
In this section we review the basic mechanism of shadowing 
in real and virtual photon-deuteron scattering at high 
energies $\nu$, or  equivalently, small $x$. 
The $\gamma^*$-deuteron cross section can be written 
as the sum of single and double scattering parts  
as illustrated in Fig.\ref{fig:mult_deut}: 
\begin{equation}
\sigma_{\gamma^* \T d} = 
\sigma_{\gamma^* {\T p}} + \sigma_{\gamma^* {\T n}}  
+ \delta \sigma_{\gamma^* {\T d}}.
\end{equation}
The first two terms describe the incoherent scattering of the  
(virtual) photon from the proton or neutron, 
while 
\begin{equation}
\delta \sigma_{\gamma^* {\T d}} = \frac{1}{2 M_{\T d}  \nu} 
\,{\T {Im}} \,{\cal A}^{(2)}_{\gamma^* {\T d}}  
\end{equation}
accounts for the coherent interaction of the projectile with both 
nucleons.

For large energies, $\nu > 3$ GeV, or small values  
of the Bjorken variable, $x < 0.1$,  
the double scattering amplitude ${\cal A}^{(2)}_{\gamma^* {\T d}}$ 
is dominated by the diffractive excitation  of  hadronic 
intermediate states (Fig.\ref{fig:mult_deut} b) described by the amplitude 
$T_{\gamma^{*} {\T {N \rightarrow XN}}}$. 
At the high  energies involved it is a good approximation  
to neglect the real part of this amplitude.
In fact, we expect 
${\rm{Re}} \,T_{\gamma^* \T{N \rightarrow XN}} \lsim 0.15\, 
{\rm{Im}} \,T_{\gamma^* \T{N \rightarrow XN}}$ 
by analogy with high-energy hadron-hadron scattering amplitudes
(see e.g. \cite{Donnachie:1992ny}). 
When including such non-zero real parts, the double scattering 
contribution changes  by less than $10\%$ \cite{Piller:1997ny}. 
We neglect the spin  and isospin dependence 
for unpolarized scattering \cite{Edelmann:1997ik}.
Of course, these degrees of freedom play a crucial role 
in polarized scattering as we will discuss in Section \ref{sec:Coh_Pol}.

Treating the deuteron target in the non-relativistic limit 
gives \cite{Gribov:1969,Gribov:1970,Bertocchi:1972,Weis:1976er}:
\begin{eqnarray} \label{eq:ds amplitude} 
{\cal A}_{\gamma^{*} {\T d}}^{(2)} &=& -\frac{2}{M} 
\int d^3r \,|\psi_{\T d}(\vec r )|^2 
\nonumber \\
&& 
\hspace*{-1cm}
\times 
\sum_{\T X}\int \frac{d^3 k}{(2\pi)^3} \,
T_{\gamma^{*} {\T{N\rightarrow XN}}}(k) \,
\frac {e^{i\vec k\cdot \vec r}}
{(q_0 - k_0)^2 - \vec k_{\perp}^2 - (q_3 - k_3)^2 - M_{\T X}^2 + i\epsilon } 
\,T_{{\T{XN}}\rightarrow \gamma^{*} {\T{N}}}(k), 
\end{eqnarray}
where $k^{\mu} = (k_0,\vec k)$ with $\vec k = (\vec k_\perp,k_3)$  
is the four-momentum transfered to the nucleon, and 
$\psi_{\T d}$ is the deuteron wave function 
normalized as $\int d^3r \,|\psi_{\T d}(\vec r)|^2 = 1$.
The sum is taken over all diffractively excited hadronic states 
with  invariant mass $M_{\T X}$ and four-momentum 
$p_{\T X} = q - k$. 
We write
\begin{equation}
\sum_{\T X} 
{|T_{\gamma^{*} {\T {N\rightarrow XN}}}|^2} = 
64\,\pi\,M^2 \nu^2
\int_{4 m_{\pi}^2}^{W^2} dM_{\T X}^2 
\,\frac{d^2\sigma^{diff}_{\gamma^{*} {\T N}}}{dM_{\T X}^2 dt}
\end{equation}
in terms of the diffractive production cross section, 
with $t = k^2$. 
The limits of integration define the kinematically permitted range 
of diffractive excitations, with their invariant mass $M_{\T X}$ 
above the two-pion production threshold and limited by the 
center-of-mass energy $W=\sqrt{s}$ of the scattering process.
We introduce the spin-averaged deuteron form factor, 
\begin{equation} \label{eq:deu_ff}
S_{\T d}(\vec k)  
= 
\int d^3 r \, e^{i\vec k\cdot \vec r} \,|\psi_{\T d}(\vec r)|^2, 
\end{equation}
perform the integration over the longitudinal 
momentum transfer in Eq.(\ref{eq:ds amplitude}) and 
then take the imaginary part of the amplitude 
${\cal A}_{\gamma^{*} {\T d}}^{(2)}$. 
Actually $k_3$ is simply fixed by energy-momentum 
conservation: 
\begin{equation} \label{eq:kz}
k_{3} \approx 
\frac{Q^2 + M_{\T X}^2}{2\nu} = \frac{1}{\lambda},
\end{equation}
which coincides with the inverse of the longitudinal propagation length 
(\ref{eq:coherence}) of the intermediate hadronic state. 
Note that the minimal momentum transfer required to 
produce  a hadronic state diffractively from a nucleon at rest 
amounts to $t_{min} \approx  - k_3^2(M_{\T X})$. 

When all steps are carried out, the result for the double scattering 
correction is \cite{Gribov:1969,Gribov:1970}
\begin{equation} \label{eq:ds_corr_full}
\delta \sigma_{\gamma^{*}{\T d}} = 
- \frac{2}{\pi} 
\int d^2 k_{\perp} \int_{4 m_{\pi}^2}^{W^2} dM_{\T X}^2  \,\, 
S_{\T d}(\vec k_{\perp},k_3 \approx \lambda^{-1}(M_{\T X}))\, \,
\frac{d^2\sigma^{diff}_{\gamma^{*} {\T N}}}{dM_{\T X}^2 dt}\,.
\end{equation}

This equation establishes the close relationship 
between shadowing in deep-inelastic scattering 
and diffractive hadron  production.
It becomes even more transparent for $x\ll 0.1$, 
i.e. large $\lambda$. 
In this limit the magnitude of shadowing  is determined 
just by the  ratio of diffractive and total $\gamma^* \T N$ cross 
sections. 
To verify this let us parametrize the  
$t$-dependence of the diffractive production cross section 
entering in Eq.(\ref{eq:ds_corr_full}) as 
\begin{equation} \label{eq:def_diff_cross_approx} 
\frac{d^2\sigma^{diff}_{\gamma^{*} {\T N}}}{dM_{\T X}^2 dt} 
=
e^{-{B} |t|}\,\left.
\frac{d^2\sigma^{diff}_{\gamma^{*} {\T N}}}{dM_{\T X}^2 dt}
\right|_{t = 0}
\approx  
e^{-{B} \vec k_{\perp}^2} \,
\left.\frac{d^2\sigma^{diff}_{\gamma^{*} {\T N}}}{dM_{\T X}^2 dt}
\right|_{t = 0},
\end{equation}
neglecting the $k_3$ dependence of $t$. 
Data from FNAL and HERA 
on diffractive photo- and leptoproduction of hadrons 
with mass $M_{\T X}^2 > 3$ GeV$^2$ give 
$B \simeq (5 \dots 7) \,$GeV$^{-2}$ 
\cite{Breitweg:1998aa,Chapin:1985,Breitweg:1997eh}. 
In the diffractive production of low mass vector mesons  
($\rho, \omega$ and $\phi$) from nucleons, a range of 
values $B \simeq (4 \dots 10) \,$GeV$^{-2}$ has been found, 
depending on $Q^2$ and on the  incident photon energy 
(for a review and references see e.g. 
\cite{Abramowicz:1998ii,Crittenden:1997yz}). 
Clearly, the soft deuteron form factor selects momenta such 
that the double scattering correction in (\ref{eq:ds_corr_full}) 
is dominated by diffractive production in the direction of the 
incident photon.

\begin{figure}[t]
\bigskip
\begin{center} 
\epsfig{file=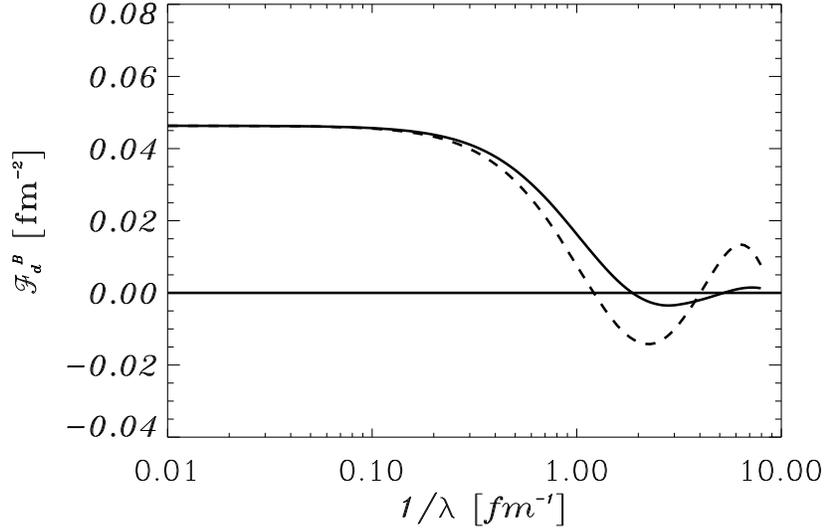,height=70mm}
\end{center}
\caption[...]{Integrated deuteron form factor 
${\cal F}_{\T d}^B$ from Eq.(\ref{eq:deuteron_ff_int}) for an 
average slope $B = 8$ GeV$^2$. The dotted curve corresponds 
to $B=0$. 
}
\label{fig:deut_ff}
\bigskip
\end{figure}

In Fig.\ref{fig:deut_ff}  
we show the deuteron form factor (\ref{eq:deu_ff}) weighted 
by the exponential $t$-dependence 
(\ref{eq:def_diff_cross_approx}) and integrated over 
transverse momentum, 
\begin{equation} \label{eq:deuteron_ff_int}
{\cal F}_{\T d}^B(\lambda^{-1}) \equiv 
{\int} \frac{d^2 k_{\perp}}{(2 \pi)^2} \,
S_{\T d}(\vec k_{\perp},k_3=\lambda^{-1})\, e^{-B  \vec k_{\perp}^2}, 
\end{equation}
as obtained with 
the Paris nucleon-nucleon potential \cite{Lacombe:1980dr} 
for a  slope  parameter $B= 8$ GeV$^{-2}$. 
We observe ${\cal F}_{\T d}^B \approx constant$ as long as the 
longitudinal propagation length $\lambda$ exceeds the deuteron size 
$\langle r^2\rangle_{\T d}^{1/2} \approx 4\,fm$. 
From Eq.(\ref{eq:cond_i}) one then finds  that 
hadronic states with an invariant mass 
\begin{equation} \label{eq:MX}
M_{\T X}^2 <M_{max}^2 =  \frac {W^2+ Q^2}
{M \langle r^2\rangle^{1/2}_{\T d}} - Q^2
\end{equation} 
contribute dominantly to double scattering. 
Combining Eqs.(\ref{eq:ds_corr_full}) and (\ref{eq:MX}) gives 
the following approximate expression for the shadowing correction 
in the limit of large longitudinal propagation length $\lambda$:  
\begin{eqnarray} \label{eq:approx}
\delta \sigma_{\gamma^{*} {\T d}} 
&\approx& 
- 8\pi  \,{\cal F}_{\T d}^B(\lambda^{-1} \rightarrow 0)
\,\left.\int_{4 m_{\pi}^2}^{M_{max}^2} dM_{\T X}^2 \,
\frac{d \sigma^{diff}_{\gamma^{*}  {\T N}}}{d M_{\T X}^2 dt} 
\right|_{t = 0} 
\approx 
- 8 \pi  {\cal F}_{\T d}^B(0) \,B \,
\sigma^{diff}_{\gamma^{*} {\T N}}.   
\end{eqnarray}
In the last step we have neglected  contributions 
to the integrated diffractive production cross section 
$\sigma^{diff}_{\gamma^{*} {\T N}}$  from  hadronic 
states with invariant masses  $M_{max} < M_{\T X} < W$. 
Since $d\sigma^{diff}_{\gamma^{*}  {\T N}}/d M_{\T X}^2$ 
drops strongly for large  $M_{\T X}$ as discussed 
in Section \ref{ssec:Diffraction}, 
this approximation is justified at large center 
of mass energies $W$ or, equivalently, at small $x$. 
For the ratio between  deuteron and  free nucleon 
structure functions we then obtain: 
\begin{equation}
R_{\T d} = \frac{F_2^{\T d}}{F_2^{\T N}} = 
\frac{\sigma_{\gamma^* \T d}}{2 \,\sigma_{\gamma^* \T N}} 
\approx 1 - 4 \pi  \,{\cal F}_{\T d}^B(0) \, B \,
\frac{\sigma^{diff}_{\gamma^{*} {\T N}}}{\sigma_{\gamma^{*} {\T N}}}.  
\end{equation}
We  use 
${\sigma^{diff}_{\gamma^{*} {\T N}}}/{\sigma_{\gamma^{*} {\T N}}} 
\approx 0.1$
for the fraction of diffractive events in 
deep-inelastic scattering from free nucleons,    
as suggested by experiment 
(see Section \ref{ssec:Diffraction}).  
Furthermore we take $B = 8$ GeV$^{-2}$. 
One  finds that shadowing at $x\ll 0.1$ in deuterium 
amounts to about $2\%$, i.e. $R_{\T d} \approx 0.98$.
The effect is small because of the large average proton-neutron 
distance in the deuteron, but the result 
agrees well with the experimental data shown in 
Fig.\ref{fig:deut_shad}.

\subsubsection{Shadowing for heavy nuclei}
\label{ssec:shad_A}

\begin{figure}[t]
\bigskip
\begin{center} 
\epsfig{file=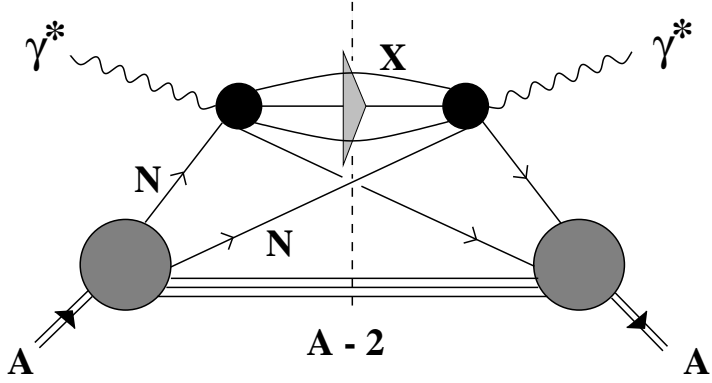,height=50mm}
\end{center}
\caption[...]{Double scattering contribution 
to deep-inelastic scattering from nuclei.
}
\label{fig:mult_double} 
\bigskip
\end{figure}

The diffractive production of hadrons from single nucleons 
also controls shadowing in heavier nuclei for which this effect 
is far more pronounced than in the deuteron.
It is an empirical fact that nuclear shadowing 
increases with the nuclear mass number $A$ of the target 
(see Section \ref{Sec:nucl_dat}). 
For $A>2$ the hadronic state which is produced in the 
interaction of the photon with one of the nucleons in the target
may scatter coherently from more than two nucleons.  
However, double scattering still dominates  
since the probability that the propagating 
hadron interacts with several nucleons along its path 
decreases with the number of scatterers.
The double scattering contribution to the total photon-nucleus cross 
section $\sigma_{\gamma^* {\T A}}$ is obtained by  
straightforward generalization 
of the deuteron result (\ref{eq:ds_corr_full}) 
\cite{Gribov:1970,Bertocchi:1972}: 
\begin{eqnarray} \label{eq:ds_A}
\sigma_{\gamma^{*}{\T A}}^{(2)} &=& 
- 8\pi 
\int d{^2 b} \int_{-\infty}^{+\infty} dz_1 
\int_{z_1}^{+\infty} dz_2 \,
\rho_{\T A}^{(2)}(\vec b,z_1;\vec b, z_2) \, \cdot 
\nonumber \\  
&&\hspace*{3cm}
 \int_{4 m_{\pi}^2}^{W^2} dM_{\T X}^2 
\cos\left[ (z_2 - z_1)/\lambda \right] 
\left. \frac{d^2\sigma^{diff}_{\gamma^{*} {\T N}}}{dM_{\T X}^2 dt}
\right|_{t\approx 0}.  
\end{eqnarray}
As illustrated in Fig.\ref{fig:mult_double} 
a diffractive state with invariant mass $M_{\T X}$ is 
produced in the interaction of the photon with  a nucleon 
located at position $(\vec b, z_1)$ in the target. 
The hadronic excitation propagates 
at fixed impact parameter $\vec b$
and then interacts  with  a second nucleon 
at $z_2$.
The probability to find two nucleons in the target 
at the same impact parameter is described by the two-body density 
$\rho_{\T A}^{(2)}(\vec b,z_1;\vec b,z_2)$ normalized as 
$\int d^3 r\,d^3 r'\,\rho_{\T A}^{(2)}(\vec r, \vec r\,') = A^2$.
The  
$\cos [ (z_2 - z_1)/\lambda]$ 
factor in Eq.(\ref{eq:ds_A}) implies that only diffractively 
excited hadrons with a longitudinal propagation length larger than 
the average nucleon-nucleon distance in the target, 
$\lambda > d \simeq 2\,\rm{fm}$, 
can contribute significantly to double scattering.

Note that nuclear short-range correlations are relevant only if the 
coherence length of the diffractively excited states
is comparable to the range of the short-range repulsive part of  
the nucleon-nucleon force, i.e. for $\lambda \lsim 0.5\,\rm{fm}$. 
In this case the shadowing effect is negligible. 
Nuclear correlations are therefore not important in the shadowing domain 
and the target can be considered as an ensemble of 
independent nucleons with  
$\rho_{\T A}^{(2)}(\vec r, \vec r\,') 
\approx \rho_{\T A}(\vec r) \rho_{\T A}(\vec r\,')$, 
where $\rho_{\T A}$ is the nuclear one-body density 
\cite{Piller:90prc,Melnitchouk:1993vc}. 

With increasing photon energies  or decreasing $x$ down to $x\ll 0.1$, 
the longitudinal propagation length of diffractively excited 
hadrons rises and eventually reaches nuclear dimensions. 
Then interactions of the excited hadronic states with several nucleons 
in the target become important. 
A simple way to account for those is a frequently used equation 
derived by Karmanov and Kondratyuk \cite{Kondratyuk:1973jept}: 
\begin{eqnarray} \label{eq:ms_A}
\delta \sigma_{\gamma^{*}{\T A}} &=& 
- 8\pi \int d{^2b} \int_{-\infty}^{+\infty} dz_1 
\int_{z_1}^{+\infty} dz_2 \,
\rho_{\T A}(\vec b,z_1)\,\rho_{\T A}(\vec b,z_2) \cdot 
\nonumber \\  
&&\hspace*{0.2cm}
\int_{4 m_{\pi}^2}^{W^2} dM_{\T X}^2 
\cos\left[ (z_2 - z_1)/\lambda \right] 
\left. \frac{d^2\sigma^{diff}_{\gamma^{*} {\T N}}}{dM_{\T X}^2 dt}
\right|_{t\approx 0} 
\exp\left[-\frac{\sigma_{{\T {XN}}}}{2} 
\int_{z_1}^{z_2} dz\,\rho_{\T A}(\vec b,z)\right].
\end{eqnarray}
The exponential attenuation factor describes the elastic 
re-scattering of the diffractively produced hadronic states 
from the remaining nucleons in the target. 
The  hadron-nucleon scattering amplitudes 
are assumed to be purely imaginary and enter in Eq.(\ref{eq:ms_A}) 
through the  cross sections $\sigma_{{\T {XN}}}$.

Equation (\ref{eq:ms_A}) has been applied in several investigations 
of nuclear shadowing using different models for the diffractive 
photoproduction cross section. The more detailed results are discussed 
in Section \ref{ssec:Shad_Model}, 
but we can get a simple estimate of nuclear shadowing at small 
Bjorken-$x$ already by just looking at the relative amount 
of diffraction in DIS from free nucleons \cite{Piller:1998cy}. 
We restrict ourselves to the double scattering correction (\ref{eq:ds_A}). 
For $x\ll 0.1$, the coherence length $\lambda$ of the hadronic states 
which dominate diffractive production in Eq.(\ref{eq:ds_A}), 
exceed the diameter of the target nucleus. 
In the limit $\lambda \rightarrow \infty$ we find: 
\begin{equation} \label{eq:shad_est_A}
\sigma^{(2)}_{\gamma^* {\T A}} \simeq - 8\pi \,B\,
\sigma^{diff}_{\gamma^* {\T N}}
\int d{^2 b} \int_{-\infty}^{+\infty} dz_1 
\int_{z_1}^{+\infty} dz_2 \,
\rho_{\T A}(\vec b,z_1) \,\rho_{\T A}(\vec b, z_2).  
\end{equation}
The slope parameter $B$ and the integrated diffractive 
production cross section $\sigma^{diff}_{\gamma^* {\T N}}$ 
have been introduced as in 
Eqs.(\ref{eq:def_diff_cross_approx}, \ref{eq:approx}).

For the nuclear densities in Eq.(\ref{eq:shad_est_A}) we use Gaussian, 
\begin{equation} \label{eq:Gauss}
\rho_{\T A}(\vec{r})
= A \,\left(\frac{3}{2 \,\pi\,\langle r^2 \rangle_{\T A} }\right)^{3/2}  \,
\exp \left( - 
\frac{3\,\vec{r}\,^2}{2\,\langle r^2 \rangle_{\T A}} 
\right),
\end{equation}
and square-well parametrizations,  
\begin{equation}
\rho_{\T A}(\vec{r}) = 
\left\{ \begin{array}{l}
A \frac{3}{4\pi} 
\left(\frac{3}{5 \,\langle r^2 \rangle_{\T A} }\right)^{3/2}  
\quad 
\mbox{for}
\quad r < \sqrt{\frac{5}{3}} \,\langle r^2 \rangle_{\T A}^{1/2}, 
\nonumber \\
0  \hspace*{3cm}\mbox{otherwise},
\end{array}\right.
\end{equation}
with  the mean square radius $\langle r^2 \rangle _{\T A} =  
\int d^3r \,r^2 \,\rho_{\T A}(r)/A$. 
For both cases the shadowing ratio 
$R_{\T A} = 
\sigma_{\gamma^* {\T A}}/
{A \sigma_{\gamma^* {\T N}}}$
is easily worked out:
\begin{equation} \label{eq:shad_est}
R_{\T A} \simeq 1 -  {\cal C} \,A\,
\left(\frac{B}{\langle r^2 \rangle _{\T A}} \right)
\frac{\sigma^{diff}_{\gamma^* {\T N}}}{\sigma_{\gamma^* {\T N}}}.
\end{equation}
For Gaussian nuclear densities one finds ${\cal C}=3$, while 
${\cal C}=2.7$ in the  square-well case.

Using again typical values for the ratio of  diffractive and total 
$\gamma^* \T N$ cross sections, 
\linebreak
${\sigma^{diff}_{\gamma^* {\T N}}}/{\sigma_{\gamma^* {\T N}}}\simeq 0.1$,  
and for the slope parameter, $B\simeq 8$ GeV$^{-2}$, 
the magnitude of $R_{\T A}$ comes out in very reasonable agreement with
experimental values as shown in Table \ref{tab:shad_est}.
\begin{table}[h]
\begin{center}
\begin{tabular}{| c | c | c | c | c | }
\hline 
                  & $^6$Li  & $^{12}$C & $^{40}$Ca & $^{131}$Xe \\ \hline     
 $R_{\T A}$       & $0.93$  & $0.84$   & $0.73$    & $0.65$    \\ \hline
 $R_{\T A}^{exp.}$& $0.94 \pm 0.07$  & $0.87 \pm 0.10$   & $0.77 \pm 0.07$ 
                   &  $0.67 \pm 0.09$ \\ \hline
\end{tabular}
\caption{
The shadowing ratio $R_{\T A}$ estimated according to Eq.(\ref{eq:shad_est}) 
in comparison to experimental data for various nuclei. 
The data are taken from 
Ref.\cite{Arneodo:1995cs,Adams:1992nf,Adams:1995is} at the smallest 
kinematically accessible values of Bjorken-$x$ 
(namely, $x \simeq 10^{-4}$). 
}
\label{tab:shad_est}
\end{center}
\end{table}
This estimate may be simple 
(in fact, higher order multiple scattering must be included 
in a more detailed analysis)
but it certainly confirms that shadowing 
in nuclear DIS is governed by the coherent interaction of diffractively 
produced states with several nucleons in the target nucleus.
A more detailed investigation of the connection between HERA data on 
diffraction and shadowing effects  measured at CERN and FNAL 
can be found in Ref.\cite{Capella:1997mn}.

Inelastic transitions between different hadronic states are neglected 
in Eq.(\ref{eq:ms_A}). 
They cannot be treated in a model-independent way. 
Estimates of such higher order diffractive dissociation 
contributions have been performed for high-energy hadron-nucleus 
scattering \cite{RamanaMurthy:1975qb,Nikolaev:1986vy}. 
In the case of neutron-nucleus collisions at  center-of-mass energies 
$s \sim 200 \,\rm{GeV^2}$ they amount to about  
$5\%$ of the total reaction  cross section.
For rising energies the 
relative importance of inelastic transitions is expected to grow
\cite{Nikolaev:1986vy}. 

\begin{figure}[t]
\bigskip
\begin{center} 
\epsfig{file=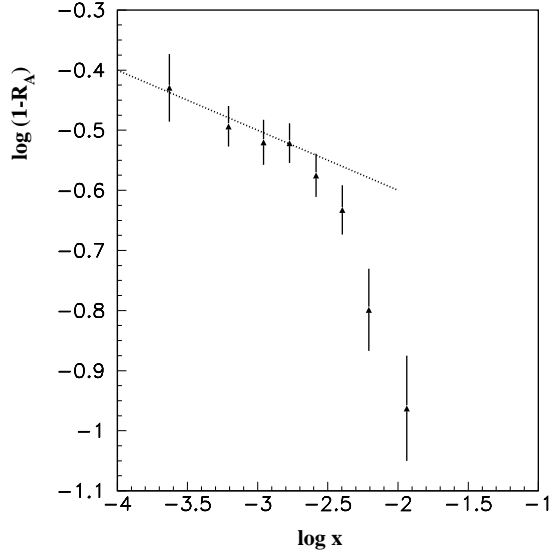,height=80mm}
\end{center}
\caption[...]{The quantity $\log \left( 1 - R_{\T A} \right)$  
as a function of $\log x$ for  
data taken on lead \cite{Adams:1995is}. 
The dashed line corresponds to the asymtotic energy dependence 
(\ref{eq:log_est}) with $\varepsilon = 0.1$. 
}
\label{fig:log1-R}
\bigskip
\end{figure}
Given the important role of diffractive production, we can now 
enter into a more detailed discussion of the $x$-dependence 
of shadowing. 
The coherence lengths $\lambda$ of hadronic states with small masses 
become comparable with nuclear dimensions for $x < 0.1$. 
As $\lambda$ increases with decreasing $x$, the shadowing effect 
grows steadily for $x\lsim 0.05$.  
At $x \ll 0.1$ it is also the energy dependence of  
the diffractive production cross section and of the hadron-nucleon cross 
section $\sigma_{\T{X N}}$ which influences 
the $x$-dependence of shadowing.  

Consider the shadowing ratio 
$R_{\T A} = {\sigma_{\gamma^* {\T A}}}/{A \sigma_{\gamma^* {\T N}}} 
= 1 - \delta\sigma_{\gamma^*\T A}/{A \sigma_{\gamma^* {\T N}}}$,  
parametrized as:
\begin{equation}  \label{eq:RA-1}
R_{\T A} - 1 = 
- c \, \left(\frac{1}{x}\right)^{\varepsilon},
\end{equation}
with a constant $c$ at small $Q^2$ where data are actually taken, 
and a characteristic exponent $\epsilon$. 
At asymptotically large energies Regge phenomenology 
suggests $\varepsilon \simeq 0.1$ (see Section \ref{ssec:Diffraction})
\footnote{
At the typical average center of mass energies 
$\overline  W < 25$ GeV 
used at experiments at CERN and FNAL 
a somewhat stronger energy dependence is expected 
through the kinematic restriction to diffractively produced 
hadronic states with masses $M_{\T X} < W$.}. 

In Fig.\ref{fig:log1-R} we show the quantity
\begin{equation}  \label{eq:log_est}
\log \left( 1 - R_{\T A} \right) = \log c  - \varepsilon \, \log x, 
\end{equation}
plotted versus $\log x$ in comparison with data taken on Pb at 
small $Q^2$. 
This plot confirms that, for $x < 3 \cdot 10^{-3}$, 
the shadowing effect indeed 
approaches the high-energy behavior expected from the 
Regge limit of diffractive production.  
Deviations from this asymptotic behavior at larger values of $x$ indicate 
how shadowing gradually builds up as the coherence length 
$\lambda \propto x^{-1}$ starts to exceed nuclear length scales for 
low mass diffractively produced states. At sufficiently high energy or small
$x$, the coherence length becomes comparable to nuclear dimensions even for 
heavy hadronic intermediate states. 
Once a major fraction of diffractively produced states 
contribute to shadowing it starts to approach its asymtotic 
high-energy behavior. 
Note this asymptotic behavior sets in when the 
coherence lengths $\lambda_{\rm   X}$ of low mass 
hadronic fluctuations of the photon exceed by far the dimension of 
the nucleus.
For example at $x = 3\cdot 10^{-3}$ and $Q^2 \simeq 0.7$ GeV$^2$,   
which corresponds to the onset of the asymptotic behavior in 
Fig.\ref{fig:log1-R},  the $\rho$ meson coherence length becomes 
$\lambda_{\rho} \simeq 36$ fm.

\subsubsection{Shadowing for real photons}
\label{sssec:shad_photon}

Data on the diffractive production of hadrons 
in high-energy photon-nucleon interactions have 
been summarized in Section 
\ref{ssec:Diffractive_photoproduction}. 
They are useful to gain insight into the relative importance of 
$\rho$, $\omega$ and $\phi$ mesons, as compared to heavier 
states, for nuclear shadowing with real photons.

Diffractive $\gamma {\T N} \rightarrow {\T {X N}}$  production 
with $M_{\T X} \lsim 1$ GeV involves primarily the light vector mesons 
$\rho$, $\omega$ and $\phi$.
Nuclear shadowing at photon energies $\nu$ up to about $200$ GeV 
is largely determined by the coherent 
multiple scattering of those diffractively produced vector mesons. 
Their propagation lengths $\lambda \simeq 2 \nu/m_{\T V}^2$ 
easily exceed nuclear dimensions as soon as $\nu > 20$ GeV. 
With rising energies additional contributions to shadowing 
from diffractively produced states with larger masses, 
$M_{\T X} > 1$ GeV, become increasingly important.

\begin{figure}[t]
\bigskip
\begin{center} 
\epsfig{file=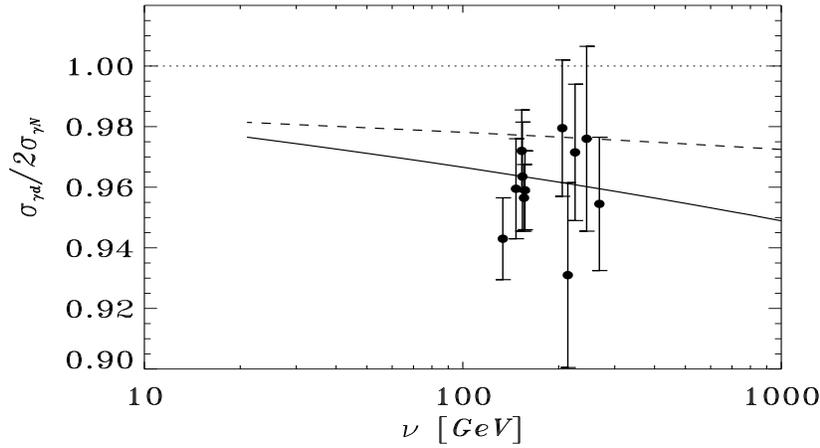,height=100mm,width=110mm}
\end{center}
\vspace*{-3cm}
\caption[...]{
The shadowing ratio $ R_{\T d} = \sigma_{\gamma \T d}/2 \sigma_{\gamma \T N}$ 
as a function of the photon energy. 
The dashed line shows the vector meson contribution. 
The experimental data are taken from the E665 collaboration 
\cite{Adams:1995sh}. (The energy values of the data have to be understood 
as average values which correspond to different $x$-bins.) 
}
\label{fig:shad_deut_photon}
\bigskip
\end{figure}

This behavior is illustrated for DIS from deuterium in 
Fig.\ref{fig:shad_deut_photon}, where  
we show the ratio of the total photon-deuteron cross section 
compared to the free photon-nucleon cross section, 
$R_{\T d} = \sigma_{\gamma {\T d}}/2 \sigma_{\gamma {\T N}}$,  from 
Ref.\cite{Piller:1997ny}. 
The empirical 
photon-proton cross section from \cite{Caldwell:1978yb} 
has been used for $\sigma_{\gamma {\T N}}$.  
The shadowing correction (\ref{eq:ds_corr_full}) 
has been calculated using a 
fit for the diffractive photon-nucleon cross section from 
Ref.\cite{Piller:1997ny}.

The observed energy dependence of shadowing in 
Fig.\ref{fig:shad_deut_photon} 
results from two sources as pointed out previously: 
the dependence of the diffractive and total photon-nucleon 
cross sections on energy implies 
$R_{\T d} - 1 \sim \nu^{0.1}$ for the shadowing 
ratio. 
An additional increase of shadowing with rising energy $\nu$ comes  
from diffractively produced states with 
large mass, $M_{\T X} > 1$ GeV, which become relevant 
at high energies.

\subsubsection{Shadowing in DIS at small and moderate $Q^2$}
\label{sssec:shad_smallQ}

So far nuclear shadowing has been measured only in 
fixed target experiments.
The kinematic conditions of such experiments imply that 
the data  for $x <  0.01$  had to be taken at small 
four-momentum transfers, 
$\overline {Q^2}\lsim 1$ GeV$^2$,   
as discussed  in Section \ref{subs:Nucl_F2}. 
The corresponding energy transfers are typically 
$50\,{\rm GeV} < \nu < 300\,{\rm GeV}$. 
The conclusions just drawn for real photons apply here too: 
nuclear shadowing as measured by  E665 and NMC  
receives major contributions from  the diffractive 
production and multiple scattering of vector mesons. 

In the  intermediate range  $0.01 < x < 0.1$, on the other hand, 
the E665 and NMC measurements 
involve momentum transfers up to $Q^2 \sim  30 \,{\T{GeV}}^2$. 
At $Q^2 > 1$ GeV$^2$ vector meson contributions  to  
diffraction and  shadowing decrease (Section \ref{ssec:Diffraction}) 
and hadronic states with 
masses $M_{\T X}^2 \sim Q^2$ become relevant. 
The data reveal that the $Q^2$-dependence of nuclear shadowing 
is very weak (Section \ref{subs:Nucl_F2}).  
This suggests that high-mass 
hadronic components of the photon which dominate    
the measured nuclear shadowing  at $Q^2 > 1$ GeV$^2$,  
interact strongly with the target, just like ordinary hadrons. 
The following section gives a schematic view of the scales 
involved, as outlined in Ref.\cite{Kopeliovich:1995ju}.

\subsection{Sizes, scales and shadowing}
\label{sssec:sizes} 

Consider DIS at small $x$ in the lab frame.
In this frame of reference the important feature is 
the nuclear interaction of hadronic fluctuations 
of the virtual  photon (see Section \ref{Sec:space_time}).  
Since the photon and its hadronic configurations 
carry high energy, the transverse separations and longitudinal 
momenta of their quark and gluon constituents are 
approximately conserved during the scattering process.
These hadronic configurations can be classified as ``small'' or ``large'', 
depending on their transverse extension. ``Large'' configurations 
have hadronic sizes of order 
$\Lambda_{\T{QCD}}^{-1} \sim 1$ fm, 
whereas ``small'' configurations are characterized by sizes which 
scale as $Q^{-1}$. 

The contribution of a given hadronic fluctuation, $\T h$, 
to the photon-nucleon interaction 
cross section is determined by its probability weight 
$w_{\gamma^*\T h}$ in the photon wave function,  
multiplied by  its cross section $\sigma_{\T {hN}}$. 
The virtual photon-nucleon cross section is:
\begin{equation}\label{eq:toy_1}
\sigma_{\gamma^* \T N} = \sum_{\T h} 
w_{\gamma^* \T h} \,\sigma_{\T {h N}}.
\end{equation}
The coherent interaction of the virtual 
photon with several nucleons behaves differently.
For example, the contribution of a hadronic fluctuation to  
double scattering, which dominates shadowing, is 
proportional to its weight in 
the photon wave  function multiplied by  the {\it square} of 
its interaction cross section. 
The double scattering correction to virtual photon-nucleus 
scattering is:  
\begin{equation} \label{eq:toy_2}
\sigma_{\gamma^* \T A}^{(2)} \sim  \sum_{\T h} w_{\gamma^*\T h} 
\,(\sigma_{\T {h N}})^2.
\end{equation}
Now, the probability to find a quark-gluon configuration of large 
size is suppressed (up to possible logarithmic terms)    
by $\Lambda_{{\T {QCD}}}^2/Q^2$ 
as compared to configurations with small transverse sizes.
On the other hand the interaction cross sections of  hadronic 
fluctuations are proportional to their squared transverse radii. 
These properties and their consequences for the cross sections 
in Eqs.(\ref{eq:toy_1}) and (\ref{eq:toy_2}) are summarized in Table 2. 
\begin{table}[h]
\begin{center}
\bigskip
\begin{tabular}{| c | c | c | c | c | }
\hline 

fluctuation h & $w_{\gamma^*\T h}$   & $\sigma_{\T{h N}}$ & 
                $w_{\gamma^*\T h}\,\sigma_{\T{h N}}$ & 
                $w_{\gamma^*\T h}\,(\sigma_{\T{h N}})^2$
\\ \hline
small size    & $ 1$  & $1/Q^2$       & $1/Q^2$  & 
                $1/Q^4$  
\\ \hline 
large size    & $\Lambda_{{\T {QCD}}}^2/Q^2$ & 
                $1/\Lambda_{{\T {QCD}}}^2$ & 
                $1/Q^2$ & $1/(\Lambda_{{\T {QCD}}}^2 Q^2) $ 
\\ \hline
\end{tabular}
\caption{Relative contributions of 
small- and large-size hadronic components of 
a virtual photon to DIS and shadowing at large $Q^2$ \cite{Kopeliovich:1995ju}.
}
\end{center}
\bigskip
\end{table}
For the scattering from individual 
nucleons one finds that both, large- and small-size 
configurations give leading contributions $\sim 1/Q^2$  to the 
photon-nucleon cross section (\ref{eq:toy_1}).
On the other hand contributions from small-size components 
to the shadowing correction $\sigma_{\gamma^* \T A}^{(2)}$ 
are suppressed by an additional power $1/Q^2$ 
as compared to large-size configurations 
(apart from contributions related to  diffractive 
production from the whole nucleus, not considered in this 
schematic picture).

In view of these scale considerations, we can now understand 
some of the previously mentioned empirical facts which, on first sight, 
seemed unrelated:
\begin{itemize}
\item [$\bullet$]
Nuclear shadowing varies only weakly with $Q^2$. 
\medskip
\item [$\bullet$]
The energy dependence of nuclear shadowing for $x\lsim 0.01$, 
as measured with  
fixed target experiments at CERN and FNAL, 
is reminiscent of hadron-nucleus collisions.
\end{itemize}
These features follow from the fact that, 
to leading order in $Q^2$,   shadowing is primarily determined 
by the interaction of large-size hadronic 
fluctuations of the exchanged photon, even at large $Q^2$.
These hadronic configurations are expected to interact like  
ordinary hadrons.

Note, those observations can be applied to diffraction as well as 
to shadowing,  given that the two phenomena are closely 
connected as established in the previous sections: 
diffraction is also a scaling effect, i.e. it survives at large $Q^2$.
Its energy dependence is expected to behave 
similarly  as in  hadron collisions.  
(For limitations to this simple picture see Sections 
\ref{ssec:Pert_Nonpert_Shad} and \ref{sec:shad_largeQ}.)

\subsection{Nuclear shadowing and parton configurations of the photon}
\label{ssec:Pert_Nonpert_Shad}

The results of the previous sections are eludicated 
by making contact with the underlying basic QCD and the parton 
structure of the virtual photon. 
The photon wave function can be decomposed in a Fock space 
expansion,
\begin{equation} \label{eq:Fock}
|\gamma \rangle = c_0 |\gamma_0 \rangle + 
                  c_{q \bar q} | q \bar q \rangle + 
                  c_{q \bar q g} |q \bar q g\rangle + \dots ,
\end{equation}
in terms of a ``bare'' photon state  $|\gamma_0 \rangle$ 
and partonic (quark-antiquark and gluonic) components. 
At large $Q^2$ the minimal Fock component $|q \bar q \rangle$
dominates the hadronic part of $|\gamma \rangle$, 
higher Fock states enter with powers of the strong coupling 
$\alpha_s$. 
Let us now have a closer look at this minimal Fock component.

\begin{figure}[t]
\bigskip
\begin{center} 
\epsfig{file=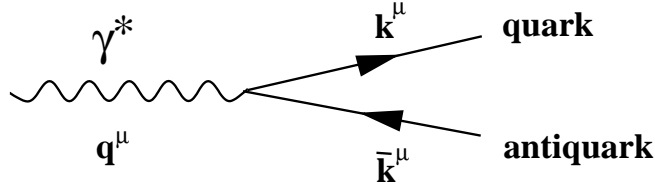,height=25mm}
\end{center}
\caption[...]{
Decomposition of a virtual photon into a quark-antiquark 
pair at large $Q^2$.
}
\label{fig:qqbar}
\bigskip
\end{figure}

Consider a virtual photon of four-momentum 
$q^{\mu} = (\nu, \vec q)$, with  $\vec q = (0, 0, q_3)$ 
and $q_3 > 0$  
defining the longitudinal direction, and $Q^2 = - q^2$. 
Let this photon split into a quark-antiquark pair as sketched in 
Fig.\ref{fig:qqbar}. 
The quark has a four-momentum $k^{\mu} = (k_0,\vec k)$ with 
$\vec k = (\vec k_{\perp},k_3)$. The fraction of the photon light-cone 
momentum carried by the quark is 
\begin{equation}
\xi = \frac{k^+}{q^+} = \frac{k_0+ k_3}{\nu + q_3}.
\end{equation}
The momentum fraction of the antiquark with 
$\bar k^{\mu} = q^{\mu} - k^{\mu}$ is obviously $1-\xi$.

For the longitudinally polarized photon, the wave function of its minimal 
$q\bar q$ fluctuation in momentum space is proportional to the 
longitudinal component of the quark pair current, multiplied by its 
propagator 
$\left[Q^2 + \frac{m_q^2 + k_{\perp}^2}{\xi (1-\xi)} \right]^{-1}$ 
where $m_q$ is the quark mass
\cite{Lepage:1980fj}. 
The quantity
\begin{equation} \label{eq:qq_mass}
M_{\T X}^2 = \frac{m_q^2 + k_{\perp}^2}{\xi (1-\xi)} 
\end{equation}
can be interpreted as the squared effective mass of the propagating 
$q\bar q$ pair.

\begin{figure}[t]
\begin{center} 
\epsfig{file=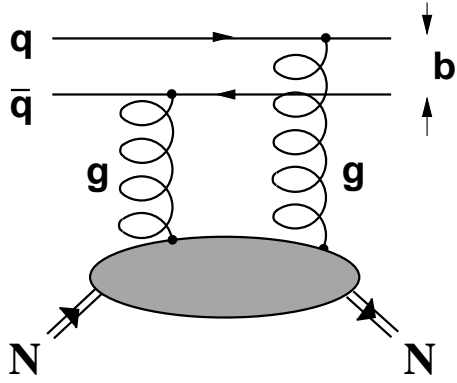,height=50mm}
\end{center}
\caption[...]{
Short distance interaction of a color singlet quark-antiquark
pair with a nucleon through two-gluon exchange.
}
\label{fig:two_gluon}
\bigskip
\end{figure}

It is useful to perform the two-dimensional Fourier transform 
with respect to 
the transverse quark momentum $\vec k_{\perp}$ conjugate to the 
transverse separation $\vec b$ of the $q\bar q$ pair. 
Neglecting the quark mass at large $Q^2$ and using generically 
one single quark flavor, the squared wave function of the 
$q\bar q$ component coupled to the longitudinally polarized 
photon becomes \cite{Bjorken:1971ah,Nikolaev:1991ja,Brodsky:1997nj}:
\begin{equation} \label{eq:qq_L}
\left|\psi_{q\bar q}^{L}(b,\xi;Q^2)\right|^2 
= 
\frac{6 \alpha}{\pi^2} \,
Q^2 \xi^2 (1-\xi)^2 \, 
K_0^2\left(b\,Q\, \sqrt{\xi (1-\xi)}\right), 
\end{equation}
where $K_n$ denotes  modified Bessel functions. 
The $\gamma^* \rightarrow q\bar q$ coupling is proportional 
to the fine structure constant $\alpha \simeq 1/137$. 
The corresponding result for a virtual photon with transverse 
polarization is:
\begin{equation}\label{eq:qq_T}
\left|\psi_{q\bar q}^{T} (b,\xi;Q^2)\right|^2 
= 
\frac{3 \alpha}{2\pi^2} \,Q^2 \xi  (1-\xi)    
\left(\xi^2 + (1-\xi)^2 \right) 
K_1^2\left(b\,Q\, \sqrt{\xi (1-\xi)}\right).
\end{equation}
Consider now DIS from a free nucleon at small Bjorken-$x$ 
and large $Q^2$. At the high energies involved 
the photon in the laboratory frame acts like a 
beam of $q\bar q$ pairs, and one can write the cross section for the 
longitudinally or transversely polarized virtual photon with the nucleon 
in the form \cite{Nikolaev:1991ja,Frankfurt:1997ri}
\begin{equation} \label{eq:sigma_LT}
\sigma_{\gamma^* \T N}^{L,T} =  \int d^2b\, \int_0^1 d\xi \, 
\left|\psi_{q\bar q}^{L,T} (b,\xi)\right|^2 \, 
\sigma_{q\bar q {\T N}}(b,x),
\end{equation}
using the wave functions of the leading $q\bar q$ fluctuations.
These wave functions as well as the $q\bar q$-nucleon cross section 
$\sigma_{q\bar q {\T N}}$
depend on the transverse separation $b$ of the quark pair. 
Since the modified Bessel functions in Eqs.(\ref{eq:qq_L}, \ref{eq:qq_T})
drop as  $K_{1,0}(y) \rightarrow e^{-y}$ at large 
$y$, the wave functions $\psi_{q\bar q}^{L,T}$ receive their dominant 
contributions from configurations with transverse size
\begin{equation}
b^2 \sim \frac{1}{Q^2 \,\xi\,(1-\xi)}\,. 
\end{equation}
Consequently, $q\bar q$ configurations at large $Q^2$ 
with comparable momenta of the quark and 
antiquark, $\xi \sim 1-\xi \sim 1/2$, 
have small transverse size, 
$b^2 \sim 1/k_{\perp}^2 \sim 1/Q^2$, or equivalently, 
large transverse momentum.
The interaction of these ``non-aligned'' configurations 
with the nucleon is therefore determined by the 
short transverse distance behavior of the cross section 
$\sigma_{q\bar q {\T N}}$  which can be calculated using 
perturbative QCD.
The reasoning goes as follows. At large $Q^2$ the leading mechanism 
responsible for the short distance interaction of the $q\bar q$ pair 
with the nucleon is two-gluon exchange (see Fig.\ref{fig:two_gluon}). 
The (color singlet) $q\bar q$ pair acts as a color dipole. Its 
interaction strength with the nucleon or any other 
(color singlet) hadron is determined by the squared color dipole 
moment, hence $\sigma_{q\bar q {\T N}}$  is proportional to $b^2$ for 
small transverse separations $b$. 
In the leading-logarithmic approximation valid at large $Q^2$ one 
derives \cite{Frankfurt:1997ri,Blaettel:1993rd}
\begin{equation} \label{eq:sigma_qq}
\sigma_{q\bar q {\T N}} (b,x) = \frac{\pi^2}{3}  \alpha_s(Q^2) 
\,b^2 \,x\,g_{\T N}(x,Q^2), 
\end{equation}
with the strong coupling constant $\alpha_s$. 
The $Q^2$ scale in (\ref{eq:sigma_qq}) is set by 
$Q^2\sim 1/b^2$. All non-perturbative effects are 
incorporated in the gluon distribution $g_{\T N}(x,Q^2)$ 
of the target nucleon. 

Small $q \bar q$ configurations interact only weakly according to 
Eq.(\ref{eq:sigma_qq}). 
This is the case for the kinematic conditions 
realized in fixed target experiments at CERN 
and FNAL  
(see Sections \ref{ssec:spin_ind_strfns} and \ref{subs:Nucl_F2}). 
The situation is  different at $x \ll 0.1$ and $Q^2 \gg  1\,\rm{GeV^2}$, 
the extreme region accessible at HERA. Here the strong increase of the 
nucleon structure function  $F_{2}^{\T N}$ at very small $x$ is 
accompanied by a correspondingly strong increase of the gluon 
distribution function. 
The gluon density becomes so large with decreasing $x$ that, even for 
small $b^2 \sim 1/Q^2$, the cross section 
$\sigma_{q\bar q {\T N}}$ can eventually reach magnitudes typical 
for ordinary hadrons \cite{Frankfurt:1996jw}. 

If either the quark or the antiquark becomes soft (that is: 
if either the momentum fraction $\xi$ or $1-\xi$ tends to zero), 
large $q\bar q$ separations contribute to the wave functions 
(\ref{eq:qq_L}, \ref{eq:qq_T}). 
In this limit perturbative QCD is not applicable. The interaction 
cross section  for such large-size configurations with small transverse 
momentum is supposed to be similar to typical hadron-nucleon cross 
sections \cite{Bjorken:1973gc}.

A detailed analysis of the ``transverse'' wave function 
(\ref{eq:qq_T}) shows that both ``small'' (non-aligned) and 
``large'' quark-antiquark configurations give 
leading $1/Q^2$ contributions to the transverse photon-nucleon cross 
section in accordance with our previous discussion. 
The situation is different for longitudinally polarized photons 
(see e.g. (\ref{eq:qq_L})). 
In this case the contributions from ``soft'' quarks 
(with $\xi \rightarrow 0$ or  $1-\xi \rightarrow 0$) are suppressed as 
$1/Q^4$ so that, to leading order in the strong coupling $\alpha_s$, 
only small size $q\bar q$ pairs contribute to $\sigma_{\gamma^*\T N}^L$. 
At next to leading 
order in $\alpha_s$, the Fock expansion (\ref{eq:Fock}) 
introduces quark-antiquark-gluon states. Large size $q\bar q g$ 
configurations are now important, and they are not suppressed by additional 
powers of $1/Q^2$ \cite{Buchmuller:1997xw}. 

At small momentum transfers, $Q^2 \lsim 1\,\rm{GeV^2}$, 
configurations of large size dominate the $q\bar q$ wave function. 
Strong interactions between quark and antiquark now lead to the formation 
of soft hadronic fluctuations including vector mesons and multi-pion states. 
Consequently, photon-nucleon cross sections at small $x$ and small $Q^2$ 
receive important contributions from the low mass vector mesons. For 
example, at $Q^2 \simeq 0.5$ GeV$^2$ almost half of the measured 
nucleon structure function $F_2^{\T N}$ comes from 
$\rho,\omega$ and $\phi$ mesons according to the calculation in
Ref.\cite{Piller:1995kh}. 

So far we have focused this discussion on free nucleons. Similar
considerations apply to deep-inelastic scattering from 
nuclei which involves the interaction of hadronic components of the 
virtual photon  with the nuclear many-body system.
To leading order in $\alpha_s$  the photon-nucleus 
cross sections are  now obtained from Eq.(\ref{eq:sigma_LT}) replacing 
$\sigma_{q\bar q \T N}$ by the corresponding $q\bar q$-nucleus  
cross section $\sigma_{q\bar q \T A}$. 
The cross section  $\sigma_{\T {h A}}$ of any hadronic fluctuation   
$\T h$  interacting with a nucleus at high energies, 
can be related to the 
cross section for the scattering from free nucleons by the 
Glauber-Gribov multiple scattering formalism 
\cite{Gribov:1970,Bertocchi:1972}. 
For a Gaussian nuclear density (\ref{eq:Gauss}) this leads to:
\begin{equation} 
\sigma_{\T{h A}}\approx
A\,\sigma_{\T{h N}}
\left[
1- \sigma_{\T{h N}} \,\frac{3}{16 \pi}\,
\frac{A-1}{\langle r^2\rangle_{\T A}}\,
\exp \left(-\frac {\langle r^2\rangle_{\T A} }{3 \,\lambda^2}\right) 
+\dots\right],\label{eq:adep}
\end{equation}
where $\lambda$ is the propagation length associated with the 
hadronic fluctuation. 
Double scattering gives  a negative correction  
proportional  to the squared cross section of the hadronic fluctuation.
Only those hadronic configurations  
with large interaction cross sections contribute significantly 
to shadowing.
Furthermore, since the nuclear mean square radius behaves 
approximately as $\langle r^2\rangle_{\T A} \sim A^{2/3}$,  
the magnitude of the double scattering correction grows for large nuclei 
with the radius of the target, i.e.  proportional to $A^{1/3}$. 
The exponential in (\ref{eq:adep})  
ensures that only hadronic fluctuations 
with propagation lengths $\lambda$ larger than  
the target dimension contribute significantly to shadowing. 
For small-sized fluctuations, interesting effects beyond 
those covered by Eq.(\ref{eq:adep}) arise from diffractive  production 
on the whole nucleus.

In accordance with our discussion in the previous section we can conclude:

\begin{itemize}

\item[i)] 
In the fixed target experiments at CERN 
and FNAL, where small values of $x < 0.01$ are accessible only at 
small average momentum transfers, $\overline Q\,^2 \lsim 1$ GeV$^2$, 
nuclear shadowing is governed by interactions of configurations 
with large transverse sizes.
Contributions 
from the vector mesons $\rho, \omega$ and $\phi$ turn out to 
be particularly important.

\medskip

\item[ii)] 
At very small $x$ together with very large $Q^2$, the growing number 
of partons in the photon-nucleus system makes them interact like 
ordinary hadrons, even if the parton configurations 
have small transverse sizes inversely proportional to $Q^2$. 
One now expects a complex interplay between soft (large-size) and hard 
(small-size) partonic components of the interacting photon which 
can no longer be classified by simple book-keeping in powers of 
$1/Q^2$. 
 
\end{itemize}

\subsection{Models} 
\label{ssec:Shad_Model}

In the following we sketch  several models which 
have been used quite successfully 
to describe nuclear shadowing as measured in  
experiments at 
CERN 
and FNAL.   
As before we restrict ourselves to lab-frame descriptions.
We do not aim for completeness but 
rather emphasize common features   
of various models and their implications 
for the underlying mechanism of nuclear shadowing.

\subsubsection{Vector mesons and aligned jets}
\label{subs:AJM}

As discussed in Section \ref{ssec:Pert_Nonpert_Shad}, the quark-antiquark 
fluctuation  of a virtual photon starts out with a 
transverse size $b^2 \sim [Q^2 \xi (1-\xi)]^{-1}$ 
where $\xi$ is the fraction of longitudinal photon momentum 
carried by one of the quarks. 
``Non-Aligned'' $q\bar q$ configurations 
with $\xi \simeq  1/2$ have small transverse size and interact 
weakly; 
``aligned'' ones with $\xi \sim 0 $ or $\xi \sim 1$ 
have large transverse size and are likely, by subsequent 
strong interactions, to turn into vector mesons if the 
$q\bar q$ invariant mass matches appropriately.

\begin{figure}[t]
\bigskip
\begin{center} 
\epsfig{file=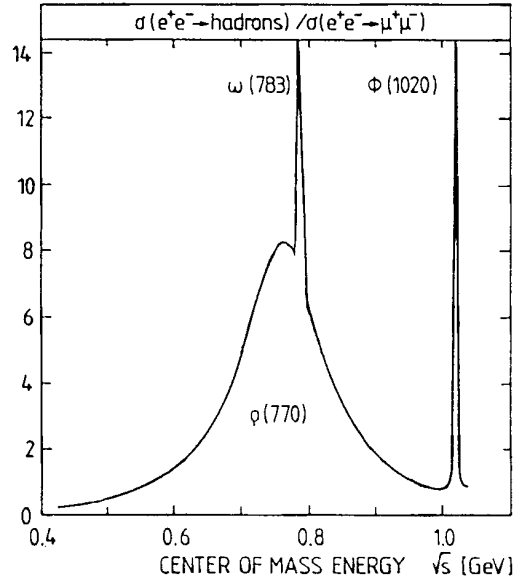,height=80mm}
\end{center}
\vspace*{1cm}
\begin{center} 
\epsfig{file=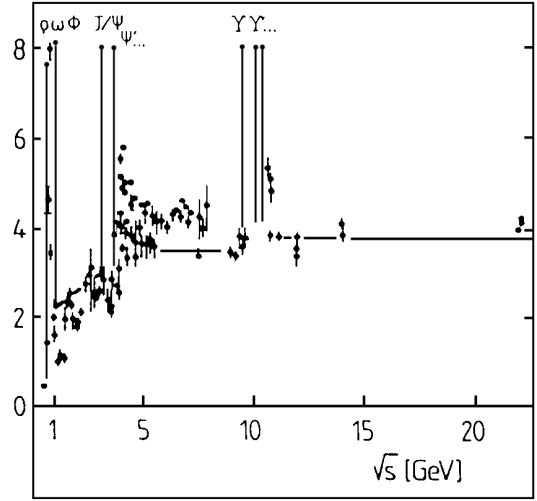,height=80mm}
\end{center}
\caption[...]{Cross section $\sigma_{e^+ e^- \rightarrow {\T {hadrons}}}/
\sigma_{e^+ e^- \rightarrow \mu^+\mu^-}$. 
}
\label{fig:ratio_epem/mupmum}
\bigskip
\end{figure}

Models which combine aspects of vector meson dominance 
and the aligned-jet picture \cite{Bjorken:1973gc}
are described in 
Refs.\cite{Piller:1995kh,Frankfurt:1989zg}. 
Their starting point is the hadronic spectrum of the virtual 
photon exchanged in the deep-inelastic scattering process.
The spectral function, $\Pi(s)$, is determined by the cross 
section for electron-positron annihilation into hadrons, 
where $s=q^2$ is the squared photon or $e^+e^-$ center-of-mass energy:
\begin{equation}  \label{eq:photon_spectral_function}
\Pi(s) = \frac{1}{12 \,\pi^2}
\frac{\sigma_{e^+ e^- \rightarrow hadrons}(s)} 
{\sigma_{e^+ e^- \rightarrow \mu^+ \mu^-}(s)}\,, 
\end{equation}
with 
\begin{equation} \label{eq:photon_spectral_function_2}
\Pi(q^2) = -\frac{1}{3\,q^2} 
\sum_{\T X} (2\pi)^3 \delta^4(q-p_{\T X}) 
\left\langle 0 \right| J_{\mu}(0) \left| {\T X} \right\rangle
\left\langle {\T X} \right| J^{\mu}(0) \left| 0 \right\rangle.
\end{equation} 
Here $J^{\mu}$ is the electromagnetic current operator. 
The sum in Eq.(\ref{eq:photon_spectral_function_2}) is taken 
over all hadronic fluctuations of the photon with 
four-momenta $p_{\T X} = q$ and squared invariant masses 
$\mu^2 \equiv p_{\T X}^2 = q^2$.
At small center-of-mass energies, 
$s\,\lsim \,1$GeV$^2$, the spectrum 
(\ref{eq:photon_spectral_function_2}) is dominated by the vector mesons 
$\rho$, $\omega$ and $\phi$ as shown  in Fig.\ref{fig:ratio_epem/mupmum}. 
The high energy spectrum at $s>1$ GeV$^2$ is characterized by 
quark-antiquark continuum plateaus together with isolated 
charmonium and upsilon resonances.

The lab frame space-time pattern of deep-inelastic scattering 
(Section \ref{Sec:space_time}) suggests that the 
nucleon structure function at small $x$ 
can be described by the following 
expression 
\cite{Gribov:1970,Frankfurt:1989zg}:
\begin{equation}  
\label{eq:F2N_AJ}
F_{2}^{\T N}(x,Q^2) =
 \frac{Q^2}{\pi}
 \int_{4 m_{\pi}^2}^{\mu_{max}^2}\,d\mu^2 
 \,
 \frac{\mu^2 \,\Pi (\mu^2)}
      {\left(\mu^2+Q^2\right)^2}\;
\int_0^1\, d \xi \, 
 \sigma_{\T{hN}}(W,\mu^2; \xi)\,.
\end{equation}
The basic idea behind this ansatz is the following. 
For $x\ll 1$, or large lab frame propagation length 
$\lambda \sim 2\nu/(Q^2 + \mu^2)$ of a 
given $q\bar q$ fluctuation of mass $\mu$, the 
vacuum spectrum $\Pi(\mu^2)$ remains more or less 
unaffected by the presence of the target nucleon. 
The high-energy virtual photon with $\nu \gg Q^2/2M$ 
behaves like a beam of hadrons with masses 
$\mu < \mu_{\max}$. 
Their maximum possible mass is determined by the condition 
that $\lambda$ must exceed the size $R\simeq 5 M^{-1}$ 
of the target nucleon, so that (roughly) 
$\mu_{\max} \sim \sqrt{\nu M}$. 
The interaction of this beam with the nucleon is described 
by the cross section $\sigma_{\T {hN}}$ which depends on 
$\mu^2$ and on the hadron/photon-nucleon center-of-mass energy 
$W = \sqrt{2 M \nu + M^2 - Q^2} \simeq \sqrt{2 M \nu} 
= Q/\sqrt x$. 
For a $q\bar q$ pair treated to leading order in $\alpha_s$, 
it also depends on the fraction $\xi$ of the photon light-cone 
momentum carried by the quark. 
The sum in Eq.(\ref{eq:F2N_AJ}) is taken over hadronic fluctuations of 
the photon with fixed invariant mass. 
The ansatz neglects contributions to the 
forward virtual photon scattering amplitude in which the mass 
$\mu$ can change during the interaction. 
\begin{figure}[t]
\bigskip
\begin{center} 
\epsfig{file=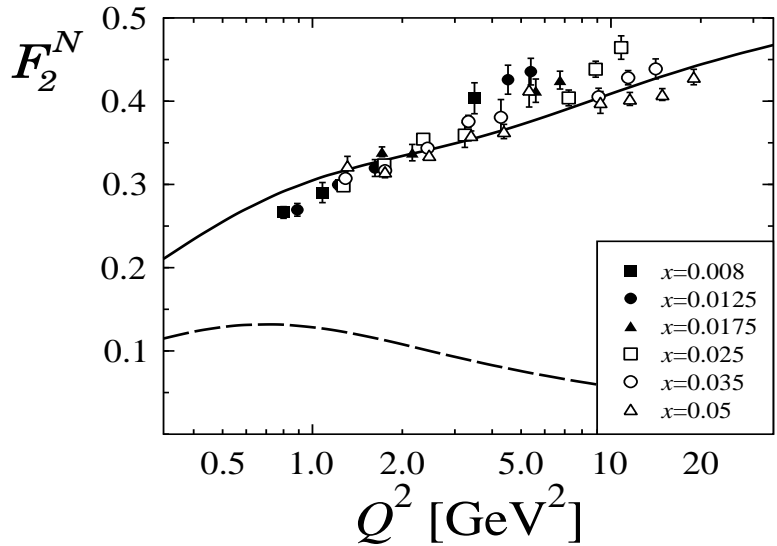,height=80mm,width=120mm}
\end{center}
\caption[...]{The nucleon structure function $F_2^{\T N}$ for small $x$ 
plotted against $Q^2$. The full line has been  obtained 
in Ref.\cite{Piller:1995kh} from Eq.(\ref{eq:F2N_AJ}).  
The dashed line indicates the contribution 
of the vector mesons $\rho$, $\omega$ and $\phi$. 
The data are from the NMC \cite{Amaudruz:1992bf}. 
}
\label{fig:fn_ajm}
\bigskip
\end{figure}

The small-$x$ structure function $F_{2}^{\T N} (x,Q^2)$ 
as given in Eq.(\ref{eq:F2N_AJ}) is governed  
by contributions from intermediate hadronic 
states with an invariant mass 
$\mu^2\sim Q^2$. 
For small momentum transfers, 
$Q^2\,\lsim\,  1\,\mbox{GeV}^2$, low mass vector mesons $\rho,\;\omega$ and 
$\phi$ are of major importance.  
Their dominance leads to the scale 
breaking behavior $F_{2}^{\T N}(x,Q^2) \sim Q^2$ for 
$Q^2 \rightarrow 0$ at small $x$.

For larger momentum transfers, 
$Q^2>m_{\phi}^2\approx 1\,\mbox{GeV}^2$,
the structure function $F_{2}^{\T N}$ is determined primarily 
by the interactions of quark-antiquark pairs from the ${q\bar q}$ continuum.
The color singlet nature of  hadronic 
fluctuations of the virtual photon implies that their interaction 
cross section is  proportional 
to their transverse size.
Quark pairs with momenta ``aligned'' along  the direction of 
the virtual photon have a large transverse size. 
Their cross sections should be comparable to typical 
hadronic cross sections.
On the other hand ``non-aligned'' quarks are characterized by 
small transverse size. 
Their cross sections should therefore be small. 
\begin{figure}[t]
\bigskip
\begin{center} 
\epsfig{file=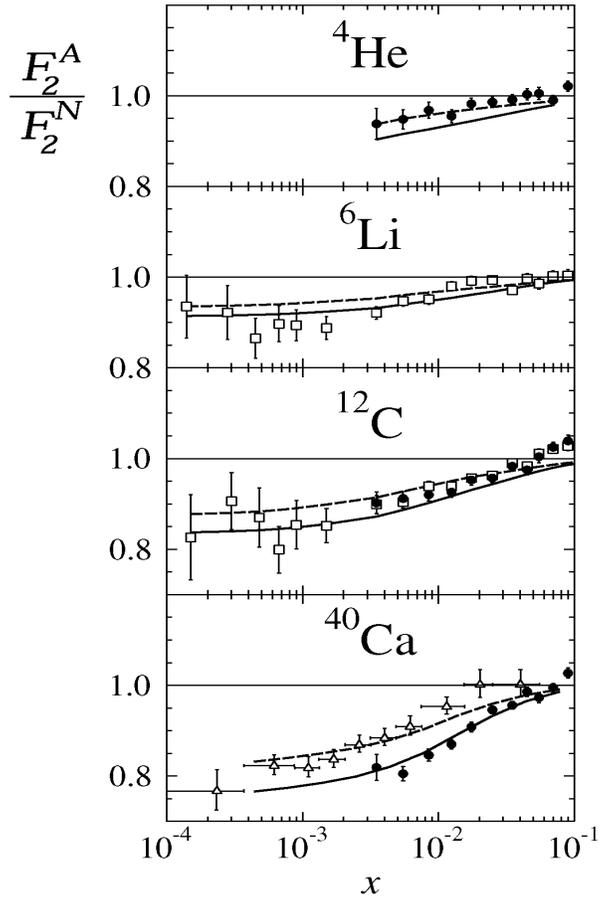,height=120mm,width=8cm}
\end{center}
\caption[...]{Results from Ref.\cite{Piller:1995kh} for the shadowing 
in He, Li, C, and Ca compared to experimental data from NMC 
(dots and squares) 
\cite{Amaudruz:1995tq,Arneodo:1995cs} and
FNAL (triangles) \cite{Adams:1995is}. 
The dashed curves show the shadowing caused by the vector mesons 
$\rho$, $\omega$ and $\phi$ only, 
the solid curves are the results including the $q\bar q$ continuum.
}
\label{fig:fa_ajm}
\bigskip
\end{figure}

With these ingredients, Eq.(\ref{eq:F2N_AJ}) gives a good 
description of the free nucleon structure function  
$F_{2}^{\T N}$ for $x < 0.1$ and moderate $Q^2$. 
A comparison with data from NMC is shown in Fig.\ref{fig:fn_ajm}. 
While the vector meson contribution vanishes as $1/Q^2$ for large $Q^2$, 
the ${q\bar q}$ continuum pairs are responsible for scaling, 
$F_{2}^{\T N}(x,Q^2) \sim \ln (Q^2)$, 
at large $Q^2$. 
Note however the importance of vector mesons at small $Q^2$. 
One finds that at $Q^2 = 1\,\mbox{GeV}^2$ almost half of 
$F_{2}^{\T N}$ at $x=0.01$ is due to vector mesons. 
At $Q^2= 10\,\mbox{GeV}^2$ they still contribute about $15\%$.
In these calculations the vector meson part of the spectrum 
$\Pi(\mu^2)$ is \cite{Bauer:1978iq}:  
\begin{equation} \label{eq:phot_vm}
\Pi^{(\T V)}(\mu^2) = 
\left(\frac{m_{V}}{g_{V}} \right)^2 
\delta(\mu^2 - m_{V}^2)
\end{equation}
with $V = \rho, \omega, \phi$, the empirical vector meson masses 
$m_{\T V}$ and the coupling constants 
$g_{\rho} = 5.0$, $g_{\omega} = 17.0$ and $g_{\phi} = 12.9$. 
The vector meson-nucleon cross sections are 
$\sigma_{\rho \T N} \approx \sigma_{\omega \T N} \approx 25$ mb,  
$\sigma_{\phi  \T N} \approx 10$ mb.

Nuclear structure functions $F_{2}^{\T A}$ for $x<0.1$  are 
expressed in an analogous way as in Eq.(\ref{eq:F2N_AJ}),  
with the hadron-nucleon cross sections $\sigma_{\T{hN}}$ 
replaced by the 
corresponding hadron-nucleus cross sections $\sigma_{\T{hA}}$. 
The relation between $\sigma_{\T{hA}}$ and $\sigma_{\T{hN}}$  
is  given by Glauber-Gribov multiple scattering theory, 
see Eq.(\ref{eq:adep}). 
In Fig.\ref{fig:fa_ajm} we present typical results for the shadowing 
ratio\footnote{It is common practice to normalize $F_2^{\T A}$ such that 
it represents the nuclear structure function {\em per nucleon}.}
$R_{\T A} = F_2^{\T A}/F_2^{\T N}$ from Ref.\cite{Piller:1995kh}. 

\begin{figure}[t]
\bigskip
\begin{center} 
\hspace*{2cm}
\epsfig{file=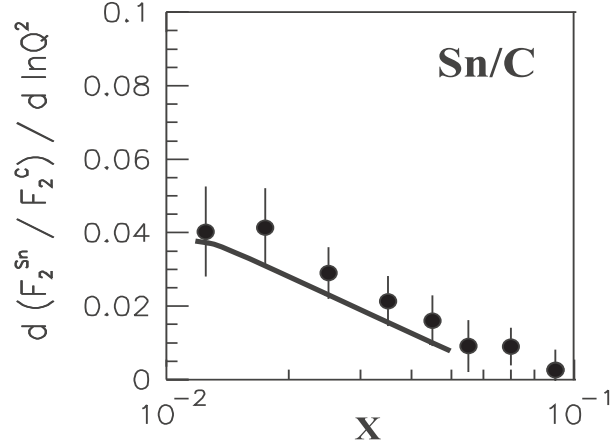,height=110mm,width=130mm}
\end{center}
\vspace*{-4cm}
\caption[...]{
The slope $b = d (F_2^{\T {Sn}}/F_2^{\T C})/d\ln Q^2$ 
indicating the $Q^2$ dependence of the shadowing ratio 
$\T{Sn}/{\T C}$. The calculation is described in   
\cite{Piller:1995kh}. Data are taken from \cite{Arneodo:1996ru}.
}
\label{fig:rqsq_ajm}
\bigskip
\end{figure}

Finally we comment on the observed weak $Q^2$-dependence 
of the shadowing effect. 
In the spectral ansatz (\ref{eq:F2N_AJ})  the given value of $Q^2$ 
selects that part of the hadron mass spectrum around $\mu^2 \sim Q^2$ 
which dominates the interaction, and hence  determines which 
cross sections $\sigma_{\T{hN}}(\mu^2)$ contribute significantly 
to the multiple scattering series. 
While the interaction cross sections decrease 
as $1/\mu^2$  with increasing mass as required by Bjorken scaling, 
pairs which are aligned with the photon momentum 
interact with large cross sections, even for large 
$\mu$, and therefore produce strong shadowing.
This is the reason for the very weak overall $Q^2$-dependence 
of shadowing  in this framework.
A comparison of results from Ref.\cite{Piller:1995kh} 
with NMC data for the slope $b$ of the ratio 
$F_2^{\T {Sn}}/F_2^{\T C} \approx a + b \ln Q^2$ 
is presented in Fig.\ref{fig:rqsq_ajm}. 
For a more detailed discussion of these issues 
including QCD corrections, see Ref.\cite{Frankfurt:1988nt}. 

\subsubsection{Vector meson dominance and pomeron exchange}

As indicated in Eqs.(\ref{eq:ds_corr_full},\ref{eq:ms_A}),
nuclear shadowing 
is directly related to the diffractive production  
cross section $d\sigma^{diff}_{\gamma^* {\T N}}/dM_{\T X}^2 \,dt$ 
or, equivalently, to the diffractive structure function 
$F_2^{D(4)}$.

Diffractive production  at $Q^2 \,\lsim\, 1\,\rm{GeV}^2$ 
is dominated by the excitation of 
the vector mesons $\rho$, $\omega$ and $\phi$. 
Their contributions can be described within the framework 
of vector meson dominance (see e.g. \cite{Bauer:1978iq}). 
Neglecting transitions between different vector mesons and 
omitting contributions from longitudinally polarized 
virtual photons one finds: 
\begin{equation} \label{eq:ms_VM}
\left.
\frac{d\sigma^{diff\,({\T V})}_{\gamma^* {\T N}} } {dM_{\T X}^2 dt}
\right|_{t\approx 0} 
= \frac{\alpha}{4}  \,
\frac{\Pi^{(\T V)}(M_{\T X}^2) \,M_{\T X}^2}{(Q^2 + M_{\T X}^2)^2}
\,\sigma_{{\T {XN}}}^2\,.
\end{equation}
Here the vector meson part (\ref{eq:phot_vm}) 
of the photon spectral function enters.   
Combining Eqs.(\ref{eq:ms_A},\ref{eq:ms_VM}) shows that 
the contribution of vector mesons to nuclear shadowing vanishes 
indeed as $1/Q^2$.

\begin{figure}[t]
\bigskip
\begin{center} 
\epsfig{file=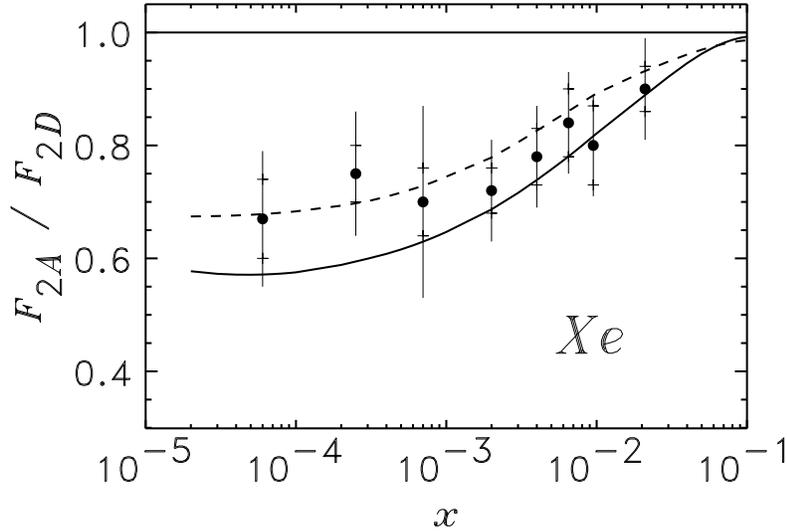,height=70mm}
\end{center}
\caption[...]{Shadowing in $\T{Xe}$. 
Details of the calculation are given in Ref.\cite{Melnitchouk:1993vc}.  
The dashed curve shows the contribution of vector mesons 
$\rho$, $\omega$ and $\phi$, while the solid curve includes 
pomeron exchange. The data 
are from the E665 collaboration \cite{Adams:1992nf}.
}
\label{fig:wally_shad_xe}
\bigskip
\end{figure}

The diffractive excitation of heavy mass states is commonly 
parametrized according to the Regge ansatz as in 
Eq.(\ref{eq:F_2D4_regge}). 
Most descriptions concentrate on the dominant contribution from 
pomeron exchange. 
The pomeron structure function $F_{2}^{\pom}$  is modeled 
in agreement with  available data on diffraction. 
At large $Q^2$ it is supposed 
to scale, i.e. it depends  at most logarithmically on $Q^2$. 
This leads to scaling for nuclear shadowing at large $Q^2$. 
On the other hand, at $Q^2 \ll 1\,\rm{GeV}^2$ one assumes 
$F_{2}^{\pom} \sim Q^2$ \cite{Kwiecinski:1988ys}  
which ensures that the shadowing 
correction to the nuclear structure function, $\delta F_{2}^{\T A}$, 
vanishes at $Q^2 = 0$, just like $F_{2}^{\T A}$ itself. 
Investigations of shadowing effects along theses lines can 
be found in 
\cite{Melnitchouk:1993vc,Kwiecinski:1988ys,Badelek:1992qa,%
Nikolaev:1991yw,Nikolaev:1992gw,Zoller:1992ns}. 
In Fig.\ref{fig:wally_shad_xe} a typical result for shadowing in $\T {Xe}$ 
from \cite{Melnitchouk:1993vc} is shown. 

\subsubsection{Generalized vector meson dominance}
\label{sssec:GVMD}

Generalized vector meson dominance models describe deep-inelastic 
lepton scattering at small $x$ purely in terms of 
hadronic degrees of freedom 
\cite{Sakurai:1972wk,Sakurai:1972zs,Schildknecht:1973gi,%
Fraas:1975gh,Devenish:1976ky,Bilchak:1988zn,%
Ditsas:1975vd,Ditsas:1976yv,Shaw:1989mn}.
For a free nucleon this leads to the following picture:
prior to its interaction with the target 
the virtual photon fluctuates
into a hadronic state with invariant mass $\mu$.
This fluctuation scatters from the target and 
converts into a hadronic state with mass $\mu'$.  
For transversely polarized photons this translates into a 
forward Compton  amplitude of the form:
%
\begin{equation}  \label{eq:A_gammaN}
{\cal A}_{\gamma^*{\T N}}^T (W,Q^2)  
\sim 
\int d\,\mu^2 \int d{\mu'}^2 
\frac{\rho(\mu,\mu';W)}{(\mu^2 + Q^2)({\mu'}^2 + Q^2)}\,, 
\end{equation}
with a double spectral distribution $\rho$ depending on the 
photon-nucleon center-of-mass energy $W$.
Here the integrals over initial and final hadronic fluctuations  
and their propagators are made explicit.

The continuum of hadronic intermediate states 
which determines the double spectral function $\rho(\mu,\mu';W)$ 
is  commonly approximated by a discrete set of narrow vector meson states  
${\T V}_n$ $(n=1,2,\dots)$.
The  resulting transverse photon-nucleon cross section is: 
%
\begin{equation} \label{eq:GVMD_sigmaT}
\sigma^T_{\gamma^* {\T N}} = \sum_{n,m} \frac{e}{g_m} 
\frac{M^2_{m}}{M^2_{m}+Q^2} \Sigma_{mn}(W)
\frac{M^2_{n}}{M^2_{n}+Q^2}\frac{e}{g_n}\,.
\end{equation}
In Refs.\cite{Fraas:1975gh,Ditsas:1976yv} the vector mesons are assumed to 
be equally spaced 
in mass,  starting with the $\rho$-meson. 
The photon-vector meson couplings $g_n$ are chosen 
to reproduce average scaling in  $e^+ e^-$ annihilation into hadrons
(see Fig.\ref{fig:ratio_epem/mupmum}).
$\Sigma_{mn}$ denotes  the imaginary part of the vector meson-nucleon 
transition amplitude, 
${\T V}_m {\T N} \rightarrow {\T V}_n \T N$, in the forward direction. 
For diagonal terms it is equal to the total ${\T V}_n$-nucleon cross 
section, $\Sigma_{nn} = \sigma({{\T V}_n \T N})$, 
which is taken to be constant.

The next step in simplification is to consider 
only diagonal ($m=n)$  and nearest off-diagonal ($m=n\pm 1$)  
contributions. 
A fine-tuned cancelation between the corresponding amplitudes $\Sigma_{mn}$  
leads to a reasonable description of the nucleon structure function 
$F_{2}^{\T N}$ at moderate momentum transfers $Q^2$.

An extension of this approach to nuclear  targets involves   
multiple scattering of hadronic fluctuations  
from  several nucleons. 
The multiple scattering process 
is described by  a coupled channel optical model 
\cite{Schildknecht:1973gi,Ditsas:1976yv} which accounts for the 
shadowing criteria in Eq.(\ref{eq:cond_i}),   
i.e. only those hadronic fluctuations with longitudinal interaction 
lengths larger than their mean free path in the nuclear medium 
contribute significantly to multiple scattering and thus to 
shadowing.

GVMD calculations applied to nuclear DIS data 
can be found in Refs.\cite{Bilchak:1989ck,Shaw:1993gx}.

\subsubsection{Vector mesons and quark scattering} 

We add a few remarks and references about approaches dealing 
with DIS in terms of quark dynamics.

The starting point in  Ref.\cite{Brodsky:1990qz} is a description of 
DIS from nucleons at large $Q^2$ and small $x$ in terms 
of quark-nucleon scattering \cite{Landshoff:1971ff}.  
The quark-nucleon scattering amplitude  is formulated 
using Regge phenomenology and constrained by the  
quark distributions of free nucleons. 
The interaction strength of quark-nucleon scattering is 
determined  by the quark-nucleon cross section, 
taken to be   about $1/3$ of the nucleon-nucleon cross section.
At center of mass energies $s \sim 200$ GeV$^2$  
one finds $\sigma_{q \T N} \approx 13$ mb \cite{Kulagin:1994fz}. 

An extension to DIS from nuclei at small $x$ 
involves the quark-nucleus scattering amplitude. 
Its connection with the amplitude for the scattering from 
free nucleons is given through the Glauber-Gribov multiple scattering series. 

In Ref.\cite{Kulagin:1994fz} the interactions of strongly correlated 
quark-antiquark pairs, i.e. vector mesons, have been added. 
One finds that vector mesons carry 
more than half of the shadowing effect measured at E665 and NMC. 
On the other hand, the interaction of uncorrelated quarks 
is also important to ensure a weak $Q^2$-dependence of shadowing.

\subsubsection{Green function methods}
\label{sssec:Green_function_methods}


The previously mentioned models have outlined in 
different ways the ingredients needed in order to understand
the physics of shadowing: the mass spectrum of 
quark-gluon fluctuations of the virtual photon, 
and the dynamics of the expanding and strongly 
interacting quark-gluon configurations in the surrounding nuclear 
system.
In most of the models the longitudinal propagation 
of hadronic fluctuations of the photon is treated by multiple 
scattering theory, 
while the transverse degrees of freedom are more or less 
``frozen'' during the passage through the nucleus. 
Several questions are faced in this context. 
The transverse size of quark-gluon fluctuations needs 
to be connected with their effective mass; 
the relationship with diffractive production must be 
elucidated; higher order terms in the multiple 
scattering series must be systematically incorporated, 
at least for heavy nuclei.

A coordinate space Green function method which permits to 
unify all those aspects has been developed in 
Refs.\cite{Kopeliovich:1998gv,Raufeisen:1998rg}. 
This work considers only quark-antiquark fluctuations of the photon. 
It turns out that 
some previous approximations can be recovered as limiting cases. 
We follow Ref.\cite{Kopeliovich:1998gv} and give here a brief summary 
of the essentials. 

\begin{figure}[t]
\bigskip
\begin{center} 
\epsfig{file=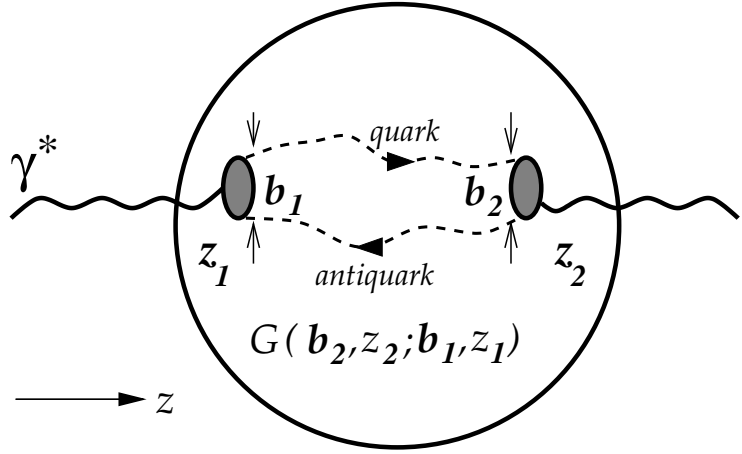,height=60mm}
\end{center}
\caption[...]{Propagation of a quark-antiquark fluctuation 
of the virtual photon $\gamma^*$ between points $z_1$ and 
$z_2$ where the pair has 
transverse separation $\vec b_1$ and $\vec b_2$. 
The Green function $G(\vec b_2,z_2;\vec b_1,z_1)$ 
sums all possible paths of the pair through the nucleus.
}
\label{fig:kopel_shad}
\bigskip
\end{figure}

Consider the scattering of a virtual photon with high energy $\nu$ 
and large squared four-momentum, $Q^2 > 1$ GeV$^2$, 
through a nucleus as illustrated in Fig.\ref{fig:kopel_shad}. 
The longitudinal ($z$-) direction is defined by the 
photon three-momentum, as usual. At point $z_1$ 
the photon produces a quark-antiquark pair with transverse 
separation $b_1$. 
Along its passage to point $z_2$ where it has a transverse separation 
$b_2$, the $q\bar q$ fluctuation experiences multiple interactions
with nucleons in the nuclear target. 
We are interested in the full Green function 
$G(\vec b_2,z_2;\vec b_1,z_1)$ which describes the propagation 
of the $q\bar q$ pair from $z_1$ to $z_2$, including 
its dynamics in the transverse space coordinate. 

This Green function enters in the shadowing part of the 
$\gamma^*$-nucleus cross section, as follows.
One writes: 
\begin{eqnarray} \label{eq:delta_sig_kop}
\delta \sigma_{\gamma^* \T{A}}(x,Q^2) 
&\equiv& \sigma_{\gamma^* \T{A}}(x,Q^2) - A \,\sigma_{\gamma^* \T{N}}(x,Q^2)
\nonumber \\
&=&
-2 \int d^2 b \int_{-\infty}^{\infty} d z_1 \,\rho_{\T A}(\vec b,z_1) 
\int_{z_1}^{\infty} d z_2 \,\rho_{\T A}(\vec b,z_2)
\,W(z_2,z_1).
\end{eqnarray}
The quantity $W(z_2,z_1)$ has dimension $(length)^4$ and 
describes the production of a $q\bar q$ fluctuation in the 
process $\gamma^* \T N \rightarrow q\bar q \,\T N$, 
its propagation from $z_1$ to $z_2$, 
and its subsequent conversion back to a virtual photon: 
\begin{eqnarray} \label{eq:W_z1_z2}
W(z_2,z_1)  = 
Re\,\int d^2 b_1 \int d^2 b_2 \int_0^1 d\xi \,
{\cal F}^*(\vec b_2,\xi)\,G(\vec b_2,z_2;\vec b_1,z_1)\, 
{\cal F}(\vec b_1,\xi)\,
e^{i\kappa(z_2 - z_1)}.
\end{eqnarray}
It involves the amplitude 
\begin{equation}
{\cal F}(\vec b,\xi) = \frac{1}{2} 
\psi_{\gamma^* \rightarrow q\bar q}(\vec b,\xi) \,\sigma(\vec b)
\end{equation}
for the $\gamma^* \T N \rightarrow q\bar q \,\T N$ process leading 
to a $q\bar q$ pair of transverse separation $\vec b$ in which the 
quark carries the fraction $\xi$ of the photon light cone momentum 
(see also Section \ref{ssec:Pert_Nonpert_Shad}). 
The color dipole cross section $\sigma(\vec b)$ has the 
characteristic color screening behavior, i.e. it vanishes 
as $b^2$ at $b\rightarrow 0$ (see also Eq.(\ref{eq:sigma_qq})), 
and the distribution of transverse separations is 
determined by the amplitude 
$\psi_{\gamma^* \rightarrow q\bar q}$. 
The normalization of $\cal F$ is such that its Fourier transform 
gives the  $\gamma^* \T N \rightarrow q\bar q \,\T N$ diffractive 
dissociation amplitude,
\begin{equation} \label{eq:dd_amp}
f(\vec k_{\perp}) = \int d^2 b\,{\cal F}(\vec b,\xi) 
e^{i\vec k_{\perp} \cdot \vec b},
\end{equation}
in plane wave impulse approximation. 
The phase factor $e^{i\kappa(z_2 - z_1)}$ involves the 
characteristic wave number of the $q\bar q$ fluctuation:
\begin{equation}
\kappa = \frac{Q^2\,\xi(1-\xi) + m_q^2}{2 \nu\,\xi (1-\xi)}
\end{equation}
where $m_q$ is the (constituent) quark mass. 
For $\xi=1/2$ the resulting 
$\kappa = \frac{Q^2 + 4 m_q^2}{2\nu} \equiv \lambda^{-1}$ 
is just the inverse coherence length of a quark and antiquark 
which travel side by side. 
(For arbitrary $\xi$ this coherence length includes the 
transverse momentum, 
$\lambda^{-1} = \kappa + \vec k_{\perp}/[2 \nu\,\xi(1-\xi)]$.) 

Let us now return to the propagation function $G$. 
It satisfies a wave equation \cite{Kopeliovich:1998gv} which can be made 
plausible by the following considerations. 
The longitudinal motion along the $z$-axis is equivalent to the 
time evolution of the $q\bar q$ fluctuation, represented by the 
operator $i\frac{\partial}{\partial z_2}$. 
The transverse dynamics has a kinetic term, 
\begin{equation} \label{eq:tkin}
t_{kin} = -\frac{\vec \nabla_{b_2}^2}{2 \nu \xi (1-\xi)}
\end{equation}
with the $2$-dimensional Laplacian acting on the 
transverse separation coordinate $\vec b_2$, and the 
denominator reflecting the effective mass of the pair. 
Interactions of the $q\bar q$ pair with the 
nuclear medium at an impact parameter $\vec b$ are introduced by an 
absorptive term, 
\begin{equation}
v(b_2,\vec b) = - \frac{i}{2} \,\sigma(b_2) 
\rho_{\T A}(\vec b,z_2).
\end{equation}
The wave equation for $G$ is then of the generic form 
$i \partial \,G/\partial z = (t_{kin} + v) \,G$ or, more precisely,
\begin{equation} \label{eq:wave_eq}
\left[i\frac{\partial}{\partial z_2} 
+ 
\frac{\vec \nabla_{b_2}^2}{2\nu\,\xi(1-\xi)} 
+ \frac{i}{2}\sigma(b_2)\,\rho_{\T A}(\vec b,z_2)\right]
G(\vec b_2,z_2;\vec b_1,z_1) = 0
\end{equation}
with the initial condition  
$G(\vec b_2,z_1;\vec b_1,z_1) = \delta^2(\vec b_2 - \vec b_1)$.
One can now discuss several interesting limits:

\underline{ i) The ``frozen'' limit }

Take the energy $\nu\rightarrow \infty$ so that the kinetic 
term (\ref{eq:tkin}) vanishes (with some extra care required 
at the kinematic corners, $\xi = 0$ and $\xi = 1$). 
Then 
\begin{equation}
G(\vec b_2,z_2;\vec b_1,z_1) = \delta^2(\vec b_2 - \vec b_1)
\,\exp\left[-\frac{\sigma(b_2)}{2} \int_{z_1}^{z_2} dz\,
\rho_{\T A}(\vec b_1,z)
\right].
\end{equation}
Inserting this expression into 
Eqs.(\ref{eq:W_z1_z2},\ref{eq:delta_sig_kop}) with 
$\lambda \rightarrow \infty$ one finds:
\begin{eqnarray}
\sigma_{\gamma^* \T{A}} &=& 
2 \int d^2 b \int d^2 b_{q\bar q} \int_0^1 d\xi\, 
\left|\psi_{\gamma^* \rightarrow q\bar q}(\vec b_{q\bar q},\xi) \right|^2 
\left\{1 - 
\exp\left[-\frac{\sigma(b_{q\bar q})}{2} \int_{-\infty}^\infty dz\,
\rho_{\T A}(\vec b,z)
\right]\right\}
\nonumber\\
\end{eqnarray}
and recovers the shadowing correction $\delta \sigma_{\gamma^* \T{A}}$ 
as in Glauber-Gribov multiple scattering theory by 
expanding the exponential. 
Note the difference compared to the standard Glauber eikonal 
approximation where the cross section $\sigma(b_{q\bar q})$ 
is averaged in the exponent.

\underline{ ii) No absorption}
 
Take the limit $\sigma \rightarrow 0$ in the wave equation  
(\ref{eq:wave_eq}). Then $G$ reduces to the free Green function 
of the $q\bar q$ pair, 
\begin{equation}
G(\vec b_2,z_2;\vec b_1,z_1) = \frac{1}{2\pi} 
\int d^2 k_{\perp}\,
\exp\left[i \vec k_{\perp}\!\cdot\!(\vec b_2 - \vec b_1) 
+ \frac{i \vec k_{\perp}^2 (z_2 - z_1)}{2 \nu\,\xi (1-\xi)} 
\right].
\end{equation}
Inserting this into Eq.(\ref{eq:W_z1_z2}) and using the diffractive 
dissociation amplitude (\ref{eq:dd_amp}) one finds 
\begin{equation} \label{eq:no_abs}
W(z_1,z_2) = \frac{1}{2\pi} 
\int_0^1 d\xi \int d^2 k_{\perp} 
\left|f(\vec k_{\perp})\right|^2\,
\cos \left[\frac{Q^2 \xi (1-\xi) + m_q^2 + \vec k_\perp^2}
{2 \nu \xi (1-\xi)} \,(z_2-z_1)\right].
\end{equation}
We identify the squared effective mass, 
$M_{\T X}^2 = (m_q^2 + \vec k_{\perp}^2)/\xi (1-\xi)$, 
of the $q\bar q$ pair as in Eq.(\ref{eq:qq_mass}) 
and introduce its 
coherence length $\lambda = 2\nu/(Q^2 + M_{\T X}^2)$. 
Inserting Eq.(\ref{eq:no_abs}) into Eq.(\ref{eq:delta_sig_kop}) one then 
recovers the double scattering result, Eq.(\ref{eq:ds_A}), 
with the factorized two-body density 
$\rho_{\T A}^{(2)}(\vec b,z_1;\vec b,z_2) = 
\rho_{\T A}(\vec b,z_1)\,\rho_{\T A}(\vec b,z_2)$. 
It is now also apparent how the additional absorption 
factor in Eq.(\ref{eq:ms_A}) is obtained, introducing an 
average cross section $\sigma_{\T{XN}}$ in the exponent.

\pagebreak

\underline{ iii) Propagation in uniform nuclear matter}

Assume that the $q\bar q$ pair moves in a nuclear medium 
of uniform density 
$\rho_{\T A}(\vec b,z) = \rho_0 = const.$ 
($\rho_0 = 0.17$ fm$^{-3}$ for normal nuclear matter). 
Suppose that the color dipole cross section is 
approximated by 
\begin{equation}
\sigma(b_{q\bar q}) = c \,b_{q\bar q}^2
\end{equation}
with a constant parameter $c$. In this case the wave equation 
(\ref{eq:wave_eq}) reduces to 
\begin{equation}
\left[i\frac{\partial}{\partial z_2} 
+ 
\frac{\vec \nabla_{b_2}^2}{2\nu\,\xi(1-\xi)} 
+ \frac{i c}{2} \,\rho_0 \,b_2^2 \right]
G(\vec b_2,z_2;\vec b_1,z_1) = 0.
\end{equation}
This is formally reminiscent of the Schr\"odinger equation 
for a harmonic oscillator with complex frequency. 
One finds \cite{Kopeliovich:1998gv} 
\begin{eqnarray}
G(\vec b_2,z_2;\vec b_1,0) = 
\frac{a}{2 \pi \sinh(\omega z)}\,
\exp\left\{-\frac{a}{2}\left[(\vec b_1^2 + \vec b_2^2) \coth (\omega z) 
- \frac{2 \vec b_1 \cdot \vec b_2}{\sinh (\omega z)}\right]\right\},
\end{eqnarray}
with
\begin{equation}
\omega^2 = \left(\frac{c\,\rho_0}{a}\right)^2 = 
i \frac{c\,\rho_0}{\nu\xi(1-\xi)}. 
\end{equation}
This is a convenient approximation to account for multiple 
scattering and absorption of the $q\bar q$ fluctuation, still 
keeping track of its transverse dynamics during its passage 
through the nuclear medium. 
Instructive results are discussed in Ref.\cite{Kopeliovich:1998gv}.

\subsubsection{Meson exchange and shadowing}
\label{sssec:shad_meson}

Up to now we have concentrated on 
diffractive contributions to nuclear  
shadowing, in which  the nucleons interacting with the 
virtual photon are  left unchanged.  
The coherent interaction of the photon 
with several nucleons in the target nucleus can also involve  
non-diffractive processes, 
in particular, reactions  in which nucleons change 
their charge.
These are commonly described by the exchange of 
mesons and sub-leading Reggeons.

Modifications to nuclear structure functions at small $x$ 
through meson exchange have been investigated in 
Refs.\cite{Melnitchouk:1993eu,Nikolaev:1997jy} for  deuterium. 
In this work significant  effects come 
from the interaction of the   
virtual photon with  pions emitted from  the target proton 
or neutron.  
Here, as in diffraction, a hadronic state X is produced which 
subsequently re-scatters from the second nucleon.
Contributions from the exchange of other mesons,  
e.g. $\rho$ and  $\omega$, turn out to be negligible.   
For the double scattering contribution through pion exchange
one finds in analogy with Eq.(\ref{eq:ds_corr_full}):
\begin{equation} \label{eq:ds_pion}
\delta \sigma_{\gamma ^* {\T d}}^{\pi} 
= 
\frac{2}{\pi} \int d^2 k_{\perp} 
\int_{4 m_{\pi}^2}^{W^2} dM_{\T X}^2 \, S_{\T d}^{\pi}(\vec k) \,
\frac{d^2 \sigma_{\gamma^* {\T N}}^{\pi}}
{dM_{\T X}^2 dt}.
\end{equation}
Here ${d^2 \sigma_{\gamma^* {\T N}}^{\pi}}/{dM_{\T X}^2 dt}$ 
is the cross section for the semi-inclusive production of 
a hadronic state with invariant mass $M_{\T X}$ from a proton or 
neutron via   pion exchange.
The form factor in Eq.(\ref{eq:ds_pion}) accounts for the 
spin-dependent response of the deuteron: 
\begin{equation} \label{eq:Sd_pi}
S_{\T d}^{\pi}(\vec k) = 
\frac{1}{3} \sum_m \int d^3 P \,
\psi_{\T d}^{m\dagger}(\vec P) \,\vec \sigma_p\cdot \hat {\vec k} \,
\,\vec \sigma_n\cdot \hat {\vec  k} \,\psi_{\T d}^m(\vec P-\vec k), 
\end{equation}
where $\vec k$ is the pion momentum and 
$\hat {\vec k} = \vec k/|\vec k|$. 
The momentum-space wave function of the deuteron with polarization $m$
is denoted by $\psi_{\T d}^m$. 
Furthermore the non-relativistic form of the 
pion-nucleon coupling is used \cite{Machleidt:1987hj}.   
Note that the energy of the exchanged pion is determined by 
$k_0 = M_{\T d} - \sqrt{M^2 + \vec P^2} - \sqrt{M^2 - (\vec P - \vec k)^2}$,  
where  $\vec P$ is the momentum of the parent nucleon. 
We denote the pion four-momentum by $k=(k_0, \vec k)$. 
For the longitudinal pion momentum one has  
$k_3  \approx y M$ with the pion light-cone momentum fraction 
$y=k\cdot q/P\cdot q$, and we introduce $t = k^2$ along with the 
usual Bjorken-$x$.

It is common to factorize the semi-inclusive differential cross 
section: 
\begin{equation}
\frac{d^2 \sigma_{\gamma^* {\T N}}^{\pi}}{dy dt}(x,Q^2;y,t) = 
f_{\pi/{\T N}}(y,t) \, \sigma_{\gamma^* \pi}(x/y, Q^2). 
\end{equation} 
Here the photon-pion cross section is related 
to the structure function of the pion by  
$F_{2}^{\pi}(x,Q^2) = 
(Q^2/4 \pi^2 \alpha) \,\sigma_{\gamma^* \pi}(x, Q^2)$,  
and the pion distribution function in the nucleon is given by: 
\begin{equation} \label{eq:f_piN}
f_{\pi/{\T N}}(y,t) = \frac{3 g_{\pi {\T {N N}}}^2}{16 \pi^2} 
\frac{\left|{\cal F}_{\pi {\T {N N}}}(t)\right|^2 (-t)}
{(t - m_{\pi}^2)^2} \,y,
\end{equation} 
with the pion-nucleon  coupling constant 
$g_{\pi {\T {NN}}}$  
and  the $\pi {\T {NN}}$ form factor ${\cal F}_{\pi {\T {N N}}}$ 
normalized to unity for on-mass-shell 
pions, i.e. ${\cal F}_{\pi {\T {N N}}}=1$ for $t = k^2 = m_{\pi}^2$.

For practical calculations  the pion structure function 
has been  approximated by that of the free 
pion,  as parametrized in 
\cite{Betev:1985pg,Gluck:1992ey} 
in accordance with Drell-Yan leptoproduction data. 
Here, however, only the region $x>0.1$ has been measured. 
An extraction of the pion structure function at small $x$ 
from  semi-exclusive  reactions at HERA has been discussed 
recently in \cite{Holtmann:1994rs,Przybycien:1996zb}.

The resulting pionic 
correction $\delta \sigma_{\gamma^* {\T d}}^{\pi}$ 
to double scattering turns out to be 
positive, i.e. it causes ``anti-shadowing''. 
The relative weight of $\delta \sigma_{\gamma^* {\T d}}^{\pi}$
decreases with decreasing $x$.
At typical values  $Q^2 =4\,\rm{GeV^2}$ and 
$0.001 < x < 0.1$ it amounts  
to around  $30\%$ of the overall shadowing correction 
\cite{Melnitchouk:1993eu}. 
In Ref.\cite{Nikolaev:1997jy} the pion correction  
$\delta \sigma_{\gamma^*{\T d}}^{\T \pi}$  has 
been found to be  negligible at large $Q^2 \,\gsim\, 10$ GeV$^2$. 
Note however that the quoted  results depend sensitively on 
the yet unknown pion structure function at small $x$, the  
deuteron wave function and the choice for the pion-nucleon 
form factor.

\subsubsection{Discussion}

The models sketched above give quite reasonable descriptions of the 
data on nuclear shadowing measured at CERN and FNAL. 
All of them support the general observation that 
nuclear shadowing as measured by 
NMC  
\cite{Amaudruz:1995tq,Arneodo:1995cs,Amaudruz:1991cc,Amaudruz:1992dj,%
Arneodo:1996rv,Arneodo:1996ru,Arneodo:1994ia} 
and E665   
\cite{Adams:1992nf,Adams:1995is,Adams:1992vm,Adams:1995sh} 
at small $x<0.01$  receives 
major contributions from the 
coherent interaction  of the vector mesons $\rho$, $\omega$ and $\phi$. 
In fact those experiments are performed 
at small average momentum transfers 
$\overline Q^2 \,\lsim\, 1$ GeV$^2$. 
On the other hand, the observed weak $Q^2$-dependence of the 
shadowing effect 
originates from  the coherent interaction of 
strongly interacting quark-antiquark fluctuations 
with large masses, $M_{\T X} > 1$ GeV.

\subsection{Interpretation of 
nuclear shadowing in the infinite momentum frame}
\label{ssec:shad_IMF}

In this section we briefly discuss how nuclear shadowing 
develops in the infinite momentum frame where  
the parton model for deep-inelastic scattering can be applied. 
We found in Section \ref{ssec:IMF_spati} 
that, in this frame, the wave functions of 
partons from different nucleons in the nucleus 
start to overlap for $x < 0.1$.
One then expects that the interaction of 
partons belonging to different nucleons increases. 
Shadowing at small $x<0.1$ is supposed to be due to the 
fusion or recombination of partons from different nucleons,
thereby effectively reducing the quark distributions of each 
nucleon.
At the same time parton fusion leads to an enhancement  
of partons at $x > 0.1$.
In Ref.\cite{Close:1989ca} modifications of parton 
distributions due to parton fusion have been derived 
and found to be proportional to $1/Q^2$. 
Therefore parton fusion processes seem to be suppressed at large momentum 
transfers but can be significant at low $Q^2$.

Procedures for modeling 
nuclear parton distributions at small $x$ have been 
proposed in Refs.\cite{Kumano:1992ef,Close:1989ca,Kumano:1994pn}.  
Recombination effects modify  these distributions 
dominantly at a low momentum scale $Q_0^2$ where   
parton fusion is calculated and incorporated in 
the initial quark and gluon distribution functions. 
Parton distributions at  $Q^2 > Q_0^2$ are then 
derived through the calculation of radiative QCD corrections 
using DGLAP evolution (see Section \ref{ssec:AP_eq}). 

To describe the measured shadowing of the NMC and E665 collaboration 
a typical scale $Q^2_0 \approx 0.8$ GeV$^2$ has been used 
in Refs.\cite{Kumano:1992ef,Kumano:1994pn}.  
As a result the empirical shadowing for $F_2^{\T A}$ can be described.
It should be mentioned that the calculation of the recombination 
effect within perturbation theory                                  
is certainly questionable at a low  momentum scale  $Q_0^2$. 
The results are strongly sensitive to  model parameters, 
such as the initial scale $Q_0^2$ and 
the input parton distributions. 

Note that the recombination effects discussed here 
involve parton distributions at a low momentum scale. 
This ``initial-state recombination'' is different from 
the ``radiative recombination'', discussed in       
Section \ref{ssec:High_parton_densities}, 
which modifies the parton evolution by recombination 
of radiatively produced partons.

\subsection{Nuclear parton distributions at small $x$}
\label{ssec:nuclear_parton_distr}

Any quantitative QCD analysis of high energy processes involving nuclei 
requires a detailed knowledge of nuclear parton distributions. 
In this section we outline the empirical information 
on their difference with respect to quark and gluon 
distribution functions  of free nucleons.

\begin{figure}[t]
\bigskip


$$
\setlength{\unitlength}{0.240900pt}
\ifx\plotpoint\undefined\newsavebox{\plotpoint}\fi
\begin{picture}(1500,900)(0,0)
\font\gnuplot=cmr10 at 10pt
\gnuplot
\sbox{\plotpoint}{\rule[-0.200pt]{0.400pt}{0.400pt}}%
\put(176.0,146.0){\rule[-0.200pt]{4.818pt}{0.400pt}}
\put(154,146){\makebox(0,0)[r]{0.8}}
\put(1416.0,146.0){\rule[-0.200pt]{4.818pt}{0.400pt}}
\put(176.0,308.0){\rule[-0.200pt]{4.818pt}{0.400pt}}
\put(154,308){\makebox(0,0)[r]{0.9}}
\put(1416.0,308.0){\rule[-0.200pt]{4.818pt}{0.400pt}}
\put(176.0,471.0){\rule[-0.200pt]{4.818pt}{0.400pt}}
\put(154,471){\makebox(0,0)[r]{1}}
\put(1416.0,471.0){\rule[-0.200pt]{4.818pt}{0.400pt}}
\put(176.0,633.0){\rule[-0.200pt]{4.818pt}{0.400pt}}
\put(154,633){\makebox(0,0)[r]{1.1}}
\put(1416.0,633.0){\rule[-0.200pt]{4.818pt}{0.400pt}}
\put(176.0,796.0){\rule[-0.200pt]{4.818pt}{0.400pt}}
\put(154,796){\makebox(0,0)[r]{1.2}}
\put(1416.0,796.0){\rule[-0.200pt]{4.818pt}{0.400pt}}
\put(176.0,113.0){\rule[-0.200pt]{0.400pt}{2.409pt}}
\put(176.0,867.0){\rule[-0.200pt]{0.400pt}{2.409pt}}
\put(234.0,113.0){\rule[-0.200pt]{0.400pt}{2.409pt}}
\put(234.0,867.0){\rule[-0.200pt]{0.400pt}{2.409pt}}
\put(282.0,113.0){\rule[-0.200pt]{0.400pt}{2.409pt}}
\put(282.0,867.0){\rule[-0.200pt]{0.400pt}{2.409pt}}
\put(322.0,113.0){\rule[-0.200pt]{0.400pt}{2.409pt}}
\put(322.0,867.0){\rule[-0.200pt]{0.400pt}{2.409pt}}
\put(357.0,113.0){\rule[-0.200pt]{0.400pt}{2.409pt}}
\put(357.0,867.0){\rule[-0.200pt]{0.400pt}{2.409pt}}
\put(388.0,113.0){\rule[-0.200pt]{0.400pt}{2.409pt}}
\put(388.0,867.0){\rule[-0.200pt]{0.400pt}{2.409pt}}
\put(415.0,113.0){\rule[-0.200pt]{0.400pt}{4.818pt}}
\put(415,68){\makebox(0,0){0.01}}
\put(415.0,857.0){\rule[-0.200pt]{0.400pt}{4.818pt}}
\put(596.0,113.0){\rule[-0.200pt]{0.400pt}{2.409pt}}
\put(596.0,867.0){\rule[-0.200pt]{0.400pt}{2.409pt}}
\put(702.0,113.0){\rule[-0.200pt]{0.400pt}{2.409pt}}
\put(702.0,867.0){\rule[-0.200pt]{0.400pt}{2.409pt}}
\put(777.0,113.0){\rule[-0.200pt]{0.400pt}{2.409pt}}
\put(777.0,867.0){\rule[-0.200pt]{0.400pt}{2.409pt}}
\put(835.0,113.0){\rule[-0.200pt]{0.400pt}{2.409pt}}
\put(835.0,867.0){\rule[-0.200pt]{0.400pt}{2.409pt}}
\put(883.0,113.0){\rule[-0.200pt]{0.400pt}{2.409pt}}
\put(883.0,867.0){\rule[-0.200pt]{0.400pt}{2.409pt}}
\put(923.0,113.0){\rule[-0.200pt]{0.400pt}{2.409pt}}
\put(923.0,867.0){\rule[-0.200pt]{0.400pt}{2.409pt}}
\put(958.0,113.0){\rule[-0.200pt]{0.400pt}{2.409pt}}
\put(958.0,867.0){\rule[-0.200pt]{0.400pt}{2.409pt}}
\put(989.0,113.0){\rule[-0.200pt]{0.400pt}{2.409pt}}
\put(989.0,867.0){\rule[-0.200pt]{0.400pt}{2.409pt}}
\put(1016.0,113.0){\rule[-0.200pt]{0.400pt}{4.818pt}}
\put(1016,68){\makebox(0,0){0.1}}
\put(1016.0,857.0){\rule[-0.200pt]{0.400pt}{4.818pt}}
\put(1197.0,113.0){\rule[-0.200pt]{0.400pt}{2.409pt}}
\put(1197.0,867.0){\rule[-0.200pt]{0.400pt}{2.409pt}}
\put(1303.0,113.0){\rule[-0.200pt]{0.400pt}{2.409pt}}
\put(1303.0,867.0){\rule[-0.200pt]{0.400pt}{2.409pt}}
\put(1378.0,113.0){\rule[-0.200pt]{0.400pt}{2.409pt}}
\put(1378.0,867.0){\rule[-0.200pt]{0.400pt}{2.409pt}}
\put(1436.0,113.0){\rule[-0.200pt]{0.400pt}{2.409pt}}
\put(1436.0,867.0){\rule[-0.200pt]{0.400pt}{2.409pt}}
\put(176.0,113.0){\rule[-0.200pt]{303.534pt}{0.400pt}}
\put(1436.0,113.0){\rule[-0.200pt]{0.400pt}{184.048pt}}
\put(176.0,877.0){\rule[-0.200pt]{303.534pt}{0.400pt}}
\put(1500,68){\makebox(0,0){$x$}}
\put(176.0,113.0){\rule[-0.200pt]{0.400pt}{184.048pt}}
\put(400,812){\makebox(0,0)[r]{$F_2^{\rm{Sn}}/F_2^{\rm{C}}$}}
\put(440,812){\makebox(0,0){$+$}}
\put(259,176){\makebox(0,0){$+$}}
\put(373,228){\makebox(0,0){$+$}}
\put(473,290){\makebox(0,0){$+$}}
\put(561,341){\makebox(0,0){$+$}}
\put(654,409){\makebox(0,0){$+$}}
\put(742,440){\makebox(0,0){$+$}}
\put(808,458){\makebox(0,0){$+$}}
\put(860,484){\makebox(0,0){$+$}}
\put(923,485){\makebox(0,0){$+$}}
\put(989,488){\makebox(0,0){$+$}}
\put(1074,490){\makebox(0,0){$+$}}
\put(1162,493){\makebox(0,0){$+$}}
\put(1255,480){\makebox(0,0){$+$}}
\put(1343,462){\makebox(0,0){$+$}}
\put(400,745){\makebox(0,0)[r]{$g_{\rm{Sn}}/g_{\rm{C}}$}}
\put(440,745){\circle{18}}
\put(440,264){\circle{18}}
\put(554,269){\circle{18}}
\put(654,432){\circle{18}}
\put(742,523){\circle{18}}
\put(835,565){\circle{18}}
\put(923,609){\circle{18}}
\put(989,617){\circle{18}}
\put(1041,584){\circle{18}}
\put(1104,625){\circle{18}}
\put(1169,555){\circle{18}}
\put(440.0,188.0){\rule[-0.200pt]{0.400pt}{36.858pt}}
\put(430.0,188.0){\rule[-0.200pt]{4.818pt}{0.400pt}}
\put(430.0,341.0){\rule[-0.200pt]{4.818pt}{0.400pt}}
\put(554.0,209.0){\rule[-0.200pt]{0.400pt}{28.908pt}}
\put(544.0,209.0){\rule[-0.200pt]{4.818pt}{0.400pt}}
\put(544.0,329.0){\rule[-0.200pt]{4.818pt}{0.400pt}}
\put(654.0,386.0){\rule[-0.200pt]{0.400pt}{21.922pt}}
\put(644.0,386.0){\rule[-0.200pt]{4.818pt}{0.400pt}}
\put(644.0,477.0){\rule[-0.200pt]{4.818pt}{0.400pt}}
\put(742.0,462.0){\rule[-0.200pt]{0.400pt}{29.149pt}}
\put(732.0,462.0){\rule[-0.200pt]{4.818pt}{0.400pt}}
\put(732.0,583.0){\rule[-0.200pt]{4.818pt}{0.400pt}}
\put(835.0,519.0){\rule[-0.200pt]{0.400pt}{21.922pt}}
\put(825.0,519.0){\rule[-0.200pt]{4.818pt}{0.400pt}}
\put(825.0,610.0){\rule[-0.200pt]{4.818pt}{0.400pt}}
\put(923.0,554.0){\rule[-0.200pt]{0.400pt}{26.499pt}}
\put(913.0,554.0){\rule[-0.200pt]{4.818pt}{0.400pt}}
\put(913.0,664.0){\rule[-0.200pt]{4.818pt}{0.400pt}}
\put(989.0,549.0){\rule[-0.200pt]{0.400pt}{32.762pt}}
\put(979.0,549.0){\rule[-0.200pt]{4.818pt}{0.400pt}}
\put(979.0,685.0){\rule[-0.200pt]{4.818pt}{0.400pt}}
\put(1041.0,502.0){\rule[-0.200pt]{0.400pt}{39.748pt}}
\put(1031.0,502.0){\rule[-0.200pt]{4.818pt}{0.400pt}}
\put(1031.0,667.0){\rule[-0.200pt]{4.818pt}{0.400pt}}
\put(1104.0,557.0){\rule[-0.200pt]{0.400pt}{32.762pt}}
\put(1094.0,557.0){\rule[-0.200pt]{4.818pt}{0.400pt}}
\put(1094.0,693.0){\rule[-0.200pt]{4.818pt}{0.400pt}}
\put(1169.0,459.0){\rule[-0.200pt]{0.400pt}{46.253pt}}
\put(1159.0,459.0){\rule[-0.200pt]{4.818pt}{0.400pt}}
\put(1159.0,651.0){\rule[-0.200pt]{4.818pt}{0.400pt}}
\sbox{\plotpoint}{\rule[-0.400pt]{0.800pt}{0.800pt}}%
\put(176,471){\usebox{\plotpoint}}
\put(176.0,471.0){\rule[-0.100pt]{303.534pt}{0.200pt}}
\sbox{\plotpoint}{\rule[-0.600pt]{1.200pt}{1.200pt}}%
\put(800,682){\makebox(0,0)[r]{$J/\psi$}}
\put(835.0,552.0){\rule[-0.050pt]{.100pt}{62.634pt}}
\put(835.0,552.0){\rule[-0.050pt]{69.138pt}{.100pt}}
\put(835.0,812.0){\rule[-0.050pt]{69.138pt}{.100pt}}
\put(1122.0,552.0){\rule[-0.050pt]{.100pt}{62.734pt}}
\put(835.0,682.0){\rule[-0.050pt]{69.138pt}{.100pt}}
\end{picture}
$$

\caption{
Results from Ref.\cite{Gousset:1996xt} for 
the ratio of the Sn and carbon gluon 
densities,  $g_{\rm Sn}(x)/g_{\rm C}(x)$,  
together with the measured ratio of structure functions  
$F_2^{\rm Sn}(x)/F_2^{\rm C}(x)$ \cite{Arneodo:1996ru}.
The box represents the extraction of the ratio of gluon distributions  
from $J/\psi$ electroproduction data \cite{Amaudruz:1992sr}.
} 
\label{fig:gousset}
\bigskip
\end{figure}
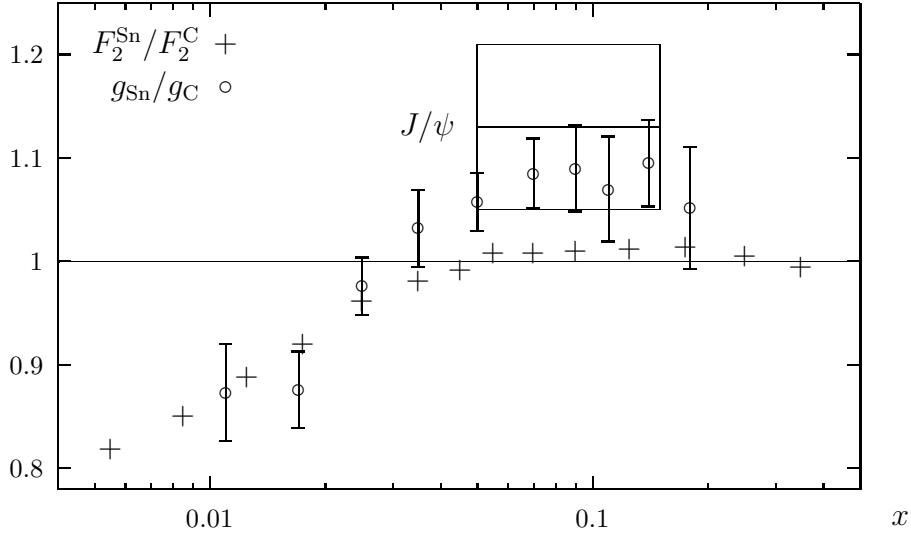

Let us first focus on  the nuclear gluon distribution.
The $Q^2$-dependence of deep-inelastic structure functions at small $x$ 
is dominated by gluon radiation. 
One can therefore extract nuclear gluon distribution 
functions from a precise analysis of scaling violations of   
the structure functions $F_2^{\T A}$.  
In leading order perturbation theory and in the limit $x \ll 0.1$ 
the DGLAP equations (\ref{eq:DGLAP},\ref{eq:DGLAP_s}) 
reduce to the simple form \cite{Prytz:1993vr}: 
\begin{equation} \label{eq:F_2_glue}
\frac{\partial F_2(x,Q^2)}{\partial \ln Q^2} 
\approx \frac{\alpha_s}{3\pi} \sum_f e_f^2 \, x \,g(2x,Q^2).
\end{equation}
This relation, with further inclusion of 
small corrections from quark contributions, 
has been used in an analysis \cite{Gousset:1996xt} of high statistics 
NMC data on 
the $Q^2$-dependence of the structure function ratio 
$F_2^{\T{Sn}}/F_2^{\T C}$ shown in Fig.\ref{fig:sn/c_a}.
The result for the corresponding ratio of nuclear gluon distributions, 
$g_{\T{Sn}}/g_{\T C}$, is shown in Fig.\ref{fig:gousset}. 
At $x<0.1$ the gluon distribution is shadowed, 
i.e.  $g_{\T{Sn}}/g_{\T C}<1$, in a similar way as
the structure function $F_2$. 
This observation is quite natural since 
$F_2$ at $x<0.1$  is dominated  
by contributions from sea quarks. 
The intimate relation between sea quarks and gluons 
through DGLAP evolution then also suggests shadowing for gluons.

At $0.05<x<0.15$ an approximate $8\%$ 
enhancement of nuclear gluons has been found.
This observation is in agreement  with an  analysis of NMC data 
for inelastic $J/\psi$-lepto\-pro\-duction \cite{Amaudruz:1992sr} 
as indicated in Fig.\ref{fig:gousset}. 
The enhancement of  nuclear gluon distributions around  
$x\simeq 0.1$ is consistent with the fact 
that the total momentum of hadrons is given by the sum of 
the momenta of its parton constituents 
\cite{Frankfurt:1990xz,Eskola:1998iy}.
The  empirical information on this sum rule applied 
to quarks has been presented in Section \ref{ssec:moments_str_fns}. 
It implies that the momentum carried by gluons is, within 
error bars, equal in nucleons and nuclei, i.e. 
\begin{equation} \label{eq:mom_SR_gluons}
\int^{1}_0 dx \,x\,
g_{\T N}(x,Q^2) \approx  
\int^{A}_0 dx \,x \, g_{\T A}(x,Q^2).
\end{equation}
Consequently, shadowing of  nuclear  gluon distributions 
at small $x$ has to be compensated by  
an enhancement at larger values of $x$. 
Assuming the latter to be located in the region 
$0.05 < x < 0.15$ leads to results similar to the 
ones shown in Fig.\ref{fig:gousset} \cite{Frankfurt:1990xz}.

Note that the close relation between shadowing and diffraction 
allows to estimate gluon shadowing using data on 
diffractive charm and dijet production from free nucleons. 
A corresponding analysis of HERA data has been carried out 
in Refs.\cite{Frankfurt:1998ym,Alvero:1998bz}.  
It suggests significantly larger shadowing for gluons than for  
quarks. 

\begin{figure}[t]
\bigskip
\hspace*{1cm}
\input{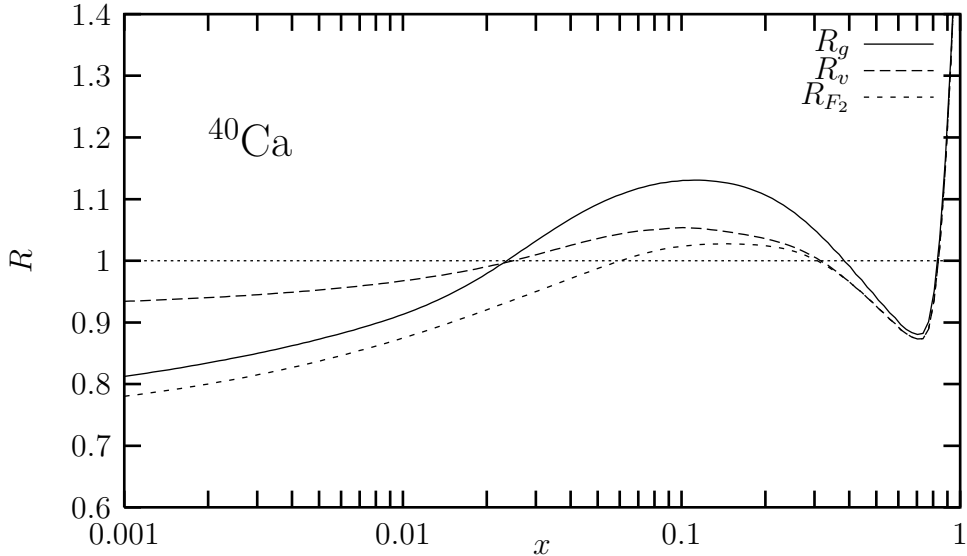}
\caption{
Momentum-space ratios from Ref.\cite{Eskola:1998iy} 
relative to the free nucleon,  
for gluon
distributions, valence quark distributions, and the $F_2$ structure
function in $^{40}$Ca at $Q^2 = 4$ GeV$^2$.
}
\label{fig:momspace-Ca}
\bigskip
\end{figure}

Nuclear effects in valence and sea quark distributions 
can be further disentangled using Drell-Yan dilepton production 
data \cite{Frankfurt:1990xz,Eskola:1998iy}. 
The E772 collaboration at FNAL has found shadowing for nuclear 
antiquark distributions at $x<0.1$ but no enhancement 
as discussed in Section \ref{sssec:DY}. 
Combining this  with the fact that the nuclear structure function
ratio $F_2^{\T A}/F_2^{\T N} \gsim 1$ for  $0.05 < x <  0.2$, 
one concludes   that nuclear valence quarks have to be enhanced 
around $x\sim 0.1$. 
From the baryon number sum rule 
\begin{equation}
\int_0^1 dx \, q_v^{\T N}(x,Q^2) =  
\int_0^A dx \, q_v^{\T A}(x,Q^2) 
\end{equation}
one then concludes that nuclear valence quark distributions, $q_v^{\T A}$,  
must be shadowed at $x<0.05$. 
Typical results from Ref.\cite{Frankfurt:1990xz} 
are shown in Fig.\ref{fig:momspace-Ca}. 

To summarize, present data on nuclear shadowing imply that all 
parton distributions are shadowed at  
$x \ll 0.1$, while only valence quarks and gluons are enhanced 
around $x\sim 0.1$. 
The kinematic range where enhancement takes place is 
related to processes which involve 
typical longitudinal distances of $1\,\rm{fm}$ in the laboratory. 
This is the region where components of the nuclear wave function 
with overlapping parton distributions should be relevant.
In Ref.\cite{Frankfurt:1990xz} it was suggested that 
at such distances inter-nucleon forces are a result 
of quark and gluon exchange  leading  to the observed 
enhancement.
In such a picture the enhancement of gluons should 
increase with the density of the nuclear target. 
A $10\%$ enhancement of glue at $x \sim 0.1$  in Sn as compared to C 
would then imply a $20\%$ increase of the gluon density 
in Pb as compared to free nucleons \cite{Frankfurt:1990xz}. 
This would imply a dramatic change of gluon fields 
in nuclear matter at distances of 
$1\,\rm{fm}$ between nucleons.

More detailed information on nuclear parton distributions 
is certainly needed. 
The shadowing region -- where nuclear effects are large -- is 
of particular interest.  
Further constraints on gluon shadowing from deep-inelastic 
scattering require data on the $Q^2$-dependence of nuclear 
structure functions at smaller values of $x$ as indicated by 
Eq.(\ref{eq:F_2_glue}).  
A more quantitative separation of nuclear effects in 
valence and sea quark distributions could be obtained  
from Drell-Yan dilepton production or neutrino scattering 
experiments with high statistics. 
On the other hand, an extraction of nuclear parton distributions 
in hadron production processes 
from nuclei, e.g.  lepto- or hadroproduction of charmonium 
or open charm (see e.g. \cite{Vogt:1991qd,Vogt:1992ki}), 
is complicated by possible final state 
interactions and higher twist corrections.

\section{Nuclear structure functions at large Bjorken-$x$}
\setcounter{section}{6}
\setcounter{figure}{0}
\label{sec:EMC}

Deep-inelastic scattering from nuclei probes the    
nuclear parton distributions.
On the other hand  conventional nuclear physics 
works well with the concept 
that  nuclei are  composed of interacting hadronic constituents, 
primarily  nucleons and pions.  
For $x>0.2$ DIS probes  
longitudinal distances smaller than  $1 \,\rm{fm}$ 
(see Section \ref{Sec:space_time}), 
less than the size of individual hadrons in nuclei. 
In this kinematic region, incoherent scattering from hadronic 
constituents of the target nucleus dominates.  
Such processes explore the quark distributions of nucleons  
bound in the nucleus.

To gain first insights suppose that the  nucleus is described 
by nucleons  moving in a mean field.  
The quark substructure of  bound nucleons   
may  differ in several respects from the quark 
distributions of  free nucleons. 
First, there is a purely kinematical effect due to the momentum distribution 
and binding energy of the bound nucleons.
This effect rescales the  energy and momentum  of the  
partonic constituents.
To illustrate this recall that for a free nucleon  
the light-cone momentum fraction of partons cannot exceed $x=1$. 
A nucleon bound in a nucleus carries a non-vanishing 
momentum which   
adds to the momenta of individual partons in that nucleon. 
As a consequence light-cone momentum fractions  up to $x = A$ 
are possible in principle, although the extreme situation in which a 
single parton 
carries all of the nuclear momentum will of course be very highly 
improbable.
On the other hand intrinsic properties of bound nucleons, 
e.g. their size, could  also change in the nuclear environment.
This may  lead to  additional, dynamical modifications of their partonic 
structure.

\subsection{Impulse approximation} 

\begin{figure}[b]
\bigskip
\begin{center} 
\epsfig{file=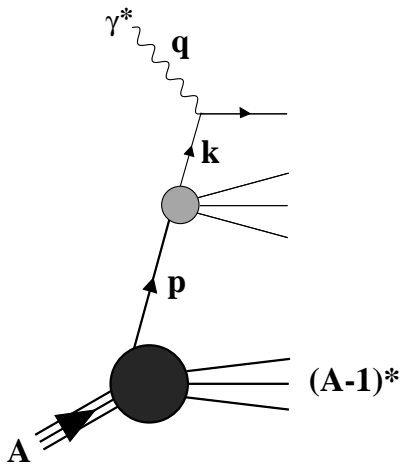,height=60mm}
\end{center}
\caption[...]{Impulse approximation for deep-inelastic scattering 
from nuclei at large $Q^2$. 
}
\label{fig:emc_ia}
\bigskip
\end{figure}

Nuclei are, in many respects, dilute systems. 
For example, in  elastic proton-nucleus 
scattering  the proton mean free path is  
of the order of $5-10\,\rm{fm}$ \cite{Schiffer:1980hb}, 
large compared to the average distance between nucleons in nuclei. 
This observation has motivated the impulse approximation 
which reduces the nuclear scattering process to 
incoherent scatterings from the individual nucleons 
(for reviews and references see 
\cite{Frankfurt:1988nt,Bickerstaff:1989ch,Arneodo:1994wf,%
Geesaman:1995yd,Jaffe:1985je,Bickerstaff:1985da}). 
Final  state interactions of the scattered hadron with the 
residual nuclear system are neglected at high energy. 
(One should note, however, that the  validity of this 
approximation, illustrated in Fig.\ref{fig:emc_ia}, 
is still under debate as discussed in 
\cite{Frankfurt:1988nt,Geesaman:1995yd,Kulagin:1994fz,%
Bickerstaff:1985da,Bickerstaff:1985ax,Hoodbhoy:1987fn,%
Saito:1994yw,Melnitchouk:1994nk} 
and references therein.)

Given the small average momenta of the weakly bound nucleons, 
their quark sub-structure is described by structure functions 
similar to those of free nucleons \cite{Kulagin:1994fz,Kulagin:1995cj}.
For a nucleon with momentum $p$ these structure functions 
depend on the scaling 
variable $x=Q^2 /2p\cdot q$ and on the squared momentum transfer 
$Q^2$.
However since the energies and momenta of bound  nucleons 
do not satisfy the 
energy-momentum  relation of free nucleons, additional 
freedom arises.

This becomes immediately obvious from the following simple kinematic 
consideration. 
In the laboratory frame deep-inelastic scattering from a nucleon 
bound in a nucleus involves the removal 
energy, $-\varepsilon_n$,  of the struck nucleon:
\begin{equation}
\varepsilon_n = M_{\T A} - M^{(n)}_{{\T{A-1}}} - M.
\end{equation}
Here $M_{\T A}$ and $M^{(n)}_{\T {A-1}}$ denote the invariant masses of the 
initial nuclear ground state and of the nuclear system,  
with a nucleon-hole state characterized by its quantum numbers $n$.
The energy of the interacting nucleon is then:
\begin{equation} \label{eq:p0_with_recoil}
p_0 = M + \varepsilon_n - T_{\T R}, 
\end{equation}
where $T_{\T R} = {\vec p^2}/{2 M^{(n)}_{\T{A-1}}}$ is the  recoil energy
of the residual nuclear system. 
We finally obtain for the squared four-momentum of this  
interacting nucleon:
\begin{equation}
p^2 = p_0^2 - \vec p^2 \simeq M^2 + 2 M (\varepsilon_n - T_{\T R} - T) 
\ne M^2,
\end{equation}
with $T=\vec p^2/2M$. The squared four-momentum of the active 
nucleon is obviously not restricted by its free invariant mass. 
It is determined by the nuclear wave function which describes  
the momentum distributions of bound target nucleons as well as the mass 
spectrum of the residual nuclear system. 
Consequently,  the structure function of a bound nucleon 
can in general depend also on $p^2$, not just on $x$ and $Q^2$.

\subsection{Corrections from binding and Fermi motion}
\label{ssec:binding_Fermi}

In the impulse approximation  deep-inelastic scattering 
from a nucleus at large $Q^2$ proceeds in two steps 
as shown  in Fig.\ref{fig:emc_ia}: 
the exchanged virtual photon scatters from a quark with   
momentum $k$. 
This quark belongs to a nucleon with momentum $p$ which is 
removed from the target nucleus. 
Treating the nucleus  as a non-relativistic bound state,  
the nuclear structure functions factorize into 
two terms \cite{Kulagin:1994fz}:
the information about  the quark and gluon substructure of the nucleons  
is included  in the  bound nucleon structure functions  
$F_{1}^{\T N}(x/y,p^2)$ and $F_{2}^{\T N}(x/y,p^2)$. 
They  depend on the fraction $x/y = k\cdot q/p\cdot q \approx k^+/p^+$ 
of the light-cone momentum\footnote{Here the photon momentum is chosen as 
$q^{\mu} = (q_0, \vec 0_\perp,q_3)$ with $q_3 < 0$.} 
of the interacting  nucleon carried 
by the quark, and 
reduce to the corresponding free nucleon structure functions 
at $p^2 = M^2$. 
Details about  nuclear structure are incorporated in the 
distribution function of nucleons with squared 
four-momentum $p^2$ and a  fraction $y$ of the 
nuclear light-cone momentum:  
\begin{eqnarray} \label{D-N/A}
D_{\rm N/A}(y,p^2) &=& \int\!\frac{d^4p'}{(2\pi)^4}\, {\cal S}(p')
        \left(1+\frac{p'_3}{M}\right)
\delta\left(y -\frac{p'^{+}}{M}\right)\,\delta(p^2 - {p'}^2).
\end{eqnarray}
Here
\begin{equation}  \label{S(p)}
{\cal S}(p)=2\pi\sum_n
\delta(p_0 - M - \varepsilon_n + T_{\T R}) 
\left|\Psi_n({\vec  p})\right|^2
\end{equation}
is the spectral function of a nucleon in the nucleus. 
It is determined by the momentum space amplitude 
$\Psi_n({\vec p})=
\langle({\T{A-1}})_n,-{\vec p}|\hat \Psi(0)|{\T A}\rangle$, 
with $\hat \Psi(0)$ 
representing the non-relativistic nucleon field operator 
at the origin ${\vec r}=0$. 
The squared amplitude $|\Psi_n({\vec p})|^2$ 
describes the probability of finding a nucleon with momentum 
$\vec p$ in the nuclear ground state $|{\T A}\rangle$, and the remaining 
${A}-1$ nucleons in a state $n$ with total momentum $-\vec p$. 
In Eq.(\ref{S(p)}) the sum over a complete set of states 
with ${A-1}$ nucleons is taken.
Note that the spectral function is normalized to 
$A$, the total number of nucleons in the nucleus. 
This leads to the proper normalization 
of the nucleon distribution function in 
Eq.(\ref{D-N/A}) (see e.g. 
Refs.\cite{Jaffe:1985je,Frankfurt:1987ui,Jung:1988jw}). 

The nuclear structure functions are then obtained by a convolution 
over the squared four-momentum of the interacting nucleons and their  
light-cone momentum fraction. 
For the structure function $F_2^{\T A}$ {\em per nucleon} 
this gives \cite{Kulagin:1994fz}: 
\begin{equation}    \label{F2A}
A \,F_2^{\T A}(x)= \int\limits_x^{A}\! dy\int\!dp^2\,
D_{\rm N/A}(y,p^2) F_2^{\T N}(x/y,p^2), 
\end{equation}
where we have suppressed the dependence on $Q^2$ for convenience.

In the following we 
examine the convolution integral (\ref{F2A}) in more detail.  
The momentum distribution of nucleons in the nuclear target,  
\begin{equation}
\int \frac{dp_0}{2\pi} {\cal S}(p) = \sum_n \left|\Psi_n({\vec  p})\right|^2 ,
\end{equation} 
falls rapidly with increasing $|\vec p|$. 
This implies that the nucleon light-cone distribution (\ref{D-N/A}) 
is strongly peaked around $y\approx 1$ and $p^2 \approx M^2$,  
with a typical width $\Delta y\sim p_F/M$ controlled by 
the Fermi momentum $p_F$. 
Expanding the bound nucleon structure function  in Eq.(\ref{F2A}) in a 
Taylor series around $y=1$ and $p^2 = M^2$,  and integrating
term by term, leads to the following expression for the nuclear
structure function per nucleon \cite{Kulagin:1994fz,Frankfurt:1987ui}, 
valid in the range $0.2 <   x  <  0.7$: 
\begin{eqnarray}    \label{expansion}
F^{\T A}_2(x) \approx F_2^{\T N}(x)
        -\frac{\langle\varepsilon\rangle}{M}\:x{F_2^{\T N}}^{\prime}(x)
        +\frac{\langle T\rangle}{3M}\:x^2{F_2^{\T N}}^{\prime\prime}(x) 
\nonumber\\
        +2\:\frac{\langle\varepsilon\rangle-\langle T\rangle}{M}
\left(p^2\frac{\partial F_2^{\T N}(x;p^2)}{\partial p^2}\right)_{p^2=M^2}, 
\end{eqnarray}
where ${F_2^{\T N}}^{\prime}(x)$ and ${F_2^{\T N}}^{\prime\prime}(x)$ 
are derivatives
of the structure function with respect to $x$. 
The mean value of the single particle energy  $\varepsilon = p_0 -M$  
is
\begin{eqnarray}    \label{E-rem}
    \langle\varepsilon\rangle &=&
    {1\over {\T A}}\int\!\frac{d^4p}{(2\pi)^4}\, {\cal S}(p)\, \varepsilon,
\label{E-kin}
\end{eqnarray}
\noindent
and
\begin{eqnarray} 
\langle T\rangle &=&
{1\over {\T A}}\int\!\frac{d^4p}{(2\pi)^4}\, {\cal S}(p)\, \frac{{\vec  p}^2}{2M}
\end{eqnarray}
represents the mean kinetic energy of  bound nucleons. 
Except for light nuclei the recoil energy $T_{\T R}$ in 
Eq.(\ref{eq:p0_with_recoil}) can be neglected. 
Then $\langle\varepsilon\rangle$ coincides with the 
separation energy.
Corrections to Eq.(\ref{expansion}) are of higher order in
$\langle\varepsilon\rangle/M$ and $\langle T\rangle/M$. 
Note that the approximate result for $F_2^{\T A}$ in Eq.(\ref{expansion}) 
is well justified in the region $0.2 <  x < 0.7$. 
Here the kinematic condition $x/y<1$ in Eq.(\ref{F2A}) 
can be ignored in accordance with the underlying expansion.

Let us briefly discuss the physical meaning of the different terms 
in Eq.(\ref{expansion}) and their implications.  
The second term on the right hand side of  Eq.(\ref{expansion}) 
involves the average separation energy of nucleons from the 
target. As such it is determined  by nuclear binding.  
In the range  $0.2 < x < 0.7$ it leads to 
a depletion of the nuclear structure function compared to the 
structure function of a free nucleon.
The third term accounts for the Fermi motion of bound 
nucleons and yields  a strong rise of the structure function 
ratio $F_2^{\T A}/F_2^{\T N}$ at large $x$. 
Finally, the fourth term in (\ref{expansion})  reflects 
the  dependence of the structure function of a bound nucleon  on 
its squared four-momentum.
Note that this contribution enters at the same order 
as binding and Fermi-motion corrections. 
Information  about the $p^2$-dependence of bound nucleon structure 
functions is rare. 
Nevertheless such effects  may lead to significant modifications of the 
EMC ratio  $F_2^{\T A} /F_2^{\T N}$ at moderate and large values of $x$.
This  has been shown for example in the framework of a simple quark-diquark 
picture for the nucleon \cite{Kulagin:1994fz}.

An important and not yet completely solved problem with respect
to the binding and Fermi-motion corrections in Eq.(\ref{expansion}) 
is a reliable calculation of
$\langle\varepsilon\rangle$ and $\langle T\rangle$.
In a simple nuclear
shell model the separation  energy is averaged over all occupied levels.  One
finds typical values $\langle\varepsilon\rangle\approx -(20-25)\;$MeV and
$\langle T\rangle\approx 18-20\;$MeV.  Correlations between nucleons change
the simple mean field picture substantially and  lead to high momentum 
components with $|\vec p| > p_F$ in the nuclear spectral 
function (\ref{S(p)}). 
This in turn causes an increase of the average separation  energy
$\langle\varepsilon\rangle$ \cite{CiofiDegliAtti:1989eg,Dieperink:1991mw}.  
In order to see this let us examine the Koltun sum rule \cite{Koltun:1972}
\begin{equation} \label{Kolsr}
\langle\varepsilon\rangle + \langle T\rangle = - 2\mu_B ,
\end{equation}
where $\mu_B$ is the binding energy per nucleon. This sum rule 
is exact if only two-body forces are present in 
the nuclear Hamiltonian.
With fixed   
$\mu_B\approx  8\,$MeV, this sum rule tells that an increase  of
$\langle T\rangle$ due to high momentum components is accompanied by an
increase of $|\langle\varepsilon\rangle|$. We refer in this
context to a calculation \cite{Benhar:1989aw} of the spectral function
of nuclear matter based on a variational method.  This calculation
shows that there is a significant probability to find nucleons with
high momentum and large separation energies. 
An integration of the
spectral function of Ref.\cite{Benhar:1989aw} gives $\langle
T\rangle\approx 38\;$MeV and $\langle\varepsilon\rangle\approx
-70\;$MeV \cite{Dieperink:1991mw}.  
In order to estimate these quantities for heavy  nuclei
one usually assumes \cite{CiofiDegliAtti:1989eg} 
that the high momentum components 
of the nucleon momentum distribution are about the same as in nuclear
matter. 
Together with Eq.(\ref{Kolsr}) this leads
to $\langle\varepsilon\rangle\approx -50\;$MeV.  

\begin{figure}[t]
\bigskip
\begin{center} 
\epsfig{file=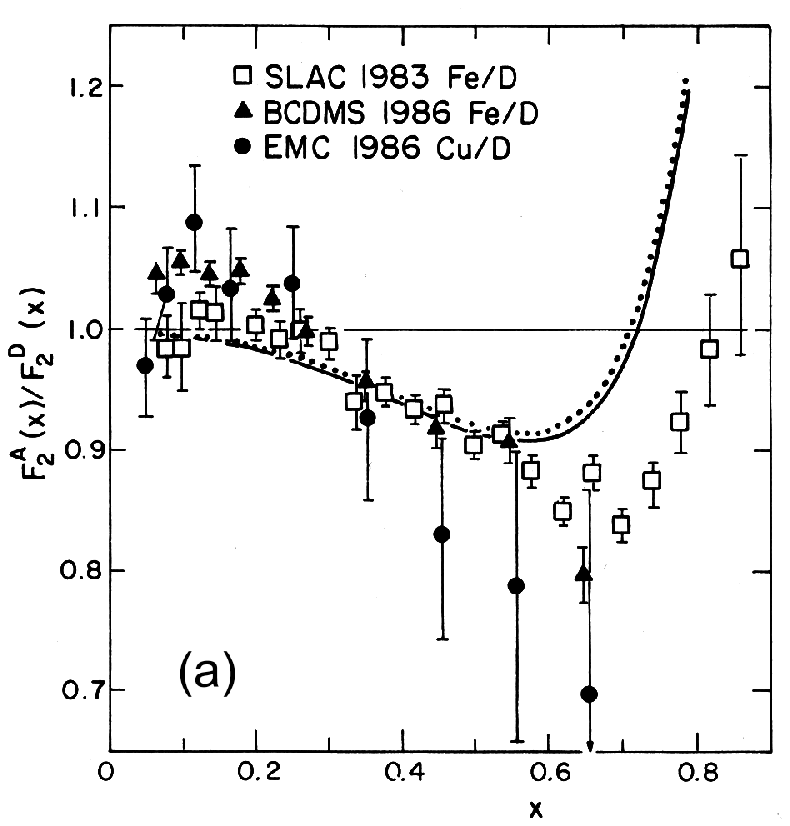,height=70mm}
\end{center}
\vspace*{1cm}
\begin{center} 
\epsfig{file=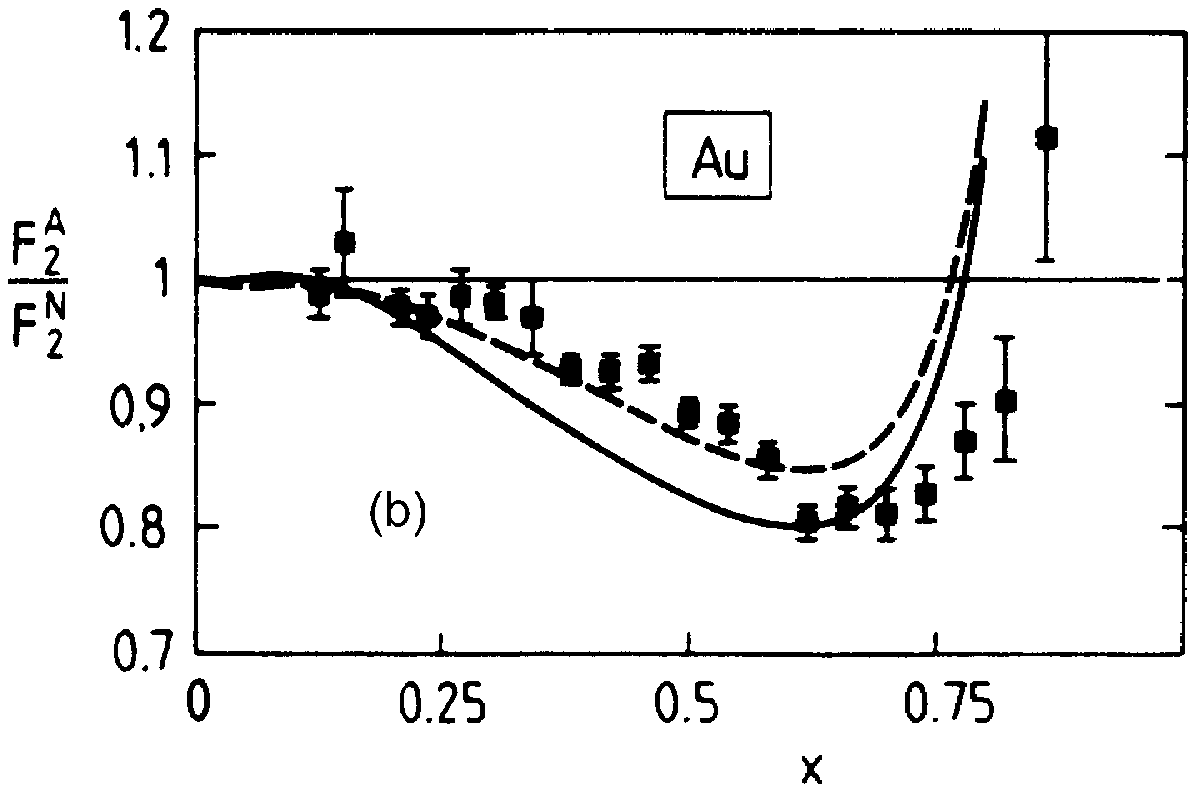,height=70mm}
\end{center}
\caption[...]{The ratio of nuclear and nucleon structure functions,   
$F_2^{\T A}/F_2^{N}$, for iron and gold taken from 
Refs.\cite{Jung:1988jw,Kulagin:1994fz}. 
(a) solid curve: calculation in Ref.\cite{Jung:1988jw}, 
dotted curve: calculation in Ref.\cite{Frankfurt:1987ui}. 
(b) results from Ref.\cite{Kulagin:1994fz}: without $p^2$-dependence 
of the bound nucleon structure function (dashed), and  
including this $p^2$-dependence 
as obtained from a simple quark-diquark picture (full). 
}
\label{fig:result_EMC}
\bigskip
\end{figure}

In Fig.\ref{fig:result_EMC} we show  typical results from 
Refs.\cite{Kulagin:1994fz,Jung:1988jw} for iron and gold.
We observe that a qualitative understanding of the EMC effect 
can indeed be reached, but 
at $x \gsim   0.5$ a more quantitative description is still lacking. 
One should note, of course, that the presentation of nuclear effects 
in terms of the ratio $F_2^{\T A}/F_2^{\T N}$ magnifies 
such effects in a misleading manner because $F_2^{\T N}$ itself 
is small in this region 
(see also the discussion in Section \ref{ssec:Co_spa_nuclei}). 

The impulse approximation picture of nuclear deep-inelastic scattering can
also be maintained in a relativistically covariant  way 
\cite{Melnitchouk:1994nk}.
Here, however, a simple factorization of nuclear structure functions 
into nuclear and nucleon parts as in Eq.(\ref{F2A}) is not possible 
any longer.  
A relativistic calculation of nuclear structure 
functions requires relativistic nuclear wave functions as well as  
a more detailed knowledge about the structure of bound nucleons. 
Nevertheless, relativistic effects seem to be small:  
in an explicit model calculation  of  the deuteron 
structure function $F_2^{\T d}$ relativistic corrections 
to the non-relativistic convolution 
(\ref{F2A}) are  less than  $2\%$  for $x<0.9$ \cite{Melnitchouk:1994rv}. 
In the region $x>1$, where nuclear structure functions are 
very small however, larger deviations are expected.

In this context a word of caution is in order. A description 
of nuclear structure functions based on nucleons alone 
is necessarily incomplete since it violates the momentum 
sum rule \cite{Frankfurt:1988nt}. 
Non-nucleonic degrees of freedom are briefly discussed in 
Section \ref{ssec:Pion}.

\subsection{Beyond the impulse approximation}

The quality of the impulse approximation has frequently been 
questioned (see e.g. \linebreak
Refs.\cite{Frankfurt:1988nt,Geesaman:1995yd,Kulagin:1994fz,%
Hoodbhoy:1987fn,Saito:1994yw,Melnitchouk:1994nk} and references therein). 
Here we give a brief summary of possible shortcomings in terms of  
models for nuclear deep-inelastic scattering which 
go beyond this  approximation.

\subsubsection{Quark exchange in nuclei}

The impulse approximation includes only incoherent 
scattering processes 
from hadronic constituents of the target nucleus. 
On the other hand, contributions involving several bound nucleons  
could also be important, and their role needs to be examined. 

One such possibility, namely quark exchange between different nucleons,  
has been investigated in 
Refs.\cite{Hoodbhoy:1987fn,Meyer:1994fg}.
The nuclear quark wave function   
which is probed in deep-inelastic scattering  must be antisymmetric 
with respect to permutations of quarks. 
This however is not realized in  the impulse approximation (\ref{F2A}).  
Antisymmetrization introduces additional quark exchange terms between 
different nucleons in the target. 

Under several simplifying assumptions a softening of the 
nuclear quark momentum distribution due to quark exchange 
has been found. 
For small nuclei the effect turned out to be significant.  
For $^4\rm{He}$ approximately $30\%$ of the 
observed depletion of the structure function ratio 
$F_2^{\T{He}}/F_2^{\T{N}}$ at $x\sim 0.6$ has been associated with 
quark exchange.
Only minor modifications have been 
found for  heavier nuclei \cite{Meyer:1994fg}.

While the estimates  based on a simple quark exchange model may 
not be reliable at a 
quantitative level, they certainly point to the fact that 
the impulse approximation is incomplete as soon as 
correlations between quarks in several nucleons come into play.

\subsubsection{Final state interactions in a mean field approach}

One of the basic assumptions of the impulse approximation 
is that interactions of the struck, highly excited nucleon 
with the residual  nuclear system can be ignored.
In general there is no solid basis for this assumption 
since the debris of the struck nucleon includes 
also low momentum fragments as seen from the target rest frame.
A proper treatment of their final state interaction  requires however a 
description of the nucleus in terms of quark and gluon degrees of 
freedom.

Investigations in this direction have been made   
starting out from a quark model for nuclear matter, 
with nucleons modeled as non-overlapping 
MIT bags \cite{Saito:1994yw,Guichon:1988jp,Saito:1992rm}.
The nucleons interact  via the exchange of scalar and vector mesons 
which couple directly to quarks.
Within the mean field approximation for the meson fields 
it is possible to describe several basic properties of nuclear 
matter, such as its compressibility and  
the binding energy per nucleon at saturation density.

This model has been applied to deep-inelastic scattering from 
finite nuclei using a local density approximation \cite{Saito:1994yw}. 
The debris of the struck  nucleon is represented by  a pair 
of spectator quarks bound in a diquark bag. 
Its interaction with the remaining nuclear system in the final 
state leads to a non-negligible  effect on nuclear structure functions:
while the full calculation including final state interactions 
allows to reproduce the structure function ratio $F_2^A/F_2^N$, 
the impulse approximation overestimates nuclear effects 
at $x\sim 0.6$ by about a factor two.
In  the framework of this model such a difference is 
expected since the binding of the nucleus 
is the result of the attractive scalar mean field experienced 
by all three constituent quarks of the interacting nucleon. 
When ignoring the binding of the spectator quark pair, as done in the 
impulse approximation, one assumes that 
the active quark which takes part in the deep-inelastic scattering 
process behaves as if it carries the binding of all three quarks, 
a feature which obviously needs to be corrected.

The mean field approach  to nuclear 
quark distributions is based on several simplifying  assumptions, 
but it nevertheless points to  the possible importance of 
final state interactions and, in more general terms, to the 
relevance of quark degrees of freedom in nuclei.

\subsection{Modifications of bound nucleon structure functions}

The intrinsic properties of nucleons bound in nuclei can 
be modified as compared to free nucleons. We summarize 
below two examples of models which deal with such 
possible changes in bound nucleon structure functions.

\subsubsection{Rescaling}

At intermediate values of the Bjorken variable, $0.2 < x < 0.7$,  
the modification 
of nuclear structure functions $F_2^{\T A}$ as compared to the 
free nucleon structure function $F_2^{\T N}$ can be described  
by a shift in the momentum scale which 
enters the structure functions. 
We briefly outline here the basic arguments 
\cite{Close:1983tn,Jaffe:1984zw,Close:1985zn,Close:1986ji,Close:1988ay}. 
Consider the  moments\footnote{For 
simplicity we use the non-singlet part only.}
\begin{equation}
M_n^{\T A}(Q^2) = \int_0^A dx\,x^{n-2} F_2^{{\T A}}(x,Q^2), 
\quad {\rm with}\;n\;{\rm{even}}. 
\end{equation}
Assume now that the moments of nuclear and nucleon structure 
functions are related by a shift of their momentum scale:
\begin{equation}
M_n^{\T A}(\mu_{\T A}^2) = M_n^{\T N}(\mu_{\T N}^2).
\end{equation}
At an arbitrary momentum transfer $Q^2$  
the  perturbative QCD evolution 
equations to leading order (see Section \ref{ssec:AP_eq}) give:
\begin{equation} \label{eq:moments_Q^2}
M_n^{\T A}(Q^2) = M_n^{\T N}\left(\xi_{\T A}(Q^2) Q^2\right),
\end{equation}
with the rescaling parameter
\begin{equation}
\xi_{\T A}(Q^2) = \left(\frac{\mu_{\T N}^2}{\mu_{\T A}^2}
\right)^{\frac{\alpha_s(\mu_{\T A}^2)}{\alpha_s(Q^2)}}.
\end{equation}
Of course Eq.(\ref{eq:moments_Q^2}) can always be satisfied 
if one allows different $\xi_{\T A}$ for different moments $n$.
However, when comparing with data it has turned out 
that the rescaling parameter is 
independent of $n$ to a good approximation.  
Consequently, the scale change for the moments 
(\ref{eq:moments_Q^2}) 
can be translated directly into a scale change for the  
structure functions themselves:
\begin{equation}
F_2^{\T A}(x,Q^2) = F_2^{\T N}\left(x,\xi_{\T A}(Q^2) Q^2\right).
\end{equation}
Good agreement with experimental data can be achieved at intermediate $x$. 
For example,  the EMC structure function data on iron    
suggest $\xi_{\rm Fe} \approx 2$ for $Q^2 = 20$ GeV$^2$ 
\cite{Close:1985zn}.

Rescaling gives a reasonable one-parameter description 
of nuclear structure functions $F_2^{\T A}$ at intermediate $x$,  
but it does not offer insights into the physical origin of 
the observed change of scale. 
One possible suggestion to explain the scale change is 
a modification of  the 
effective confinement scale for quarks in the 
nucleus as compared to free nucleons 
\cite{Close:1983tn,Jaffe:1984zw,Close:1985zn}.  

Scale changes are not simply related to possible ``swelling'' 
of nucleons inside nuclei which 
is constrained by inclusive electron-nucleus 
scattering data in the quasielastic region.
The experimentally observed $y$-scaling 
indicates a rather small increase of  the charge radius   
for bound nucleons. 
For example, the study of \cite{Sick:1985a,Sick:1985b}  comes to the 
conclusion that any increase of the nuclear radius in 
nuclei should be less than $6\%$ of its free radius.

Related discussions and a comparison with nuclear DIS data 
can be found in  
Refs.[202,216,\linebreak 231--234].
\nocite{Kumano:1994pn,CiofiDegliAtti:1989eg,%
Nachtmann:1984py,Bickerstaff:1985mp,Mulders:1985ec,%
Bickerstaff:1987ck}

\subsubsection{Color screening in bound nucleons}

The scenario of Refs.\cite{Frankfurt:1988nt,Frankfurt:1985cv,Frank:1996pv} 
assumes that the 
dominant contribution to the structure function $F_2^{\T N}$ 
at large $x \sim 0.6$ is given by small size (pointlike) 
parton configurations  in the nucleon.
In a nuclear environment such configurations interact only 
weakly with other nucleons due to the screening of their color charge.
It is argued that the probability for  
pointlike configurations is reduced in bound nucleons. 
In fact, the probability to find parton configurations 
of {average} size in the nucleon should actually 
be enhanced in nuclei since they experience the attraction of 
the nuclear mean field. 
Then the variational principle with normalization condition on 
the wave function implies that {small}-size configurations 
should indeed be suppressed.

An estimate of such deformations in the wave function of nucleons 
bound in heavy nuclei gives  
for $x \sim 0.5$ \cite{Frankfurt:1985cv}:  

\begin{equation} \label{eq:str_fn_ratio_FS}
\frac{F_2^{\T A}}{F_2^{\T N}} 
\sim 1 + \frac{4 \, \langle U \rangle}{E^*} 
\sim 0.7 \,\mbox{--} \,0.8.
\end{equation}
Here $\langle U \rangle$ is the average potential energy 
per nucleon, $\langle U \rangle \sim - 40$ MeV, 
and $E^* \sim 0.5$ GeV is  the typical energy scale for 
excitations of the nucleon. 
Since $\langle U \rangle$ scales with the nuclear 
density, the nuclear dependence of the structure function ratio  
(\ref{eq:str_fn_ratio_FS}) is roughly consistent with data.

It should be mentioned that the proposed suppression of 
rather rare pointlike configurations in  bound nucleons does not 
necessarily imply a substantial change of average properties
of a bound nucleon, such as  
its electromagnetic radius \cite{Frankfurt:1988nt}.

\subsection{Pion contributions to nuclear structure functions}
\label{ssec:Pion}

In conventional nuclear physics 
meson exchange is responsible for the binding of nucleons in the nucleus. 
Therefore  deep-inelastic scattering from mesons 
present in the nuclear wave function should 
give additional contributions 
to nuclear structure functions.
Pions, which are responsible for most of the intermediate- 
and long-range nucleon-nucleon force, are supposed to play 
the prominent role (see e.g. 
\cite{Ericson:1983um,Berger:1985na,%
Birbrair:1989hb,Kulagin:1989mu,Jung:1990pu}).

The framework is the Sullivan process \cite{Sullivan:1972}. 
Its contribution to the nucleon structure function 
$F_{2}^{\T N}$ reads:
\begin{equation} \label{eq:Sullivan_F}
\delta^{\pi} F_{2}^{\T N}(x) = \int_x^1 {\T d} y 
\,f_{{\T{\pi/ N}}}(y)\, 
F_{2}^{{\T\pi}}(x/y), 
\end{equation}
where 
\begin{equation}\label{eq:pion_LC_N}
f_{\T{\pi/N}}(y) = \frac{3 g^2_{\T{\pi N N}}}{16 \pi^2}
\int^{t_{min}}_{-\infty} {\T d} t 
\frac{ - t |{\cal F}_{{\T{\pi N N}}}(t)|^2}{(t-m_{{\T \pi}}^2)^2}
\,y
\end{equation}
specifies  the distribution of pions with light-cone momentum 
fraction $y$ in the nucleon, 
while $F_{2}^{{\T \pi}}$ is the pion structure function.  
Equation (\ref{eq:Sullivan_F}) describes deep-inelastic scattering 
from a pion emitted from its nucleon source. 
The nucleon receives a momentum transfer 
equal to the pion momentum $k^{\mu} = (\omega, \vec k)$.
The minimal squared momentum transfer $t=k^2$ required 
for pion emission is $t_{min} = -M^2 y^2/(1-y)$. 
One finds that $\delta^{\T{\pi}} F_2^{\T N}$ gets its dominant 
contributions from pions with  momenta 
$|\vec k|\simeq 300 -400$ MeV. 
Pions with smaller momenta are suppressed by  the explicit 
factor $y$ in Eq.(\ref{eq:pion_LC_N}), while pions with large 
momenta are suppressed by the pion propagator and 
the $\T{\pi N N}$ form factor ${\cal F}_{\T{\pi NN}}$ 
\cite{Ericson:1983um}.

The convolution ansatz 
in Eq.(\ref{eq:Sullivan_F}) suffers from similar problems as  
convolution for nuclear structure functions discussed 
in Section \ref{ssec:binding_Fermi}: 
the interacting pion is not on its mass
shell, i.e. $k^2 \approx - \vec k^2 \ne m_{\pi}^2$. 
Therefore the pion structure function depends also on $k^2$. 
Furthermore, final state interactions of the pion 
debris with the recoil nucleon are neglected. 

The detailed treatment of pionic effects in nuclei includes 
the pion propagation in the medium with $\Delta$ resonance 
excitation, Pauli effects and short range spin-isospin 
correlations. 
All these effects are incorporated in the pion-nuclear 
response function $R(\vec k,\omega)$ 
which determines the spectrum of pionic excitations in the nuclear medium.
The resulting distribution function of pions in a nucleus is 
\cite{Ericson:1983um}
\begin{equation} \label{eq:pion_LC_A}
f_{\T \pi/{\T A}} (y) = \frac{3 g^2_{\T{\pi N N}}}{16 \pi^2} y 
\int_{M^2 y^2}^\infty {\T d} |\vec k|^2 \int_0^{|\vec k|-My} {\T d} \omega 
\,\frac{ \vec k^2 |{\cal F}_{{ \T{\pi N N}}}(\vec k^2)|^2}
{(t-m_{{\T \pi}}^2)^2}
\,R(|\vec k|,\omega), 
\end{equation}
where $t = \omega^2 - \vec k^2$. 
Using the the Sullivan description (\ref{eq:Sullivan_F}) 
the contribution of excess pions to the nuclear structure function 
$F_{2}^{\T A}$ can be calculated according to: 
\begin{equation}
\delta^{\pi} F_{2}^{{\T A}}(x) = 
\int_x^1 {\T d} y 
\left(f_{{\T{\pi/ A}}}(y)- f_{{\T{\pi/ N}}}(y)\right) 
F_{2}^{ {\T\pi}}(x/y).
\end{equation}
In the original work in Ref.\cite{Ericson:1983um}, 
using the empirical pion structure function,  
a significant enhancement of the 
ratio $F_{2}^{\T A}/F_{2}^{\T N}$ was found at $x<0.3$. 
This observation was in agreement with the early 
EMC data \cite{Aubert:1983}.
Later data on nuclear 
structure functions showed only a minor enhancement 
around $x\simeq 0.15$ (see Section \ref{subs:Nucl_F2}).
In addition Drell-Yan  data from E772 \cite{Alde:1990im}
have demonstrated that the antiquark distribution in nuclei 
is not significantly enhanced as compared to free nucleons,  
in disagreement with the first pion model 
calculations. 
However, as already emphasized in Ref.\cite{Ericson:1983um}, the 
pion contribution to nuclear structure functions is 
very sensitive to the 
strength of repulsive short range spin-isospin correlations in nuclei. 
Variations of this strength  by $15\%$ can easily lead to 
$30\%$ changes in $\delta^{\T{\pi}} F_{2}^{\T A}$. 
While it is not difficult to accommodate the 
very small observed pionic enhancement within such uncertainties, 
it is still a challange to arrive at a consistant overall picture 
of nuclear DIS which rigorously satisfies the requirements of 
the momentum sum rule.

\subsection{Further notes}

Related studies of pion field effects as well as other nuclear medium 
corrections and their implications on nuclear DIS have been performed in 
Refs.\cite{Brown:1995su,Miller:1996qg,Dieperink:1997iv,Koltun:1997py}. 
These studies include calculations within the delta-hole model 
\cite{Dieperink:1997iv}, the role of NN correlations and the energy dependence 
of nuclear response functions \cite{Koltun:1997py}, possible effects of 
``Brown-Rho scaling'' on nuclear structure functions \cite{Brown:1995su}, 
and implications of low-energy pion-nucleus scattering data for nuclear 
deep-inelastic scattering and Drell-Yan production 
\cite{Miller:1996qg}.

Some further investigations use a relativistic many-body approach 
to treat mesonic and binding corrections to 
reproduce nuclear effects in the EMC and Drell-Yan measurements 
\cite{Marco:1996vb,Marco:1997xb}. 
From the point of view of nuclear many-body theory, the best nuclear 
wave functions have been employed in 
Refs.\cite{Benhar:1997emc,Benhar:1998gb}, 
treating both short and long range correlations in nuclear matter 
and helium at the most advanced level. 
Two-nucleon correlations turn out to be important in 
nuclear DIS. 
It should be mentioned, however, that the results of  
\cite{Marco:1996vb,Benhar:1997emc,Benhar:1998gb} have met with some 
debate concerning the proper choice of the ``flux factor''. 
Questions of rigorous baryon number conservation 
\cite{Frankfurt:1988nt} 
have also been raised.

It has been suggested to investigate  intrinsic properties of 
bound nucleons  in semi-inclusive deep-inelastic scattering 
from nuclei 
\cite{Frankfurt:1985cv,Melnitchouk:1996vp,Simula:1996xk,CiofidegliAtti:1993ep}.
Measuring the scattered lepton in coincidence with 
the residual nuclear system should provide detailed 
information on changes in bound nucleon structure functions.
Possible experiments are discussed at HERMES 
\cite{Ingelman:1996ge}.

\section{Deep-inelastic scattering from polarized nuclei}
\setcounter{section}{7}
\setcounter{figure}{0}
\label{sec:Pol_DIS}

Understanding the spin structure of the proton and the neutron 
is a central issue in QCD. Both the polarized neutron and proton 
structure functions, $g_1^{\rm n}$ and $g_1^{\rm p}$, are needed
in the investigation of flavor singlet quark spin distributions 
(see e.g. Ref.\cite{Anselmino:1995gn}), and in the experimental 
test of the fundamental Bjorken sum rule (\ref{eq:Bj-SR}). 

Since free neutron targets are not available one must 
resort to polarized nuclei, such as the deuteron and $^{3}\rm{He}$, 
where the neutron spin plays a well defined role in building up 
the total polarization of the nuclear target. 
Polarized deep-inelastic scattering from deuterium 
\cite{Adeva:1998vv,Anthony:1999rm,Anthony:1999py,Abe:1998wq} 
and 
$^{3}\rm{He}$ 
\cite{Abe:1997qk,Ackerstaff:1997ws,Abe:1997cx}
has been studied with 
high precision. 
In order to deduce accurate information about the individual 
nucleon spin structure functions from these data, it is essential 
to correct for genuine nuclear effects. 
In addition, the presence of the tensor interaction between 
nucleons in nuclei creates specific spin effects which are of interest 
in their own right.

\subsection{Effective polarization}

As discussed in Section \ref{Sec:space_time}, nuclear structure functions at 
Bjorken-$x > 0.2$ are dominated by the incoherent scattering 
from bound nucleons. 
For polarized nuclei, the non-trivial spin-orbit structure 
of the wave functions causes new effects. 
Bound nucleons can carry orbital angular momentum, so 
their polarization vectors need not be aligned with the total polarization of
the target nucleus: 
depolarization effects occur. 

In order to describe such nuclear depolarization phenomena 
it is useful to introduce effective polarizations for 
nucleons bound in the nucleus. 
Let $|A \uparrow\rangle$ represent a nuclear state polarized 
in the $z$-direction. Then the effective polarization of 
protons or neutrons in that nucleus is 
\begin{eqnarray}
\label{eq:nucleon_polarizations}
{\cal P}_{\T p}^{\T A} &=& 
\left \langle A \uparrow \right| \sum_{i=protons} \sigma_z (i)
\left| A \uparrow \right \rangle, 
\\ 
{\cal P}_{\T n}^{\T A} &=& 
\left \langle A \uparrow \right| \sum_{i=neutrons} \sigma_z (i)
\left| A \uparrow \right \rangle. 
\end{eqnarray}
When the nuclear depolarization effects are described entirely 
in terms of these effective polarizations of bound nucleons, 
the nucleon spin structure functions 
(say, $g_1^{\T A}$ and $g_2^{\T A}$) have the following simple additive 
form: 
\begin{equation} \label{eq:g1A_depol}
g_{1,2}^{\T A}(x,Q^2) = {\cal P}_{\T p}^{\T A} \, g_{1,2}^{\T p}(x,Q^2)
+ {\cal P}_{\T n}^{\T A} \, g_{1,2}^{\T n}(x,Q^2).
\end{equation}
Nuclear depolarization effects are important over the whole 
kinematic range of recent measurements. 
Within the impulse approximation these effects exceed 
by far the influence of nuclear binding and Fermi motion 
at $0.2 < x < 0.7$ (see Section \ref{ssec:pol_DIS_largex}).

\subsection{Depolarization in deuterium and 
$^{3}$He}

In case of the deuteron the effective proton 
and neutron polarizations are simply determined by 
the D-state admixture in the deuteron wave function, 
induced by the tensor interaction between proton and neutron 
in the spin triplet state. One finds:
\begin{equation} \label{eq:nucleon_polarization_d}
{\cal P}_{\T p}^{\T d} = {\cal P}_{\T n}^{\T d} =  1-\frac{3}{2} P_{\T D} 
\end{equation}
where $P_{\T D}$ is the D-state probability. 
The numerical values of 
${\cal P}_{\T {p,n}}^{\T d}$ range between $0.91$ and $0.94$ 
using deuteron wave functions calculated with 
the Paris \cite{Lacombe:1980dr} or Bonn \cite{Machleidt:1987hj} 
nucleon-nucleon potential, respectively.

Apart from the interest in neutron spin structure functions 
the deuteron with its triplet spin structure 
is of interest all by itself. 
Its spin-$1$ property leads to additional structure functions as 
given in Eq.(\ref{eq:hadten_A}). 
In particular, the new spin structure functions $b_1$ and $b_2$ 
are accessible in deep-inelastic scattering from polarized 
deuterons and can be investigated in forthcoming 
HERMES measurements \cite{Jackson94}.

Polarized  $^3\T{He}$ can be viewed, to a first approximation, 
as a polarized neutron target, with 
the proton-neutron subsystem in a spin singlet configuration 
and the surplus neutron carrying the spin of the three-body 
system. 
Corrections to this picture come from the admixture of 
S$'$- and D-wave components to the 
$^3\T{He}$ wave function. 
The consequence is that the effective neutron polarization 
is reduced from unity, 
and the effective proton polarization does not vanish: 
\begin{eqnarray} \label{eq:pn_He}
{\cal P}_{\T n}^{\T{He}} &=& 1  -\frac{2}{3} P_{\T S'} - 
\frac{4}{3} P_{\T D} 
= 0.86 \pm 0.02, 
\\
\label{eq:pp_He}
{\cal P}_{\T p}^{\T{He}} &=& - \frac{2}{3} 
\left(P_{\T D} - P_{\T S'} \right)  
= - 0.056 \pm 0.008. 
\end{eqnarray}
These results are obtained from three-body calculations using 
realistic nucleon-nucleon interactions \cite{CiofidegliAtti:1993zs}, 
omitting however effects from meson exchange 
currents (see Section \ref{subs:Helium})

\subsection{Nuclear coherence effects in polarized deep-inelastic scattering}
\label{sec:Coh_Pol}

Coherence phenomena such as shadowing at small Bjorken-$x$ are dominated 
by the interaction of diffractively excited hadronic states with 
several nucleons over large longitudinal distances in the 
target nucleus. 
The characteristic space-time properties of DIS are independent of 
the target or beam polarization. 
Therefore, nuclear coherence effects are also expected in polarized 
scattering. 
We explore such effects in the following for deuterium and $^3\T{He}$.

\subsubsection{Polarized single and double scattering in the deuteron}

Consider the deuteron spin structure functions $g_1^{\T d}$ and $b_1$  
at small values of the Bjorken variable, $x<0.1$. 
Following the discussion in Section \ref{ssec:nucl_str_fns}, these structure 
functions can be expressed in terms of virtual photon-deuteron helicity 
amplitudes. 
At large $Q^2$ in the Bjorken limit which we keep throughout this section, 
only the helicity conserving amplitudes enter. 
As usual, we choose a right-handed, transversely polarized (virtual) 
photon (index ``$+$'') for reference. 
We denote the helicity conserving $\gamma^*\T d$ amplitudes 
by ${\cal A}^{\gamma^*\T d}_{+ {H}}$, where ${H}=0,+,-$
refers to the helicity state of the polarized deuteron, and we choose the 
direction defined by the photon momentum $\vec q$ as quantization axis. 
The spin structure functions of interest are then expressed as 
(\ref{eq:g1A_hel},\ref{eq:b1_hel})
\begin{eqnarray}
\label{eq:g1d_hel}
g_1^{\mathrm d} &=& \frac{1}{4 \pi e^2} \T{Im} \,\left(
{\cal A}^{\gamma^*\mathrm{d}}_{+-}-{\cal A}^{\gamma^*\mathrm {d}}_{++}
\right),\\
\label{eq:b1d_hel}
b_1 &=& \frac{1}{4 \pi e^2} \T{Im} \,\left(
2 {\cal A}^{\gamma^* \mathrm{d}}_{+0}-
{\cal A}^{\gamma^*\mathrm{d}}_{++}-
{\cal A}^{\gamma^*\mathrm{d}}_{+-}
\right).
\end{eqnarray}
Let us now decompose ${\cal A}^{\gamma^*\T d}_{+ {H}}$ into incoherent, 
single scattering terms and a coherent double scattering contribution. 
We will use the non-relativistic deuteron wave function, 
\begin{eqnarray} \label{psi}
\psi_{H}(\vec{r})&=&\frac{1}{\sqrt{4\pi}}\left[\frac{u(r)}{r}+
\frac{v(r)}{r}\frac{1}{\sqrt{8}}\hat{S}_{12}(\hat{\vec{r}})
\right]\chi_{{H}},
\end{eqnarray}
where  $r=|\vec r|$, and $\chi_{H}$ denotes the 
$S=1$ spin wave function of the deuteron. 
The tensor operator 
$\hat{S}_{12}(\hat{\vec {r}}) = 
3 (\vec \sigma_{\mathrm p}\cdot \vec r) 
(\vec \sigma_{\mathrm{n}}\cdot \vec r)/r^2 - 
\vec \sigma_{\mathrm{p}}\cdot \vec \sigma_{\mathrm{n}}$, 
and $u(r)$, $v(r)$ are the  S- and D-state radial wave functions 
normalized as $\int_0^{\infty} {\T d} r \,[u^2(r)+v^2(r)] =1$. 
The   D-state probability is 
$P_{\mathrm D} = \int_0^{\infty} {\T d} r \,v^2(r)$.
We have $P_{\T D} \simeq  5.8\%$ for the Paris potential 
\cite{Lacombe:1980dr} 
and  $P_{\T D} \simeq 4.3\%$ for the Bonn potential 
\cite{Machleidt:1987hj}.

In the polarized deuteron, the proton or neutron can have 
their spins either parallel or antiparallel with respect to the 
$z$-axis defined by $\vec q/|\vec q|$. 
Let the corresponding projection operators be 
$P^{\mathrm{p},\mathrm{n}}_{\uparrow}$ and 
$P^{\mathrm{p},\mathrm{n}}_{\downarrow}$, respectively. 
The amplitude for single scattering of the virtual photon from a proton 
in the polarized deuteron is:
\begin{eqnarray} \label{eq:A_1wf}
{\cal A}^{\gamma^* \mathrm{p}}_{+ {H}}  
&=&\int {\T d}^3r \,
\psi^{\dag}_{{H}}(\vec{r})  
\left( P^{\mathrm{p}}_{\uparrow}{A}^{\gamma^*\mathrm{p}}_{+\uparrow}+
P^{\mathrm{p}}_{\downarrow}{A}^{\gamma^*\mathrm{p}}_{+\downarrow} \right)
\,\psi_{{H}}(\vec{r}), 
\end{eqnarray}
with the helicity conserving $\gamma^*$-proton amplitudes
${\cal A}^{\gamma^*\mathrm{p}}_{+ \uparrow}$ and 
${\cal A}^{\gamma^*\mathrm{p}}_{+ \downarrow}$. 
The analogous amplitudes for single scattering from the neutron 
are obtained by the replacement  
$[\T p \leftrightarrow \T {n}]$. 
We then have 
${\cal A}^{\gamma^*\mathrm{d}}_{+{H}} = 
{\cal A}^{\gamma^*\mathrm{p}}_{+{H}}+
{\cal A}^{\gamma^*\mathrm{n}}_{+{H}}$ at the 
single scattering level. 
Combining with Eqs.(\ref{eq:g1_helicity},\ref{eq:g1d_hel}) one finds 
%
\begin{equation} \label{eq:g_1d_single}
g_1^{\T d}=\left(1-\frac{3}{2}P_{\T D}\right)(g_1^{\T p}+g_1^{\T n}) 
=2 {\cal P}^{\T d} \,g_1^{\T N},
\end{equation}
%
where ${\cal P}^{\T d}$ is the effective nucleon 
polarization (\ref{eq:nucleon_polarization_d}) in the deuteron and 
$g_1^{\T N} = (g_1^{\T p}+g_1^{\T n})/2$. 
Of course, $b_1 = 0$ at the single scattering level since 
nucleons as spin $1/2$ objects do not have a structure function $b_1$.

Next we concentrate on the coherent double scattering amplitude, 
\begin{equation}
\delta {\cal A}^{\gamma^*\mathrm{d}}_{+{H}} 
= {\cal A}^{\gamma^*\mathrm{d}}_{+{H}} - 
\left(
{\cal A}^{\gamma^*\mathrm{p}}_{+{H}}+
{\cal A}^{\gamma^*\mathrm{n}}_{+{H}}
\right)
\end{equation}
which simultaneously involves both the proton and the 
neutron. 
At $x <  0.1$ this amplitude is dominated, as in 
the unpolarized case, by diffractive production and 
rescattering of intermediate hadronic states, 
but now with the polarized target nucleons excited by 
polarized virtual photons. 
We introduce the diffractive production amplitudes
$T^{\T{XN}}_{+\uparrow}(k)$ 
and  
$T^{\T{XN}}_{+\downarrow}(k)$ 
which  describe the diffractive production
process $\gamma^* \T N \rightarrow \T X \T N$  
for right handed photons on polarized nucleons, 
with momentum transfer  $k$. 
Following steps similar to those described in 
Section \ref{ssec:shad_deu}, one finds 
\cite{Edelmann:1997ik,Edelmann:1997qe}:
\begin{eqnarray} \label{eq:A_2_ds}
\delta{\cal A}^{\gamma^* \mathrm{d}}_{+H}&=&\frac{i}{2M \nu}\sum_{\T X}
\int\frac{d^2k_{\perp}}{(2\pi)^2}\int d^2 b\,
e^{i\vec{k}_{\perp}\cdot\vec{b}}
\int_{-\infty}^0dz\,
e^{i {z}/{\lambda}}\nonumber\\
&&\psi^{\dag}_{H}(\vec{r})
\left(
P^{\T p}_{\uparrow}T^{\T{Xp}}_{+\uparrow}(k)+P^{\T p}_{\downarrow}
T^{\T{Xp}}_{+\downarrow}(k)\right)\,\left(
P^{\T n}_{\uparrow}T^{\T{Xn}}_{+\uparrow}(k)+P^{\T n}_{\downarrow}
T^{\T{Xn}}_{+\downarrow}(k)\right)
\psi_{H}(\vec{r}),
\end{eqnarray}
with the longitudinal propagation length 
$\lambda = 2\nu \,(M_{\T X}^2 + Q^2)^{-1}$ of the diffractively 
produced intermediate system. 
We recall from Section \ref{ssec:DIS_SPTH} that a hadronic fluctuation of mass 
$M_{\T X}$ contributes to coherent double scattering only if its propagation 
length $\lambda$ exceeds the deuteron diameter,  
$\langle r^2 \rangle_{\T d} ^{1/2} \simeq 4\,\T{fm}$.

In the following we approximate the dependence of the 
diffractive production amplitudes on the 
momentum transfer $t = k^2 \approx - \vec k_{\perp}^2$ by: 
\begin{equation} \label{eq:T_ampl_approx}
T^{\T {XN}}(k) \approx e^{- B \,\vec k_{\perp}^2 /2 }  
\,T^{\T {XN}}, 
\end{equation}
with the forward amplitude $T^{\T{XN}} \equiv T^{\T{XN}}(\vec k = 0)$. 
Various data on diffractive leptoproduction at  $Q^2\,\lsim \,3\,\T{GeV}^2$ 
suggest  an average slope $B \simeq (5 \dots 10) \,\T{GeV}^{-2}$ 
(for references see e.g. \cite{Crittenden:1997yz,Abramowicz:1998ii}).
We then define  the integrated (longitudinal) form factor  
\begin{equation} \label{eq:F_H}
{\cal F}_{H}^B(\lambda^{-1})=
\int\frac{d^2k_{\perp}}{(2\pi)^2}\, 
\,S_H(\vec k_{\perp},\lambda^{-1}) \, e^{-B \,\vec k_{\perp}^2}, 
\end{equation}
where  
\begin{equation} \label{eq:FF_H}
S_{H}(\vec k)=\int
d^3r\,|\psi_{H}(\vec r )|^2 e^{i\vec k \cdot \vec r} 
\end{equation}
is the conventional helicity-dependent deuteron form factor. 
Next we introduce helicity dependent diffractive production 
cross sections for transversely polarized virtual photons by:
\begin{equation}
\frac {1}{8 M^2 \nu^2}\sum _{\T X} 
\left(|T^{\T X \T p}_{+\uparrow}|^2 + |T^{\T X \T n}_{+\uparrow}|^2 \right) 
= 
16\pi \int_{4m_{\pi}^2}^{W^2} {\T d} M_{\T X}^2 
\left.\frac{d^2\sigma_{\uparrow}^{\gamma^*_\perp {\T N}}}{dM_{\T X}^2 dt}
\right|_{t = 0},  
\end{equation}
and a corresponding expression for 
$d^2\sigma_{\downarrow}^{\gamma^*_\perp {\T N}}$, 
with the center-of-mass energy $W$ of the $\gamma^* \T N$ system. 
The resulting coherent double scattering correction to the 
spin structure function $g_1^{\T d}$ becomes \cite{Edelmann:1997ik}: 
\begin{eqnarray} \label{eq:delta_g1}
\delta g_1^{\T d}(x,Q^2) &=&  
-\frac{2 \,Q^2}{e^2 x}\int_{4 m_\pi^2}^{W^2} {\T d} M_X^2
\left[
\frac{d^2\sigma^{\gamma^*_\perp \T N}_{\downarrow}}
{dM_{\T X}^2 dt} -
\frac{d^2\sigma^{\gamma^*_\perp \T N}_{\uparrow}}
{dM_{\T X}^2 dt} 
\right]_{t = 0} 
{\cal F}^B_+(\lambda^{-1}). 
\end{eqnarray}
Similarly one obtains for $b_1$ 
from Eqs.(\ref{eq:b1d_hel},\ref{eq:A_2_ds})
\cite{Edelmann:1997ik,Nikolaev:1997jy,Edelmann:1997qe}:  
\begin{equation} \label{eq:b1_ds}
b_1=\frac{2 \,Q^2}{e^2 x}
\int_{4 m_{\pi}^2}^{W^2} dM_X^2
\,\left.\frac{d^2\sigma^{\gamma^*_T \T N}}
{dM_X^2 dt}\right|_{t=0}
\,\left(
{\cal F}^B_+(\lambda^{-1}) + {\cal F}^B_-(\lambda^{-1}) 
- 2 {\cal F}^B_0(\lambda^{-1}) \right).
\end{equation}

\subsubsection{Shadowing in $\lowercase{g}_1^{\lowercase{\T d}}$}
\label{sssec:shad_in_g1d}

The difference of polarized diffractive virtual photoproduction cross 
sections which enters in Eq.(\ref{eq:delta_g1}) has so far 
not yet been measured.
Nevertheless it is possible to estimate the shadowing correction 
$\delta g_1$ to an accuracy which is sufficient for a reliable 
extraction of the neutron structure function $g_1^{\T n}$ from 
current experimental data. 
With inclusion of 
shadowing and the effective nucleon polarization 
in the deuteron one finds:
\begin{equation} \label{eq:g1n_est}
g_1^{\T n} \approx  
\frac{ {\cal P}^{\T d} - \delta g_1^{\T d}/(2 g_1^{\T N})}
{({\cal P}^{\T d})^2} 
\, g_1^{\T d} - g_1^{\T p}. 
\end{equation}
The measured spin structure functions have the property 
$|g_1^{\T d}| < |g_1^{\T p}|$, at least for $x > 0.01$ \cite{Adeva:1998vw}. 
This implies that, at the present level of data accuracy,  
shadowing effects 
and  uncertainties in the deuteron D-state probability do not 
play a major role in the extraction of $g_1^{\T n}$.

To estimate the amount of shadowing in  $g_1^{\T d}$ and its influence on 
the extraction of $g_1^{\T n}$ one can study 
the double scattering contribution   (\ref{eq:delta_g1}) 
in the framework of a simple model. 
In the laboratory frame at small $x < 0.1$   
the exchanged virtual photon first converts to a  
hadronic state X  which then interacts with the target 
(see Section \ref{ssec:DIS_SPTH}), 
dominant contributions coming from hadronic states 
with invariant mass $M_{\T X}^2 \sim  Q^2$. 
Consider therefore a single effective hadronic state with 
a coherence length $\lambda \sim 1/(2Mx)$. 
Comparing shadowing  for  unpolarized and 
polarized structure functions  gives \cite{Edelmann:1997ik}:
\footnote{Note that shadowing corrections  
in unpolarized structure functions,   
$\delta F_{1,2}^{\T d} =  F_{1,2}^{\T d} -  
(F_{1,2}^{\T p} -  F_{1,2}^{\T n})/2$,  
are defined {\em per nucleon},  as $F_{1,2}^{\T d}$ themselves.} 
\begin{eqnarray} \label{eq:g1_F1}
\frac{\delta g_1^{\T d}}{2 g_1^{\T N}} \approx {\cal R}_{g_1} 
\frac{\delta F_1^{\T d}}{F_1^{\T N}} 
\approx 
{\cal R}_{g_1} \frac{\delta F_2^{\T d}}{F_2^{\T N}}, 
\quad \mbox{with}\quad 
{\cal R}_{g_1} = 2 
\frac{{\cal F}^{B}_{+}(2Mx)}{{\cal F}^{B}(2Mx)}.
\end{eqnarray}
At small $x$ and $B = 7$ GeV$^{-2}$ 
one finds ${\cal R}_{g_1} = 2.2$ for both the Paris 
and Bonn nucleon-nucleon potentials 
\cite{Lacombe:1980dr,Machleidt:1987hj}. 
Although shadowing for $g_1^{\T d}$ turns out to be approximately 
twice as large as for the unpolarized structure function $F_2^{\T d}$, it 
still leads to negligible effects on the extraction of $g_1^{\T n}$,   
at least at the present level of experimental accuracy. 
Using the experimental data on shadowing for $F_2^{\T d}$ 
\cite{Adams:1995sh}  
one finds that the shadowing correction in (\ref{eq:g1n_est}) 
amounts at $x \sim 0.01$ to less than $5\%$ of the experimental error on 
$g_1^{\T n}$ for the SMC analysis \cite{Adams:1997hc}.

\subsubsection{The tensor structure function ${\lowercase{ b}_1}$ 
               at small $\lowercase{x}$}

The shadowing correction for the unpolarized 
structure function $\delta F_{1}^{\T d}$ and 
the deuteron tensor structure function $b_1$ 
are directly related. 
In order to see this, note again that 
the propagation lengths (\ref{eq:coherence}) of 
diffractively produced  hadrons exceed the deuteron size
$\lambda > \langle r^2\rangle_{\T d} \approx 4\,\T{fm}$ 
at small $x$. 
The deuteron form factors become approximately constant, 
i.e. ${\cal F}^{B}_{H} (\lambda^{-1} < 1/4\,\T{fm}) 
\approx {\cal F}^{B}_{H} (0)$, 
and a comparison with the double scattering correction for the 
unpolarized structure function (\ref{eq:approx}) gives 
\cite{Edelmann:1997ik}:
\begin{equation} \label{eq:b1_est}
b_1= {\cal R}_{b_1} \,\delta F_1, 
\quad \mbox{with}\quad 
{\cal R}_{b_1} = 2 \,\frac{{\cal F}^{B}_{0}(0)-{\cal F}^{B}_{+}(0)}
{{\cal F}^{B} (0)}.
\end{equation}
With $B = 7$ GeV$^{-2}$ we find from the 
Paris nucleon-nucleon potential \cite{Lacombe:1980dr}
${\cal R}_{b_1} = -0.66$, while the Bonn 
one-boson-exchange potential \cite{Machleidt:1987hj} leads to 
${\cal R}_{b_1} = -0.58$.
Using data for $F_2^{\T d}/F_2^{\T N}$ 
\cite{Adams:1995sh}  one can estimate $b_1$ at small $x$ 
and finds that it reaches 
about   $2\%$ of the unpolarized structure function 
$F_1^{\T N}$ at $x\ll 0.1$ \cite{Edelmann:1997ik,Strikman:1996YALE}. 

\begin{figure}[t]
\bigskip
\begin{center} 
\epsfig{file=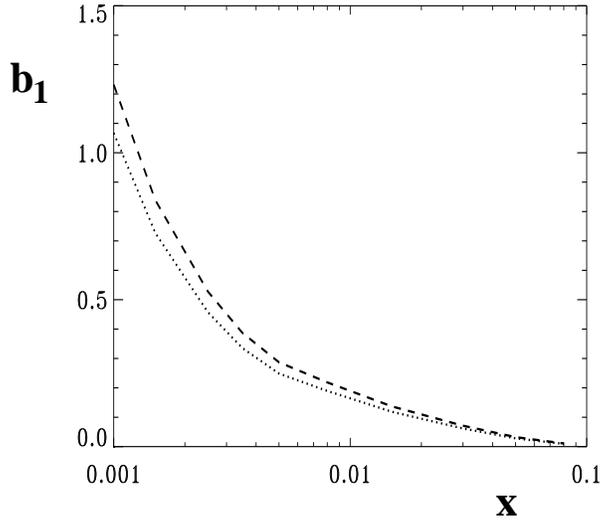,height=70mm}
\end{center}
\caption[...]{
Double scattering contribution to the 
tensor structure function $b_1$ from \cite{Edelmann:1997ik} . 
The dotted and dashed curves correspond to the Bonn \cite{Machleidt:1987hj}  
and Paris potential \cite{Lacombe:1980dr}, respectively. 
}
\label{fig:b1}
\bigskip
\end{figure}

In Fig.\ref{fig:b1} we present $b_1$ as obtained from Eq.(\ref{eq:b1_est}).  
The result shown here corresponds to the kinematics of E665 
\cite{Adams:1995sh}.
Estimates of $b_1$ at large $Q^2\gg 1$GeV$^2$  and small $x\ll 0.1$ 
can be found in \cite{Nikolaev:1997jy}. 
It should be mentioned that the magnitude of $b_1$ 
at $x<0.1$ exceeds estimates from previous model 
calculations   
which are applicable at moderate and large $x$,  
by several orders of magnitude 
(see e.g. \cite{Khan:1991qk,Umnikov:1997qv}). 
Unfortunately, the effect of $b_1$ in the observable asymmetry, 
which is proportional to $b_1/F_1^{\T d}$, is only of the order 
of $10^{-2}$, as already mentioned. 

\subsubsection{The $^3$He case} 
\label{subs:Helium}

Although $^3\T{He}$ would appear to be an ideal neutron target 
because of its large effective neutron polarization (\ref{eq:pn_He}), 
the extraction of neutron spin structure functions 
nevertheless requires dealing with significant nuclear effects.
We concentrate again on the structure function $g_1$ at small 
values of $x$.  
As for deuterium two types of corrections are relevant:  
higher angular momentum components of the $^3\T{He}$ wave 
function lead to effective proton and neutron polarizations 
(\ref{eq:pn_He},\ref{eq:pp_He}). 
Furthermore, the coherent interaction of the 
virtual photon 
with several nucleons causes shadowing.
Accounting for both effects the neutron structure function 
can be deduced from $g_1^{\T{He}}$ as follows: 
\begin{equation} \label{eq:g_1^n(He)_a}
g_1^{\mathrm n} \approx 
\frac{{\cal P}_{\mathrm n}^{\T{He}} - 
\delta g_1^{\T He}/g_1^{\mathrm n}}
{({\cal P}_{\mathrm n}^{\T{He}})^2} \,
g_1^{\mathrm{He}} -  
\frac{{\cal P}_{\mathrm p}^{\T{He}}}
{{\cal P}_{\mathrm n}^{\T{He}}} \,g_1^{\mathrm p}. 
\end{equation}
Since $|{\cal P}_{\mathrm p}^{\T{He}}| \ll 
|{\cal P}_{\mathrm n}^{\T{He}}|$ 
the proton contribution in (\ref{eq:g_1^n(He)_a}) is 
indeed suppressed. 

Uncertainties in the effective nucleon polarizations which 
may influence  the extraction of $g_1^{\T n}$,   
result also from non-nucleonic degrees of freedom, e.g. mesons and 
$\Delta$-isobars, present in $^3$He. 
In Ref.\cite{Frankfurt:1996nf} this is demonstrated for 
the non-singlet nucleon and $\T A=3$ structure functions:  
\begin{eqnarray} 
\label{eq:g1ns_1}
&&
\Delta g_1^{\T A = 1} (x,Q^2) = g_1^{\T p}(x,Q^2) -  g_1^{\T n}(x,Q^2),
\\
\label{eq:g1ns_3}
&&
\Delta g_1^{\T A = 3} (x,Q^2) = 
g_1^{\T{H}}(x,Q^2) -  g_1^{\T{He}}(x,Q^2), 
\end{eqnarray}
where  $g_1^{\mathrm{H}}$ is the triton spin structure function. 
From the effective nucleon polarizations 
(\ref{eq:pn_He},\ref{eq:pp_He}) and 
isospin symmetry of the three-body nuclear wave function 
one obtains:
\begin{equation} \label{eq:g13/g11}
\Delta g_1^{\T A = 3} = 
\left(1 -\frac{4}{3} P_{\T S'} - \frac{2}{3} P_{\T D}\right) 
\Delta g_1^{\T A = 1}.
\end{equation}
Applying the Bjorken sum rule (\ref{eq:Bj-SR}) 
to the $\T A = 1$ and $\T A = 3$ systems and taking their  
ratio leads to:
\begin{equation} \label{eq:Bj_A=1/3}
\frac
{\int_0^1 dx 
\left[ g_1^{\mathrm{He}}(x,Q^2) - g_1^{\mathrm{H}}(x,Q^2) 
\right]}
{\int_0^1 dx 
\left[ g_1^{n}(x,Q^2) - g_1^{p}(x,Q^2) 
\right]}
= \frac{G_A(^3\mathrm{H})}{G_A(\mathrm{n})}.
\end{equation}
The axial vector coupling constants
$G_A(^3\mathrm{H})$ and $G_A(\mathrm{n})$ are 
measured in the $\beta$-decay of tritium and 
the neutron, respectively, with $G_A(\T n) \equiv g_A = 1.26$ 
\cite{Caso:1998tx}. 
If one considers incoherent scattering from 
individual nucleons and accounts only for the effective 
nucleon polarizations  one finds from 
Eqs.(\ref{eq:g13/g11},\ref{eq:Bj_A=1/3}):  
\begin{equation}
\frac{G_A(^3\mathrm{H})}{g_A} = 
1 -\frac{4}{3} P_{\T S'} - \frac{2}{3} P_{\T D} 
=0.922 \pm 0.006.
\end{equation}
This result is however at variance with the empirical ratio  
${G_A(^3\mathrm{H})}/{g_A}\approx 0.963\pm 0.003$ 
\cite{Budick:1991zb,Caso:1998tx}.
One  concludes that simply using effective 
nucleon polarizations from realistic three-nucleon calculations 
would lead to a violation of the Bjorken sum rule 
by approximately $4\%$ \cite{Frankfurt:1996nf,Kaptari:1990qt}. 
On the other hand, 
it is known from nuclear $\beta$-decay and Gamov-Teller transitions 
that  axial coupling constants in nuclei are renormalized by 
meson exchange currents and $\Delta$-isobars \cite{Ericson:1988gk}.  
The possible influence of these non-nucleonic degrees of freedom 
on the extraction of $g_1^{\T n}$ has to be carefully 
investigated \cite{Frankfurt:1996nf}.

At small values $x<0.1$ 
of the Bjorken variable, 
coherent multiple scattering 
from several nucleons in the target leads to the 
shadowing correction $\delta g_1$. 
Assuming again, as in Section \ref{sssec:shad_in_g1d}, 
that the photon-nucleus scattering 
at small $x$ can be represented by the interaction of 
one effective hadronic fluctuation  
with invariant mass $M_{\T X}^2 \sim Q^2$, one finds \cite{Frankfurt:1996nf}:
\begin{equation} \label{eq:shad_g1He_approx}
\frac{g_1^{\T{He}}}{g_1^{\mathrm{n}}} 
\approx 
\left(2 \,\frac{F_2^{{\mathrm{He}}}}{F_2^{\mathrm N}} - 1\right), 
\end{equation}
i.e. shadowing in $g_1^{\T {He}}$ is about 
twice as large as  for the unpolarized structure function 
$F_2^{\mathrm{He}}$. 
This is also true for  shadowing  in  the non-singlet structure 
function $\Delta g_1^{\T A = 3}$ \cite{Frankfurt:1996nf}. 
A similar result has been found for the polarized 
deuterium case, see e.g. (\ref{eq:g1_F1}). 

In Ref.\cite{Frankfurt:1996nf} 
the shadowing correction (\ref{eq:shad_g1He_approx}) 
has been combined with the nuclear Bjorken sum rule (\ref{eq:Bj_A=1/3}). 
In the nuclear case, shadowing  
reduces the small-$x$ contribution to the Bjorken integral.
Consequently   the non-singlet nuclear structure function 
must  be enhanced, i.e. $\Delta g_1^{\T A=3}  > \Delta g_1^{\T A=1}$, 
somewhere in the region $x \,\gsim \,0.1$.  
This is suggested to occur around $x\sim 0.1$, where 
the projectile may still interact  with two nucleons inside the 
target.  
As a consequence a significant anti-shadowing is obtained 
in this kinematic region.
The dynamical origin for such an 
enhancement is supposed to be independent of  
the flavor channel considered, so that this 
is also expected to occur for $g_1^{\T{He}}$.

To summarize this discussion we emphasize that a precise 
extraction of the neutron spin structure function $g_1^{\T n}$ 
from $^3\T{He}$ data at $x \lsim 0.2$ requires a careful 
analysis of nuclear effects due to non-nucleonic 
(e.g. meson and $\Delta$-isobar) degrees of freedom 
in the $^3\T{He}$ wave function  and a detailed understanding of 
shadowing and anti-shadowing effects.
The use of deuterium as a target may have advantages since non-nucleonic 
admixtures are supposed to be smaller
due to weaker nuclear binding. 
Furthermore, especially at small $x$ the influence of 
shadowing on the extraction of $g_1^{\T n}$ is less 
pronounced for deuterium as compared to $^3$He.

\subsection{Polarized deep-inelastic scattering from nuclei at 
            $x>0.2$}
\label{ssec:pol_DIS_largex}

At moderate and large values of $x$, the distances probed by 
DIS from nuclei are smaller than $2$ fm as outlined in 
Section \ref{Sec:space_time}. 
Incoherent interactions of the virtual photon with hadronic 
constituents of the nucleus dominate, 
and the usual starting point is a description based on the impulse 
approximation including nucleon degrees of freedom only 
\cite{Kulagin:1995cj,CiofidegliAtti:1993zs,Woloshyn:1985,Schulze:1997rz}. 
This approach is not without question, but not much 
progress has so far been made beyond that level.


In the following we first discuss general features of the nuclear 
spin structure functions $g_1^{\T A}$ and 
$g_T^{\T A}= g_1^{\T A}+ g_2^{\T A}$
in the framework of the impulse approximation. 
The nuclear target is treated non-relativistically. 
This allows, as in the unpolarized case, to factorize  nucleon and nuclear 
degrees of freedom, introducing structure functions of 
nucleons as bound quasi-particles.
The nuclear structure functions are then given as two-dimensional 
convolutions of the bound nucleon structure functions and the nucleon 
light-cone momentum distributions \cite{Kulagin:1995cj}: 
\begin{eqnarray} 
\label{eq:g_1^A_conv}
g_1^{\T A}(x)
&=& 
\sum_{{i = 1}}^A \int\!dp^2\int\limits_x\frac{dy}y\,
    D_1^{i}(y,p^2)\,g_1^{i}\!\left(\frac{x}y,p^2\right), \\
\label{eq:g_T^A_conv}
g_T^{\T A}(x)
&=& 
\sum_{{i = 1}}^A \int\!dp^2\int\limits_x\frac{dy}y\,
    \left[ D_T^{i}(y,p^2)\,g_T^{i}\!\left(\frac{x}y,p^2\right)
           + D_{T2}^{i}(y,p^2)\,g_2^{\T i}\!\left(\frac xy,p^2\right)
           \right].                           
\end{eqnarray}
Here we have suppressed the $Q^2$-dependence for simplicity. 
The expressions in 
Eqs.(\ref{eq:g_1^A_conv}, \ref{eq:g_T^A_conv}) 
resemble the result for the unpolarized 
nuclear structure function in Eq.(\ref{F2A}): 
the exchanged virtual photon scatters from quarks 
which carry a fraction  $x/y$ of  the light-cone momentum 
of their parent nucleons, which in turn   
have  a fraction $y = p^+/M$  of the nuclear 
light-cone momentum and a squared four-momentum $p^2$.

The nucleon distribution functions in 
Eqs.(\ref{eq:g_1^A_conv},\ref{eq:g_T^A_conv}) 
are given by: 
\begin{eqnarray}
\label{eq:nucl_distr_pol,1}
D_1(y,p^2)
&=& \int\!\frac{{\T d}^4{p'}}{(2\pi)^4}\
    {\rm tr}
    \left[ {\cal S}_{\|}(p')
           \left( \widehat{\Sigma}_0 + \widehat{\Sigma}_z\right)
    \right]
    \delta \left( y - \frac{p'^{+}}{M} \right)
    \delta \left( p^{2} - p'^{2} \right),    \\
\label{eq:nucl_distr_pol,T}
D_T(y,p^2)
&=& \int\!\frac{{\T d}^4 p'}{(2\pi)^4}\
    {\rm tr}
    \left[ {\cal S}_{\perp}(p')\
           \widehat{\Sigma}_\perp
    \right]
    \delta \left( y - \frac{p'^+}{M} \right)
    \delta \left( p^2 - p'^2 \right),    \\
\label{eq:nucl_distr_pol,T2}
D_{T2}(y,p^2)
&=& \int\!\frac{{\T d}^4{p'}}{(2\pi)^4}\
    {\rm tr}
    \left[ {\cal S}_{\perp}(p')\
           \widehat{\cal T}_2
    \right]
    \delta \left( y - \frac{p'^+}{M} \right)
    \delta \left( p^2 - p'^2 \right),
\end{eqnarray}
with the polarized nucleon spectral function:
\begin{equation}\label{eq:spectral_pol}
{\cal S}_{\sigma\sigma'}(p)
=  2\pi \sum_n
   \delta \left(p_0 - M - \varepsilon_n + T_R \right)
   \psi_{n,\sigma}({\vec  p}) \psi^{*}_{n,\sigma'}({\vec p}). 
\end{equation}
Here the summation is performed over the complete set of 
states with $A-1$ nucleons.
The functions $\psi_{n,\sigma}({\vec p})
= \left\langle (\T A-1)_n,-{\vec p}|\psi_\sigma(0)|\T A \right\rangle$
are the probability amplitudes to find  a nucleon    
with polarization $\sigma$  in the nuclear ground state
and the remaining $A-1$
nucleons in a state $n$ with total momentum $-{\vec p}$.
The separation energy $\varepsilon_n$ and the recoil energy  
of the residual nuclear system $T_R$ enter in Eq.(\ref{eq:spectral_pol}) 
as for  the unpolarized case (\ref{S(p)}).  
For $g_1$ the target nucleus is chosen to be polarized 
parallel to the photon momentum as indicated in 
Eq.(\ref{eq:nucl_distr_pol,1}) by the subscript $\|$. 
For the transverse structure function $g_T$ 
the target polarization is taken perpendicular to the momentum 
transfer and denoted by $\perp$. 

The nucleon spin operators which multiply the spectral 
functions in 
Eqs.(\ref{eq:nucl_distr_pol,1},\ref{eq:nucl_distr_pol,T},%
\ref{eq:nucl_distr_pol,T2}) 
refer to the active nucleon. They read:
\begin{eqnarray} \label{eq:spin_op}
\widehat{\Sigma}_0
&=& \frac{\vec \sigma \cdot \vec p}M,   
\quad \quad \quad \quad
\widehat{\Sigma}_j
=
\left( 1-\frac{{\vec  p}^2}{2M^2} \right)
\sigma_j
 + \frac{\vec \sigma \cdot \vec p}{2M^2}\ p_j,
\nonumber \\
\widehat{\cal T}_2
&=& - { {\vec  p}_\perp \cdot {\vec  S}_\perp \over M }
      \left(\frac{ \vec \sigma \cdot \vec p }{M}
          + \sigma_z \left( 1 - \frac{p_z}{M} \right)
    \right),
\end{eqnarray}
where $j$ denotes spatial indices. 
Furthermore, ${\vec p}_\perp$ is the transverse component of the nucleon
three-momentum,  ${\vec  p} = ({\vec  p}_\perp, p_z)$, and 
${\vec S}_\perp$ determines the transverse spin quantization
axis relative to the photon momentum which is taken along the $z$-direction.

From Eqs.(\ref{eq:g_1^A_conv},\ref{eq:g_T^A_conv}) we observe 
that $g_1^{\T A}$ is expressed entirely in terms of 
the corresponding nucleon structure function $g_1^{\T N}$. 
This is different for $g_T^{\T A}$ which  
receives contributions from $g_T^{\T N}$ as well as from $g_2^{\T N}$.

If the bound nucleon structure functions in  
Eqs.(\ref{eq:g_1^A_conv},\ref{eq:g_T^A_conv})
are replaced by free ones, 
one ends up with the conventional one-dimensional 
convolution ansatz for nuclear structure functions 
\cite{Kulagin:1995cj,CiofidegliAtti:1993zs}.
Relativistic contributions  which lead beyond the convolution 
formula (\ref{eq:g_1^A_conv},\ref{eq:g_T^A_conv})
have been investigated  in 
Refs.\cite{Melnitchouk:1995tx,Piller:1996mf}, 
and corrections have been estimated within a quark-diquark 
model for the bound nucleon. 
Deviations from non-relativistic convolution were generally found 
to be small, except at  very large $x>0.9$.

Given that the nuclear spectral functions, polarized as well 
as unpolarized, receive their major contributions from small nucleon 
momenta, systematic expansions of $g_1^{\T A}$ and $g_T^{\T A}$ 
for $x < 0.7$ can be performed around $y =1$ and the mass-shell 
point $p^2 = M^2$, keeping terms of order $\varepsilon/M$ 
and $\vec p^2/M^2$. 
Results and 
applications to spin structure functions of the deuteron 
and $^3$He are discussed in 
Refs.\cite{Kulagin:1995cj,CiofidegliAtti:1993zs,%
Schulze:1997rz,Kaptari:1995di}. 
As a general rule, the structure functions at $x < 0.7$ 
are well described using simply the effective 
nucleon polarizations 
(\ref{eq:nucleon_polarization_d},\ref{eq:pn_He},\ref{eq:pp_He}).

\section{Further developments and perspectives}
\setcounter{section}{8}
\setcounter{figure}{0}
\label{sec:perspectives}

We close this review with a short summary of the key physics points 
together with  an outlook on several selected 
topics for which investigations are 
still actively under way.
We comment on exclusive vector meson production from nuclei, 
questions of shadowing at large $Q^2$ and the issue of high 
parton densities in nuclear systems.

\subsection{Coherence effects in DIS and in the exclusive electroproduction 
           of vector mesons}

Nuclear shadowing in inclusive deep-inelastic lepton scattering is a prime 
source of information on coherently propagating hadronic or quark-gluon 
fluctuations of the virtual photon in a nuclear medium. 
By selecting different kinematic cuts in $Q^2$ and energy transfer 
$\nu$, one can focus on different components of the photon's 
Fock space wave function. 
An even more stringent selection of such components can be 
achieved in exclusive photo- and electroproduction processes,
and in particular in high-energy diffractive vector meson production. 
Data on vector meson production from nuclei have become available 
in recent years at 
FNAL (E665) \cite{Adams:1995bw}, CERN (NMC)  
\cite{Arneodo:1994qb,Arneodo:1994id}, and 
DESY (HERMES) \cite{Ackerstaff:1999wt}, 
and further experiments are under discussion at TJNAF 
\cite{Dytman:1998} and 
DESY (HERA, HERMES) \cite{Ingelman:1996ge}.

Depending on energy and momentum transfer, the mechanism of 
vector meson formation can be quite different. In the range 
$\nu >  3$ GeV and $Q^2 \,\lsim \,1$ GeV$^2$  
the production process is well described using the vector 
meson dominance picture (see e.g. \cite{Bauer:1978iq}): 
in the lab frame the photon converts into a vector meson 
prior to scattering from the target.  
On the other hand, at large $Q^2 \gg 1$ GeV$^2$ 
perturbative QCD calculations show that the 
photon-nucleon interaction produces 
an initially small-sized, color singlet quark-antiquark wave packet 
\cite{Frankfurt:1996jw,Brodsky:1994kf}. 
At high photon energies the finally observed vector meson is then 
formed at a much later stage.

The transition from small to large $Q^2$ interpolates between 
non-perturbative hadron formation and perturbative quark-antiquark-gluon 
dynamics, a question of central importance in QCD. 
Nuclear targets are particular helpful at this point because they serve 
as analyzers for the coherent interaction of the produced $q\bar q$-gluon 
system with several nucleons \cite{Brodsky:1988xz}. 
The distance between two nucleons provides the ``femtometer stick'' 
which can be used to measure the relevant coherence lengths 
(for reviews and references see 
\cite{Nikolaev:1992si,Frankfurt:1994hf,Jain:1996dd}). 

The characteristic scales for this discussion have  been 
encountered several times in previous sections. First, there is the 
typical longitudinal distance (propagation length) 
\begin{equation}\label{coh}
\lambda \approx \frac{2\nu}{m^2 + Q^2}\,.
\end{equation}
It represents the distance over which a hadronic fluctuation 
of invariant mass $m$ propagates in the lab frame when 
induced by a photon of energy $\nu$ and virtuality $Q^2$. 
At large $Q^2$ the initially produced wave packet is characterized 
by its transverse size $b$. For longitudinally 
polarized photons, 
\begin{equation}\label{ej}
b  =  \frac{const}{Q}.
\end{equation}
In perturbative QCD the minimal Fock space component has 
$const \sim 4$--$5$ at $Q^2 \gsim 5 \,\T{GeV}^2$ 
\cite{Frankfurt:1996jw,Kopeliovich:1994pw}. 
Thus for $Q^2 = 5\,\T{GeV}^2$  and $const = 4$, the 
transverse size of the initial wave packet is $b\sim 0.4$ fm, 
a small fraction of the diameter of a fully developed $\rho$ meson.

Recent measurements at HERMES \cite{Ackerstaff:1999wt} have observed effects 
related to the coherence length $\lambda$ in 
$\rho^0$ electroproduction on hydrogen, deuterium, 
$^3$He, and $^{14}$N. 
The range of energy and momentum transfers covered by the 
experiment is $9 \,\T{GeV} < \nu < 20 \, \T{GeV}$ and 
$0.4 \,\T{GeV}^2 < Q^2 < 5 \, \T{GeV}^2$. 
This implies coherence lengths in the range 
$0.6\,\T{fm} \,\lsim\,\lambda < 8\,\T{fm}$
covering scales from individual nucleons up to and beyond nuclear dimensions.
(The interesting upper section of the available $Q^2$ interval,  
$Q^2\,\gsim\,  4$ GeV$^2$, has been accessible only  for small 
energies with  $\nu$ with $\lambda \,\lsim\,1$ fm 
in these measurements.)
\begin{figure}[t]
\bigskip
\begin{center} 
\epsfig{file=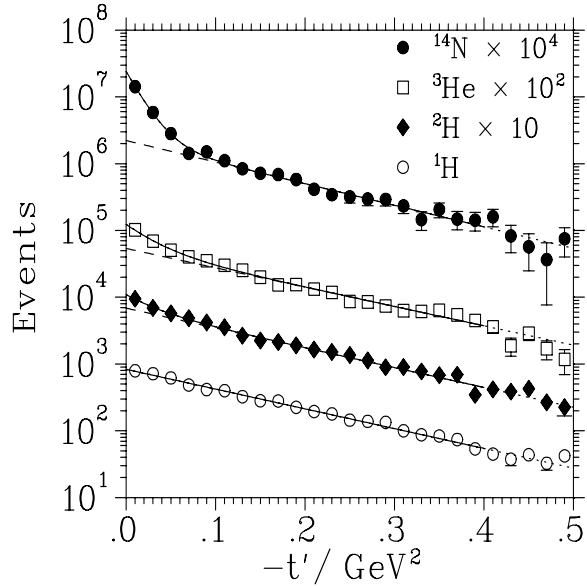,height=80mm}
\end{center}
\caption[...]{
The rates of produced $\rho^0$'s plotted 
against $t' = t - t_0$ for hydrogen, deuterium, 
$^3$He, and $^{14}$N. 
The solid lines show fits to the data, 
the dashed lines are inferred incoherent contributions 
(for details see Ref.\cite{Ackerstaff:1999wt}). 
}
\label{fig:vm_prod_A}
\bigskip
\end{figure}

Given the four-momenta $q$ and $k$ of the incoming virtual photon 
and the produced $\rho$ meson, 
the $t$-channel momentum transfer to the nucleon is $t = (q - k)^2$.
In Fig.\ref{fig:vm_prod_A} the rate of produced $\rho^0$'s is plotted 
against $t' = t - t_0$, 
the squared momentum transfer above threshold  
($|t_0| \simeq \lambda^{-2}$).
At $|t'| \ll 0.1$ GeV$^2$ 
coherent production dominates, leaving the nucleus as a whole in 
the ground state. Such coherent processes fall off rapidly 
with the nuclear form factor, so that at  $|t'|\,\gsim\,0.1$ 
GeV$^2$ mostly incoherent $\rho$ production from individual nucleons 
remains.

\begin{figure}[t]
\vspace*{-5cm}
\begin{center} 
\hspace*{-2cm}
\epsfig{file=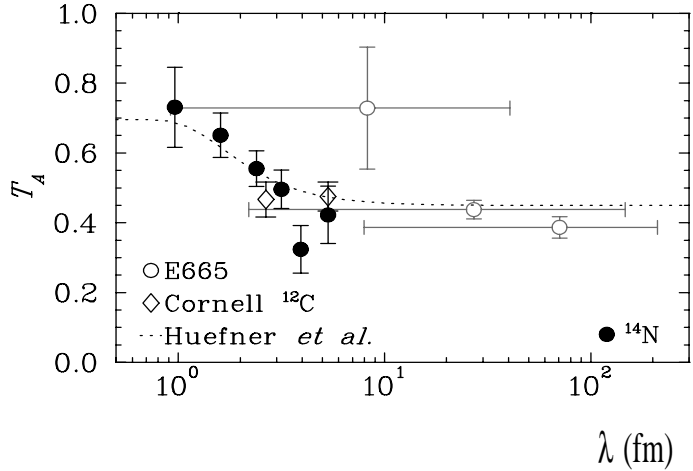,height=120mm,width=100mm}
\end{center}
\caption[...]{The nuclear transparency $T_{\T A}$ 
as a function of the propagation length $\lambda$ 
for $^{14}$N. 
HERMES data \cite{Ackerstaff:1999wt} are indicated by full dots. 
The open symbols represent data from previous experiments 
\cite{Clellan:1969,Adams:1995bw}. 
The dashed curve shows a Glauber calculation from Ref.\cite{Hufner:1996dr}
}
\label{fig:trans}
\bigskip
\end{figure}

Consider now the incoherent production of vector mesons 
from nuclei.
In the absence of coherent rescattering processes the nuclear 
$\rho$ production cross section  $\sigma_{\T A}$ would simply 
be $A$ times the production cross section $\sigma_{\T N}$ 
on a free nucleon. 
Nuclear effects are conveniently discussed in terms of the 
transparency ratio, 
$T_{\T A} =\sigma_{\T A}/({A}  \sigma_{\T N})$. 
The measured ratio for $^{14}$N is plotted as a 
function of the longitudinal propagation length $\lambda$ 
in Fig.\ref{fig:trans}. 
The deviation of   $T_{\T A}$ from unity for $\lambda \,\lsim\,1$ fm 
simply reflects the ``trivial'' final state rescattering of the 
$\rho$ meson after being produced on one of the nucleons.
More interesting effects are visible when $\lambda$ exceeds the average 
nucleon-nucleon distance of about $2$ fm. 
Now the hadronic fluctuations of the photon can scatter coherently 
on several nucleons also prior to the production of the final state 
vector meson, and the transparency ratio $T_{\T A}$ 
systematically decreases until it exceeds the nuclear diameter. 
The dashed curve in Fig.\ref{fig:trans} shows a theoretical prediction 
calculated within the vector meson dominance model \cite{Hufner:1996dr}. 
Its agreement with data suggests that the production process is 
dominated, given the relatively low $Q^2$ involved, 
by hadronic fluctuations which interact about as strongly as the produced 
$\rho$ meson. 
Further systematic investigations of such coherence length effects, 
especially its detailed dependence on the momentum transfer $t$, 
are discussed at TJNAF \cite{Dytman:1998,Frankfurt:1998vx}. 

It is interesting to push these observations to more extreme 
regions of very large $Q^2$ and $\nu$. 
Once the energy transfer exceeds several tens of GeV, 
a further scale enters. At large $Q^2$ the $q\bar q$ fluctuation 
of the photon starts out initially as a small-sized wave packet. 
The time it takes for this wave packet to develop into the 
final vector meson is called the formation time $\tau_f$ 
\cite{Nikolaev:1992si,Frankfurt:1994hf,Jain:1996dd}.
To be specific, let the observed vector meson again be a $\rho^0$. 
The small initial wave packet is generally not an eigenstate of the strong 
interaction Hamiltonian. Now consider expanding the wave packet in 
hadronic eigenstates. Clearly, for a wave packet with a size small 
compared to typical hadronic dimensions, several such eigenstates 
are necessary to represent the wave function of the packet. 
Let one of those hadronic eigenstates in the expansion be the 
$\rho$ meson itself (mass $m_{\rho}$), let 
another one be a neighboring state with larger mass 
(say, $m_{\rho'} = m_{\rho} + \delta m$). 
The characteristic propagation length of the $\rho$ component of the 
$q\bar q$-gluon wave packet is $\lambda \simeq 2\nu/(m_{\rho}^2 + Q^2)$, 
that of the neighboring state is 
$\lambda' \simeq 2\nu/(m_{\rho}^2 + Q^2 + 2 m_{\rho} \,\delta m)$. 
The phase difference between both states is determined by their wave 
numbers $1/\lambda$ and $1/\lambda'$. 
The time it takes to filter out all but the $\rho$ meson component 
when passing through the nucleus is then 
\begin{equation}\label{fmt}
\tau_f \sim \frac{\nu}{m_{\rho}\,\delta m}\,.
\end{equation}
Thus when $\nu$ is sufficiently large so that $\tau_f$ reaches 
nuclear dimensions {\em and} $Q^2$ is large, a small-sized 
$q\bar q$ wave packet induced by the photon has a chance to travel 
over large distances inside nuclei and interact weakly, its cross 
section being proportional to $b^2 \sim 1/Q^2$. 

This phenomenon is commonly named color coherence or 
color (singlet) transparency 
\cite{Brodsky:1988xz,Nikolaev:1992si,Frankfurt:1994hf,Jain:1996dd}. 
It has been addressed in exclusive vector meson production 
experiments at FNAL (E665) \cite{Adams:1995bw}  and CERN (NMC) 
\cite{Arneodo:1994id}. 
The interpretation of data in terms of color transparency is 
still under debate (see e.g. \cite{Kopeliovich:1995bj,Abramowicz:1995hb}). 
Possible future developments with nuclei at HERA 
\cite{Ingelman:1996ge} could offer 
an enormous extension of the accessible kinematic range.

\subsection{Nuclear shadowing in DIS at large $Q^2$}
\label{sec:shad_largeQ}

Experimental information on nuclear shadowing 
in inclusive DIS is available, up to now, 
only from fixed target experiments, with the 
kinematic range restricted 
to $Q^2\, \lsim \, 1$ GeV$^2$ at $x\ll 0.1$.
Although there are currently no data on nuclear shadowing 
at large $Q^2 \gg 1$ GeV$^2$, it is nevertheless instructive 
to investigate what one should expect in comparison with the 
previous results at smaller $Q^2$. 
Such a study is possible due to recent data on 
diffractive production from the HERA collider and 
has been performed in Refs.\cite{Piller:1997ny,Frankfurt:19966nx,%
Barone:1997ij}.

One finds that contributions from vector mesons 
are negligible at large momentum transfers  since diffractive vector 
meson production is strongly suppressed at $Q^2 > 10$ GeV$^2$. 
Furthermore, the  ZEUS data \cite{Breitweg:1998gc} on the ratio of 
diffractive to total photon-nucleon cross sections (Fig.\ref{fig:diff_tot}) 
suggest that diffractively produced states with large mass are 
relevant at large $Q^2$. 
Therefore  at large $Q^2$ shadowing probes 
the coherent interaction of quark-gluon configurations with large 
invariant mass. 
This is complementary to fixed target experiments 
at FNAL (E665)  and  CERN (NMC) where the coherent interaction 
of low mass vector mesons played a dominant role 
(see Sections \ref{sssec:shad_smallQ} and \ref{ssec:Shad_Model}).

In Section \ref{sssec:sizes}  we have argued that to leading order 
in $1/Q^2$ shadowing is dominated by the interaction of large-size 
hadronic fluctuations of the exchanged photon. 
This suggests a weak energy dependence of nuclear shadowing. 
However, at very small $x$ together with very large $Q^2$, 
the steadily growing number of partons in the photon-nucleon system 
makes quark-gluon configurations of the photon interact like 
ordinary hadrons, even if they have small transverse size 
proportional to $1/Q^2$ (see e.g. Eq.(\ref{eq:sigma_qq})). 
As a consequence one expects a more rapid energy dependence of shadowing 
as compared to the case of small $Q^2$.
On similar grounds the diffractive leptoproduction cross 
section at $x \ll 0.1$ and $Q^2 \gg 1$ GeV$^2$ 
should rise  more strongly than suggested by Regge phenomenology.
Possible indications  for this  behavior have been 
found  at HERA \cite{Adloff:1997sc,Breitweg:1998gc}, 
signaling the onset of a new kinematic regime with 
a complex interplay between soft (large size) and hard 
(small size) partonic components of the interacting photon.
A systematic investigation of strong interaction dynamics in this kinematic 
region is a major challenge. 
Here electron-nucleus collider experiments could give important 
new insights \cite{Ingelman:1996ge}.

\subsection{Physics of high parton densities}
\label{ssec:High_parton_densities}

An elementary QCD  treatment of radiative corrections 
in nucleon and nuclear structure functions, or
equivalently, in quark and gluon distributions 
is provided by the DGLAP evolution equations 
(see Section \ref{ssec:AP_eq}).
In leading logarithmic approximation 
one sums over  $\alpha_s \ln Q^2/\Lambda^2$ terms,
each of which represents a quark radiating a gluon or a 
gluon splitting into a $q\bar q$ or gluon pair. 
A contribution  $(\alpha_s \ln Q^2/\Lambda^2)^n$ 
is associated with the radiation from $n$ partons 
in a physical gauge.
Due to radiation partons loose momentum. 
Therefore  DGLAP evolution leads to a rise of  
quark and gluon distribution functions  at small $x$:
strongly increasing numbers of partons carry smaller and smaller 
fractions of the total momentum. 
For example, consider gluons which dominate  
the dynamics of parton distributions at small $x$. 
Suppose that we are given an initial gluon distribution, 
prior to radiative QCD corrections, at a momentum scale $Q_0^2$. 
This initial distribution can be non-singular (finite) 
at $x\to 0$. Now turn on QCD radiation and DGLAP evolution. 
The resulting asymptotic behavior of the gluon distribution function 
at $x \ll 1$ and large $Q^2$ is (see e.g. \cite{Gribov:1984tu}):
\begin{equation}
x g(x,Q^2) \sim \exp\left\{\frac{12}{5} 
\sqrt{\ln \left(\frac{\ln(Q^2/\Lambda^2)}{\ln(Q_0^2/\Lambda^2)}\right)  
\ln(1/x)}
\right\}.
\end{equation}
One observes a strong rise of $x g(x,Q^2)$ at $x \ll 1$.
This increase of the gluon density with decreasing $x$ can however 
not continue  indefinitely.  
At very small $x$ the density  of gluons becomes 
so large that they interact with each other reducing their density 
through annihilation.
Thus one enters a new regime of very high parton densities where  
standard methods of perturbation theory are inappropriate  
despite small values for the coupling   $\alpha_s$, 
and resummation techniques to infinite order have to be applied 
(for references see e.g. 
\cite{McLerran:1994ni,Jalilian-Marian:1997xn,Ayala:1997em,Kovchegov:1997dm,%
Mueller:1999wm,Kovchegov:1999yj}). 

For further illustration let us view the scattering 
process in a frame where the target 
momentum is very large, $P = |\vec P| \rightarrow \infty$.
A  measurement at 
specific values of $Q^2$ and $x$ probes   partons 
over a longitudinal distance  
$\Delta z \sim 1/(P x)$ and with a transverse size 
$\Delta b \sim 1/\sqrt{Q^2}$.
Thus at small values of $x$ and moderate $Q^2$ 
parton wave functions overlap, leading to high parton densities, 
so that the rate of parton-parton annihilation processes  
increases. 
This rate 
involves the probability of finding at least 
two partons per unit area in the nucleon \cite{Gribov:1984tu,Mueller:1986wy}.
In this respect parton recombination can be regarded as a  
``non-linear'' correction to radiation processes 
described by standard DGLAP evolution, where parton distribution 
functions enter linearly (see e.g. Eq.(\ref{eq:DGLAP})).

The existing HERA data on free nucleon structure functions 
at   $Q^2\,\gsim\,1$ GeV$^2$ and $x \,\gsim \,10^{-4}$ 
do not show a clear sign for the need of additional parton fusion 
corrections in the evolution equations 
(see e.g. \cite{Cooper-Sarkar:1997jk}).  
However, phenomena related to high parton densities 
at small $x$ should be magnified in nuclei  
since nuclear parton densities are enhanced.  
For example, if shadowing is ignored and 
the nuclear gluon distribution is assumed to be 
the sum of the gluon distributions of  
individual nucleons,  
$g_{\T A}(x,Q^2) = {A}\, g_{\T N}(x,Q^2)$, 
then the ratio of gluon densities in a nucleus and a nucleon per (transverse) 
area  is \cite{Frankfurt:1998eu}:
\begin{equation} \label{eq:gN/gA_incoherent_sum}
\left.
\frac{g_{\T A}(x,Q^2)}{\pi R_{\T A}^2} \right/ 
\frac{g_{\T N}(x,Q^2)}{\pi r_{\T N}^2}
\approx 
\frac{{\T A} \,r_{\T N}^2} {R_{\T A}^2}\approx \,0.5 \,{\T A}^{1/3},   
\end{equation}
where $r_{\T N}\approx 0.8\,\rm{fm}$ and 
$R_{\T A}\approx r_0 A^{1/3} \approx 1.1\,{\rm fm} \,{\T A}^{1/3}$ 
have been used for nucleon and nuclear radii, respectively.
Effects beyond DGLAP evolution should therefore be amplified 
in nuclei and set in already at a larger value of $x$ as compared 
to  free nucleons.

In Ref.\cite{Mueller:1986wy,Qiu:1987wh} corrections to DGLAP evolution for 
nuclear parton distributions  have been  calculated in the 
leading logarithmic approximation taking into account   
corrections due to parton-parton recombination. 
The nuclear gluon  distribution  is written in two parts: 
$x\,g_{\T A}(x,Q^2) = x\,\T A \,g_{\T N}(x,Q^2) + 
\delta \left(\,x\,g_{\T A}(x,Q^2)\right)$. 
The first term is associated with independent nucleons and evolves 
accordingly. The second term is the 
correction of interest here and 
describes the interaction of gluons from different nucleons. 
Its evolution  due to gluon-gluon recombination reads  
\cite{Mueller:1986wy,Qiu:1987wh}:  
\begin{equation} \label{eq:non-lin.G}
Q^2 \frac{\partial}{\partial Q^2} 
\frac{\delta \left(\,x\,g_{\T A}(x,Q^2)\right)}{{\T A}} = 
-\frac{81}{16}\frac{{\T A}^{1/3}}{Q^2 \,r_0^2} \,\alpha_s^2(Q^2)
\int_x^1 \frac{dx'}{x'}\left[x' g_{\T N}(x',Q^2)\right]^2.
\end{equation}
Clearly gluon-gluon recombination is enhanced in nuclei. 
One finds \cite{Arneodo:1996qa,Strikman:acta96} 
that the $x$-range where non-linear effects 
(\ref{eq:non-lin.G}) become significant, differs for heavy 
nuclei and free nucleons by more than two orders 
of magnitude, assuming $x g_{\T N}(x) \sim x^{-0.2}$.

Investigations of unitarity constraints in hard two-body amplitudes 
\cite{Frankfurt:1996jw} also suggest that non-linear effects at high 
parton densities can be significant in nuclei at $Q^2 \sim 10$ GeV$^2$ 
and $x \sim 10^{-4}$. 

While the kinematic bounds for the applicability of ``normal'' DGLAP 
evolution are quite well defined, the dynamical mechanisms responsible 
for slowing down the rapid increase of parton distributions 
at large $Q^2$ and very small $x$ are not yet clear. 
One is entering a  new domain of QCD, dealing with partonic systems 
of high density, which presents new challenges.
 
\bigskip
\bigskip

{\bf \large Acknowledgments}

We gratefully acknowledge many discussions and conversations with 
S. Brodsky, L. Frankfurt, P. Hoyer, B. Kopeliovich, S. Kulagin, L. Mankiewicz, 
W. Melnitchouk, G.A. Miller, N. Nikolaev, K. Rith, M. Sargsian, M. Strikman, 
A.W. Thomas and M. V\"anttinen.  

\bibliographystyle{h-elsevier}
\bibliography{report}

\begin{thebibliography}{100}

\bibitem{Frankfurt:1988nt}
L.L. Frankfurt and M.I. Strikman,
\newblock Phys. Rept. 160 (1988) 235.

\bibitem{Jaffe:1985je}
R.L. Jaffe,
\newblock Relativistic Dynamics and Quark Nuclear Physics, edited by M.B.
  Johnson and A. Picklesimer, Wiley, New York, USA, 1986.

\bibitem{Bickerstaff:1989ch}
R.P. Bickerstaff and A.W. Thomas,
\newblock J. Phys. G15 (1989) 1523.

\bibitem{Arneodo:1994wf}
M. Arneodo,
\newblock Phys. Rept. 240 (1994) 301.

\bibitem{Geesaman:1995yd}
D.F. Geesaman, K. Saito and A.W. Thomas,
\newblock Ann. Rev. Nucl. Part. Sci. 45 (1995) 337.

\bibitem{Roberts:1990ww}
R.G. Roberts,
\newblock The Structure of the Proton: Deep Inelastic Scattering (Cambridge
  University Press, Cambridge, UK, 1990).

\bibitem{Muta:1987mz}
T. Muta,
\newblock Foundations of Quantum Chromodynamics (World Scientific, Singapore,
  Singapore, 1987).

\bibitem{Cheng:1984}
T.P. Cheng and L.F. Li,
\newblock Gauge Theory of Elementary Particle Physics (Oxford University Press,
  Oxford, UK, 1984).

\bibitem{Cooper-Sarkar:1997jk}
A.M. Cooper-Sarkar, R.C.E. Devenish and A. {De Roeck},
\newblock Int. J. Mod. Phys. A13 (1998) 3385.

\bibitem{Badelek:1996ss}
B. Badelek et~al.,
\newblock J. Phys. G22 (1996) 815.

\bibitem{Badelek:1996rmp}
B. Badelek and J. Kwiecinski,
\newblock Rev. Mod. Phys. 68 (1996) 445.

\bibitem{Adams:1996gu}
E665, M.R. Adams et~al.,
\newblock Phys. Rev. D54 (1996) 3006.

\bibitem{Aid:1996au}
H1, S. Aid et~al.,
\newblock Nucl. Phys. B470 (1996) 3.

\bibitem{Adloff:1997mf}
H1, C. Adloff et~al.,
\newblock Nucl. Phys. B497 (1997) 3.

\bibitem{Derrick:1996ef}
ZEUS, M. Derrick et~al.,
\newblock Z. Phys. C69 (1996) 607.

\bibitem{Derrick:1996hn}
ZEUS, M. Derrick et~al.,
\newblock Z. Phys. C72 (1996) 399.

\bibitem{Breitweg:1998dz}
ZEUS, J. Breitweg et~al.,
\newblock Eur. Phys. J. C7 (1999) 609.

\bibitem{Arneodo:1997qe}
NMC, M. Arneodo et~al.,
\newblock Nucl. Phys. B483 (1997) 3.

\bibitem{Whitlow:1990dr}
L.W. Whitlow,
\newblock PhD thesis, Stanford University, 1990,
\newblock SLAC Report 357.

\bibitem{Benvenuti:1989rh}
BCDMS, A.C. Benvenuti et~al.,
\newblock Phys. Lett. B223 (1989) 485.

\bibitem{Donnachie:1992ny}
A. Donnachie and P.V. Landshoff,
\newblock Phys. Lett. B296 (1992) 227.

\bibitem{Collins:1977jy}
P.D.B. Collins,
\newblock An Introduction to Regge Theory and High-Energy Physics (Cambridge
  University Press, Cambridge, UK, 1977).

\bibitem{Abe:1994xx}
CDF, F. Abe et~al.,
\newblock Phys. Rev. D50 (1994) 5518.

\bibitem{Apel:1979sp}
Serpukhov-CERN, W.D. Apel et~al.,
\newblock Nucl. Phys. B154 (1979) 189.

\bibitem{Bauer:1978iq}
T.H. Bauer, R.D. Spital, D.R. Yennie and F.M. Pipkin,
\newblock Rev. Mod. Phys. 50 (1978) 261.

\bibitem{Brodsky:1995kg}
S.J. Brodsky, M. Burkardt and I. Schmidt,
\newblock Nucl. Phys. B441 (1995) 197.

\bibitem{Whitlow:1990gk}
L.W. Whitlow et~al.,
\newblock Phys. Lett. B250 (1990) 193.

\bibitem{Benvenuti:1990fm}
BCDMS, A.C. Benvenuti et~al.,
\newblock Phys. Lett. B237 (1990) 592.

\bibitem{Benvenuti:1987zj}
BCDMS, A.C. Benvenuti et~al.,
\newblock Phys. Lett. B195 (1987) 91.

\bibitem{Berge:1991hr}
J.P. Berge et~al.,
\newblock Z. Phys. C49 (1991) 187.

\bibitem{Altarelli:1978tq}
G. Altarelli and G. Martinelli,
\newblock Phys. Lett. B76 (1978) 89.

\bibitem{Adloff:1997yz}
H1, C. Adloff et~al.,
\newblock Phys. Lett. B393 (1997) 452.

\bibitem{Adeva:1998vv}
SMC, B. Adeva et~al.,
\newblock Phys. Rev. D58 (1998) 112001.

\bibitem{Adeva:1998vw}
SMC, B. Adeva et~al.,
\newblock Phys. Rev. D58 (1998) 112002.

\bibitem{Abe:1997qk}
E154, K. Abe et~al.,
\newblock Phys. Lett. B404 (1997) 377.

\bibitem{Airapetian:1998wi}
HERMES, A. Airapetian et~al.,
\newblock Phys. Lett. B442 (1998) 484.

\bibitem{Ackerstaff:1997ws}
HERMES, K. Ackerstaff et~al.,
\newblock Phys. Lett. B404 (1997) 383.

\bibitem{Anthony:1999rm}
E155, P.L. Anthony et~al.,
\newblock (1999), hep-ex/9904002.

\bibitem{Anthony:1999py}
E155, P.L. Anthony et~al.,
\newblock (1999), hep-ex/9901006.

\bibitem{Abe:1997dp}
E154, K. Abe et~al.,
\newblock Phys. Lett. B405 (1997) 180.

\bibitem{Abe:1997cx}
E154, K. Abe et~al.,
\newblock Phys. Rev. Lett. 79 (1997) 26.

\bibitem{Anthony:1996mw}
E142, P.L. Anthony et~al.,
\newblock Phys. Rev. D54 (1996) 6620.

\bibitem{Abe:1998wq}
E143, K. Abe et~al.,
\newblock Phys. Rev. D58 (1998) 112003.

\bibitem{Anselmino:1995gn}
M. Anselmino, A. Efremov and E. Leader,
\newblock Phys. Rept. 261 (1995) 1.

\bibitem{Lampe:1998eu}
B. Lampe and E. Reya,
\newblock (1998), hep-ph/9810270.

\bibitem{Adams:1994id}
SMC, D. Adams et~al.,
\newblock Phys. Lett. B336 (1994) 125.

\bibitem{Abe:1996dc}
E143, K. Abe et~al.,
\newblock Phys. Rev. Lett. 76 (1996) 587.

\bibitem{Wandzura:1977qf}
S. Wandzura and F. Wilczek,
\newblock Phys. Lett. B72 (1977) 195.

\bibitem{Abramowicz:1998ii}
H. Abramowicz and A. Caldwell,
\newblock DESY-98-192.

\bibitem{Adloff:1997sc}
H1, C. Adloff et~al.,
\newblock Z. Phys. C76 (1997) 613.

\bibitem{Ahmed:1995ns}
H1, T. Ahmed et~al.,
\newblock Phys. Lett. B348 (1995) 681.

\bibitem{Breitweg:1998gc}
ZEUS, J. Breitweg et~al.,
\newblock Eur. Phys. J. C6 (1999) 43.

\bibitem{Breitweg:1998aa}
ZEUS, J. Breitweg et~al.,
\newblock Eur. Phys. J. C1 (1998) 81.

\bibitem{Derrick:1996ma}
ZEUS, M. Derrick et~al.,
\newblock Z. Phys. C70 (1996) 391.

\bibitem{Derrick:1995wv}
ZEUS, M. Derrick et~al.,
\newblock Z. Phys. C68 (1995) 569.

\bibitem{Ingelman:1993qf}
G. Ingelman and K. Prytz,
\newblock Z. Phys. C58 (1993) 285.

\bibitem{Goulianos:1995wy}
K. Goulianos,
\newblock Phys. Lett. B358 (1995) 379.

\bibitem{Chapin:1985}
T.J. Chapin et~al.,
\newblock Phys. Rev. D31 (1985) 17.

\bibitem{Breitweg:1997eh}
ZEUS, J. Breitweg et~al.,
\newblock Eur. Phys. J. C2 (1998) 237.

\bibitem{Breitweg:1997za}
ZEUS, J. Breitweg et~al.,
\newblock Z. Phys. C75 (1997) 421.

\bibitem{Adloff:1997mi}
H1, C. Adloff et~al.,
\newblock Z. Phys. C74 (1997) 221.

\bibitem{Aid:1995bz}
H1, S. Aid et~al.,
\newblock Z. Phys. C69 (1995) 27.

\bibitem{Derrick:1994dt}
ZEUS, M. Derrick et~al.,
\newblock Z. Phys. C63 (1994) 391.

\bibitem{Crittenden:1997yz}
J.A. Crittenden,
\newblock Tracts in Modern Physics, Volume 140, Springer  (1997).

\bibitem{Goulianos:1983vk}
K. Goulianos,
\newblock Phys. Rept. 101 (1983) 169.

\bibitem{Piller:1998cy}
G. Piller, L. Ferreira and W. Weise,
\newblock Eur. Phys. J. A4 (1999) 287.

\bibitem{Hoodbhoy:1989am}
P. Hoodbhoy, R.L. Jaffe and A. Manohar,
\newblock Nucl. Phys. B312 (1989) 571.

\bibitem{Sather:1990bq}
E. Sather and C. Schmidt,
\newblock Phys. Rev. D42 (1990) 1424.

\bibitem{Jackson94}
H.E. Jackson,
\newblock AIP Conf. Proc. 339 (1994).

\bibitem{Edelmann:1997ik}
J. Edelmann, G. Piller and W. Weise,
\newblock Phys. Rev. C57 (1998) 3392.

\bibitem{Amaudruz:1995tq}
NMC, P. Amaudruz et~al.,
\newblock Nucl. Phys. B441 (1995) 3.

\bibitem{Gomez:1994ri}
J. Gomez et~al.,
\newblock Phys. Rev. D49 (1994) 4348.

\bibitem{Benvenuti:1987az}
BCDMS, A.C. Benvenuti et~al.,
\newblock Phys. Lett. B189 (1987) 483.

\bibitem{Aubert:1983}
EMC, J.J. Aubert et~al.,
\newblock Phys. Lett. B123 (1983) 275.

\bibitem{Arneodo:1995cs}
NMC, M. Arneodo et~al.,
\newblock Nucl. Phys. B441 (1995) 12.

\bibitem{Adams:1992nf}
E665, M.R. Adams et~al.,
\newblock Phys. Rev. Lett. 68 (1992) 3266.

\bibitem{Adams:1995is}
E665, M.R. Adams et~al.,
\newblock Z. Phys. C67 (1995) 403.

\bibitem{Adams:1992vm}
E665, M.R. Adams et~al.,
\newblock Phys. Lett. B287 (1992) 375.

\bibitem{Amaudruz:1991cc}
NMC, P. Amaudruz et~al.,
\newblock Z. Phys. C51 (1991) 387.

\bibitem{Amaudruz:1992dj}
NMC, P. Amaudruz et~al.,
\newblock Z. Phys. C53 (1992) 73.

\bibitem{Arneodo:1996rv}
NMC, M. Arneodo et~al.,
\newblock Nucl. Phys. B481 (1996) 3.

\bibitem{Arneodo:1996ru}
NMC, M. Arneodo et~al.,
\newblock Nucl. Phys. B481 (1996) 23.

\bibitem{Heynen71}
V. Heynen et~al.,
\newblock Phys. Lett. B34 (1971) 651.

\bibitem{Brookes73}
G.R. Brookes et~al.,
\newblock Phys. Rev. D8 (1973) 2826.

\bibitem{Caldwell:1973bu}
D.O. Caldwell et~al.,
\newblock Phys. Rev. D7 (1973) 1362.

\bibitem{Michalowski:1977eg}
S. Michalowski et~al.,
\newblock Phys. Rev. Lett. 39 (1977) 737.

\bibitem{Arakelian:1978rc}
E.A. Arakelian et~al.,
\newblock Phys. Lett. B79 (1978) 143.

\bibitem{Caldwell:1979ik}
D.O. Caldwell et~al.,
\newblock Phys. Rev. Lett. 42 (1979) 553.

\bibitem{Bianchi:1994ax}
N. Bianchi et~al.,
\newblock Phys. Lett. B325 (1994) 333.

\bibitem{Weise:1993}
W. Weise,
\newblock Phys. Rept. 13 (1974) 53.

\bibitem{Adams:1995sh}
E665, M.R. Adams et~al.,
\newblock Phys. Rev. Lett. 75 (1995) 1466.

\bibitem{Arneodo:1996kd}
NMC, M. Arneodo et~al.,
\newblock Nucl. Phys. B487 (1997) 3.

\bibitem{Bari:1985ga}
BCDMS, G. Bari et~al.,
\newblock Phys. Lett. B163 (1985) 282.

\bibitem{Benvenuti:1994bb}
BCDMS, A.C. Benvenuti et~al.,
\newblock Z. Phys. C63 (1994) 29.

\bibitem{Vakili:1999qt}
CCFR, M. Vakili et~al.,
\newblock (1999), hep-ex/9905052.

\bibitem{Arrington:1996hs}
J. Arrington et~al.,
\newblock Phys. Rev. C53 (1996) 2248.

\bibitem{Rock:1992jy}
S. Rock et~al.,
\newblock Phys. Rev. D46 (1992) 24.

\bibitem{Bosted:1992fy}
P. Bosted et~al.,
\newblock Phys. Rev. C46 (1992) 2505.

\bibitem{Filippone:1992iz}
B.W. Filippone et~al.,
\newblock Phys. Rev. C45 (1992) 1582.

\bibitem{Day:1987az}
D.B. Day et~al.,
\newblock Phys. Rev. Lett. 59 (1987) 427.

\bibitem{Nachtmann:1974aj}
O. Nachtmann,
\newblock Nucl. Phys. B78 (1974) 455.

\bibitem{Tao:1996uh}
E140X, L.H. Tao et~al.,
\newblock Z. Phys. C70 (1996) 387.

\bibitem{Dasu:1988ru}
S. Dasu et~al.,
\newblock Phys. Rev. Lett. 60 (1988) 2591.

\bibitem{Amaudruz:1992wn}
NMC, P. Amaudruz et~al.,
\newblock Phys. Lett. B294 (1992) 120.

\bibitem{Dasu:1994vk}
S. Dasu et~al.,
\newblock Phys. Rev. D49 (1994) 5641.

\bibitem{Field:1989uq}
R.D. Field,
\newblock Applications of Perturbative QCD (Addison-Wesley, Redwood City, USA,
  1989).

\bibitem{Alde:1990im}
D.M. Alde et~al.,
\newblock Phys. Rev. Lett. 64 (1990) 2479.

\bibitem{Frankfurt:1990xz}
L.L. Frankfurt, M.I. Strikman and S. Liuti,
\newblock Phys. Rev. Lett. 65 (1990) 1725.

\bibitem{Berger:1981ni}
E.L. Berger and D. Jones,
\newblock Phys. Rev. D23 (1981) 1521.

\bibitem{Amaudruz:1992sr}
NMC, P. Amaudruz et~al.,
\newblock Nucl. Phys. B371 (1992) 553.

\bibitem{Guy:1987us}
WA25, J. Guy et~al.,
\newblock Z. Phys. C36 (1987) 337.

\bibitem{Allport:1989vf}
BEBC WA59, P.P. Allport et~al.,
\newblock Phys. Lett. B232 (1989) 417.

\bibitem{Kitagaki:1988wc}
E745, T. Kitagaki et~al.,
\newblock Phys. Lett. B214 (1988) 281.

\bibitem{Guy:1989iz}
BEBC, J. Guy et~al.,
\newblock Phys. Lett. B229 (1989) 421.

\bibitem{Brodsky:1997de}
S.J. Brodsky, H.C. Pauli and S.S. Pinsky,
\newblock Phys. Rept. 301 (1998) 299.

\bibitem{Vanttinen:1998iz}
M. V{\"a}nttinen, G. Piller, L. Mankiewicz, W. Weise and K.J. Eskola, 
\newblock Eur. Phys. J. A3 (1998) 351.

\bibitem{Ioffe:1969kf}
B.L. Ioffe,
\newblock Phys. Lett. B30 (1969) 123.

\bibitem{LlewellynSmith:1985pv}
C.H. Llewellyn-Smith,
\newblock Nucl. Phys. A434 (1985) 35c.

\bibitem{Hoyer:1996nr}
P. Hoyer and M. V{\"a}nttinen,
\newblock Z. Phys. C74 (1997) 113.

\bibitem{Braun:1995jq}
V. Braun, P. Gornicki and L. Mankiewicz,
\newblock Phys. Rev. D51 (1995) 6036.

\bibitem{Collins:1982uw}
J.C. Collins and D.E. Soper,
\newblock Nucl. Phys. B194 (1982) 445.

\bibitem{Balitskii:1988/89}
I.I. Balitskii and V.M. Braun,
\newblock Nucl. Phys. B311 (1988/89) 541.

\bibitem{Lai:1997mg}
H.L. Lai et~al.,
\newblock Phys. Rev. D55 (1997) 1280.

\bibitem{Mankiewicz:1996ep}
L. Mankiewicz and T. Weigl,
\newblock Phys. Lett. B380 (1996) 134.

\bibitem{Weigl:1996ii}
T. Weigl and L. Mankiewicz,
\newblock Phys. Lett. B389 (1996) 334.

\bibitem{Brodsky:1991gn}
S.J. Brodsky and I.A. Schmidt,
\newblock Phys. Rev. D43 (1991) 179.

\bibitem{Eskola:1998iy}
K.J. Eskola, V.J. Kolhinen and P.V. Ruuskanen,
\newblock Nucl. Phys. B535 (1998) 351.

\bibitem{Frankfurt:1998ym}
L.L. Frankfurt and M.I. Strikman,
\newblock (1998), hep-ph/9812322.

\bibitem{Alvero:1998bz}
L. Alvero, L.L. Frankfurt and M.I. Strikman,
\newblock Eur. Phys. J. A5 (1999) 97.

\bibitem{Nikolaev:1975vy}
N.N. Nikolaev and V.I. Zakharov,
\newblock Phys. Lett. B55 (1975) 397.

\bibitem{Close:1988xw}
F.E. Close and R.G. Roberts,
\newblock Phys. Lett. B213 (1988) 91.

\bibitem{Kumano:1992ef}
S. Kumano,
\newblock Phys. Rev. C48 (1993) 2016.

\bibitem{RamanaMurthy:1975qb}
P.V.R. Murthy et~al.,
\newblock Nucl. Phys. B92 (1975) 269.

\bibitem{McLerran:1994ni}
L. McLerran and R. Venugopalan,
\newblock Phys. Rev. D49 (1994) 2233.

\bibitem{Jalilian-Marian:1997xn}
J. Jalilian-Marian, A.  Kovner, L. McLerran and H. Weigert, 
\newblock Phys. Rev. D55 (1997) 5414.

\bibitem{Ayala:1997em}
A.L. Ayala, M.B.G. Ducati and E.M. Levin,
\newblock Nucl. Phys. B493 (1997) 305.

\bibitem{Kovchegov:1997dm}
Y.V. Kovchegov, A.H. Mueller and S. Wallon,
\newblock Nucl. Phys. B507 (1997) 367.

\bibitem{Mueller:1999wm}
A.H. Mueller,
\newblock (1999), hep-ph/9904404.

\bibitem{Kovchegov:1999yj}
Y.V. Kovchegov,
\newblock (1999), hep-ph/9901281.

\bibitem{Miller:1997xh}
G.A. Miller,
\newblock Phys. Rev. C56 (1997) 8.

\bibitem{Miller:1997cr}
G.A. Miller,
\newblock Phys. Rev. C56 (1997) 2789.

\bibitem{Burkardt:1998bt}
M. Burkardt and G.A. Miller,
\newblock Phys. Rev. C58 (1998) 2450.

\bibitem{Miller:1998tp}
G.A. Miller and R. Machleidt,
\newblock Phys. Lett. B455 (1999) 19.

\bibitem{Blunden:1999hy}
P.G. Blunden, M. Burkardt and G.A. Miller,
\newblock Phys. Rev. C59 (1999) 2998.

\bibitem{Miller:1999ap}
G.A. Miller and R. Machleidt,
\newblock (1999), nucl-th/9903080.

\bibitem{Blunden:1999gq}
P.G. Blunden, M. Burkardt and G.A. Miller,
\newblock (1999), nucl-th/9906012.

\bibitem{Piller:1997ny}
G. Piller, G. Niesler and W. Weise,
\newblock Z. Phys. A358 (1997) 407.

\bibitem{Gribov:1969}
V.N. Gribov,
\newblock Sov. Phys. JETP 29 (1969) 483.

\bibitem{Gribov:1970}
V.N. Gribov,
\newblock Sov. Phys. JETP 30 (1970) 709.

\bibitem{Bertocchi:1972}
L. Bertocchi,
\newblock Nuovo Cim. 11A (1972) 45.

\bibitem{Weis:1976er}
J.H. Weis,
\newblock Acta Phys. Polon. B7 (1976) 851.

\bibitem{Lacombe:1980dr}
M. Lacombe et~al.,
\newblock Phys. Rev. C21 (1980) 861.

\bibitem{Piller:90prc}
G. Piller and W. Weise,
\newblock Phys. Rev. C42 (1990) 1834.

\bibitem{Melnitchouk:1993vc}
W. Melnitchouk and A.W. Thomas,
\newblock Phys. Lett. B317 (1993) 437.

\bibitem{Kondratyuk:1973jept}
V.A. Karmanov and L.A. Kondratyuk,
\newblock JETP Lett. 18 (1973) 266.

\bibitem{Capella:1997mn}
A. Capella,  A. Kaidalov, C. Merino, D. Pertermann and J. Tran Thanh Van,
\newblock Eur. Phys. J. C5 (1998) 111.

\bibitem{Nikolaev:1986vy}
N.N. Nikolaev,
\newblock Z. Phys. C32 (1986) 537.

\bibitem{Caldwell:1978yb}
D.O. Caldwell et~al.,
\newblock Phys. Rev. Lett. 40 (1978) 1222.

\bibitem{Kopeliovich:1995ju}
B.Z. Kopeliovich and B. Povh,
\newblock Z. Phys. A356 (1997) 467.

\bibitem{Lepage:1980fj}
G.P. Lepage and S.J. Brodsky,
\newblock Phys. Rev. D22 (1980) 2157.

\bibitem{Bjorken:1971ah}
J.D. Bjorken, J.B. Kogut and D.E. Soper,
\newblock Phys. Rev. D3 (1971) 1382.

\bibitem{Nikolaev:1991ja}
N.N. Nikolaev and B.G. Zakharov,
\newblock Z. Phys. C49 (1991) 607.

\bibitem{Brodsky:1997nj}
S.J. Brodsky, A. Hebecker and E. Quack,
\newblock Phys. Rev. D55 (1997) 2584.

\bibitem{Frankfurt:1997ri}
L.L. Frankfurt, A.V. Radyushkin and M.I. Strikman,
\newblock Phys. Rev. D55 (1997) 98.

\bibitem{Blaettel:1993rd}
B. Blaettel, G. Baym, L.L. Frankfurt and M.I. Strikman,
\newblock Phys. Rev. Lett. 70 (1993) 896.

\bibitem{Frankfurt:1996jw}
L.L. Frankfurt, W. Koepf and M.I. Strikman,
\newblock Phys. Rev. D54 (1996) 3194.

\bibitem{Bjorken:1973gc}
J.D. Bjorken and J.B. Kogut,
\newblock Phys. Rev. D8 (1973) 1341.

\bibitem{Buchmuller:1997xw}
W. Buchm{\"u}ller, M.F. McDermott and A. Hebecker,
\newblock Nucl. Phys. B487 (1997) 283.

\bibitem{Piller:1995kh}
G. Piller, W. Ratzka and W. Weise,
\newblock Z. Phys. A352 (1995) 427.

\bibitem{Frankfurt:1989zg}
L.L. Frankfurt and M.I. Strikman,
\newblock Nucl. Phys. B316 (1989) 340.

\bibitem{Amaudruz:1992bf}
NMC, P. Amaudruz et~al.,
\newblock Phys. Lett. B295 (1992) 159.

\bibitem{Kwiecinski:1988ys}
J. Kwiecinski and B. Badelek,
\newblock Phys. Lett. B208 (1988) 508.

\bibitem{Badelek:1992qa}
B. Badelek and J. Kwiecinski,
\newblock Nucl. Phys. B370 (1992) 278.

\bibitem{Nikolaev:1991yw}
N.N. Nikolaev and B.G. Zakharov,
\newblock Phys. Lett. B260 (1991) 414.

\bibitem{Nikolaev:1992gw}
N.N. Nikolaev and V.R. Zoller,
\newblock Z. Phys. C56 (1992) 623.

\bibitem{Zoller:1992ns}
V.R. Zoller,
\newblock Phys. Lett. B279 (1992) 145.

\bibitem{Sakurai:1972wk}
J.J. Sakurai and D. Schildknecht,
\newblock Phys. Lett. B40 (1972) 121.

\bibitem{Sakurai:1972zs}
J.J. Sakurai and D. Schildknecht,
\newblock Phys. Lett. B42 (1972) 216.

\bibitem{Schildknecht:1973gi}
D. Schildknecht,
\newblock Nucl. Phys. B66 (1973) 398.

\bibitem{Fraas:1975gh}
H. Fraas, B.J. Read and D. Schildknecht,
\newblock Nucl. Phys. B86 (1975) 346.

\bibitem{Devenish:1976ky}
R. Devenish and D. Schildknecht,
\newblock Phys. Rev. D14 (1976) 93.

\bibitem{Bilchak:1988zn}
C.L. Bilchak, D. Schildknecht and J.D. Stroughair,
\newblock Phys. Lett. B214 (1988) 441.

\bibitem{Ditsas:1975vd}
P. Ditsas, B.J. Read and G. Shaw,
\newblock Nucl. Phys. B99 (1975) 85.

\bibitem{Ditsas:1976yv}
P. Ditsas and G. Shaw,
\newblock Nucl. Phys. B113 (1976) 246.

\bibitem{Shaw:1989mn}
G. Shaw,
\newblock Phys. Lett. B228 (1989) 125.

\bibitem{Bilchak:1989ck}
C. Bilchak, D. Schildknecht and J.D. Stroughair,
\newblock Phys. Lett. B233 (1989) 461.

\bibitem{Shaw:1993gx}
G. Shaw,
\newblock Phys. Rev. D47 (1993) 3676.

\bibitem{Brodsky:1990qz}
S.J. Brodsky and H.J. Lu,
\newblock Phys. Rev. Lett. 64 (1990) 1342.

\bibitem{Landshoff:1971ff}
P.V. Landshoff, J.C. Polkinghorne and R.D. Short,
\newblock Nucl. Phys. B28 (1971) 225.

\bibitem{Kulagin:1994fz}
S.A. Kulagin, G. Piller and W. Weise,
\newblock Phys. Rev. C50 (1994) 1154.

\bibitem{Kopeliovich:1998gv}
B.Z. Kopeliovich, J. Raufeisen and A.V. Tarasov,
\newblock Phys. Lett. B440 (1998) 151.

\bibitem{Raufeisen:1998rg}
J. Raufeisen, A.V. Tarasov and O.O. Voskresenskaya,
\newblock (1998), hep-ph/9812398.

\bibitem{Melnitchouk:1993eu}
W. Melnitchouk and A.W. Thomas,
\newblock Phys. Rev. D47 (1993) 3783.

\bibitem{Nikolaev:1997jy}
N.N. Nikolaev and W. Schafer,
\newblock Phys. Lett. B398 (1997) 245.

\bibitem{Machleidt:1987hj}
R. Machleidt, K. Holinde and C. Elster,
\newblock Phys. Rept. 149 (1987) 1.

\bibitem{Betev:1985pg}
NA10, B. Betev et~al.,
\newblock Z. Phys. C28 (1985) 15.

\bibitem{Gluck:1992ey}
M. Gl{\"u}ck, E. Reya and A. Vogt,
\newblock Z. Phys. C53 (1992) 651.

\bibitem{Holtmann:1994rs}
H. Holtmann, G. Levman, N.N. Nikolaev, A. Szczurek and J. Speth,
\newblock Phys. Lett. B338 (1994) 363.

\bibitem{Przybycien:1996zb}
M. Przybycien, A. Szczurek and G. Ingelman,
\newblock Z. Phys. C74 (1997) 509.

\bibitem{Arneodo:1994ia}
NMC, M. Arneodo et~al.,
\newblock Phys. Rev. D50 (1994) 1.

\bibitem{Close:1989ca}
F.E. Close, J. Qiu and R.G. Roberts,
\newblock Phys. Rev. D40 (1989) 2820.

\bibitem{Kumano:1994pn}
S. Kumano,
\newblock Phys. Rev. C50 (1994) 1247.

\bibitem{Prytz:1993vr}
K. Prytz,
\newblock Phys. Lett. B311 (1993) 286.

\bibitem{Gousset:1996xt}
T. Gousset and H.J. Pirner,
\newblock Phys. Lett. B375 (1996) 349.

\bibitem{Vogt:1991qd}
R. Vogt, S.J. Brodsky and P. Hoyer,
\newblock Nucl. Phys. B360 (1991) 67.

\bibitem{Vogt:1992ki}
R. Vogt, S.J. Brodsky and P. Hoyer,
\newblock Nucl. Phys. B383 (1992) 643.

\bibitem{Schiffer:1980hb}
J.P. Schiffer,
\newblock Nucl. Phys. A335 (1980) 339.

\bibitem{Bickerstaff:1985da}
R.P. Bickerstaff, M.C. Birse and G.A. Miller,
\newblock Phys. Rev. D33 (1986) 3228.

\bibitem{Bickerstaff:1985ax}
R.P. Bickerstaff, M.C. Birse and G.A. Miller,
\newblock Phys. Rev. Lett. 53 (1984) 2532.

\bibitem{Hoodbhoy:1987fn}
P. Hoodbhoy and R.L. Jaffe,
\newblock Phys. Rev. D35 (1987) 113.

\bibitem{Saito:1994yw}
K. Saito and A.W. Thomas,
\newblock Nucl. Phys. A574 (1994) 659.

\bibitem{Melnitchouk:1994nk}
W. Melnitchouk, A.W. Schreiber and A.W. Thomas,
\newblock Phys. Rev. D49 (1994) 1183.

\bibitem{Kulagin:1995cj}
S.A. Kulagin, W. Melnitchouk, G. Piller and W. Weise, 
\newblock Phys. Rev. C52 (1995) 932.

\bibitem{Frankfurt:1987ui}
L.L. Frankfurt and M.I. Strikman,
\newblock Phys. Lett. B183 (1987) 254.

\bibitem{Jung:1988jw}
H. Jung and G.A. Miller,
\newblock Phys. Lett. B200 (1988) 351.

\bibitem{CiofiDegliAtti:1989eg}
C. {Ciofi Degli Atti} and S. Liuti,
\newblock Phys. Lett. B225 (1989) 215.

\bibitem{Dieperink:1991mw}
A.E.L. Dieperink and G.A. Miller,
\newblock Phys. Rev. C44 (1991) 866.

\bibitem{Koltun:1972}
D.S. Koltun,
\newblock Phys. Rev. Lett. 28 (1972) 182.

\bibitem{Benhar:1989aw}
O. Benhar, A. Fabrocini and S. Fantoni,
\newblock Nucl. Phys. A505 (1989) 267.

\bibitem{Melnitchouk:1994rv}
W. Melnitchouk, A.W. Schreiber and A.W. Thomas,
\newblock Phys. Lett. B335 (1994) 11.

\bibitem{Meyer:1994fg}
H. Meyer, P.J. Mulders and W.F.M. Spit,
\newblock Nucl. Phys. A570 (1994) 497.

\bibitem{Guichon:1988jp}
P.A.M. Guichon,
\newblock Phys. Lett. B200 (1988) 235.

\bibitem{Saito:1992rm}
K. Saito, A. Michels and A.W. Thomas,
\newblock Phys. Rev. C46 (1992) 2149.

\bibitem{Close:1983tn}
F.E. Close, R.G. Roberts and G.G. Ross,
\newblock Phys. Lett. B129 (1983) 346.

\bibitem{Jaffe:1984zw}
R.L. Jaffe, F.E. Close, R.G. Roberts and G.G. Ross,
\newblock Phys. Lett. B134 (1984) 449.

\bibitem{Close:1985zn}
F.E. Close, R.L. Jaffe, R.G. Roberts and G.G. Ross,
\newblock Phys. Rev. D31 (1985) 1004.

\bibitem{Close:1986ji}
F.E. Close, R.G. Roberts and G.G. Ross,
\newblock Phys. Lett. B168 (1986) 400.

\bibitem{Close:1988ay}
F.E. Close, R.G. Roberts and G.G. Ross,
\newblock Nucl. Phys. B296 (1988) 582.

\bibitem{Sick:1985a}
I. Sick,
\newblock Phys. Lett. B157 (1985) 13.

\bibitem{Sick:1985b}
I. Sick,
\newblock Nucl. Phys. A434 (1985) 677c.

\bibitem{Nachtmann:1984py}
O. Nachtmann and H.J. Pirner,
\newblock Z. Phys. C21 (1984) 277.

\bibitem{Bickerstaff:1985mp}
R.P. Bickerstaff and G.A. Miller,
\newblock Phys. Lett. B168 (1986) 409.

\bibitem{Mulders:1985ec}
P.J. Mulders,
\newblock Phys. Rev. Lett. 54 (1985) 2560.

\bibitem{Bickerstaff:1987ck}
R.P. Bickerstaff and A.W. Thomas,
\newblock Phys. Rev. D35 (1987) 108.

\bibitem{Frankfurt:1985cv}
L.L. Frankfurt and M.I. Strikman,
\newblock Nucl. Phys. B250 (1985) 143.

\bibitem{Frank:1996pv}
M.R. Frank, B.K. Jennings and G.A. Miller,
\newblock Phys. Rev. C54 (1996) 920.

\bibitem{Ericson:1983um}
M. Ericson and A.W. Thomas,
\newblock Phys. Lett. B128 (1983) 112.

\bibitem{Berger:1985na}
E.L. Berger and F. Coester,
\newblock Phys. Rev. D32 (1985) 1071.

\bibitem{Birbrair:1989hb}
B.L. Birbrair, E.M. Levin and A.G. Shuvaev,
\newblock Nucl. Phys. A491 (1989) 618.

\bibitem{Kulagin:1989mu}
S.A. Kulagin,
\newblock Nucl. Phys. A500 (1989) 653.

\bibitem{Jung:1990pu}
H. Jung and G.A. Miller,
\newblock Phys. Rev. C41 (1990) 659.

\bibitem{Sullivan:1972}
J.D. Sullivan,
\newblock Phys. Rev. D5 (1972) 1732.

\bibitem{Brown:1995su}
G.E. Brown, M. Buballa, Zi Bang Li  and J. Wambach,
\newblock Nucl. Phys. A593 (1995) 295.

\bibitem{Miller:1996qg}
G.A. Miller,
\newblock Proc. LAMPF Symposium: 20 Years of Meson Factory Physics, edited by
  B.F. Gibson et~al., 1996.

\bibitem{Dieperink:1997iv}
A.E.L. Dieperink and C.L. Korpa,
\newblock Phys. Rev. C55 (1997) 2665.

\bibitem{Koltun:1997py}
D.S. Koltun,
\newblock Phys. Rev. C57 (1998) 1210.

\bibitem{Marco:1996vb}
E. Marco, E. Oset and P. {Fernandez de Cordoba},
\newblock Nucl. Phys. A611 (1996) 484.

\bibitem{Marco:1997xb}
E. Marco and E. Oset,
\newblock Nucl. Phys. A645 (1999) 303.

\bibitem{Benhar:1997emc}
O. Benhar, V.R. Pandharipande and I. Sick,
\newblock Phys. Lett. B410 (1997) 79.

\bibitem{Benhar:1998gb}
O. Benhar, V.R. Pandharipande and I. Sick,
\newblock JLAB-THY-98-12.

\bibitem{Melnitchouk:1996vp}
W. Melnitchouk, M. Sargsian and M.I. Strikman,
\newblock Z. Phys. A359 (1997) 99.

\bibitem{Simula:1996xk}
S. Simula,
\newblock Phys. Lett. B387 (1996) 245.

\bibitem{CiofidegliAtti:1993ep}
C. {Ciofi degli Atti} and S. Simula,
\newblock Phys. Lett. B319 (1993) 23.

\bibitem{Ingelman:1996ge}
G. Ingelman, A. {De Roeck} and R. Klanner, editors,
\newblock Proc. Workshop Future Physics at HERA, DESY, Hamburg, Germany, 1996.

\bibitem{CiofidegliAtti:1993zs}
C. {Ciofi degli Atti}, S. Scopetta, E. Pace and G. Salme,
\newblock Phys. Rev. C48 (1993) 968.

\bibitem{Edelmann:1997qe}
J. Edelmann, G. Piller and W. Weise,
\newblock Z. Phys. A357 (1997) 129.

\bibitem{Adams:1997hc}
SMC, D. Adams et~al.,
\newblock Phys. Lett. B396 (1997) 338.

\bibitem{Strikman:1996YALE}
M.I. Strikman,
\newblock Proc. Symposium on Spin Structure of the Nucleon, Yale, USA, 1994.

\bibitem{Khan:1991qk}
H. Khan and P. Hoodbhoy,
\newblock Phys. Rev. C44 (1991) 1219.

\bibitem{Umnikov:1997qv}
A.Y. Umnikov,
\newblock Phys. Lett. B391 (1997) 177.

\bibitem{Frankfurt:1996nf}
L.L. Frankfurt, V. Guzey and M.I. Strikman,
\newblock Phys. Lett. B381 (1996) 379.

\bibitem{Caso:1998tx}
C. Caso et~al.,
\newblock Eur. Phys. J. C3 (1998) 1.

\bibitem{Budick:1991zb}
B. Budick, J.S. Chen and H. Lin,
\newblock Phys. Rev. Lett. 67 (1991) 2630.

\bibitem{Kaptari:1990qt}
L.P. Kaptari and A.Y. Umnikov,
\newblock Phys. Lett. B240 (1990) 203.

\bibitem{Ericson:1988gk}
T.E.O. Ericson and W. Weise,
\newblock Pions and Nuclei (Oxford University Press, Oxford, UK, 1988).

\bibitem{Woloshyn:1985}
R.M. Woloshyn,
\newblock Nucl. Phys. A495 (1985) 749.

\bibitem{Schulze:1997rz}
R.W. Schulze and P.U. Sauer,
\newblock Phys. Rev. C56 (1997) 2293.

\bibitem{Melnitchouk:1995tx}
W. Melnitchouk, G. Piller and A.W. Thomas,
\newblock Phys. Lett. B346 (1995) 165.

\bibitem{Piller:1996mf}
G. Piller, W. Melnitchouk and A.W. Thomas,
\newblock Phys. Rev. C54 (1996) 894.

\bibitem{Kaptari:1995di}
L.P. Kaptari, A.Yu Umnikov, C. {Ciofi degli Atti}, S. Scopetta and 
K.Yu Kazakov,
\newblock Phys. Rev. C51 (1995) 52.

\bibitem{Adams:1995bw}
E665, M.R. Adams et~al.,
\newblock Phys. Rev. Lett. 74 (1995) 1525.

\bibitem{Arneodo:1994qb}
NMC, M. Arneodo et~al.,
\newblock Phys. Lett. B332 (1994) 195.

\bibitem{Arneodo:1994id}
NMC, M. Arneodo et~al.,
\newblock Nucl. Phys. B429 (1994) 503.

\bibitem{Ackerstaff:1999wt}
HERMES, K. Ackerstaff et~al.,
\newblock Phys. Rev. Lett. 82 (1999) 3025.

\bibitem{Dytman:1998}
S. Dytman, H. Fenker and P. Ross, editors,
\newblock Proc. Jefferson Lab Physics and Instrumentation with $6$--$12$ GeV
  Beams, TJNAF, Newport News, USA, 1998.

\bibitem{Brodsky:1994kf}
S.J. Brodsky, L.L. Frankfurt, J.F. Gunion, A.H. Mueller and M.I. Strikman,
\newblock Phys. Rev. D50 (1994) 3134.

\bibitem{Brodsky:1988xz}
S.J. Brodsky and A.H. Mueller,
\newblock Phys. Lett. B206 (1988) 685.

\bibitem{Nikolaev:1992si}
N.N. Nikolaev,
\newblock Comments Nucl. Part. Phys. 21 (1992) 41.

\bibitem{Frankfurt:1994hf}
L.L. Frankfurt, G.A. Miller and M. Strikman,
\newblock Ann. Rev. Nucl. Part. Sci. 44 (1994) 501.

\bibitem{Jain:1996dd}
P. Jain, B. Pire and J.P. Ralston,
\newblock Phys. Rept. 271 (1996) 67.

\bibitem{Kopeliovich:1994pw}
B.Z. Kopeliovich, J. Nemchick, N.N. Nikolaev and B.G. Zakharov,
\newblock Phys. Lett. B324 (1994) 469.

\bibitem{Hufner:1996dr}
J. H{\"u}fner, B.Z. Kopeliovich and J. Nemchik,
\newblock Phys. Lett. B383 (1996) 362.

\bibitem{Frankfurt:1998vx}
L.L. Frankfurt, G. Piller, M. Sargsian and M.I. Strikman,
\newblock Eur. Phys. J. A2 (1998) 301.

\bibitem{Clellan:1969}
G.N. McClellan et~al.,
\newblock Phys. Rev. Lett. 23 (1969) 554.

\bibitem{Kopeliovich:1995bj}
B.Z. Kopeliovich and J. Nemchik,
\newblock (1995), nucl-th/9511018.

\bibitem{Abramowicz:1995hb}
H. Abramowicz, L.L. Frankfurt and M.I. Strikman,
\newblock Surveys in High Energy Physics 11 (1997) 51.

\bibitem{Frankfurt:19966nx}
L.L. Frankfurt and M.I. Strikman,
\newblock Phys. Lett. B382 (1996) 6.

\bibitem{Barone:1997ij}
V. Barone and M. Genovese,
\newblock Phys. Lett. B412 (1997) 143.

\bibitem{Gribov:1984tu}
L.V. Gribov, E.M. Levin and M.G. Ryskin,
\newblock Phys. Rept. 100 (1983) 1.

\bibitem{Mueller:1986wy}
A.H. Mueller and J. Qiu,
\newblock Nucl. Phys. B268 (1986) 427.

\bibitem{Frankfurt:1998eu}
L.L. Frankfurt and M.I. Strikman,
\newblock Proc. Int. Conf. on Deep Inelastic Scattering and QCD (DIS 98),
  Brussels, Belgium, 1998.

\bibitem{Qiu:1987wh}
J. Qiu,
\newblock Nucl. Phys. B291 (1987) 746.

\bibitem{Arneodo:1996qa}
M. Arneodo et~al.,
\newblock Proc. Workshop Future Physics at Hera, DESY, Hamburg, Germany, 1996.

\bibitem{Strikman:acta96}
M.I. Strikman,
\newblock Acta Phys. Pol. B27 (1996) 3431.

\end{thebibliography}

\end{document}